\documentclass[10pt,prb,aps,floatfix,twocolumn,superscriptaddress]{revtex4-2}
\usepackage{graphicx}
\usepackage{amssymb}
\usepackage{amsmath}
\usepackage{xcolor}
\begin{document}

\author{Jun Takahashi$^\dagger$}
\affiliation{Center for Quantum Information and Control, University of New Mexico, Albuquerque, New Mexico 87131, USA}
\affiliation{Institute for Solid State Physics, The University of Tokyo, 5-1-5 Kashiwanoha, Kashiwa, Chiba 277-8581, Japan}

\author{Hui Shao$^\dagger$}
\affiliation{School of Physics and Astronomy, Beijing Normal University, and \\ Key Laboratory of Multiscale Spin Physics (Beijing Normal University),
Ministry of Education, Beijing 100875, China}

\author{Bowen Zhao}
\affiliation{Department of Physics, Boston University, 590 Commonwealth Avenue, Boston, Massachusetts 02215}

\author{Wenan Guo}
\affiliation{School of Physics and Astronomy, Beijing Normal University, and \\ Key Laboratory of Multiscale Spin Physics (Beijing Normal University),
Ministry of Education, Beijing 100875, China}

\author{Anders W. Sandvik}
\email{Corresponding author. E-mail: sandvik@bu.edu}
\affiliation{Department of Physics, Boston University, 590 Commonwealth Avenue, Boston, Massachusetts 02215}

\title{SO(5) multicriticality in two-dimensional quantum magnets}

\begin{abstract}
We resolve the long-standing problem of the nature of the quantum phase transition between a N\'eel antiferromagnet and a spontaneously dimerized valence-bond
solid in two-dimensional spin-1/2 magnets---the putative deconfined quantum-critical point.
We study a class of $J$-$Q$ models, in which the standard Heisenberg exchange $J$ competes with multi-spin interactions
$Q_n$ formed by products of $n$ singlet projectors on adjacent parallel links of the lattice. Our large-scale quantum Monte Carlo (QMC) calculations provide
unambiguous evidence for first-order transitions in these models, with the strength of the discontinuities increasing with $n$. In the case of the widely studied
$n=2$ and $n=3$ models, the first-order signatures are very weak but visible in high-precision correlation functions on large lattices. On intermediate length scales
(up to a hundred or more lattice constants, depending on the observable) we can extract well-defined scaling dimensions (critical exponents) that are common to the
models with small $n$, indicating close proximity to a universal quantum critical point. By combining two different $Q$ terms, specifically $J$-$Q_2$-$Q_6$,
the transition can be continuously tuned from weak to stronger first-order. In the plane $(Q_2,Q_6)$, with $J=1-Q_2$, the two coexisting order parameters
on the first-order line scale with an unusually large exponent $\beta \approx 0.85$. This exponent and others coincide closely with known bounds for an SO($5$)
symmetric conformal field theory (CFT), including a relevant SO($5$) singlet operator that destabilizes the generic deconfined critical point. We quantitatively
characterize the emergent SO($5$) symmetry by computing the scaling dimensions of its leading irrelevant perturbations. The large $\beta$ value and likewise large
correlation length exponent, $\nu \approx 1.4$, explain why the transition remains near-critical on the first-order line even quite far away (parametrically) from
the critical point and in many different models without fine-tuning. Beyond the CFT scaling dimensions, the exponent that had previously been identified with the
divergent correlation length does not have a corresponding level in the CFT spectrum. We explain this emergent ``pseudocritical'' length scale as originating from
bubbles of one phase inside the other that are compressed in size by repulsive interactions between the two order parameters. This length scale is reflected in
various crossover behaviors of observables when traversing the weak first-order transition. The separation between the gaps (masses) associated with the bubbles and the
amplitude (``Higgs'') mode caused by a dangerously irrelevant perturbation is an analogy to the mass hierarchy problem in particle physics. We further argue that
the multicritical point is most likely the tip of a gapless spin liquid phase in frustrated Heisenberg models into which the $J$-$Q$ models
can be continuously deformed. Our results are at variance with the prevailing ``walking'' scenario of a generic deconfined quantum critical
point described by a complex CFT. The multicritical point should exist within real Hamiltonians, though perhaps only outside the regime amenable to sign-free
QMC simulations.
\end{abstract}

\date{\today}

\maketitle

\tableofcontents

\section{Introduction}
\label{sec:intro}

Quantum phase transitions out of the two-dimensional (2D) N\'eel antiferromagnetic (AFM) state have been of central interest in condensed matter physics
since the early days of high-$T_{\rm c}$ superconductivity in the cuprates \cite{anderson87,chakravarty89,dagotto89,manousakis91,chubukov94}, and soon thereafter
also in the context of frustrated quantum magnetism \cite{diep04,savari17}. Two classes of non-magnetic states have been of particular interest;
quantum spin liquids (QSLs) \cite{fezekas74,wen91,savari17}, which have no conventional long-range order (but should have topological order), and
crystalline valence-bond-solid (VBS) states \cite{dagotto89,read89,read90,sachdev08} in which lattice symmetries are spontaneously broken by the formation
of some pattern of modulated singlet density on the lattice links. Theories of quantum magnets often place AFM, QSL, and VBS ground states within the
same phase diagram \cite{hermele04,hermele05,hermele08,xu09,thomson17,song19,yu20,zou21,schackleton21,schackleton22},
but it has proven difficult to understand precisely how these states of matter relate to each other in specific quantum spin models
and materials, and what the nature is of the quantum phase transitions between the phases.

Theories involving the hypothesized deconfined quantum critical point (DQCP) \cite{senthil04a,senthil04b,levin04,tanaka05,senthil06} as a ``beyond Landau''
scenario for the generic AFM--VBS transition have remained unsettled and controversial. Monte Carlo studies of 2D quantum spin Hamiltonians \cite{sandvik07}
and related 3D classical systems \cite{nahum15a,sreejith19} have identified tantalizing signatures of the DQCP, e.g., emergent symmetries
\cite{sandvik07,jiang08,lou09,nahum15b,sreejith19,zhao19,serna19,takahashi20}, exponents for SU($N$) symmetric models agreeing with field theory calculations for
large $N$ \cite{kaul08,kaul12,block13,dyer15}, and apparent manifestations of deconfined spinon excitations \cite{shao16,ma18,yang25}. Other observed features appear
to be at odds with a true DQCP, e.g., scaling violations that have been interpreted either as a weak first-order transition \cite{jiang08,kuklov08a,kuklov08b,chen13}
or unusual scaling behaviors not anticipated at the DQCP \cite{sandvik10a,nahum15a,shao16}. These counter-indicators may require only minor modifications of
the original theory, e.g., scenario of a near-critical ``walking'' first-order transition associated with a complex conformal field theory (CFT), which has dominated
the theoretical discourse during the past several years \cite{wang17,gorbenko18a,gorbenko18b,ma20,nahum20,he21}. Alternatively, a more drastic overhaul of the
theory of AFM--VBS transitions will be required, e.g., if a continuous transition is realized only as a fine-tuned multicritical point \cite{zhao20,lu21,chester24}.
A resolution of the problem is pressing, given the prominent position the DQCP has taken in field theories of condensed matter and beyond, despite its unsettled
nature, and considering also that experimental platforms targeting DQCP physics are under active development \cite{zayed17,guo20,cui23}.

\subsection{Reconciling theory and numerics}

\begin{figure}[t]
\includegraphics[width=84mm]{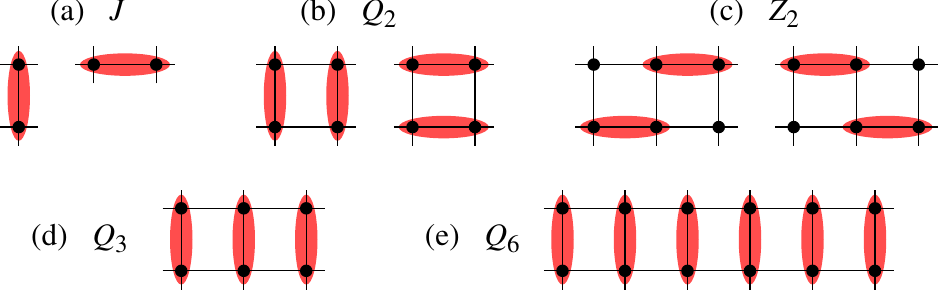}
\caption{Interactions between $S=1/2$ spins on the square lattice. Red ellipses indicate singlet projectors
$({\bf S}_i \cdot {\bf S}_j -1/4)$, and two or more ellipses in the same lattice cell correspond to products.
  The original $J$-$Q$ (here referred to as $J$-$Q_2$) model \cite{sandvik07} combines the conventional AFM Heisenberg exchange in (a)
  at strength $J$ with the four-spin interaction in (b) at strength $Q_2$. Other arrangements of the projectors can also be considered, such
  as the $Z_2$ operator in (c). Interactions $Q_n$ consist of stacks of $n$ projectors, e.g., with $n=3$ in (d), while $Z_n$ has
  $n$ projectors in a staircase formation. The largest interaction considered here is the twelve-spin case in (e). In (c)-(e), the $90^\circ$
  rotated interaction patterns are also included in the summation over the full lattice, so that all square-lattice symmetries are preserved.
  In addition to the $J$-$Q_n$ models, we also consider Hamiltonians with three different couplings; $J$-$Q_2$-$Z_2$, $J$-$Q_2$-$Z_3$,
  and $J$-$Q_2$-$Q_6$.}
\label{jqterms}
\end{figure}

Here we will present unambiguous evidence for a line of first-order AFM--VBS transitions ending at a multicritical point with emergent SO($5$) symmetry.
We apply quantum Monte Carlo (QMC) methods to study ground states of a class of spin-$1/2$ $J$-$Q$ models (those introduced in Refs.~\cite{sandvik07,lou09} as well
as generalizations not studied previously), in which multi-spin interactions favoring locally correlated singlets on the 2D square lattice are added to the
conventional Heisenberg two-spin exchange $J$. The models are illustrated and further explained in Fig.~\ref{jqterms}.

The $J$-$Q$ models and their generalizations permit QMC studies of the transition between the AFM state and a spontaneously dimerized columnar VBS state in great
detail without approximations other than well controlled small statistical errors and finite lattice size. In some cases we here take the linear size as large
as $L=1024$, which is unprecedented in ground state simulations. We consider several different $Q$ terms, with multi-spin interactions $Q_n$ consisting
of $n$ singlet projectors on adjacent parallel lattice links, i.e., arranged as stacks of length or height $n$, as illustrated for some cases in Fig.~\ref{jqterms}.
Focusing on the AFM--VBS transition, we compare results for several of these $J$-$Q_n$ models, anticipating strongly first-order transitions for large $n$
\cite{takahashi20}. For $n=2$ and $n=3$ we find first-order transitions with very small coexisting order parameters. There is a significant range of length
scales (up to a hundred or more lattice spacings) on which the systems exhibit robust quantum critical scaling, which we study using several different correlation
functions and observables probing emergent symmetries.

We also investigate a model combining two different $Q$ terms, in order to realize an entire line of AFM--VBS transitions in a phase diagram with two axes.
Specifically, we use a $J$-$Q_2$-$Q_6$ model to access a line of AFM--VBS transitions in the plane of $Q_2$ and $Q_6$ (with $Q_2+J=1$). Here we observe
critical scaling of the growth of the coexisting order parameters upon moving further into the first-order line. We also study the first-order line with the
$J$-$Q_2$-$Z_2$ and $J$-$Q_2$-$Z_3$ models, where the $Z_2$ interaction, illustrated in Fig.~\ref{jqterms}(c), has a staircase arrangement of two singlet projectors
and $Z_3$ is an extension to three projectors. Comparisons of the different models provide information on the relevant (continuum field) operator content of the
lattice interactions.

\begin{table}[t]
  \caption{Scaling dimensions obtained here for the $J$-$Q$ models compared with values from the
    SO($5$) CFT bootstrap \cite{chester24} and fuzzy-sphere exact diagonalization with 10 orbitals
    (from Table II of Ref.~\cite{zhou23}). The subscripts correspond to the order parameter ($\Delta_\phi$), the relevant SO($5$)
    singlet operator ($\Delta_s$), the rank-$2$ traceless symmetric tensor ($\Delta_t$), the conserved current operator ($\Delta_j$), and the
    leading SO($5$) perturbation; a rank-$4$ dangerously irrelevant operator ($\Delta_4$). Star superscripts indicate necessary or
    assumed values; $\Delta_\phi=0.63$ was used as input in the CFT numerical bootstrap calculation \cite{chester24} and the other exponents
    depend on this value, while $\Delta_j=2$ must hold in the CFT and was imposed for calibration of the level spectrum in the fuzzy
    sphere calculation \cite{zhou23}. Our convention for statistical errors here and henceforth is that the digit(s) within ()
    correspond to the one standard deviation error of the preceding digit, i.e., $3.723(11)$ means $3.723 \pm 0.011$.}
\centering
\begin{tabular}{l|lllll}
  \toprule
    ~~~ & ~$\Delta_\phi$ & ~$\Delta_s$ & ~$\Delta_t$ & ~$\Delta_j$ & ~$\Delta_4$\cr
\hline
  This work & 0.607(4) & 2.271(3) & 1.417(7) & 2.01(3) & 3.723(11) \cr
  \hline
  SO($5$) CFT & 0.630$^*$ & 2.359 & 1.519 & 2$^*$ & 3.884 \cr
\hline
  Fuzzy sphere & 0.585 & 2.831 & 1.458 & 2$^*$ & 3.895 \cr
\botrule
\end{tabular}
\label{dtable}
\end{table}

\begin{figure*}[t]
\includegraphics[width=147mm]{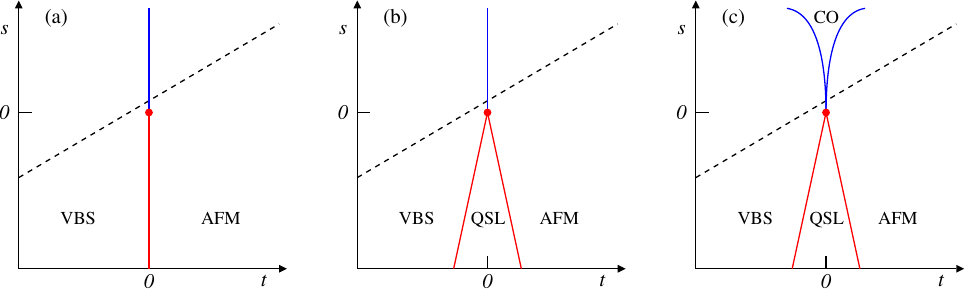}
\caption{Schematic phase diagrams containing transitions between VBS and AFM phases in a space of two scaling fields $(s,t)$, corresponding to the
relevant singlet and traceless symmetric tensor in the multicritical SO($5$) CFT \cite{chester24}. The red circles indicate a multicritical point
in all three phase diagrams, classified as tricritical in (a), bicritical in (b), and tetracritical in (c). The blue vertical lines in (a) and (b)
indicate first-order transitions ending at the multicritical point, below which there is a line of continuous DQCP transitions in (a) and a QSL phase
intervening between the VBS and AFM phases in (b). The blue curves in (c) are the phase boundaries of an extended AFM--VBS coexistence phase ending
at the tetracritical point, which is also the tip of a QSL phase as in (b). The slanted dashed lines show a possible path taken with the models
studied here when a single parameter is tuned, e.g., the ratio $J/Q_n$ in a $J$-$Q_n$ model. The field $t$ changes the symmetry of the order parameter
from ${\mathbb Z}_4$ in the VBS [with emergent U($1$) symmetry close to criticality] to SO(3) in the AFM, while $s$ does not violate the SO($5$)
symmetry of the multicritical point. Case (a) corresponds to the conventional DQCP scenario, where a QSL phase (not shown) may also in principle connect
to a lower end point of the generic continuous transition. Both (a) and (b) with multicritical points hosting emergent SO($5$) symmetry are consistent with our
results, though when including other arguments (b) is more likely. We can positively exclude the coexistence phase in (c). The placement of the dashed
lines is in accord with the $J$-$Q$ models in the regime without QMC sign problem, where only the first-order line can be crossed---though sufficiently
close to the multicritical point so that its scaling dimensions can be reliably determined.}
\label{phases}
\end{figure*}

Our multifaceted analysis of the AFM--VBS transition in the $J$-$Q$ models and new theoretical insights allow us to tie together many analytical and numerical
results that previously appeared to be in conflict with each other. The perceived discrepancies are largely consequences of intricate scaling
behaviors for a system not located exactly at the critical point but hosting a near-critical ground state with coexisting weak AFM and VBS orders. We completely
resolve the nature of the phase transition, demonstrating a line of first-order transitions terminating at an SO($5$) multicritical point. We determine several
scaling dimensions of this critical point, including relevant ones (i.e., those related to the conventional critical exponents) as well as some irrelevant
ones; specifically, those characterizing the leading perturbations of the SO($5$) symmetry.

Although emergent SO($5$) symmetry has been part of the SU($2$) DQCP story (as a special case of the general SU($N$) theory)
almost from the beginning \cite{tanaka05,senthil06,lee15,wang21} and has been observed in previous
numerical studies \cite{nahum15b,sreejith19,takahashi20}, it was assumed that the continuous transition is generic, described by a conformal field theory (CFT) without
any relevant SO($5$) symmetric (singlet) perturbations. In the multicriticality scenario there is a relevant singlet perturbation in addition to the relevant operator
that is tuned to bring a model between the AFM and VBS phases.

As shown in Table \ref{dtable}, both relevant and irrelevant scaling dimensions agree reasonably well
with recent results for a multicritical SO($5$) CFT \cite{chester24}, obtained using a variant of the numerical bootstrap method \cite{poland19}. Strictly speaking,
the bootstrap results in this case reflect only bounds on the scaling dimensions, some of which had been obtained previously \cite{nakayama16,li18}. In O($N$) models,
the true scaling dimensions typically fall exactly on the boundary of the CFT allowed region of scaling dimensions, and in general it is believed that they must at
least be close to the boundary. The small discrepancies between our values and those of Ref.~\cite{chester24} likely stem from an input value in the bootstrap
calculation, the scaling dimension $\Delta_\phi$ of the order parameters, that was not known precisely and to some extent impacts the output values of the other
scaling dimensions. Our refined value of $\Delta_\phi$ in Table \ref{dtable} differs slightly from the previously best estimate, and using it in the bootstrap
calculation indeed improves the agreement with all the other values \cite{chester24b}. Our values of $\Delta_\phi$, $\Delta_t$, and $\Delta_s$, coincide at a
level of uncertainty of only $1-2$\% with the CFT permissible boundaries in the planes $(\Delta_\phi,\Delta_t)$ and $(\Delta_\phi,\Delta_s)$ computed previously
\cite{li18}.

In addition to the operators in Table \ref{dtable}, we also identify the next primary operator in the same sector as the $t$ field driving the AFM--VBS
transition (rank-$2$, parity-even, Lorentz spin $\ell=0$), with scaling dimension $\Delta_{t'} = 4.711(4)$. This operator will play an important role in our scenario
for unusual scaling behaviors in the neighborhood of the weak first-order transition.

Concrete evidence for a multicritical point in the context of the DQCP was to our knowledge first proposed in Ref.~\cite{zhao20} by three of us (though
with, in hindsight, a misidentification of one of the scaling dimensions), and attempts to construct a corresponding field theory were made in Ref.~\cite{lu21}.
The good agreement with the CFT calculations \cite{li18,chester24}, in particular showing the relevance of an SO($5$) singlet operator, firmly establishes the
existence of the multicritical point. This scenario is not specific to the $J$-$Q$ models studied here, but should be universal for quantum magnets with
AFM--VBS transitions as well as other systems with analogous order parameters.

The CFT calculations did not address the nature of the phase diagram in any particular microscopic model. The phases and transitions generated when subjecting the
SO($5$) CFT to a specific relevant microscopic perturbation are not automatically known. The critical point was assumed to be of the tricritical variant \cite{chester24},
though the scaling dimensions alone cannot rule out some other multicritical point. Three putative phase diagrams relevant to generic 2D quantum magnets are
presented schematically in a space of two relevant scaling fields in Fig.~\ref{phases}. Our $J$-$Q$ results are in principle consistent with both the cases
depicted in Fig.~\ref{phases}(a) and \ref{phases}(b), where a first-order line terminates at a multicritical point; tricritical in (a) and bicritical in (b). In
Fig.~\ref{phases}(a) there is a generic line of DQCP transitions below the multicritical point, which corresponds to the original DQCP scenario, specifically
the theory with SO($5$) symmetry built in explicitly \cite{tanaka05,senthil06} and with no relevant singlet assumed (and where the nature of a putative end point
of the critical line was not addressed). In contrast, in Fig.~\ref{phases}(b) the multicritical point is also the tip of a gapless QSL phase that should inherit
aspects of the critical
point.

Figure \ref{phases}(c) depicts a tetracritical point scenario, which within a Landau-type theory would require effectively attractive
interactions between the AFM and VBS order parameters \cite{bruce75}. Studying the way in which the SO($5$) symmetry is violated for large system sizes on the
first-order line, we can conclude that the interactions actually must be repulsive, thus excluding an extended coexistence phase.

While we only directly study the first-order line, both in extreme proximity to the multicritical point and further away from it, several works
\cite{jiang12,gong14,morita15,wang18,ferrari20,nomura21,yang22,wang22,liu22,liu23,viteritti23,corboz26} have pointed to a QSL phase intervening between AFM and VBS phases
in frustrated quantum magnets. Fig.~\ref{phases}(b) is also the phase diagram proposed in Ref.~\cite{yang22} in the context of a narrow QSL phase that could
be close to vanishing at a multicritical point. The $s$ and $t$ fields would then correspond to specific microscopic model parameters
that individually drive either a weak first-order transition, as in the $J$-$Q$ models studied here, or a gapless QSL phase of the type realized in
frustrated Heisenberg models. Many of the standard frustrated models can in principle be continuously deformed into the $J$-$Q$ models studied here by
adding $Q$ terms (which has not yet been done).

As shown on the third row in Table \ref{dtable}, the scaling dimensions that we determine within the $J$-$Q$ models also agree approximately with values extracted
from the spectrum of an electronic SO($5$) model on a ``fuzzy sphere'' \cite{zhou23}, though the results of the latter are not yet fully converged with respect to
the number of orbitals $n_f$, especially in the case of $\Delta_s$. The results for the largest number, $n_f=10$, is given in Table \ref{dtable} but
extrapolating the available $\Delta_s(n_f)$ data with a $1/n_f$ correction actually produces rather good agreement with our result. However, the interpretation
of the fuzzy sphere results---in particular $\Delta_s$ flowing to more relevant values versus $n_f$---was that of a DQCP described by a
non-unitary CFT \cite{zhou23}, where the SO($5$) singlet of the complex CFT is presumed irrelevant at the fixed point despite the apparent
flow to a stable value below $3$ for small $n_f$. In our opinion, the complex CFT scenario, in which the true DQCP only exists below (but presumably close to) 2+1
dimensions \cite{wang17,gorbenko18a,gorbenko18b,nahum20,ma20}, is superfluous in light of the close agreement between the SO($5$) multicritical unitary
CFT and our findings here.

Given the evolution of the coexisting order parameters on the first-order line that we examine in detail here, and the prospect of further reducing the already
very small ordered moments by frustrating (sign problematic in QMC) terms, a natural scenario is that a disordered QSL phase opens on the other side of the
multicritical point. This would be an analogy with the paramagnetic phase above the critical temperature in a 3D classical O($N$) model, except for the
fact that emergent O($N_{\rm A}+N_{\rm B}$) symmetry is possible in conventional classical spin models only at the junction of transitions in models
with $N_{\rm A}=N_{\rm B}=1$ \cite{pelissetto02,eichhorn13}. This does not, however, exclude the possibility of emergent SO($5$) symmetry in the quantum spin
models studied here, since field theories with topological terms will fall into different universality classes. The QSL phase also should have non-trivial
topological properties, unlike a conventional classical paramagnet. Because of the remaining uncertainty on the nature of the phase diagram, Fig.~\ref{phases}(a),
Fig.~\ref{phases}(b), or some more exotic scenario, we will refer to the critical point as multicritical. However, we regard Fig.~\ref{phases}(b) as the most
plausible, with support coming also from fuzzy sphere calculations including SO($5$) violating terms \cite{chen23,liao26}. Scaling dimensions were not
determined precisely there, but a disordered phase was argued to emerge from a point that, in light of Ref.~\cite{chester24} and our present work, should
be the SO($5$) multicritical point.

Our results not only establish beyond reasonable doubt a first-order line ending at a multicritical SO($5$) point, but we also have obtained new insights into the nature
of the weak first-order transition, including the key question of why many models seem to almost automatically harbor such transitions without fine-tuning. A
major reason for the latter is that the exponent $\beta$ governing the growth of the order parameters on the coexistence line is unusually large,
$\beta \approx 0.85$, and the exponent controlling the correlation length is likewise large; $\nu_s \approx 1.4$ (both these exponents are related to the scaling
dimensions in Table \ref{dtable} in the standard way). Thus, the order parameters are small and the correlation length is large on the first-order line even
quite far away (in the parameter space of a typical lattice model) from the critical point. It is therefore not unlikely for a typical model to have a transition
inside the near-critical window. A further reason could be that typical lattice interactions are dominated by the relevant operator with the smaller scaling
dimension $\Delta_t$. From our observations of the effects of different $Q_n$ interactions, it appears that an operator must in practice have a large spatial extent
for it to overlap substantially with the $s$ field and drive a more strongly first-order transition, like the $n=6$ operator Fig.~\ref{jqterms}(e).

\subsection{Massive excitations and crossover scales}
\label{sub:crossover}

At a conventional weakly first-order transition close to a critical point (often called fluctuation driven first-order transitions), when approaching
the transition point there is a crossover from nearly critical fluctuations to a coexistence state with long-range order and finite correlation
length. In the present context, this correlation length $\xi_s$ in the close proximity of the first-order line is controlled by the SO($5$) symmetric
field $s$ in the phase diagrams in Fig.~\ref{phases}, with a divergence $\xi_s \sim s^{-{\nu_s}}$ upon approach to the multicritical point along
the first-order line. Thus, when tuning the parameter $t$ that drives the VBS--AFM transition, the corresponding correlation length $\xi_t \sim t^{-\nu_t}$
initially can grow large when $s>0$ is very small, until a crossover $\xi_t \to \xi_s < \infty$ occurs very close to $t=0$.

\begin{figure*}[t]
\includegraphics[width=115mm]{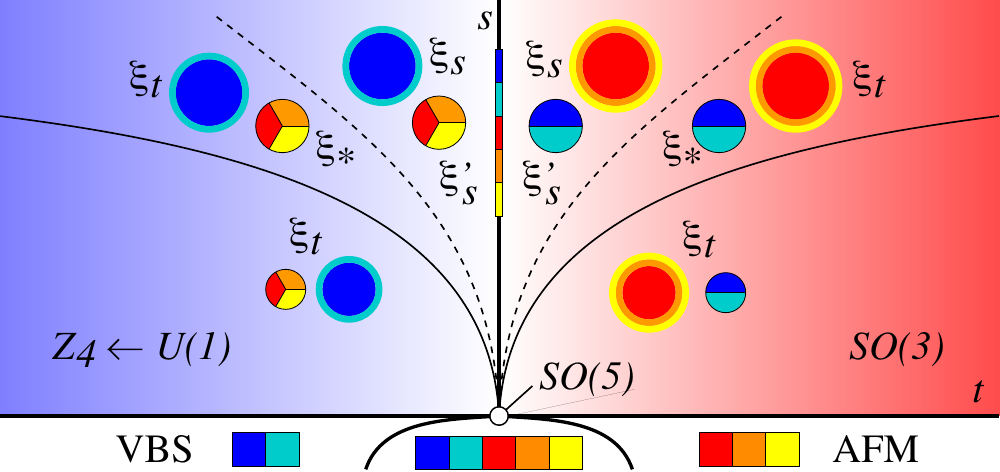}
\caption{Symmetries and excitations in the neighborhood of the weak first-order line for small $s>0$ in the phase diagrams in Fig.~\ref{phases}(a) or \ref{phases}(b).
The five colors illustrate the components $(v_1,\ldots,v_5)$ of a unified order parameter ${\bf v}$, with the VBS corresponding to $(v_1,v_2)$, indicated by blue
and cyan, and the AFM components are $(v_3,v_4,v_5)$, shown as red, orange, and yellow. The VBS order parameter is shown as symmetry-broken into the $v_1$ (blue)
component and the AFM is ordered in its $v_3$ (red) component. The circles indicate the different types of defects (excitations), whose associated length
scales in the different $(t,s)$ regimes are indicated. A filled circle surrounded by $n=1$ or $n=2$ rings corresponds to the gapped amplitude mode
of the order parameter and the $n$ associated gapless Goldstone modes. The circles with $n=2$ or $n=3$ sectors depict $n$ massive bubble fluctuations into the
opposite phase. Far away from the SO($5$) point the bubbles and amplitude modes are governed by the same scaling dimension $\Delta_t$ and their length scales only
differ by an order-one factor. On approach to the first-order line, the order parameter repulsion becomes relevant (in the neighborhood of the solid curves fanning out
from the multicritical point) and further impedes fluctuations into the components of ${\bf v}$ not associated with the phase, thus reducing their typical size to
$\xi_* < \xi_t$ and increasing the corresponding mass (of order the inverse length scale). The length scale $\xi_t$ eventually crosses over to $\xi_s$ governed
by the SO($5$) singlet dimension $\Delta_s$ in the close neighborhood (indicated by the dashed curves) of the first-order transition, whence the bubble size
converges to $\xi'_s \propto \xi_s^{\nu_*/\nu} < \xi_s$. On the first-order line, the system breaks SO($5$) symmetry equivalently to the AFM or VBS sector, up to a
very large length scale associated with the (dangerously) irrelevant perturbation of the SO($5$) symmetric multicritical point. There is likely a gapless spin
liquid phase below the SO($5$) symmetric multicritical point, as in Fig.~\ref{phases}(b).}
\label{defects}
\end{figure*}

In the near-critical $J$-$Q$ model, another length scale $\xi_*$ appears which is shorter than the conventional correlation length $\xi_t$, being associated with
a ``pseudocritical'' exponent $\nu_* < \nu_t <\nu_s$. We will demonstrate that the scale $\xi_*$ appears as a consequence of effectively repulsive interactions
between the AFM and VBS order parameters. This repulsion disfavors SO($5$) symmetry, thus altering the nature of the near-critical fluctuations in a rather
complex way illustrated schematically in Fig.~\ref{defects}.

Consider the five-component unified order parameter ${\bf v}=(v_1,\ldots,v_5)$, with $D=(v_1,v_2)$ and $M=(v_3,v_4,v_5)$ corresponding to the VBS and AFM, respectively.
Let us assume that the system is in the VBS phase with the symmetry broken so that $v_1$ is the ordered component. There is then a gapless Goldstone mode related to
$v_2$ fluctuations and a massive amplitude (``Higgs'') mode of the dimer order parameter $D$. Sufficiently close to the multicritical point, the O($3$) AFM order parameter
$M$ also forms but only as ``bubbles''; defects of a finite typical size $\xi_t$ governed by the scaling dimension $\Delta_t$. The amplitude mode must also be governed
by this same scaling dimension, under the assumption that $s$ is so small that $t$ is effectively the only relevant perturbation and all divergent length scales
are controlled by it. However, upon moving very close to the weak first-order transition at $s>0$, for $|t|$ between the solid black curves in Fig.~\ref{defects},
the repulsive interactions between the two order parameters also become relevant, exceeding the energy scale associated with the $t$ field. Then bubble
fluctuations into any of the components $(v_3,v_4,v_5)$ become energetically more expensive, since a domain wall is required that has coexisting AFM and
VBS order. We will derive an exponent $\nu_*$ governing the typical size $\xi_* \sim|t|^{-\nu_*}$, with $\nu_* < \nu_t$, of the bubbles in this regime. At least
to leading order, the VBS amplitude mode is not affected by the repulsive interactions. Thus, there are two different growing length scales as $|t|$ is reduced,
until eventually the scale $\xi_t$ of the amplitude fluctuations saturates at $\xi_s$ and the bubble scale converges to $\xi'_s \propto \xi_s^{\nu_*/\nu_s} < \xi_s$.
This converged regime corresponds to $|t|$ falling between the dashed curves in Fig.~\ref{defects}. The same set of arguments can be applied for a system in the
AFM phase as well, then with two emergent bubble excitations instead of three in the VBS phase, and, correspondingly, two remaining Goldstone modes instead of just one.

It should be noted that the VBS order parameter breaks $\mathbb{Z}_4$ symmetry, not U($1$), on the largest length scales---even larger than $\xi_t$ and $\xi_*$.
We will demonstrate this explicitly within the formalism of dangerously irrelevant operators causing $\mathbb{Z}_q$ to U($1$) crossovers, familiar in the context of 3D
classical clock models \cite{amit82,oshikawa00,lou07,okubo15,leonard15,shao20}. Similarly, because the multicritical point only has emergent SO($5$) symmetry,
when increasing $s$ on the first-order line the symmetry will be split, SO($5$) $\to$ SO($3$)$\times$U($1$) and finally $\to$ SO($3$)$\times$$\mathbb{Z}_4$, but
this breakdown of the SO($5$) symmetry takes place on a length scale larger than $\xi_s$ and does not affect the scenario laid out above. Effectively, the system
on the first-order line breaks SO($5$) symmetry in the standard way, with four Goldstone modes, as if the symmetry were exact, up to a length scale larger than
$\xi_s$. We stress that $\xi_*$ is not a conventional length scale induced by a dangerously irrelevant operator
\cite{amit82,oshikawa00,lou07,okubo15,leonard15,shao20}, though it does depend on a specific dangerously irrelevant SO($5$) violating CFT operator
(related to order-parameter repulsion) that we identify.

\begin{figure}[t]
\includegraphics[width=83mm]{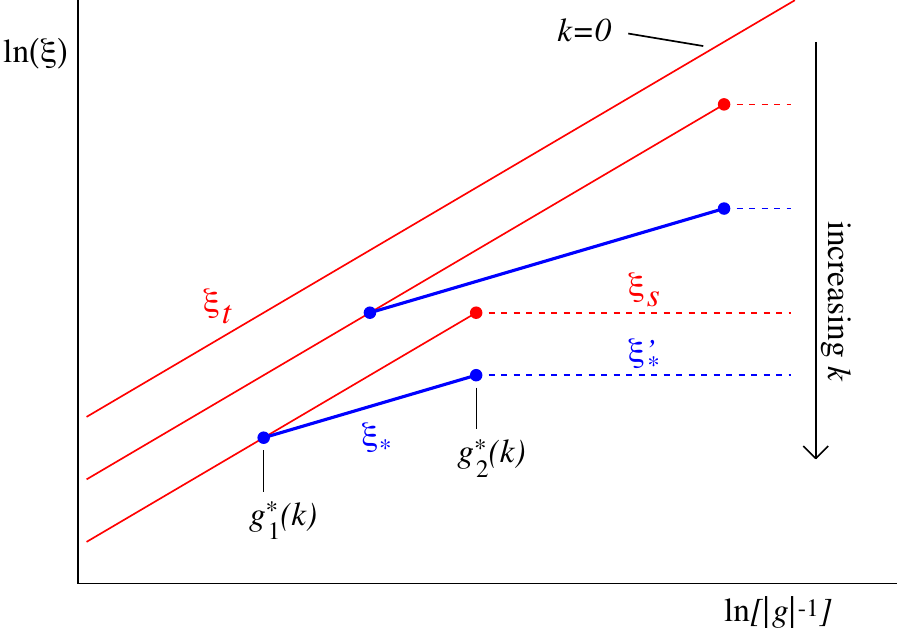}
\caption{Illustration of crossover from the conventional correlation length $\xi_t(g,k)$ to a slower growing bubble size $\xi_*(g,k)$ (the AFM bubbles in the
  VBS or vice versa) when the weak first-order transition is approached in a model space $(g,k)$. Here changing $g$ at fixed $k$ corresponds to moving on the dashed
  line in the space of relevant CFT fields ($s,t$) in Fig.~\ref{phases}(b), and increasing $k\ge 0$ corresponds to shifting the line higher in $s$. The critical
  point is at $g=k=0$ (where also $s=t=0$), where the correlation length $\xi=\xi_t \sim g^{-\nu_t}$ that governs both the bubbles and the amplitude fluctuations
  of the order parameter applies all the way to $g=0$. For $k>0$, a slower growth of the bubble size, $\xi_* \sim g^{-\nu_*}$, commences at $g=g^*_1(k)$ and persists
  until $g=g^*_2(k)$, where $\xi_t$ (which still corresponds to the amplitude mode) saturates to $\xi_s \sim k^{-\nu_s}$ and $\xi_*$ converges to
  $\xi'_s=\xi_s^{\nu_*/\nu_s}$. The crossover point $g^*_1(k)$ moves to the right as $k \to 0$ and the range of $\xi_*$ expands because of
  the strongly diverging $\xi_s(k \to 0)$. In reality the crossovers are smooth, without the sharp kinks drawn here.}
\label{xi}
\end{figure}

Figure \ref{xi} illustrates the evolution of the different length scales in a model with two tuning parameters. There are three stages of growth, two different
near-critical power-law behaviors and finally convergence to two finite scales in the neighborhood of the transition. This first crossover of the correlation
length was in fact observed, but not explained, in the 3D loop model by Nahum et al.~\cite{nahum15a}, and was also in hindsight present in results for the $J$-$Q$
model \cite{shao16,sandvik20}. Here we will study the crossover of the correlation length in detail and also use it to explain anomalous finite-size scaling
behaviors of various quantities where the scaling function is controlled by two arguments. Such scaling was first suggested in \cite{shao16} but a complete
understanding of the ground state and the physical meaning of the length scales was lacking.

We will explicitly demonstrate the presence of repulsion between the order parameters
and its irrelevant nature at the multicritical point using a specific probe of the SO($5$) symmetry
violations in the $J$-$Q$ models. The repulsive interaction also rules out the phase diagram with a coexistence phase and tetracritical point in Fig.~\ref{phases}(c).
The bubble scale $\xi_*$ is most directly visible in the way the emergent SO($5$) symmetry is violated by order parameter repulsion in the neighborhood
of the critical point, and we use finite-size scaling of a quantity characterizing the deviations from perfect symmetry to extract a precise value
of the exponent $\nu_*$. This exponent can account for the growth of the observed correlation length after the crossover (between the solid and dashed curves in
Fig.~\ref{defects} and between $g_1^*$ and $g_2^*$ in Fig.~\ref{xi}), as well as scaling behaviors with crossovers of many other physical observables. We derive
the relationship,
\begin{equation}\label{nustardeltas}
\frac{1}{\nu_*} = (3-\Delta_s)(\Delta_{t'}-\Delta_t),
\end{equation}
between $\nu_*$, the relevant scaling dimensions $\Delta_s$ and $\Delta_t$ in Table \ref{dtable}, and the irrelevant scaling dimension $\Delta_{t'} \approx 4.7$ of
a primary CFT operator in the same symmetry sector as $t$ that we argue is contained in the repulsive interaction; this scaling dimension is also obtained from
finite-size scaling of a measure of SO($5$) deformation. Eq.~(\ref{nustardeltas}) is consistent with all numerical data and actually produces our best estimate
for $\Delta_s$ on the basis of our estimates of $\nu_*$, $\Delta_t$, and $\Delta_{t'}$.

The length scale at which $\xi_t$ and $\xi_*$ become clearly distinguishable depends on the crossing distance $s$ (at $t=0$), which in a model is
determined by a microscopic parameter that we denote by $k$ in Fig.~\ref{xi}; the crossover takes place versus the tuning parameter $g$ at $g^*_1(k)$.
The closer the system is to the critical point, the longer the conventional correlation length persists, i.e., $g^*_1(k) \to 0$ when $k \to 0$. When
approaching the first-order line at $g=0$, the defects can grow only up to a size that is also dictated by the distance $s \sim k$ to the critical point.
Thus, there is a second crossover to an essentially saturated bubble size at $g^*_2(k) < g^*_1(k)$ and this size diverges when $k\to 0$. We will argue that
the range of bubble sizes $\xi_*$ realized between $g^*_1(k)$ and $g^*_2(k)$ diverges when $k \to 0$. Thus, the pseudocritical length scale is a well
defined critical property at the AFM--VBS transition, not just a conventional first-order crossover.

Though we are not aware of any previous explanation of the emergent pseudocritical length scale, our derivation of $\nu_*$, Eq.~(\ref{nustardeltas}) is
inspired by an RG analysis by Bruce and Aharony \cite{bruce75} of attractive order-parameter interactions, which lead to a coexistence phase extending
out from a tetracritical point, as in Fig.~\ref{phases}(c). If the tetracritical point has emergent O($N$) symmetry, the boundaries of the coexistence
phase are also in part governed by the scaling dimension of the irrelevant perturbation of the symmetry. In analogy, we argue here that the repulsive
order-parameter interaction at AFM--VBS transition is associated with a region of influence that is very similar, in a scaling sense, to the coexistence
phase. Results of Ref.~\cite{bruce75} can therefore be adapted to the present case. We then take the additional step of identifying the corresponding
length scale $\xi_*$ and deriving its exponent, Eq.~(\ref{nustardeltas}). Moreover, this analogy with Ref.~\cite{bruce75} is what allows us to conclusively
show that there is no coexistence phase, which is not possible from just a phase diagram extracted from numerical results for the order parameters,
because a coexistence phase close to a multicritical point is typically extremely narrow and easily mistaken for a direct first-order transition.
The key to making the distinction is the way in which the SO($5$) symmetry is violated, which we also illustrate using a simple classical O($5$)
deformation model.

The defects (bubbles and amplitude fluctuations) correspond to virtual particles in the ground state (vacuum) and should have their standard counterparts
as real excitations (particles). We do not study the mass gaps of these excitations here (which can be done with imaginary-time correlations and analytic
continuation to real frequency, as in Refs.~\cite{ma18,yang25}) but they will be proportional to the inverse length scales that we study. Our scenario then
has aspects loosely resembling electroweak symmetry breaking and the Higgs mechanism in the standard model of particle physics. There the starting point is
a combined SU($2$)$\times$U($1$) field (massless $W_1,W_2,W_3$ gauge bosons and hypercharge $B$) and an SO($4$) Higgs field. The coupling between these leads
to the massive Higgs boson, massive $W^\pm$ and $Z$ vector ($\ell=1$) bosons, and the massless photon. In our case, the starting point is a field of higher
symmetry, SO$(5$), and a Higgs-like mechanism produces massive bubbles (scalar pseudo-Goldstone soliton-like configurations) in two stages due to two perturbations
separating the AFM and VBS order parameters. The Higgs-like boson was already there as the amplitude mode of the SO($2$) (VBS) or SO($3$) (AFM) symmetry-broken
long-range order. Initially, relatively far away from the phase transition, only the tuning field $t$ is relevant and governs both the Higgs and bubble masses.
However, when $|t|$ is very small the dangerously irrelevant operator that we identify controls the bubbles, separating their mass $\propto \xi_*^{-1}$ from
the smaller Higgs mass $\propto \xi_t^{-1}$, with the extreme separation taking place in the immediate neighborhood of the first-order line whence
$\xi_t \to \xi_s$ and $\xi_* \to \xi'_s = \xi_s^{\nu_*/\nu_s}$.

The analogy to particle physics is stronger with alternatives to the standard Higgs theory, e.g., minimal composite Higgs, \cite{bellazzini14}, little Higgs,
\cite{schmaltz05}, and technicolor \cite{hill03}, where the starting point is also typically a single field of higher symmetry, e.g., SO($5$) or SO($6$). While
the analogy between the Higgs boson and the amplitude mode in antiferromagnets has been investigated previously \cite{podolsky11,podolsky12,lohofer15,lohofer17,qin17},
the generation of bubble excitations with their own mass brings the analogy much closer to a true Higgs mechanism. This analogy with particle physics should of
course not be taken too literally, but rather as the spin model providing an interesting example of generation of a mass hierarchy in a different setting.

There is of course also an analogy to the Anderson-Higgs mechanism in superconductors \cite{anderson63,pekker15}, where the starting point is two U($1$) fields;
the photon and the superconducting order parameter that already has its massive amplitude mode. Upon coupling, the two transverse photon modes become gapped along
with a likewise gapped longitudinal polarization mode, giving rise to the Meissner effect. This situation may be almost realizable in quantum spin models in close
proximity to an SO($4$) multicritical point \cite{zhao19,serna19,lee19}. However, the superconductor is charged and coupled to the photon field through the Coulomb
interaction, and this ``eats'' the would-be Goldstone mode. In the spin models, there is instead a gapless Goldstone mode left, as in the VBS part of the
illustration in Fig.~\ref{defects} [where in the SO($4$) case, there would be two instead of three gapped bubble excitations].

The closest analogy of our findings in condensed matter is with Zhang's SO($5$) theory of $d$-wave superconductivity \cite{zhang97}, which exactly has the structure
of a splitting into SO($3$) AFM and U($1$) superconducting sectors and whose $\pi$-resonance is our bubble excitation (though a length scale analogous
to our $\xi_*$ was not discussed there). In the context of high-$T_c$ superconductivity, this mechanism of doping-driven SO($5$) transition was never confirmed,
and experimentally it was essentially ruled out \cite{demler04}. However, the multicritical SO($5$) scenario that we present here in the context of 2D quantum
spin models at $T=0$ should also apply to the transition between AFM and superconducting ground states at half-filling \cite{assaad97,chen26}.

\subsection{Paper outline}

The rest of the paper is organized as follows:

In Sec.~\ref{sec:status} we provide a more detailed account of the background and current status of the 2D
AFM--VBS transition and its persistent controversies.

In Sec.~\ref{sec:overview} we briefly describe the QMC methods used and the different ways in which we analyze the AFM--VBS transition. We also explain in some
more detail how these calculations collectively lead us to the coherent picture of SO($5$) multicriticality outlined above.

In Sec.~\ref{sec:order}, after showing evidence for larger first-order discontinuities with increasing $n$ in the $J$-$Q_n$ models, we present detailed
results and analysis of correlation functions delivering the common order-parameter scaling dimension $\Delta_\phi$.

Other correlation functions giving estimates of $\Delta_t$ and $\Delta_j$ are presented in Sec.~\ref{sec:scaledims}.

We proceed in Sec.~\ref{sec:jq2q6} to study lines of first-order transitions
in the $J$-$Q_2$-$Z_2$, $J$-$Q_2$-$Z_3$, and $J$-$Q_2$-$Q_6$ models. Detailed studies of the latter model, in particular, show that the growth of the coexisting
order parameters with increasing $Q_6$ is fully compatible with the exponents of the multicritical point, i.e., the exponent $\beta$, which is related to the
scaling dimensions $\Delta_\phi$ and $\Delta_s$ in Table \ref{dtable}.

We characterize the emergent SO($5$) symmetry in Sec.~\ref{sec:symm}, going beyond previous works \cite{nahum15b,sreejith19,takahashi20} by extracting irrelevant
scaling dimensions, $\Delta_4$ and $\Delta_{t'}$, of two different perturbations of the symmetry. We also extract the pseudocritical crossover exponent
$\nu_*$ from one of the indicators of SO($5$) violation. Moreover, we present a simple O($5$) deformation that provides further insights into
repulsive interactions as the source of the irrelevant asymmetry between the AFM and VBS components of the SO($5$) order parameter.

All the above results provide us with the necessary components for deriving and testing the relationship Eq.~(\ref{nustardeltas}) between $\nu_*$ and
CFT scaling dimensions in Sec.~\ref{sec:dark}, which also allows us to obtain the most precise value of the SO($5$) singlet dimension $\Delta_s$.

In Sec.~\ref{sec:anomalous} we present a unified framework for how all three length-scale exponents, $\nu_t$, $\nu_s$, and $\nu_*$, impact scaling of physical
observables, with Binder cumulant crossings and the spin stiffness as specific examples.

In Sec.~\ref{sec:discuss} we further discuss our main results and their broader impact in the context of the DQCP
phenomenon and beyond.

\section{The DQCP enigma}
\label{sec:status}

Our current understanding of the AFM--VBS transition in 2D quantum magnets, as reported in this paper, builds on decades of work by a large number of
researchers using many different theoretical and computational techniques. While the proliferation of ideas and results has in some cases been
difficult to reconcile, in the end it is the synergistic combination of different results and approaches that has allowed us to reach a solid conclusion
of the nature of the transition and its broader implications in quantum many-body physics. In this section we provide further relevant background to our
present work, up to the rather confusing status of the field prior to the developments reported here, mainly up until the first version of the paper
(available as preprint arXiv:2405.06607v1), with a few more recent developments deferred to Sec.~\ref{sec:discuss} as they strongly relate to our findings.
After this summary of the prior status of the field, we preview our results and resolution of the previously puzzling aspects of the transition in
Sec.~\ref{sec:overview}.

\subsection{AFM--VBS transition and putative DQCP}

The AFM ground state of the 2D $S=1/2$ Heisenberg model is well described by spin wave theory \cite{anderson52,manousakis91}. Though there is no
analytical proof (there is for $S=1$ and higher \cite{neves86,dyson87}), the numerical evidence of robust long-range order from QMC simulations is
overwhelming \cite{reger88,liang90,sandvik10c,jiang11,sandvik26}. The very large ordered moment, about 60\% of the value in the absence of quantum
fluctuations, is also in the end the reason why spinwave theory works so well in this case.

When the AFM order is reduced by some kind of competition
between interactions (still keeping the spin rotational invariance) and the ordered moment is reduced, spinwave theory fails and all analytical
descriptions become challenging when the amplitude fluctuations cannot be neglected \cite{chubukov95}. However, the continuum field theory (the nonlinear
$\sigma$-model) is well understood \cite{chakravarty89,chubukov94} and can describe the quantum phase transition into a non-degenerate featureless (with
no topological order) paramagnetic state through a Wilson-Fisher O($3$) [strictly speaking SO($3$)] critical point. QMC simulations perform well, as
exemplified by numerous studies of the transition in Heisenberg systems with imposed coupling patterns (dimer or plaquette modulation)
\cite{sandvik94,troyer96,matsumoto01,lohofer15,ma18b}.

Studies of VBS (spontaneously dimerized) ground states of deformed Heisenberg models also have a long history---here again we limit the discussion to
spin-rotationally invariant interactions. The discrete symmetry breaking associated with the formation of a pattern of varying singlet density turned out to
be much more challenging than the AFM phase and its SO($3$) symmetry breaking. In quantum field theory, the VBS state on the lattice can be understood as
the confined phase of certain topological defects of a compact U($1$) gauge field coupled to matter in the form of $S=1/2$ objects commonly referred to as spinons
\cite{haldane88,read89,read90,murthy90}.

In a more intuitive picture, the spinons in the ground state of a strongly ordered VBS are bound into short bipartite singlets---the valence bonds forming a
four-fold degenerate columnar pattern on the square lattice. An $S=1$ excitation of the VBS involves the
presence of a triplet bond, or, in a basis of singlet bonds and unpaired spins, two bound $S^z=1/2$ objects in a background of singlets. When the order is
reduced, longer valence bonds are required to describe the ground state, which corresponds to a longer confinement scale. The energy required to excite one
valence bond---the spin gap---should then be reduced as well. Spinon bound states can be directly observed in QMC simulations of $J$-$Q$ models in the
valence bond basis \cite{tang11a,tang11b,shao16}.

In the simplistic picture based on valence bonds, a continuous transition at which the VBS melts would imply spinon proliferation and deconfinement by
unbinding of spinon pairs when the distribution of valence-bond lengths becomes very broad. The AFM phase forms on the other side of the DQCP by condensation
of spinons. In the basis of valence bonds, the AFM ground state is also characterized by a broad distribution \cite{liang88}, with the probability of a bond
of length $l$ decaying as $l^{-3}$ \cite{lou07b,beach09b}, but here some fraction of the spinons have formed a condensate, reflecting the N\'eel order. Once
the SO($3$) symmetry is broken in the thermodynamic limit (or by a weak external staggered magnetic field in a finite system), the singlet character of the
(bipartite) valence bonds is lost \cite{beach09b} as the spins previously forming singlets now experience effectively opposite magnetic fields from the
staggered magnetic order.

When elevating the symmetry of the spins from SU($2$) to SU($N$) (with a specific representation suitable for bipartite systems), the Heisenberg model spontaneously
dimerizes into a four-fold degenerate columnar state for large $N$ through a mean-field-like mechanism. Corrections have been studied with $1/N$ expansions
\cite{read89,read90} and the so predicted value of $N$ at which dimerization takes place, $N \approx 4.5$, has been confirmed by QMC simulations \cite{harada03,beach09}.
However, it proved difficult to obtain reliable results for the dimerization (VBS) transition in the frustrated spin-$1/2$ SU($2$) Heisenberg Hamiltonians that were
typically the focus of research in the field of quantum magnetism.

The main computational obstacle here is that QMC simulations are hindered by the ``sign problem'', i.e., our inability to formulate the problem with positive
definite sampling weights (except in special cases) \cite{chandra99,henelius00}. Early studies of frustrated Heisenberg models were therefore limited to small
lattices accessible with exact diagonalization \cite{dagotto89,schulz92,einarsson95,misguich99,mambrini06} or series expansions around some solvable limit
\cite{singh99}. It is only recently that methods targeting larger systems, such as the density matrix renormalization group (DMRG) \cite{white92,gong14,he17,wang18}
as well as more general matrix-product and tensor-product based methods \cite{liu22,liu23,corboz26}, have been able to approach the necessary degree of reliability for
2D Hamiltonians with VBS (and also QSL) ground states. Machine learning is also making rapid inroads in producing good variational wave functions
\cite{nomura21,viteritti23,chen24,rende25,moss25,viteritti26}. Still, with all existing methods it is difficult to reliably study quantum phase
transitions in the traditional frustrated spin models.

The AFM--VBS transition in 2D quantum magnets came into renewed focus with the DQCP proposal \cite{senthil04a}, according to which the transition is generically
continuous. In contrast, a generic first-order transition between two ordered phases is expected within the conventional Landau-Wilson-Ginzburg (LGW) framework, with
continuous transitions realized only at fine-tuned multicritical points. The essential non-Landau aspect of the DQCP proposal is that the AFM and VBS order parameters
are not the fundamental degrees of freedom in a low-energy description, but they emerge from the same spinon field by either confinement or condensation.

In the DQCP theory the spinons at the critical point are coupled to a non-compact (defect-free) U($1$) gauge field \cite{motrunich04,senthil04a,senthil04b},
originally in a CP$^{1}$ theory extended to CP$^{N-1}$ for SU($N$) spins. Working with SU($N$) spinons \cite{murthy90}, Senthil et al.~showed that DQCPs are
realized for large $N$, and, motivated at least in part on prior suggestive numerical results for related lattice models \cite{assaad97,sandvik02}, conjectured
that such non-LGW continuous transitions persist all the way to $N=2$ \cite{senthil04a,senthil04b}. While subsequent numerical work for spin Hamiltonians with
rather large values of $N$ match well with results from $1/N$ expansions \cite{kaul12,block13,dyer15}, the fate of the transition for small $N$ has remained
uncertain, in particular for $N=2$, on which most of the work has focused.

In principle, the putative DQCP universality class should also be accessible in 3D classical Heisenberg models (or the nonlinear $\sigma$-model in
the continuum) with ``hedgehog'' defects suppressed in analogy with the non-compact gauge field in the theory \cite{motrunich04}. Though the symmetry
of the order parameter is the same as in the conventional model, where hedgehogs proliferate at the phase transition, the absence of topological defects is
believed to change the universality from that of the Wilson-Fisher O($3$) transition. Though encouraging results for unconventional critical exponents were
obtained by Monte Carlo simulations of such a defect-suppressed model \cite{motrunich04}, in practice the complete suppression of defects on the lattice
is difficult, as the singular nature of hedgehogs strictly appears only in the continuum.

Numerical results for quantum spin systems and other related lattice models, to be detailed below, have shown that the AFM--VBS transition in the SU($2$)
case is associated with emergent SO($5$) symmetry of the combined order parameter with three AFM components and two VBS components. DQCP theories starting
from such a higher symmetry had already been put forward \cite{tanaka05,senthil06,lee15}; the nonlinear sigma model with a topological $\theta$ term,
which for SO($5$) symmetry is the WZW level $k=1$ CFT. This is essentially a generalization to two spatial dimensions
of the much better understood case of the WZW $k=1$ theory in one space dimension, which describes the dimerization transition in spin chains with half-odd
integer $S$ \cite{affleck87}. A number of other SO($5$) field theories have also been proposed and investigated in the context of a ``web of dualities''
that links seemingly different theories to the same critical point \cite{seiberg16,wang17}. Similar dualities have been demonstrated explicitly for low-energy
theories of spin chains \cite{roberts19}. Whether or not these theories properly describe the 2D AFM--VBS transition, in lattice models as well as real
materials \cite{cui23}, is a key question whose answer may have ramifications also for other proposed non-LGW transitions \cite{bi18,jiang23,christos23,christos24}.
The only currently realistic way to answer this question is through unbiased numerical studies of lattice models.

\subsection{Simulations of lattice models}
\label{sec:lattmodels}

To circumvent the numerical difficulties associated with traditional frustrated quantum magnets that may harbor AFM--VBS transitions, the $S=1/2$ $J$-$Q$ model was
introduced specifically to study this transition without QMC sign problems on large lattices \cite{sandvik07}. This simplest version of the model is defined in terms
of nearest-neighbor singlet projectors $P_{ij}=\frac{1}{4} - {\bf S}_i \cdot {\bf S}_j$ by the Hamiltonian
\begin{equation}\label{hamjq}
H = -J\sum_{\langle ij\rangle}P_{ij} - Q_2 \sum_{\langle ijkl\rangle}P_{ij}P_{kl},
\end{equation}
where the lattice bonds $ij$ and plaquettes $ijkl$ are illustrated in Figs.~\ref{jqterms}(a) and \ref{jqterms}(b), respectively. The $J$ term is just the standard
AFM Heisenberg exchange and the $Q_2$ interaction can be extended \cite{lou09} to $n$ projectors, as also illustrated in Fig.~\ref{jqterms}. For negative coupling
constants, the $Q_n$ interaction clearly favors columnar alignment of singlets for large $n$. Even in the case of $n=2$, which was studied first,
there is a columnar VBS phase for large coupling $Q_2$, with the transition into the AFM state taking place at Heisenberg interaction strength $J/Q_2 \approx 0.045$.
The initial QMC results \cite{sandvik07} indeed supported a continuous transition of the predicted DQCP type, e.g., with emergent U($1$)
VBS order parameter and rather large values of the anomalous dimensions $\eta_{\rm AFM} \approx \eta_{\rm VBS} \approx 0.26$ (much larger than $\eta \approx 0.03$
at the O($3$) transition) and overall scaling behaviors at finite temperature similar to the large-$N$ CP$^{N-1}$ DQCP theory \cite{melko08,kaul08}.

Scaling anomalies on larger lattices were subsequently found in several works and were taken as evidence either of a rather conventional weak first-order
transition \cite{jiang08,chen13}, unusually strong scaling corrections \cite{sandvik10a,harada13}, or behaviors influenced by two divergent length scales
\cite{shao16}. Similar scaling violations were also found in lattice versions of the proposed field theory \cite{motrunich08,kuklov08a,kuklov08b}, and later
in a 3D classical loop model inspired by the $J$-$Q$ models \cite{nahum15a,nahum15b}. Many results have also been reported for 3D classical dimer models
\cite{powell08,powell09,charrier08,chen09,sreejith14,sreejith19}, which were argued to realize the same DQCP transition in the presence of appropriate
interactions. Some exponents match reasonably well those obtained with $J$-$Q$ and loop models \cite{sreejith14,sreejith19}, though also in this case large
scaling corrections were pointed out.

Fermion models with putative DQCP transitions between AFM and $d$-wave superconducting ground states have been studied extensively
by determinant based QMC simulations \cite{assaad16,gazit18,li19,gotz22,sato23}, but the system sizes accessible here are much smaller, making it
difficult to obtain reliable results for critical exponents or to reach the system sizes where the scaling anomalies become significant.

As mentioned, the predicted emergent U($1$) symmetry of the critical (and near-critical) VBS order parameter was found in early studies of $J$-$Q$ models
\cite{sandvik07,jiang08,lou09}, and later SO($5$) symmetry was detected in the 3D loop \cite{nahum15b}, dimer \cite{sreejith19}, and fermion \cite{gazit18}
models. The close resemblance of the loop and $J$-$Q$ models suggested that this higher symmetry should emerge also in the latter. Indeed, the very
similar values that had already been found for the AFM and VBS anomalous dimensions (which must be the same if the higher symmetry exists) \cite{sandvik07,sandvik12}
in hindsight had its explanation in emergent SO($5$) symmetry. Indications of the higher symmetry were also observed in emergent degeneracies of the excited
level spectrum of the $J$-$Q_2$ model \cite{suwa16}.

A version of the $J$-$Q$ model with plaquette-singlet ground state (where the singlets form on units of four spins in a checker-board like pattern instead of the
columnar dimers) was later constructed which exhibited SO($5$) symmetry on large length scales despite the transition being clearly first-order \cite{takahashi20}.
The strong influence of the higher symmetry even rather far away from the presumed DQCP suggests that the exponent governing the emergent SO($5$) symmetry must
be very large. However, no quantitative results for such exponents were available. In variants of the $J$-$Q$ \cite{zhao19} and loop \cite{serna19} models with
two-fold degenerate VBS states, similarly persistent emergent O($4$) symmetry was found.

The close connection between the 3D loop and $J$-$Q$ models deserves further discussion. The similarity is apparent in the (2+1)D configuration space
to which spin-$1/2$ models are mapped for the purpose of QMC simulations (for reviews, see Refs.~\cite{evertz03,sandvik10b}), where efficient system updates are
carried out by constructing space-time loops of spins \cite{evertz93,sandvik99}. The spins can in fact be completely integrated out, which results in improved
estimators for observables expressed solely with the system partitioned into a set of close-packed loops \cite{evertz03,beach06,sandvik10c}. Thus, the $J$-$Q$
model is mapped to a close-packed loop model, and the $Q$ term corresponds closely to a tunable loop-loop interaction in the loop model \cite{sandvik10b}.
The main difference between the $J$-$Q$ and the classical 3D loop model is that the latter explicitly builds in space-time symmetry (as any of the three space
dimension can be regarded as the time dimension), while in the $J$-$Q$ model Lorentz invariance is emergent. The required dynamic exponent $z=1$ was
indeed found early on in the $J$-$Q_2$ model \cite{sandvik07}, and, given also the very similar scaling behaviors observed in the loop and $J$-$Q$ model
\cite{nahum15a,sandvik20} (and to some extent also the 3D dimer model \cite{powell08,sreejith19}), emergent Lorentz invariance in the $J$-$Q$ model
is not in doubt.

In general, for the different types of lattice models investigated, the anomalous dimension(s) extracted from spin and dimer correlation functions have been
rather stable over time and across models, with already the first $J$-$Q_2$ study resulting in $\eta=0.26(3)$ \cite{sandvik07}, common to both order parameters.
Later, much larger systems still gave a compatible exponent, $\eta= 0.27(1)$ \cite{sandvik12}, from system size up to $L\approx 100$. The often cited ``best'' estimate
up until now (prior to the results presented here) is from the 3D loop model, $\eta_{\rm AFM} = 0.259(6)$ and $\eta_{\rm VBS} = 0.25(3)$ \cite{nahum15a}, where
system sizes up to $L=512$ were studied. Results for the 3D classical dimer model are also consistent with these $\eta$ values \cite{sreejith14}.

In contrast, estimates of the correlation length exponent $\nu$ in the $J$-$Q$ model have changed over time, from $\nu = 0.78(3)$ in the first study
\cite{sandvik07} and then gradually decreasing in later works \cite{melko08,sandvik10a,harada13,block13}, to the most precise recent estimate $\nu = 0.455(2)$
\cite{sandvik20}. In the 3D loop model, large drifts were observed as a function of the system size, with $\nu \approx 0.62$ holding over a range of
moderate system sizes before drifting toward $\nu \approx 0.46$ (i.e., consistent with the $J$-$Q$ model) for the largest system sizes
\cite{nahum15a}---in a way very similar to what we have illustrated qualitatively in Fig.~\ref{xi}. Such drifts versus the system size were also previously
observed in the $J$-$Q$ model \cite{shao16,sandvik20}, though the crossover was not analyzed extensively.

The larger exponent, $\nu \approx 0.62$, extracted for the loop model at the smaller length scales \cite{nahum15b}, also agrees with a rough estimate
obtained from a short-distance correlation function delivering $\Delta=3-1/\nu$ in the same model \cite{nahum15b}. The value is also in good agreement
with rather stable values of $\nu$ observed in the 3D dimer model \cite{sreejith14} (where the system sizes were smaller and a potential crossover of
the exponent may therefore not have been visible). These results raise the question as to why $\nu$ drifts so much with the system size and which value,
$\nu \approx 0.62$ or $\nu \approx 0.45$, is correct (if any). Note that still smaller values of $\nu$ should result from finite-size scaling at a first-order
transition, with $1/\nu \to d+z'=4$ (with $z'=2$ coming from the $\propto L^{-2}$ finite-size gaps of the quantum rotor states associated with
AFM long-range order at the transition) \cite{zhao19}. Thus, it appears that the observed behavior is in some way related to criticality, even if the
flow eventually should tend to a different behavior characterizing the first-order transition.

Another puzzling finite-size drift that was pointed out early on in a study of the $J$-$Q$ model was an anomalous behavior of the spin stiffness $\rho_s$,
which should scale as $\rho_s \sim L^{-z}$ at a critical point with dynamic exponent $z$, here with $z=1$. Instead, a somewhat slower decay was observed
\cite{jiang08}, but, as mentioned above, interpretations relying on $z \not =1$ can be excluded (see also Ref.~\cite{sandvik11}). The anomalous behavior
of $\rho_s$ in the $J$-$Q$ model \cite{jiang08} (and similar quantities in lattice field theories \cite{kuklov08a,kuklov08b}) had initially prompted
suggestions of a first-order transition, spurring further work to demonstrate it more clearly \cite{chen13}. Anomalous logarithmic scaling corrections
\cite{sandvik10a} or two diverging length scales was also proposed \cite{shao16}. A weak first-order transition was more recently indicated in simulations of
$J$-$Q$ models including appropriate order parameter fields, which make it easier to observe what appears to be discontinuities \cite{demidio23b}---though
perhaps still with some room for anomalous scaling.

The bipartite entanglement entropy of the $J$-$Q_3$ model for a corner-less bipartition is consistent with four Goldstone modes, as expected for an AFM--VBS
coexistence state near an SO($5$) DQCP \cite{deng23}, while the corner contributions for a particular bipartition of the $J$-$Q_2$ model have a near-critical
SO($5$) CFT form \cite{demidio23a} (also on the honeycomb lattice \cite{kumar26}). In contrast, another series of results on entanglement entropy in $J$-$Q_2$
and $J$-$Q_3$ models, using different bipartitions, were interpreted as a complete inconsistency with the CFT description \cite{zhao22,liao23,song23a}. The
expected CFT behavior of the corner contributions was observed only for SU($N$) variants of the model \cite{kaul12} with $N \agt 7$ \cite{song23b}.
The profound dependence on the way the entanglement subregion is defined \cite{demidio23a,kumar26} in this case needs to be better understood and
some steps in this direction have already been taken \cite{zhu26}.

As mentioned above, many different quantum field theories are believed to describe the DQCP through the web of dualities \cite{wang17}. Though all
theories may be equivalent, when treated within approximate schemes (e.g., mean-field theories) they are typically different. One approach
then is to use lattice numerics to try to identify which theory within which approximation can best describe available data. One such example is
Ref.~\cite{ma18}, where dynamic spectral functions of a planar (spin anisotropic) variant of the $J$-$Q$ model were computed using stochastic analytic
continuation \cite{shao23} of QMC computed imaginary-time correlation functions. The standard isotropic model was also considered in the same way
recently \cite{yang25}. In both models, the spinon dispersion
relation matches very well that of a parton mean field treatment, the square-lattice $\pi$-flux state, of four-fermion QED$_3$, and
interactions included within the random-phase-approximation further improved the agreement with the QMC computed spectral functions. This result makes
it clear that spinon deconfinement takes place at the AFM--VBS transition on very large length scales, despite the transition likely being ultimately
first-order (as is now certain for all the $J$-$Q_n$ models, as will be shown in Sec.~\ref{sub:corrL}).

It has become apparent that the AFM--VBS and related transitions in the lattice models are sufficiently close to a (possibly inaccessible) critical point for its
scaling properties to be realized up to rather large length scales. However, based on the existing literature it is not yet clear how reliable the so far extracted
scaling dimensions are, i.e., to what degree their values have been affected by the ultimate crossover to first-order behavior.

\subsection{Conformal bootstrap method and the complex CFT scenario}

The numerical conformal bootstrap \cite{rattazzi08,poland19} has emerged as a central method in studies of CFTs in three dimensions. Here the
constraints imposed by conformal invariance, along with other applicable symmetries, are used to systematically exclude combinations of scaling
dimensions, in some cases leading to small regions (``islands'') of possible scaling dimensions. After initial successes with Ising and O($N$)
models \cite{kos16}, the method was applied to the DQCP \cite{nakayama16}, including the SO($5$) case \cite{li18,poland19}. While the applied
constraints were not sufficient to produce small islands in the space of scaling dimensions, useful bounds were found. The bound on the scaling
dimension of the order parameter corresponds to $\eta \agt 0.6$ \cite{li18,poland19}, far exceeding the numerically found value $\eta \approx 0.26$
(though $\eta \approx 0.6$ was claimed in one of the fermion models \cite{li19}). Moreover, the correlation length exponent $\nu \approx 0.45$
\cite{nahum15a,sandvik20} violates $\nu > 0.51$ \cite{nakayama16}---a bound that is applicable to any CFT, including the presumed generic DQCP, with
only a single relevant operator obeying all the symmetries of the Hamiltonian. It should be noted, however, that the larger value mentioned above,
$\nu \approx 0.62$ obtained in some calculations or within ranges of smaller system sizes, does satisfy the bootstrap bound.

The discouraging conclusions of the bootstrap calculation, in combination with a prevailing perception that all the models studied had similar
weak first-order transitions, spurred speculations that the DQCP must not be fundamentally reachable. Instead, it may exist only outside
3D space-time \cite{wang17}, below some critical dimensionality. Specifically, based on $\epsilon$-expansions for the level-$k$
WZW nonlinear $\sigma$-model, $d_c<3$ was found for all $k$ by Nahum \cite{nahum20} and $d_c=2.77$ was obtained for $k=1$ (the
case presumed to be applicable to the DQCP) by Ma and Wang \cite{ma20}. A similar $d_c$ value was found using the CFT bootstrap \cite{he21}.
The critical dimensionality may then be sufficiently close to $d=3$ for ``walking'' behavior of a near-critical system to be observed
\cite{gorbenko18a,gorbenko18b}. This type of slow evolution under renormalization group (RG) flow would then be responsible for the gradual
drift of some of the critical exponents when the system slowly crosses over to the first-order fixed point.

Walking-type near-criticality can in some cases be described by a non-unitary CFT, with complex scaling dimensions. The perhaps most well
studied case is the five-state Potts model in two dimensions, which has a phase transition where the correlation length is about 2500 lattice
spacings \cite{buffenoir91}. There is no corresponding unitary CFT, and a real Hamiltonian with the 5-state full permutation symmetry maintained
can never be driven to a continuous transition. A complex CFT arising after collision of two real fixed points versus the number of states
$q$ (in this case the Ising and tricritical Ising fixed points coinciding for $q=4$) can, however, describe observed crossover behaviors
\cite{gorbenko18b,ma19}. A complex fixed point was even explicitly demonstrated in a non-Hermitian variant of the quantum 5-state Potts model
\cite{tang24}. Though there are no direct indications from the lattice calculations that this type of scenario can quantitatively explain any
numerical data for the AFM--VBS transition, it has been widely accepted as the correct theory.

The bounds on scaling dimensions from the CFT bootstrap assumed a generic critical point with only one relevant operator; in commonly
used CFT language the rank-$2$ traceless symmetric tensor, with field strength denoted by $t$ on the vertical axes in Fig.~\ref{phases}. In classical
statistical physics the corresponding exponent $\nu_t$ (or its inverse ratio with the conventional correlation exponent \cite{fisher74})
is often called the crossover exponent, as the transition involves an explicit change in symmetry of the Hamiltonian when crossing an O($N$)
critical point by turning on perturbations of the form $\phi_i^2$ of some of the components $i=1,\ldots,N$ of the order parameter $\boldsymbol{\phi}$.
At the DQCP, the subset of these perturbations that take the system from the AFM to the VBS phase is still symmetric in the
microscopic lattice Hamiltonian, e.g., in the $J$-$Q$ models the symmetries of the Hamiltonian do not change when tuning coupling ratios.
However, these perturbations are not symmetric under the putative emergent SO($5$) symmetry, since the order parameter changes from SO($3$)
in the AFM phase to ${\mathbb Z}_4$ in the VBS phase, with the latter enlarged to U($1$) as the critical point is approached, then combining with
the AFM order parameter to form the unified SO($5$) symmetric order parameter.

All fully symmetric perturbations---singlet operators under the SO($5$) symmetry---were assumed to be irrelevant at the DQCP \cite{nahum15b,wang17}.
This is in sharp contrast to the classical O($N$) models, where changing the temperature from the critical $T_c$ is the relevant symmetric perturbation
that takes the system from a paramagnet above $T_c$ to an ordered state with broken O($N$) symmetry below $T_c$. A generic SO($5$) DQCP transition
would correspond to Fig.~\ref{phases}(a) below the special multicritical point, with the entire line consisting of AFM--VBS transitions in the same
universality class, presumably satisfying the bounds on $\nu$ and $\eta$ found in Refs.~\cite{nakayama16,li18,poland19}, unless the complex
CFT description is necessary here.

The original DQCP scenario \cite{senthil04a,senthil04b} did not involve an emergent SO($5$) symmetry but an explicitly SO($5$) symmetric theory
was later constructed as an alternative to the CP$^{N-1}$ field theory for $N=2$ (while for $N>2$ no higher symmetry is expected) \cite{senthil06}.
The original CP$^{N-1}$ scenario for $N=2$ could therefore in principle also correspond to the phase diagram in Fig.~\ref{phases}(a) below the multicritical
point, but now with only emergent SO($5$) symmetry on the critical line, as in the case of lattice models within this scenario. The SO($5$) singlet field $s$
would be a redundant (or irrelevant) direction. The critical line can still be supplanted by a first-order transition
for some sufficiently large values of $s$, and a special (multicritical) point would then separate the two types of transitions.

Emergent SO($5$) symmetry has been observed numerically on large length scales \cite{nahum15b,takahashi20}, and within the generic (i.e., not multicritical) DQCP
scenario this symmetry in combination with signs of weak discontinuities (walking behavior) in lattice models has been interpreted as line of pseudocritical
transitions controlled by a nearby complex SO($5$) WZW $k=1$ fixed point. This scenario of a generic complex CFT was presented as the solution to the perceived
fine-tuning problem \cite{wang17}; the generic weak first-order transitions with apparent SO($5$) symmetry observed in many lattice models with a single
tuning parameter.

The fuzzy sphere calculation \cite{zhu23,zhou23} involves a continuum electronic model coupled to a central magnetic monopole, previously introduced by Haldane
in the context of the quantum Hall effect \cite{haldane83}. The spherical symmetry is optimal for numerical studies of 3D CFTs, in the sense that the level spectrum
of a Hamiltonian is directly related to the spectrum of scaling dimensions in an underlying low-energy CFT description. Remarkably, even with a very small
number of electronic (Landau level) orbitals (10 or less), a spectrum in very close agreement with the 3D Ising CFT was found in a model with ${\mathbb Z}_2$
(Ising ferromagnetic) symmetry breaking \cite{zhu23}. By subsequently studying a fermionic SO($5$) WZW theory \cite{lee15}, likewise regularized in the basis
of Landau orbitals, Zhou et al.~\cite{zhou23} found scaling dimensions (reproduced here in Table \ref{dtable}) that are in approximate agreement with those of
the multicritical CFT. However, the presence of a singlet operator that becomes relevant with increasing number of orbitals was interpreted as
pseudocriticality, signaling an eventual RG flow toward a first-order transition controlled by a complex CFT \cite{zhou23}.

\subsection{Multicriticality scenario}

An order--order transition involving unrelated broken symmetries
should normally be expected to be continuous only at fine-tuned multicritical points. The conjecture that there is no relevant
SO($5$) singlet operator is an important ``beyond-Landau'' aspect of the DQCP scenario, implying a generic continuous transition between the AFM and VBS
states when an interaction implicitly violating the SO($5$) symmetry is tuned, favoring either the three AFM components or the two VBS components of the
five-dimensional order parameter. However, there is now mounting evidence that the AFM--VBS transition in the lattice models studied so far actually is continuous
only at a multicritical point. In Ref.~\cite{zhao20}, a scaling dimension that we now identify as $\Delta_t \approx 1.4$ was found within the $J$-$Q_2$ model. Since
it was believed that the correlation length exponent $\nu\approx 0.45$ (in both the $J$-$Q$ and loop models) should correspond to a relevant operator with
all symmetries of the Hamiltonian, the finding of another symmetric relevant operator associated with a different exponent, $\nu_t = (3-\Delta_t)^{-1} \approx 0.62$,
would imply a multicritical point.

Though similar values of $\nu_t$ had also been found in some earlier works, as detailed in Sec.~\ref{sec:lattmodels} above, the mismatch between $\nu$ and $\Delta_t$,
even within the same model (the loop model \cite{nahum15a,nahum15b}) had not been addressed previously. We now know (Sec.~\ref{sec:dark}) that the above exponent
$\nu \approx 0.45$ ($\Delta \approx 0.80$), which we will denote as $\nu_*$ henceforth (and for which we find a slightly smaller value, $\nu_* \approx 0.42$),
does not control the conventional correlation length $\xi=\xi_t$ when approaching the critical point exactly from one of the phases. Instead, it governs the slower
growing bubble size when the weak first-order transition is approached; see Figs.~\ref{defects} and \ref{xi}. Regardless of their physical interpretation
\cite{shao16}, the existence of two length-scale exponents $\nu_t$ and $\nu_*$ is not consistent with the generic DQCP scenario. Indeed, we will show in
Sec.~\ref{sec:dark} that the exponent $\nu_*$ depends on both $\nu_s$ and $\nu_t$, and also on the scaling dimension of the leading irrelevant SO($5$) perturbation.

Adding to the rise of the multicriticality scenario, some years ago numerical evidence started to build for gapless QSL phases located between AFM and VBS phases
in frustrated spin-$1/2$ models \cite{gong14,morita15,wang18,ferrari20,nomura21,yang22,wang22,liu22,corboz26} (in contrast to a gapped phase claimed earlier
\cite{jiang12}) that can be continuously deformed into the $J$-$Q$
models with direct AFM--VBS transitions. This potential to tune a model from a QSL to a direct AFM--VBS transition prompted the suggestion \cite{yang22}
that the generic phase diagram of frustrated models with QSL, AFM, and VBS phases is like that in Fig.~\ref{phases}(b). Such a phase diagram was indeed
found in a $J_1$-$J_2$-$J_3$ Heisenberg model \cite{liu23}, though it was not clear, because of the numerical challenges for conventional frustrated spin
models, whether the direct transition is continuous or first-order---a generic DQCP line was suggested as most likely. Field theories to incorporate
the QSL and SO($5$) multicriticality within the DQCP scenario are also now under active investigations \cite{lu21,schackleton21,schackleton22,liu22}.

If the AFM--VBS transitions in the lattice models indeed are situated on the first-order side and very close to a multicritical point, then the previous
bounds on critical exponents from the CFT bootstrap are no longer of concern in this context \cite{zhao20}. The conclusion of the first DQCP bootstrap
calculations \cite{nakayama16,li18} was colored by the desire to test the generic unitary CFT scenario, which could be ruled out when $\Delta_t$ did not
agree with numerical estimates under the presumed constraint $\Delta_s>3$, thus spawning the complex CFT walking scenario. In principle, a generic DQCP
line with $\Delta_s>3$ could still exist, as in Fig.~\ref{phases}(a), but no numerics so far has been able to access such a line in a spin model (i.e., there
are no exponent estimates pertaining to it), while the QSL phase in Fig.~\ref{phases}(b) has been observed in many models and the weak
first-order transitions likewise in several (other) models.

Very recently, Chester and Su \cite{chester24} applied a modified bootstrap method to the SO($5$) CFT, using one input scaling dimension, $\Delta_\phi=0.63$,
from a previous lattice calculation to predict other scaling dimensions agnostically (i.e., not insisting $\Delta_s > 3$), leading to the results
listed on the second row of Table \ref{dtable}. This calculation produced a relevant SO($5$) singlet operator, with a rather large scaling dimension
$\Delta_s \approx 2.36$, thus indicating a multicritical point (assumed tricritical). The procedure actually corresponded to the identification of scaling
dimensions located on the boundary of the CFT allowed region, and similar boundaries had also previously been computed (with less CFT constraints) for the
most important pairs of scaling dimensions; $(\Delta_\phi,\Delta_t)$ and $(\Delta_\phi,\Delta_s)$ \cite{li18}. It is believed that the true scaling dimensions
fall on or very close to the allowed boundaries, as is the case for the simpler O($N$) models \cite{poland19} (where alternatively small ``islands'' of allowed
regions can also be produced that indeed fall on the exclusion boundaries). The value $\Delta_s \approx 2.36$ \cite{chester24} (in our calculation slightly
smaller, as seen in Table \ref{dtable}) corresponds to a correlation function decaying with an exponent $2\Delta_s$ close to $5$, which may explain
why it had remained undetected (or mistaken for a scaling correction) in the lattice calculations.

It would now appear that multicriticality is the simplest explanation of all the previous lattice numerics. In addition to the conformal bootstrap calculation
supporting it, at face value the recent work on the fuzzy sphere also seems consistent with this conclusion, though the authors of Ref.~\cite{zhou23} instead
take their results as support of the non-unitary CFT scenario. In the multicritical scenario the observation of the singlet switching from irrelevant to weakly
relevant versus the number of orbitals $n_f$ simply would mean that $\Delta_s$ converges more slowly than some of the other scaling dimensions. However, it was
argued that the specific dependence on $n_f$ instead supports the complex CFT scenario \cite{zhou23}, where $\Delta_s$ also has an imaginary part that was
not accessed.

The non-unitary CFT description was originally proposed \cite{wang17} in order to try to reconcile the generic DQCP scenario and interpretations of the
numerical lattice data available at the time. With the current knowledge that a multicritical unitary CFT is possible \cite{li18,chester24}, the scenario of
non-unitary CFT seems obsolete---violating the principle of Occam's Razor. Indeed, in another application of the fuzzy sphere method \cite{chen23,liao26}, it
was found that the CFT spectrum is only stable within a narrow range of an interaction parameter in the electronic model. The previous calculation had
assumed that this parameter, which regulates the strength of a scalar operator, only tunes scaling corrections and that there is an optimal point at
which these corrections are minimized, similar to classical models in which scaling behavior can be optimized by tuning away the leading scaling correction
\cite{hasenbusch10}. Moreover, by moving away from the line of exact SO($5$) symmetry, another phase between the AFM and VBS phases was found \cite{chen23,liao26},
thus potentially realizing a phase diagram such as the one in Fig.~\ref{phases}(b), conjectured previously in Ref.~\cite{yang22}. All these developments make the
multicritical SO($5$) point and the phase diagram in Fig.~\ref{phases}(b) the most likely scenario for the AFM--VBS transition, setting the stage for our
study producing comprehensive evidence for it and uncovering its broader consequences.

\section{Overview of Calculations}
\label{sec:overview}

Given the length of the paper and the large number of interrelated results and insights, we here provide a concise summary of our methodology
and how we have drawn the conclusions that were already outlined above in Sec.~\ref{sec:intro}. The purpose here is to give some more concrete pointers
to specific numerical results and reasoning supporting these conclusions. As such, this section provides a guide to the very detailed analysis presented in
the subsequent sections and how the different aspects of the problem form a coherent picture of the weak AFM--VBS transition and its multicritical
end point.

\subsection{Quantum Monte Carlo methods}
\label{sub:qmc}

We study the $J$-$Q_2$ model defined in Eq.~(\ref{hamjq}) and its extensions illustrated in Fig.~\ref{jqterms} using two types of QMC simulation algorithms:

(1) Stochastic series expansion (SSE), where the partition function $Z={\rm Tr}\{{\rm e}^{-\beta H}\}$
is sampled in the form of strings of the local operators of the Hamiltonian appearing in a series expansion of ${\rm e}^{-\beta H}$, along with basis
states---eigenstates of the $S^z_i$ operators---to perform the trace operation stochastically. To obtain results suitable for finite-size scaling
at a $z=1$ quantum critical point, we take the inverse temperature as $\beta=aL$. Any factor of proportionality $a$ works asymptotically for large
$L$, and we use either $a=1$ or in some cases (the $J$-$Q_2$ model) take $a=1/c$, where $c \approx 2.38$ is the estimated velocity of excitations
\cite{suwa16} of the near-critical $J$-$Q_2$ model.

(2) Depending on the quantity studied, it is some times more practical to instead project out the ground state from a translationally invariant
singlet state $|\Psi\rangle$ (an amplitude-product state \cite{liang88}), using either the operator ${\rm e}^{-\beta H/2}$, which can be series expanded
as in the SSE method, or $(-H)^{pN/2}$, where $N$ is the total number of spins and $p \propto L$ should be expected for convergence (in the
case of criticality with $z=1$). The computational effort, which scales with the length of the operator string, is then similar to the expansion of the
exponential, where $\beta \propto L$ is required for convergence to the ground state and the corresponding mean string length is $\propto \beta N$.
Though a singlet state $|\Psi \rangle$ is used as the ``trial state'', the sampling of the normalization of the projected state,
$\langle \Psi|{\rm e}^{-\beta H}|\Psi\rangle$ or $\langle \Psi|(-H)^{pN}|\Psi\rangle$, is still carried out in the $S^z$ basis. Because of the special
properties of the valence-bond basis, it is easy to switch to it and evaluate expectation values in the space of transition graphs of left and right
projected states, which restores rotational invariance and reduces the statistical errors.

The implementation of this
projector QMC (PQMC) method is overall very similar to the SSE method---essentially the two methods only use different boundary conditions in imaginary time.
However, the computational effort to completely reach the ground state can be less in PQMC,
given that only singlet states at zero momentum are included in the sampling space, thus eliminating many of the low-lying excitations that affect the
$T \to 0$ convergence of $T>0$ QMC simulations. Moreover, operator expectation values can be conveniently expressed using transition graphs \cite{liang88}
of valence-bond configurations, i.e., when measuring observables the basis is switched back to that of valence bonds, which is an essentially trivial
task. To converge to the ground state the parameter $\beta$ or $p$ typically must only be a small multiple of $L$, and we have checked for full convergence of
all results presented here.

Both the above QMC methods are well established, and for implementation details we refer to Ref.~\cite{sandvik10b} for SSE and Ref.~\cite{sandvik10c}
for PQMC. Some useful PQMC operator estimators are further discussed in Refs.~\cite{beach06} and \cite{tang11c}. The methods are both
exact in the sense that, for a given system size, there are no approximations beyond statistical errors.

\subsection{Specific results and synthesis}

The scenario that we have arrived at for the AFM--VBS transition consists of many separate parts and concepts that are difficult to discuss in a linear
manner, as they all depend on each other in some way and the complete picture emerges only when all the components are put together. We therefore often
refer to concepts and results that have not yet been discussed in detail but that motivate our numerical analysis in the different sections. The main
numerical work is roughly organized in the order of the conventional relevant scaling dimensions first, then numerical results and insights
pertaining to the emergent SO($5$) symmetry, followed by results on the first-order line. It is only after all these aspects of the transition have
been discussed that the pseudocritical scaling and crossover close to the first-order line can be fully appreciated and explained.

In Sec.~\ref{sec:order} we first study spin and dimer correlations (corresponding to the AFM and VBS order parameters) of several $J$-$Q_n$ models and
establish beyond doubt that all these models have first-order AFM--VBS transitions. Nevertheless, for $n=2$ and $n=3$ the coexistence states are sufficiently
weakly ordered for the properties of the nearby critical point to be well established; specifically, two-point correlation functions of many operators exhibit
stable power laws up to distances of 100 lattice spacings or more. The exponents are common to the $J$-$Q_2$ and $J$-$Q_3$ models, and in some cases the $J$-$Q_4$
model also produces very similar power-law decays, though its first-order tendencies are much stronger.

We further analyze the correlation functions of the $J$-$Q_2$ model and extract the scaling dimension $\Delta_\phi$ to unprecedented precision. We implement
the trick of taking the distance ($r$) derivative $C'(r)$ of a correlation function $C(r)$ as a ``filter'' \cite{nahum15a} to minimize the impact of the eventual
long-range order at the weak first-order transitions. With the PQMC method we have converged such calculations to the ground state for system sizes up to $L=1024$,
where critical scaling is observed up to distances close to $200$. We also fit an appropriate function directly to $C(r)$ for $L$ up to $512$, with scaling
corrections included, and the results of this approach also agree well with those from the derivatives. All these results lead us to the firm conclusion that
$\Delta_\phi$ is about 4\% smaller than the previously used input value for the CFT bootstrap calculation (see Table \ref{dtable}), which may appear to be a minor
discrepancy but is an important one when it comes to establishing agreement with the SO($5$) CFT. The individual values for $\Delta_\phi$ obtained from spin and
dimer correlations agree with each other to within a statistical error of about $1$\% and we therefore assume that they really are the same, as required by the
SO($5$) CFT description.

In Sec.~\ref{sec:scaledims} we use a symmetric multi-spin (singlet) correlator not containing the order parameters in attempts to extract the scaling dimensions
$\Delta_t$ and $\Delta_s$ of the two relevant symmetric perturbations of the critical point. However, while the value of $\Delta_t$ is relatively easy to extract
from the long-distance correlation decay, giving the value listed in Table \ref{dtable}, it is difficult to distinguish the subleading contributions governed
by $\Delta_s$ from other expected scaling corrections. We obtain $\Delta_s$ in other ways, as explained further below. We further
argue that a spatial bond ($J$-term) modulation used to induce a helical VBS phase in Ref.~\cite{zhao20} contains the conserved current
operator, which allows us to extract its scaling dimension $\Delta_j$. This scaling dimension was already obtained approximately in Ref.~\cite{zhao20},
but without noting the role of the conserved current operator and with a numerical value $1.90(2)$; five error bars off the expected value $2$. With improved
results for the corresponding correlation function, we now find a value consistent with the general requirement $\Delta_j=2$ for a Noether current operator.
This result is important in that it demonstrates explicitly that a scaling dimension with a known CFT value can be extracted despite the fact that the
systems we study cannot reach the critical point exactly.

The most natural scenario for a multicritical point is that it is attached to a first-order line on which the relevant singlet field $s$ varies and the
SO($5$) violating field vanishes ($t=0$), as in either of the scenarios in Fig.~\ref{phases}(a) and Fig.~\ref{phases}(b). To study such a first-order
line in a 2D parameter space, in Sec.~\ref{sec:jq2q6} we study the phase diagram of the $J$-$Q_2$-$Q_6$ model, keeping $J+Q_2=1$ and extracting transition
values $Q_{2c}$ for several values of $Q_6$. As expected, we find that the transition for increasing $Q_6$ develops gradually stronger first-order signatures,
which we use to further test the multicriticality scenario. Using finite-size scaling of a Binder cumulant, we show consistency with the value in
Table \ref{dtable} of the correlation-length exponent $\nu_s=(3-\Delta_s)^{-1}$ pertaining to the first-order line. Moreover, we extrapolate the order parameters
to infinite size, using the common value $m^2(L)$ of the two squared order parameters where they cross each other, to ensure that the system is properly in the
coexistence state for each $L$. We find power-law scaling of the form $m^2 \sim (Q_6-Q_{6c})^{2\beta}$ with $\beta \approx 0.87(5)$, which agrees with the value
computed from the independently extracted scaling dimensions in Table \ref{dtable}; $\beta = \Delta_\phi/(3-\Delta_s) \approx 0.84$. The fitted critical $Q_6$
value is negative, $Q_{6c}= -0.07(2)$, and cannot be studied directly with sign-free QMC simulations. However, there is now no specific objective reason to doubt
that this multicritical point exists in the real model space, in contrast to the common view that only a critical point described by a complex CFT is possible
\cite{wang17,ma20,nahum20,he21,zhou23}.

In Sec.~\ref{sec:jq2q6} we also carry out similar analyses of the first-order lines of the $J$-$Q_2$-$Z_2$ and $J$-$Q_2$-$Z_3$ models, which reveal no detectable changes
in the order parameters when $Z_2$ or $Z_3$ is increased, even though the transition point evolves. This behavior implies that the $Z_2$ and $Z_3$ operators are
dominated by their $t$ field content, which points to one plausible reason why many models show weak first-order AFM--VBS transitions (beyond the
fact that the critical exponents $\beta$ and $\nu_s$ imply large critical scaling regions): the overlap of the $s$ field with typical
lattice interactions is small. Extreme operators such as the spatially large $Q_6$ terms are required to shift the transition significantly
into the coexistence line (increasing the $s$ field).

In addition to the scaling dimensions of the relevant CFT operators listed in Table \ref{dtable}, in Sec.~\ref{sec:symm} we extract the scaling dimensions of leading
irrelevant perturbations of the SO($5$) symmetry, using quantities probing the symmetry deformations
inherent to the Hamiltonian (instead of analyzing corresponding correlation
functions that decay very fast for irrelevant operators, thus being affected by large statistical errors). The leading lattice-induced ${\mathbb Z}_4$ deformation of
the emergent U($1$) symmetric VBS order parameter is detected in the size dependent distribution $P(D_x,D_y)$ and corresponds in the SO($5$) CFT to an $\ell=0$
(Lorentz spin), parity-even, rank-$4$ (internal symmetry) dangerously irrelevant primary operator, for which we obtain the scaling dimension denoted $\Delta_4$.

Surprisingly, we find that the inherent symmetry deformation involving both AFM and VBS components, detected via $P(M_z,D_x)$, is not governed by $\Delta_4$ but
by a larger scaling dimension $\Delta_{2,3} \approx 4.71$ which satisfies $\Delta_{2,3} \approx \Delta_4+1$. Here the subscripts $2,3$ refer to violation of the
symmetry between the $3$-component AFM order parameter and the $2$-component VBS order parameter. An exact relationship $\Delta_{2,3} = \Delta_4+1$ would correspond
to the first descendant of the rank-$4$ primary, which, however, is not possible because the symmetries of the Hamiltonian should not allow the presence of this
operator. Given the exact SO($3$) symmetry of the AFM order parameter ${\bf M}$ and the $\mathbb{Z}_4$ or U($1$) symmetry of the VBS order parameter ${\bf D}$, the
probed perturbation is really one involving amplitudes, in the simplest form $D^2M^2$. Decomposing this into its allowed SO($5$) operators, the most likely scenario is
that the deformation is dominated by the first irrelevant primary operator $t'$ in the same symmetry sector as the $t$ field driving the transition (the imposed
deformation that is tuned to zero at the critical point), i.e., an irrelevant $\ell=0$, parity-even, rank-$2$ operator. Such an operator with scaling dimension
$\approx 4.35$ was identified in the fuzzy sphere calculation \cite{zhou23}, which, given the discrepancies in some of the other scaling dimensions in Table \ref{dtable},
is not that far off from our value. We thus posit that actually $\Delta_{2,3} \equiv \Delta_t'$. The rank-$4$ operator with smaller scaling dimension $\Delta_4 \approx 3.7$
($\approx 3.9$ in the fuzzy sphere calculation) may be expected to eventually govern the behavior at the longest length scales but must have a comparably very small
amplitude in the $J$-$Q$ models or, possibly, is completely absent. We will argue that the physical source of the deformation $t'$ indeed is an effectively repulsive
interaction $D^2M^2$ between the AFM and VBS order parameters, which we also illustrate using a simple classical O($5$) deformation model, which rules out the phase
diagram with coexistence phase and tetracritical point in Fig.~\ref{phases}(c) (which would require an attractive interaction).

Moreover, the scaling dimension $\Delta_{t'}$ is related by Eq.~(\ref{nustardeltas}) to the observed emergent exponent $\nu_*$ of the pseudocritical (bubble)
scale. As discussed above in Sec.~\ref{sec:lattmodels}, a glaring inconsistency between previous results for $J$-$Q$ and loop models and potential CFT descriptions
of the DQCP phenomenon has been the anomalously small correlation length extracted from various quantities, e.g., Binder cumulants \cite{nahum15a,sandvik20}.
The value $\nu \approx 0.45$ violates a generic CFT bound $\nu > 0.51$ under the assumption of only one relevant perturbation that maintains all the
symmetries of the Hamiltonian, i.e., the absence of relevant SO($5$) singlet operators \cite{nakayama16}. This bound is no longer of concern if there is a
relevant singlet in the CFT spectrum, in which case the transition is fine-tuned as proposed in Refs.~\cite{zhao20,yang22}.
However, as seen in Table \ref{dtable}, there is no scaling dimension $\Delta_* = 3-1/\nu \approx 0.80$ of the right symmetry in the CFT spectrum
computed in Ref.~\cite{chester24} and also no operator close to this value in the fuzzy sphere calculation \cite{zhou23}. In Ref.~\cite{sandvik20} an
apparent scaling dimension $\Delta_* \approx 0.80$ was actually identified in the form of a particular correlation function decaying as
$r^{-2\Delta_*}$. This value of $\Delta_*$ matched very well the correlation length exponent with its estimated value at the time, $\nu_* \approx 0.45$,
through the standard relationship $1/\nu_* = 3-\Delta_* \approx 2.2$. However, we now have a better way of estimating the pseudocritical exponent, from
a scaling behavior of the emergent SO($5$) away from the transition, which we investigate in Sec.~\ref{sub:so5} and find $1/\nu_* \approx 2.40$. We believe that
$\Delta_*$ (whose value stays close to $0.80$ in the improved analysis here) is also associated in some way with the phenomenon of pseudocriticality,
but we do not know the exact relationship (if any) between $\Delta_*$ and $\nu_*$. $\Delta_*$ has no impact on any other results in this work beyond presenting
a lingering mystery. Presumably, a perturbation theory of the SO($5$) CFT on the weak first-order line could answer this question, but this undertaking
is beyond the scope of the present paper.

A major result of our work is that $\nu_*$ can be derived from a generalization of results for a classical tetracritical point with emergent O($N$) symmetry by
Bruce and Aharony \cite{bruce75}. They showed that the shape of phase boundaries of the coexistence phase of two order parameters, schematically drawn in
Fig.~\ref{phases}(c), is given by a combination of the crossover exponent, $\nu_t$ in the present context, and the scaling dimension of the leading irrelevant
perturbation of the O($N$) symmetry. The tetracritical point with its adjacent coexistence phase forms in Landau theory as a result of attractive interactions
between two order parameters \cite{bruce75}, but as already mentioned we have ruled out a coexistence phase in the $J$-$Q$ model.

In Sec.~\ref{sec:dark} we will argue that, while there is no new phase induced by repulsive interactions, an associated crossover region nevertheless surrounds
the first-order transition; schematically between the dashed and solid curves fanning out from the SO($5$) point in Fig.~\ref{defects}. This region is associated with
a length scale that we can derive by starting from scaling arguments previously developed for the attractive case. We arrive at the following expression, already
presented in a different form as Eq.~(\ref{nustardeltas}):
\begin{equation}
\frac{1}{\nu_*} = \frac{1}{\nu_s}\left ( \frac{1}{\nu_t} + |y_{t'}| \right ),
\label{nustarform}
\end{equation}
where $y_{t'}=3-\Delta_{t'}$. Given our explicit determination of the scaling dimension of the inherent dangerously irrelevant SO$(5)$ perturbation (which becomes
relevant on the first-order line) causing the asymmetry between the critical AFM and VBS order parameters, we will assume that $\Delta_{t'}$ is to be used here,
not the smaller dimension $\Delta_4$, at least on the length scales accessible in practice close to the multicritical point (and note that the divergence of $\xi_*$
is anyway cut off in a manner similar to conventional first-order transitions when moving even closer to the transition, inside the fan marked by dashed curves
in Fig.~\ref{defects}). The value of $\nu_*$ extracted in Sec.~\ref{sub:so5} from a quantity corresponding to the size of the window within which emergent SO($5$)
symmetry is manifested (the regime where the repulsive order parameter interactions are relevant) is indeed fully consistent with this expression.

By using the extracted value of $\nu_*$ together with our estimates for $\nu_t$ and $y_{t'}$, we can alternatively solve Eq.~(\ref{nustarform}) for
$1/\nu_s=3-\Delta_s$, which leads to our best value of $\Delta_s$ in Table \ref{dtable}. The value is close to the predicted CFT bootstrap value in Table
\ref{dtable} and, in combination with our values of $\Delta_\phi$ and $\Delta_t$, falls very close to previously computed SO($5$) CFT bounds \cite{li18} (which
are believed to be close to the true exponents \cite{chester24}). In Sec.~\ref{sec:jq2q6} we show that our $\Delta_s$ value is also fully compatible with
the growth of the coexisting order parameters when moving along the first-order line of the $J$-$Q_2$-$Q_6$ model, i.e., the exponent $\beta$ that we
extract from the order parameters on the first-order line to a precision of a few percent.

In Sec.~\ref{sec:dark} we show that a detailed understanding of the observed correlation length (as obtained from correlation functions
or from critical data collapse procedures) requires all the relevant length scales, i.e., the three exponents $\nu_t$, $\nu_s$, and $\nu_*$. We further explain
how the scaling crossover of the correlation length varies as the first-order line is crossed at different distances from the multicritical point as in
Fig.~\ref{xi}, using the $J$-$Q_2$-$Q_6$ model with different values of $Q_6$ (corresponding to $k$ in Fig.~\ref{xi}). The impact of the bubble length scale
governed by $\nu_*$ fades away as the critical point is approached, becoming truly extinct at the critical point.

The three length scales governed by $\nu_t$, $\nu_s$, and $\nu_*$ impact also other physical observables that intrinsically are related to
the correlation length. In Sec.~\ref{sec:anomalous} we discuss the previous ansatz with two length scales (corresponding to $\nu_t$ and $\nu_*$)
\cite{shao16} in light of our present understanding, which implies a reinterpretation of the physical nature of the length scales and their exponents
assumed in Ref.~\cite{shao16} (but with the same scaling forms). While $\nu_t$ and $\nu_*$ often dominate scaling properties, accounting for the full
finite-size dependence of many observables also requires that $\nu_s$ is taken into account when the transition is crossed as in Figs.~\ref{phases}(a)
and \ref{phases}(b), where both the fields $s$ and $t$ are varied. We demonstrate the resulting intricate behavior using Binder cumulants and the spin
stiffness $\rho_s$.

\section{Order parameters}
\label{sec:order}

Here we begin by precisely defining spin and dimer correlations and their critical exponents in Sec.~\ref{sec:corrdef}.
In Sec.~\ref{sub:corrL} we present PQMC results for these correlations at distance $r=L/2$ on lattices of size up to $L=384$
for the $J$-$Q_2$ model and somewhat smaller, $L=320$, for the $J$-$Q_3$ and $J$-$Q_4$ models. The main result here is that the $J$-$Q_2$
model undergoes a very weak first-order transition, which is clearly detectable in the correlation functions of the order parameters
only with high-quality data (small error bars) for large lattices. For increasing $n$, the $J$-$Q_n$ models host transitions of increasing
first-order character, as seen in the magnitude of the coexisting AFM and VBS order parameters.

In Sec.~\ref{sub:corr}, we examine the distance dependence of the correlation functions $C_{s,d}(r)$ and their $r$ derivatives. In the presence
of the weak coexisting long-range orders at the transition point, the derivatives show power-law behaviors over longer distances and are useful
for extracting the scaling dimension $\Delta_\phi$, as was previously found with the 3D loop model \cite{nahum15a}. In the case of the $J$-$Q_2$
model, we present ground-state converged PQMC results for derivatives up to $r=256$ for a system as large as $L=1024$, revealing a significant
window of $r$ for which a scaling dimension $\Delta_\phi \approx 0.61$ effectively governs the behavior of both the spin and dimer derivatives.
To complement and confirm the validity of this approach, we also analyze the correlation functions themselves with scaling corrections included,
using system sizes up to $L=512$, which is the largest size for which the correlations are well converged to the ground state (i.e., the derivatives
converge faster). The two approaches deliver compatible results for the AFM and VBS values of $\Delta_\phi$, with the individual scaling dimensions
being the same to within statistical errors of less than 1\%. Our best estimate of the common value is $\Delta_\phi=0.607(4)$.

\subsection{Spin and dimer correlation functions}
\label{sec:corrdef}

We define the staggered spin correlation functions as
\begin{equation}
C_{\rm s}({\bf r}) = \langle {\bf S}_{\bf 0}\cdot {\bf S}_{\bf r}\rangle \epsilon_{\bf 0}\epsilon_{\bf r},
\label{cafmdef}
\end{equation}
where $\epsilon_{\bf r}=(-1)^{r_x+r_y}$ is the AFM phase factor for the spin at ${\bf r} = (r_x,r_y)$. The dimer correlation function is
defined as
\begin{equation}
  C_{\rm d}({\bf r}) = [\langle ({\bf S}_{\bf 0}\cdot {\bf S}_{\bf \hat x})({\bf S}_{\bf r}\cdot {\bf S}_{\bf r+\hat x})\rangle
  -\langle {\bf S}_{\bf 0}\cdot {\bf S}_{\bf \hat x}\rangle^2]\theta_{\bf 0}\theta_{\bf r}
\label{cvbsdef}
\end{equation}
where $\hat x = (1,0)$ is the lattice vector in the $x$ direction and $\theta_{\bf r}=(-1)^{r_x}$ is the appropriate phase factor for
a columnar dimer pattern in the ordered state, corresponding to the wave-vector of the dominant critical fluctuations; $q=(\pi,0)$ and
$(0,\pi)$ (the latter with $x \leftrightarrow y$ in the above definitions). In all cases, we average translationally over the reference spin
$r=0$ and also use the reflection and rotation symmetries of the square lattice as appropriate.

In a system with long-range order of a given kind, the corresponding long-distance correlation function should converge to a nonzero value
representing the square $m^2$ of the ordered moment, while in a critical system a power law decay of the form
\begin{equation}
C(r) \propto r^{-2\Delta_\phi} \equiv r^{-(1+\eta_\phi)}
\label{cretaform}
\end{equation}
should be observed asymptotically. Here $\Delta_\phi$ is the scaling dimension of the order parameter considered, and its indicated relationship
to the critical exponent $\eta_\phi$ (the anomalous dimension) is
\begin{equation}
2\Delta_\phi = d+z-2+\eta_\phi
\end{equation}
for spatial dimensionality $d=2$ and dynamic exponent $z=1$. Evidence for $z=1$ was presented already in the first study of the $J$-$Q_2$ model
\cite{sandvik07} and later, e.g., in Ref.~\cite{suwa16}. The similarities between the $J$-$Q_2$ model and the manifestly space-time invariant
classical 3D loop model \cite{nahum15a,nahum15b} also supports $z=1$, as we already discussed in Sec.~\ref{sec:lattmodels}.

\begin{figure*}[t]
\includegraphics[width=170mm]{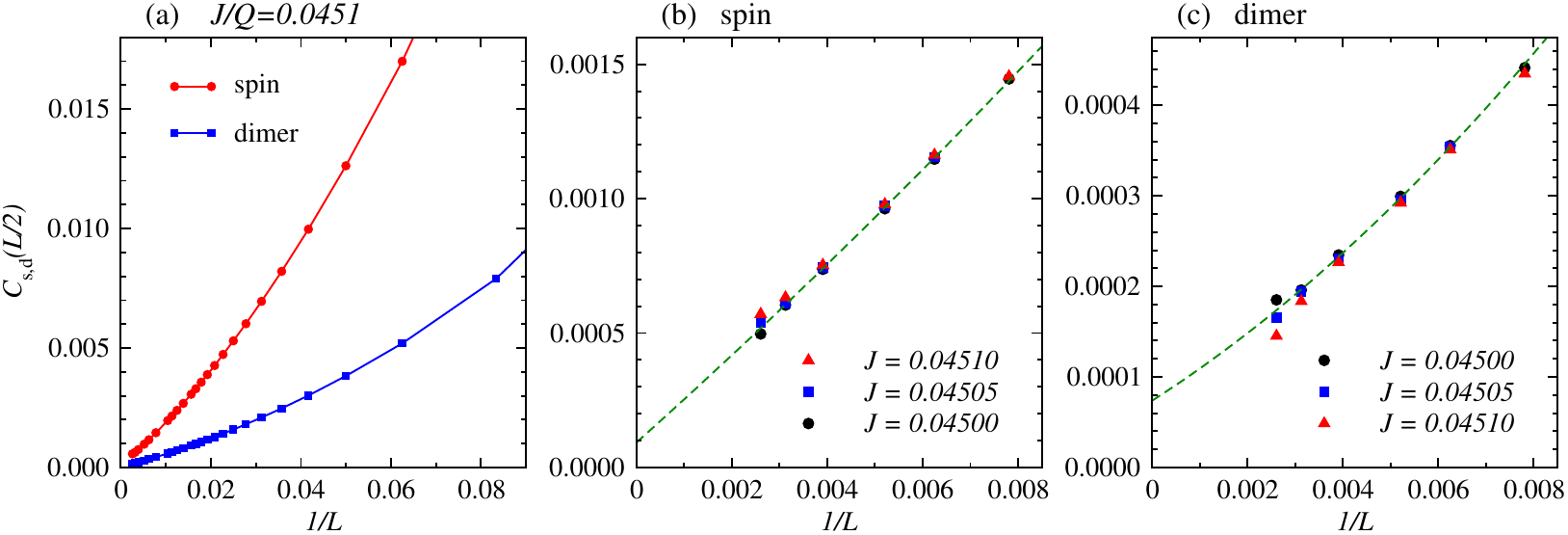}
\caption{Spin and dimer correlation functions at distance $L/2$ vs the inverse system size in the near-critical $J$-$Q_2$ model, obtained with
ground-state converged PQMC calculations. In (a), both correlation functions are shown vs $1/L$, with $L$ from $16$ (spin) or $12$ (dimer)
up to $L=384$. In (b) and (c) the correlation functions at three values of the coupling $J$ (at fixed $Q_2=1$) are zoomed in on the points for
$L=128$, $160$, $192$, $256$, $320$, and $384$. The error bars for the largest system in (c) are about half the symbol
size, while for all other cases they are much smaller. The green dashed curves are cubic collective fits to the data for $J=0.04500$ and
$J=0.04505$ (with several smaller system sizes also included in the fitting).}
\label{corL_jq2}
\end{figure*}

Below we will show both the spin and dimer correlations only for separations along the line ${\bf r}=(r,0)$. We have also computed the correlations along
the lines $(r,r)$ and $(0,r)$, the latter of which is not equivalent to $(r,0)$ in the case of $C_d$ for the $x$-oriented dimer operator.
We find the same type of behavior (including the decay exponent that we extract) in all cases in both spin and dimer correlations.

\subsection{Phase coexistence in $J$-$Q_n$ models}
\label{sub:corrL}

Long-distance correlation functions in finite systems can be studied either by considering distances $r \ll L$ for large $L$ or by taking one of the
longest distances in a periodic system, e.g., $r=L/2$. The former approach delivers the true correlation function in the thermodynamic limit, while
$C(L/2)$ for large $L$ extrapolates to the correct squared order parameter in a system with long-range order. In a critical CFT system, the same scaling
exponent governs the correlations versus $r$ and versus $L$, while the overall factor is different because of boundary enhancements when $r \approx L/2$.
For the purpose of detecting weak long-range order, investigating $C(L/2)$ versus $L$ is preferable, as only the behavior versus the single parameter
$L$ has to be monitored.

Figure \ref{corL_jq2}(a) shows both spin and dimer correlations at $r=L/2$ versus $1/L$ at the value of the critical coupling ratio of the $J$-$Q_2$ model
obtained from Binder cumulant crossings in Ref.~\cite{sandvik20}. It is certainly clear that any nonzero values for $L \to \infty$ must be very small,
suggesting a critical point or very nearly critical ground state. For system sizes up to about $L=100$, power laws can describe the behaviors reasonably
well, e.g., in Ref.~\cite{sandvik12} the exponent $1+\eta \approx 1.27$ was found this way for both spin and dimer correlations. However, the exponent
drifts significantly for larger systems. For the 3D loop model, system sizes up to $L \approx 500$ were considered \cite{nahum15a} and it was found that
$\eta$ extracted under the assumption of the critical form Eq.~(\ref{cretaform}) eventually tends toward negative values, while for shorter distances
$\eta$ matches well the value in the $J$-$Q$ model. Such drifting behavior is also found with the present results in Fig.~\ref{corL_jq2}(a).

In Figs.~\ref{corL_jq2}(b) and \ref{corL_jq2}(c) we zoom in on the largest system sizes for three different values of $J$ (here keeping $Q_2=1$) close to the
transition point. Comparing the variations in the spin and dimer correlations in this small window of couplings leaves little doubt that there is
one value of $J$, or possibly a very small range of values, for which both order parameters extrapolate to nonzero values. Thus, instead of a critical
ground state, AFM and VBS long-range orders coexist, as expected at a first-order transition or in a coexistence phase. We will later, in Sec.~\ref{sub:so5},
provide strong evidence for no finite coexistence phase, only a first-order AFM--VBS coexistence line (on which we will move by adjusting
another coupling in Sec.~\ref{sec:jq2q6}). The present data indicate that the transition value of $J$ is between the values $0.04500$ and $0.04505$;
slightly lower than the estimate in Ref.~\cite{sandvik20} but close to an older estimate based on correlation lengths in Ref.~\cite{sandvik10b}.

The dashed curves in Figs.~\ref{corL_jq2}(b) and \ref{corL_jq2}(c) are cubic polynomial fits to the data at both $J=0.04500$ and $0.04505$ (i.e.,
one function was collectively fitted to two data sets) in order for the curve to roughly correspond to the behavior at the transition point. While these
curves likely provide reasonably good estimates for the coexistence state located between these two couplings, they are here intended only to
illustrate the main qualitative trends. More detailed analysis of the nonzero $L \to \infty$ order parameters will be deferred to Sec.~\ref{sec:jq2q6},
where we move further into the first-order coexistence state by introducing a third tunable interaction ($Q_6$).

A leading linear correction in $1/L$ of the order parameter is expected in general for a system breaking O($N>1$) symmetry, reflecting gapless Goldstone
modes with their associated transverse correlations decaying as $1/r$ in two spatial dimensions. The AFM order parameter is O($3$) symmetric [strictly speaking
SO($3$), but the lack of reflection symmetry of the AFM state is irrelevant here] and the ultimately ${\mathbb Z}_4$ symmetric VBS order parameter has emergent
U($1$) symmetry up to some length scale (as discussed further in Sec.~\ref{sub:u1}) in a near-critical VBS phase. Moreover, the two order parameters
develop emergent SO($5$) symmetry (discussed in Sec.~\ref{sub:so5}), and so the $1/L$ behavior with corrections in the form of higher powers in
Figs.~\ref{corL_jq2}(b) and \ref{corL_jq2}(c) is expected for both order parameters over some range of system sizes, both inside the respective ordered
phases and at the transition point. As will be shown in detail in Sec.~\ref{sub:corr}, the distance dependence of the correlations is characterized
by a crossover from critical power-law decay to the nonzero squared ordered moment of the coexistence state.

\begin{figure}[t]
\includegraphics[width=75mm]{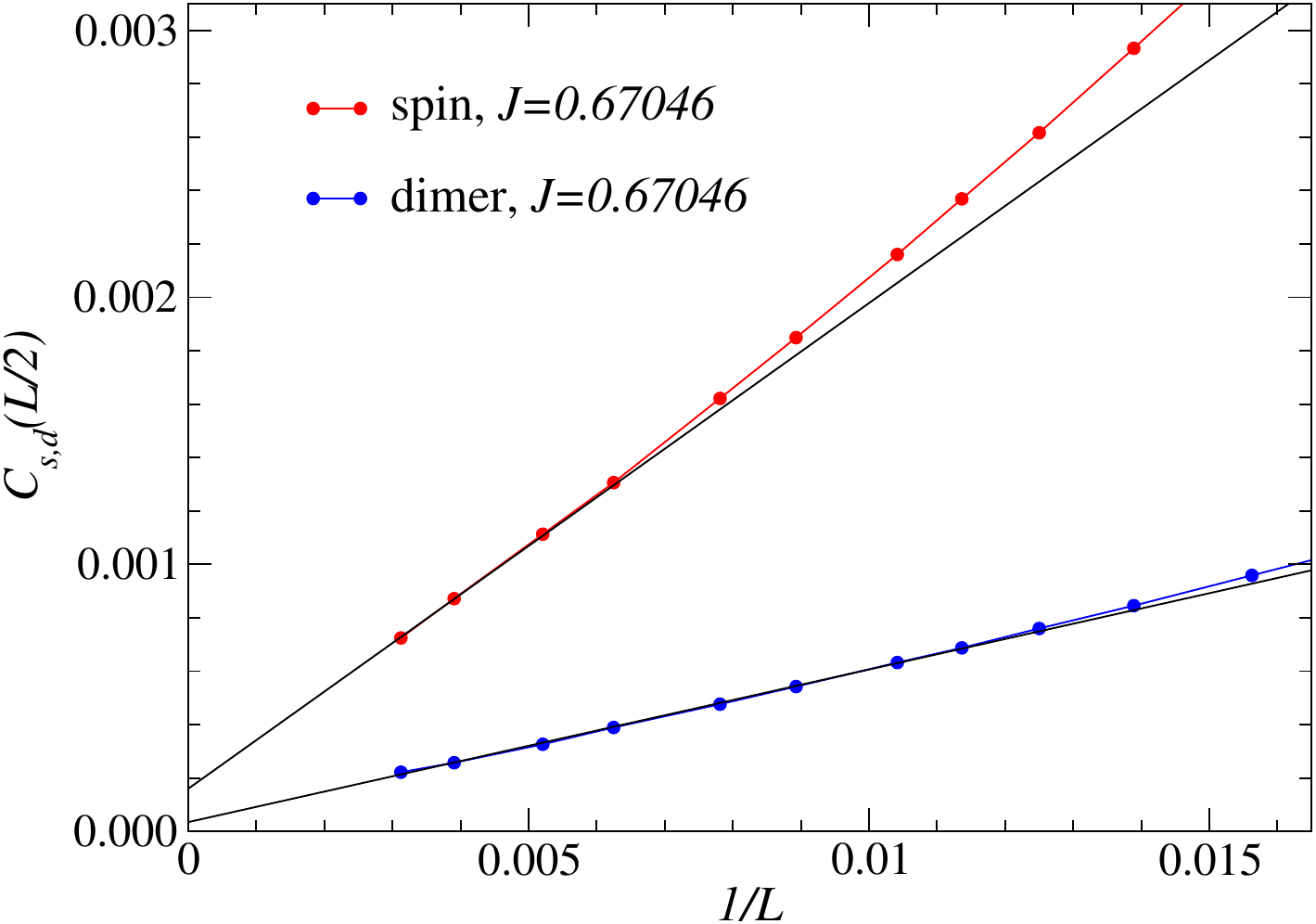}
\caption{Long-distance ($r=L/2$) spin and dimer correlations vs the inverse system size in the $J$-$Q_3$ model close to its AFM--VBS transition
  point (with $Q_3=1$ in all cases). The results for the largest system sizes in each case fall on almost straight lines with nonzero intercept,
  indicating a first-order transition with coexisting AFM and VBS long-range orders (located at $J$ slightly above the value used here).}
\label{corL_jq3}
\end{figure}

In Fig.~\ref{corL_jq3} we show results for the $J$-$Q_3$ model close to its AFM--VBS transition. Here we only draw straight lines in $1/L$
through the data points for the largest systems. We note that the dimer correlation for the largest size ($L=320$) deviates slightly upward
from the linear behavior, and there is also a less apparent downtrend in the spin correlation. These behaviors suggest that the value of $J$ used
here is marginally inside the VBS phase. Focusing on the AFM order parameter, where linear behavior exactly at the transition should persist when
$L \to \infty$ both inside the AFM phase and on the coexistence line (which is in practice close enough to the $J$ value used here), we see that
the extrapolated correlation function is roughly twice that of the $J$-$Q_2$ model in Fig.~\ref{corL_jq2}(b), i.e., the first-order discontinuity
is stronger in the $J$-$Q_3$ model, though not very dramatically so. The VBS order parameter actually appears to be a bit smaller, but it should
be kept in mind that the extrapolation is likely not quantitatively reliable.

\begin{figure}[t]
\includegraphics[width=80mm]{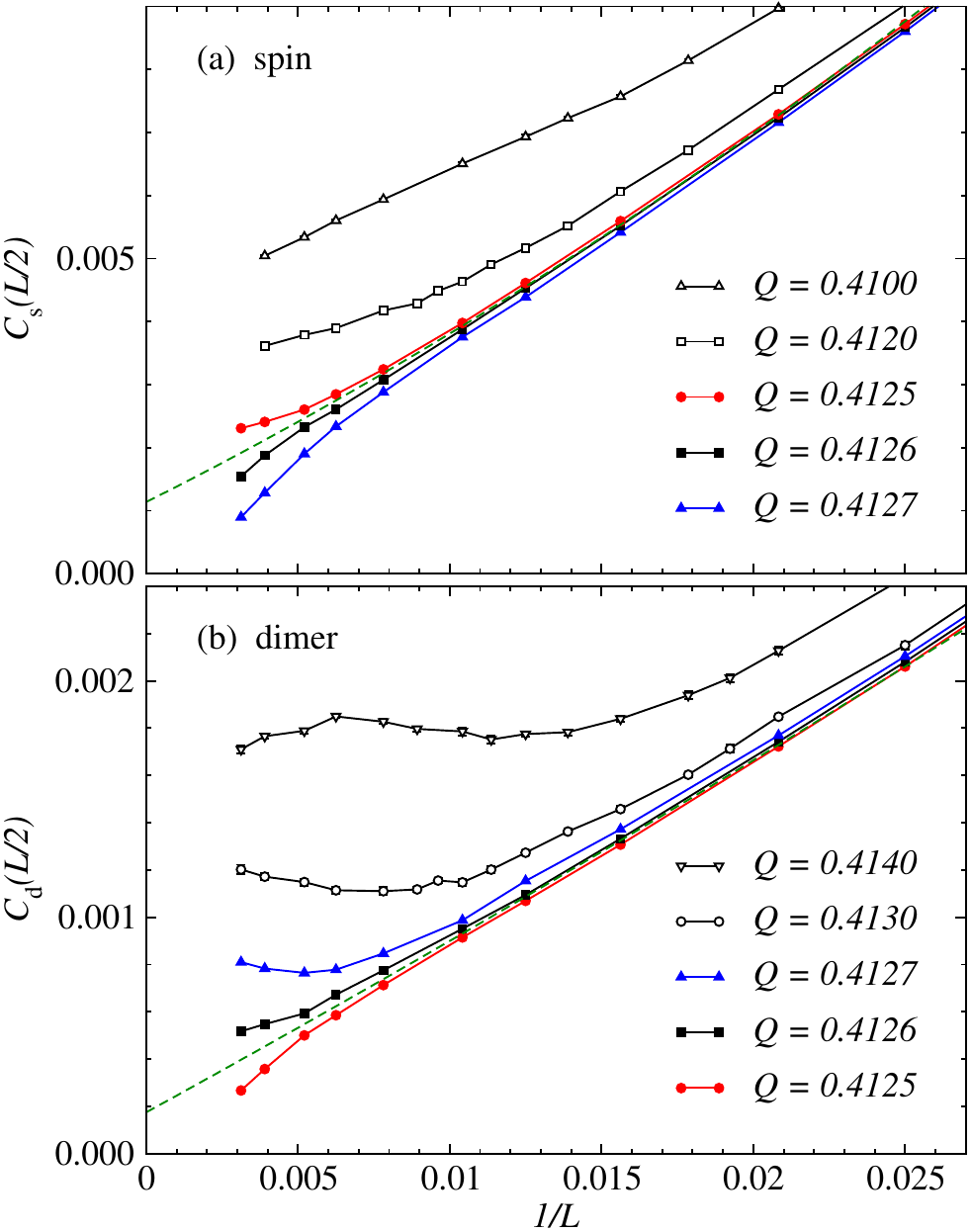}
\caption{Spin (a) and dimer (b) correlation functions in the $J$-$Q_4$ model at distance $L/2$ vs the inverse system. Results in both (a) and (b)
are shown with red, blue, and black solid symbols for values of the coupling $Q$ (with $J+Q=1$) close to the transition point. Other values of $Q$
(open black symbols) correspond to systems further inside the AFM phase in (a) and the VBS phase in (b). Error bars are at most of the order of
the symbol size. The green dashed curves are cubic collective fits to the data for $Q=0.41251$ and $Q=0.41260$, but do not necessarily
extrapolate reliably to infinite size because of possible crossover behavior for systems larger than available here.}
\label{corL_jq4}
\end{figure}

Moving on to the $J$-$Q_4$ model, in Fig.~\ref{corL_jq4} we show data for values of the $Q$ coupling (here with $J+Q=1$) close to the transition
as well as further inside the ordered phases. The roughly extrapolated order parameters of the coexistence state are clearly significantly
higher than those of the $J$-$Q_2$ and $J$-$Q_3$ models.

In Fig.~\ref{corL_jq4}(b) we observe nonmonotonic behavior of the dimer order parameter inside the VBS phase, close to the transition point. Such behavior
is natural in the neighborhood of a first-order transition (another example was discussed in Ref.~\cite{sandvik06}), where a small system effectively
is in the coexistence state and the order parameter that eventually is stabilized is averaged together with its much smaller value pertaining when the system
fluctuates into the other phase. Once the system is large enough, it no longer fluctuates between the two phases, and the computed order parameter therefore
represents, in the limit of infinite system size, the actual value in the ordered phase. The nonmonotonic behavior is only completely observable
in Fig.~\ref{corL_jq4}(b) at $Q=0.4140$, but the upward trend is clear also at $Q=0.4130$ and $Q=0.4127$. A precursor to the increasing behavior is
seen at $Q=0.4126$. The maximum seen at $L \approx 150$ at $Q=0.4140$ gradually diminishes as $Q$ is increased further and eventually (data not shown
here) a completely monotonic behavior sets in, with $C_d(L/2)$ rapidly approaching (on account of the gapped VBS state with finite correlation length)
its $L\to \infty$ value from above.

In Fig.~\ref{corL_jq4}(a) the spin correlations do not exhibit nonmonotonicity for the available system sizes, though likely a maximum followed
by convergence from above would also set in here for larger system sizes. Precursors similar to those preceding the upturn with increasing $L$
in Fig.~\ref{corL_jq4}(b) are seen, and the same arguments for nonmonotonicity apply in the AFM phase as well. The less pronounced nonmonotonic
behavior here should follow because the AFM order parameter has three components with exact SO($3$) symmetry, representing a larger phase space compared
to the eventually discrete four-fold degenerate VBS pattern. The fluctuations out of the AFM phase, even relatively close to the transition point, will
therefore be weaker.

Clearly, in both phases, the nonmonotonic behaviors will shift to larger systems upon moving closer to the transition point, and exactly at the transition
the crossover out of the coexistence phase never takes place; thus a monotonic behavior is expected at the transition point. It is very difficult
(computationally expensive) to locate the transition point to sufficient precision to follow the coexisting order parameters for large $L$. The
transition window shrinks rapidly with increasing system size---in classical $d$-dimensional systems as $L^{-d}$ and, correspondingly in quantum systems
as $L^{-d-z}$, where $z$ is the dynamic exponent (which likely should be taken as $z=2$ for long-ranged order parameters with continuous symmetry \cite{zhao19}).
The relative probability of either of the two phases changes smoothly from close to $0$ to close to $1$ when traversing this narrow window.

The curves fitted in Fig.~\ref{corL_jq4} can represent the size dependence when the coexisting order parameters have developed the emergent SO($5$) symmetry.
This symmetry must eventually break down (as discussed further in Sec.~\ref{sub:so5}), which should be accompanied by a crossover of the order parameters
versus $1/L$ also at the transition point. The SO($3$) symmetric AFM order parameter should still approach its infinite-size limit as $1/L$, but with a
different constant of proportionality. Due to its ultimate discrete nature, the VBS order parameter will asymptotically converge exponentially.
Such crossovers, which require very large sizes to be observed in these models, make it essentially impossible to extract the discontinuities in the
order parameters at the transition. For scaling purposes, we have nevertheless found a practically useful method for extrapolating the strength of the
coexisting order parameters, which we do on the first-order line of the $J$-$Q_2$-$Q_6$ model in Sec.~\ref{sec:jq2q6}.

\subsection{Scaling dimension of the order parameters}
\label{sub:corr}

In the first study of the $J$-$Q_2$ model \cite{sandvik07}, in PQMC simulations of system sizes up to $L=32$ it was found that the critical decay of the
AFM and VBS order parameters was similar, with an exponent $\eta = 0.26(3)$, or $\Delta_\phi \approx 0.630(15)$ by Eq.~(\ref{cretaform}). Later studies for
larger lattices found consistent results, e.g., in Ref.~\cite{sandvik12} $\eta \approx 0.27$ described both
order parameters for $L$ up to $96$. These results were based on the distance integrated correlation functions, which scale with the system size in
the same way as the $r \ll L$ correlator with $r$. For a generic correlation function $C(r)$ the squared order parameter is
\begin{equation}
m^2 = \frac{1}{N}\sum_{\bf r} |C({\bf r})|,
\end{equation}
and if $C(r) \sim r^{-2\Delta_\phi}$ then $m^2(L) \sim L^{-2\Delta_\phi}$.

In the 3D loop model \cite{nahum15a}, a study of much larger system sizes revealed that $\eta$ extracted from correlation functions at $r=L/2$
drifts toward smaller values with increasing $L$, eventually turning negative. For small to moderate values of $r$, up to $r \approx 20$, the
above quoted values of $\eta$ still describe the behavior, however, thus lending support to the loop and $J$-$Q$ models hosting the same kind of critical
fluctuations. The behavior of $\eta$ turning negative (with $1+\eta$ eventually tending to $0$ at the first-order transition that is now without doubt)
can clearly be understood as arising from the small ordered moment of a coexistence state, which dominates the correlation function for large distances;
$C(r) \to m^2$ for $r \to \infty$, while not affecting much the critical fluctuations for small $r$.

\begin{figure*}[t]
\includegraphics[width=150mm]{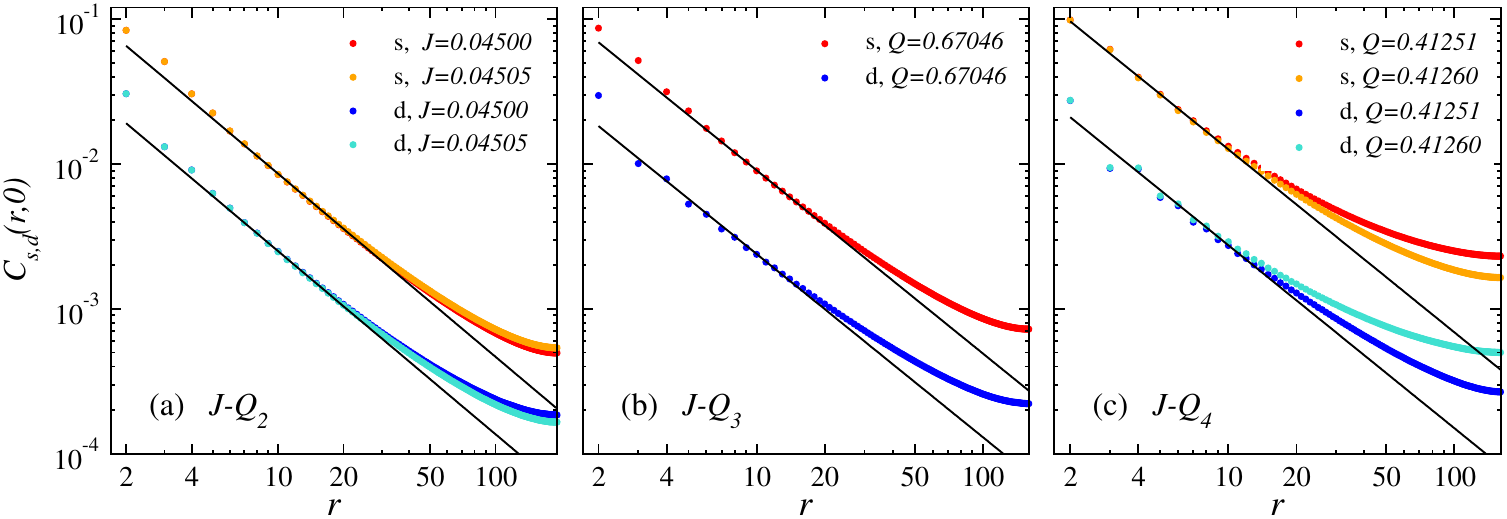}
\caption{Spin and dimer correlations of the $J$-$Q_2$ (a), $J$-$Q_3$ (b), and $J$-$Q_4$ (c) model at points close to their respective
AFM--VBS transition points. The lines correspond to the critical form $C_{s,d} \propto x^{-2\Delta_\phi}$, where $\Delta_\phi=0.63$ and the
amplitude has been adjusted for good fits for $x\approx [5,20]$ in the $J$-$Q_2$ model and smaller ranges in the other two models. The
system size is $L=384$ in (a) and $320$ in both (b) and (c).}
\label{sdcor}
\end{figure*}

\begin{figure*}
\includegraphics[width=150mm]{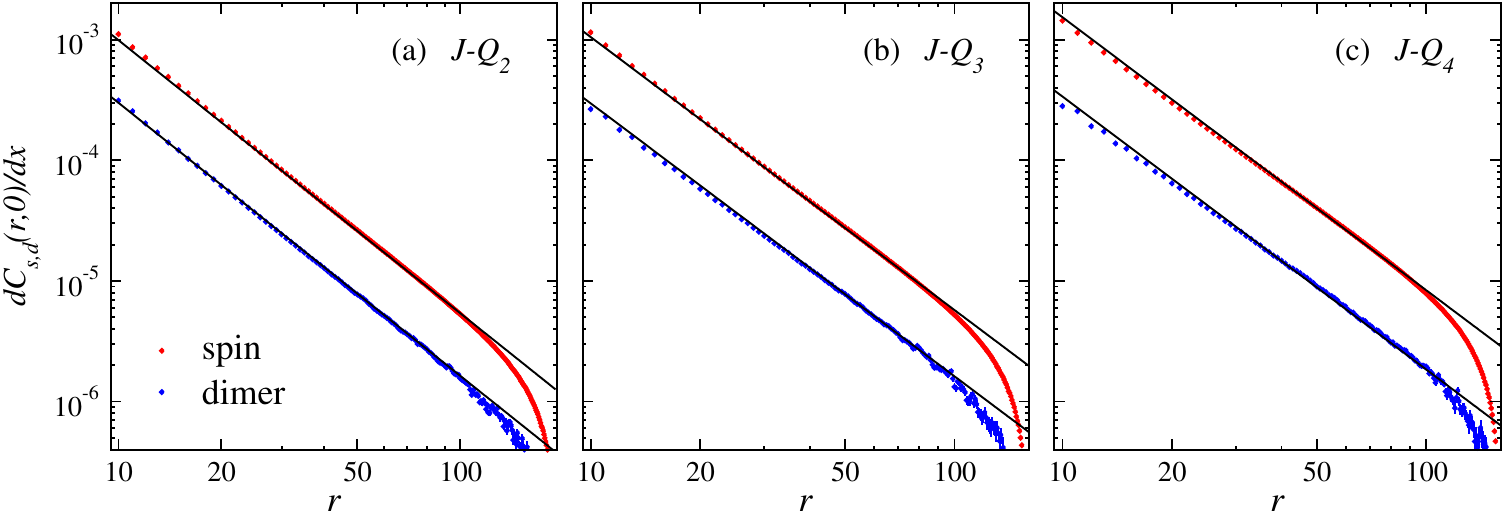}
\caption{First derivative, based on Eq.~(\ref{cderdef}), of the spin (red symbols) and dimer (blue symbols) correlation functions of the $J$-$Q_2$ (a),
$J$-$Q_3$ (b), and $J$-$Q_4$ (c) model. In (a) and (c), the results represent averages of the two couplings stated in Figs.~\ref{sdcor}(a) and \ref{sdcor}(c).
The black lines all have slope $-2.26$ and have been adjusted for best fit to the data in the range $x \approx [30,80]$.}
\label{sdder}
\end{figure*}

Nahum et al.~\cite{nahum15a} further made a very interesting and useful observation: the first discrete $r$ derivative of the correlation
functions exhibited good power-law scaling, $C'(r) \sim r^{-2-\eta}$, up to much larger distances than $C(r) \sim r^{-1-\eta}$. For $L=512$,
power law decays were observed for distances up to $r \approx 200$, with $\eta_{s} = 0.256(6)$ and $\eta_{d} = 0.25(3)$, respectively,
for the AFM and VBS correlations (with the larger error bar on the VBS result explained by much larger statistical errors of the dimer
correlations). This much extended scaling behavior of the derivatives can be explained by writing the correlation function
(spin or dimer) for a finite system as
\begin{equation}
C(r,L)=C_1(r)+C_2(L)+C_3(r,L),
\label{c123}
\end{equation}
which clearly can always be done, with non-uniqueness circumvented by (in principle) minimizing $C_3$.
In the case at hand, one would suspect that $C_1(r)$ is close to the critical power law described by
the exponent $1+\eta$ and $C_2(L)$ is essentially the small size dependent magnetization squared, i.e., a function similar to $C_{s,d}(L/2)$
in Figs.~\ref{corL_jq2}, \ref{corL_jq3}, and \ref{corL_jq4}. The mixed term $C_3(r,L)$ can of course not completely vanish, but if it is relatively
small until $r$ approaches $L/2$ then $C'(r,L)=\partial C(r,L)/ \partial r$ will be close to $d C_1(r)/dr$, which, in turn, should be close
to the critical form $\propto r^{-2-\eta}$. The results of Ref.~\cite{nahum15a} clearly showed this to be the case for the AFM and VBS
correlation functions of the 3D loop model.

We here test the derivative method for the $J$-$Q$ models. We also fit the correlation function itself to a form including appropriate
scaling corrections. Both methods produce compatible values of the scaling dimension, $\Delta_\phi \approx 0.61$ from both spin and dimer
correlations, slightly smaller than most previous estimates---in particular the SO($5$) CFT input value $\Delta_\phi = 0.63$ in Table \ref{dtable}.

\subsubsection{Correlation functions and derivatives}
\label{sec:corandderiv}

We first examine the $r$ dependent spin and dimer correlation functions on relatively large systems, $L=384$ for the $J$-$Q_2$ model and $L=320$ for the
$J$-$Q_3$ and $J$-$Q_4$ models. Ground state results for $C_{s,d}(r)$ from PQMC simulations are shown in Fig.~\ref{sdcor}. For the $J$-$Q_2$ model the two
couplings that we previously determined as brackets of the transition point (Fig.~\ref{corL_jq2}) are used in Fig.~\ref{sdcor}(a). The results are
indistinguishable in the plot for short distances but slightly split off from each other for longer distances. The lines show the decay expected for a critical
system with the scaling dimension $\Delta_\phi=0.63$ assumed in the CFT bootstrap calculation \cite{chester24}, which is within the error bars of the values
obtained with the loop model \cite{nahum15a} and is also consistent with earlier results for the $J$-$Q$ model \cite{sandvik07,sandvik12}. The form indeed
describes reasonably well both the AFM and VBS data up to $r \approx 20$, though corrections are also clearly present. It is not possible
to determine a reliable exponent from these results from a fit to a single power law, primarily because of the crossover to a much slower decaying
form reflecting the weak long-range order, but also because of the short-distance corrections. The results overall look very similar to those for
the 3D loop model \cite{nahum15b}.

For the $J$-$Q_3$ model, in Fig.~\ref{sdcor}(b) we show results for the same near-critical coupling as in Fig.~\ref{corL_jq2}. Again the
short-distance behaviors, beyond the regime where corrections are apparent, are reasonably well described by a common exponent $\Delta_\phi = 0.63$.
This behavior applies up to somewhat shorter distances than for the $J$-$Q_2$ model, which again indicates a slightly more ordered coexistence state.
The same trend of diminishing range of the critical decay, due to more pronounced long-range order, is seen for the $J$-$Q_4$ model in
Fig.~\ref{sdcor}(c), where we show results for the two couplings bracketing the transition point according to Fig.~\ref{corL_jq4}.

Next we study the derivatives. To minimize discretization errors, we use the fourth-order approximant of the first $r$ derivative and define
$C'(r)$ with an overall minus sign in order to have a positive quantity;
\begin{eqnarray}
  C'(r) & = & \frac{2}{3} \left [ C(r-1)-C(r+1) \right ] \nonumber \\
   && ~~- \frac{1}{12} \left [ C(r-2)-C(r+2) \right ].
\label{cderdef}
\end{eqnarray}
Results for all three models and the same lattice sizes as above are shown in Fig.~\ref{sdder}. The derivatives are not as sensitive as
the correlation functions themselves to the exact value of the coupling in the neighborhood of the transition point. To reduce the
statistical errors further, we have therefore taken the average over the two couplings used in Fig.~\ref{sdcor}(a) and Fig.~\ref{sdcor}(c)
for the $J$-$Q_2$ and $J$-$Q_3$ model, respectively, which also corresponds to an interpolation with the two data sets to a value closer
to the transition point than the two individual values.

For all three models, we find a very substantial range of distances over which the decay is closely described by $\propto r^{-1-2\Delta_\phi}$,
with $\Delta_\phi \approx 0.63$. The behavior is very similar to that in the 3D loop model \cite{nahum15a}. It is remarkable that even the
$J$-$Q_4$ model exhibits this critical behavior up to distances $r \approx 100$, though with larger corrections than the $J$-$Q_2$ and
$J$-$Q_3$ models. This critical behavior is manifested even though the model has a clearly ordered coexistence state according to the
size dependent correlation functions in Fig.~\ref{corL_jq4}---with the AFM correlations extrapolating to a value about 10 times larger
than in the $J$-$Q_2$ model in Fig.~\ref{corL_jq2}. The good agreement between these results for all three $J$-$Q_n$ models, as well as the
previously studied 3D loop model \cite{nahum15a}, should leave no doubt that the behavior reflects universal scaling originating from a
nearby critical point.

We use the $J$-$Q_2$ model, the transition of which is the closest to the critical point, to extract the most reliable value of $\Delta_\phi$. Based on
just the results for $L=384$ in Fig.~\ref{sdder}(a), it appears that the data agree with the previous estimates of the scaling dimension,
$\Delta_\phi \approx 0.63$. However, the indicated scaling behavior is statistically good only over distance windows of a few tens of lattice
spacings, and the best exponent also depends to some extent on how the window is chosen. We therefore proceed to study the evolution of the derivatives
$C'_{s,d}(r)$ with increasing system size and also take corrections to the leading power law behavior into account in fits to $C_{s,d}(r)$.

As we have seen, the derivatives show much better scaling than the correlation functions themselves, for reasons that can be understood
based on Eq.~(\ref{c123}) when there is weak long-range order. We have observed a related favorable aspect of the PQMC simulation algorithm:
the derivatives at large $r$ converge significantly faster with the projection power than the correlations themselves.
The variant of the PQMC method used here (Sec.~\ref{sub:qmc}) samples contributions to the normalization of the ground state projected
out of a singlet state, $\langle \Psi|(-H)^{pN}|\Psi\rangle$, where $p$ typically has to be larger than the system length $L$ for complete
convergence to the ground state (we often use $p=2L$). In contrast, the derivatives of the spin correlations are well converged already for
$p\approx L/2$, which allows us to reach larger system sizes when targeting the derivative $C'(r)$, even though $C(r)$ is not yet fully converged
to the ground state. The derivative of the dimer correlation function appears to converge somewhat slower and is also noisier. We therefore
only analyze the derivative of the spin correlation function here, using system sizes up to $L=1024$.

\begin{figure}[t]
\includegraphics[width=75mm]{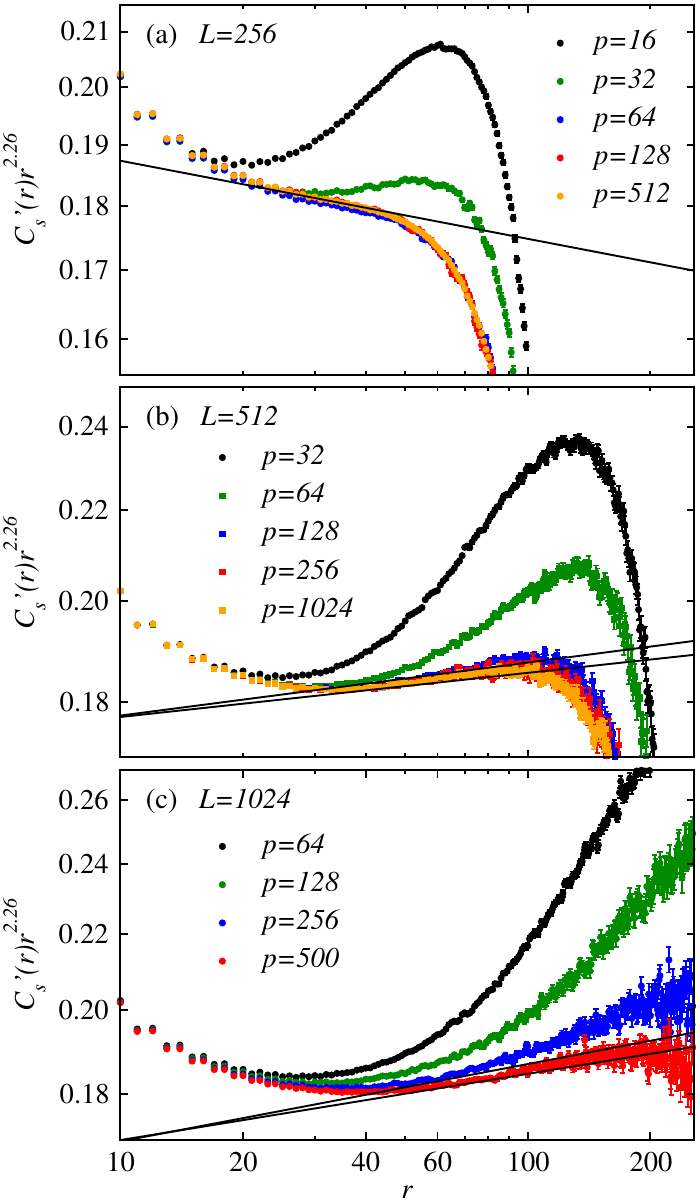}
\caption{Discrete derivative, Eq.~(\ref{cderdef}), of the spin correlation function in the near-critical $J$-$Q_2$ model on lattices of
size $L=256$ (a), $512$ (b), and $1024$ (c). The results were obtained by PQMC simulations with a total of $Np$ operators in the string, with
$p$ ranging from $L/16$ to $2L$ for $L=256$ and $L=512$, and up to $p=500 \approx L/2$ for $L=1024$. The results have been multiplied by $r^{2.26}$,
so that the previously presumed scaling dimension $\Delta_\phi=0.63$ would give an asymptotically flat behavior for a large critical system.
Lines are drawn in the regions of approximate power-law behavior to provide size dependent estimates of $\Delta_\phi$. In (a), the single line
has slope $-0.030$, corresponding to $\Delta_\phi=0.645$ for $L=256$. In (b) and (c), the two lines correspond to $\Delta_\phi \in [0.618,0.620]$
for $L=512$ and $\Delta_\phi \in [0.609,0.612]$ for $L=1024$.}
\label{d1024}
\end{figure}

The faster convergence of $C'(r)$ can likely be traced to the role of the long-range ordered component of $C(r)$. The value of the ordered moment,
which should most strongly affect the $r$-independent term $C_2(L)$ in Eq.~(\ref{c123}), converges relatively slowly, while the power-law contribution dominating
$C_1(r)$ at the transition converges faster. In Fig.~\ref{d1024} we show results for system sizes $L=256$, $512$, and $1024$, in each case with the projection
parameter $p=L/16$, $L/8$, $L/4$, and $L/2$. In the case of $L=1024$, the largest $p$ is slightly smaller, $p=500$ instead of $512$, because
this is the largest value of $p$ for which our implementation of the PQMC method works with four-byte integers; with an address space of
$2^{32}$ elements of a linked list of vertices (see Refs.~\cite{sandvik10b} and \cite{sandvik10c} for technical details on the
sampling procedures). For $L=256$ and $L=512$ we also include results for $p=2L$, where also $C(r)$ is fully converged. In Fig.~\ref{d1024} we
have multiplied the derivatives by $r^{2.26}$, which would give $r$ independent values (over some range of distances) if the scaling dimension
takes the previously assumed value $\Delta_\phi=0.63$.

For $L=256$ and $L=512$, already $p=L/4$ gives well converged results for $C'(r)$. For $1024$, there are still some differences between the
results for $p=256$ and $500$. For $L=512$, convergence is explicitly confirmed by the essentially perfect agreement between the results for
$p=L/2$ and $p=2L$. Given that reasonably good convergence is also seen for $L=512$ already with $p=L/4$ and the rather small differences
between $p=L/4$ and $p\approx L/2$ for $L=1024$, the latter calculation should also be well converged within error bars.

These results for three large system sizes allow us to refine the value of $\Delta_\phi$.
We have fitted power laws to a range of distances where such behavior appears to hold. Though the range of distances with approximative power-law
behavior is limited, it grows with increasing system size and an effective size dependent scaling dimension can be defined to within a
rather small level of uncertainty.

In Fig.~\ref{d1024}(a), we see that the effective scaling dimension is larger than $0.63$ for $L=256$, i.e., the power-law decay is faster than $r^{-2.26}$.
However, for both $L=512$ and $L=1024$ the decay for large distances is slightly slower, corresponding to $\Delta_\phi$ smaller than $0.63$. The decay
exponent adjusted by the multiplier $r^{2.26}$ is in the range $[2.236,2.240]$ for $L=512$ and $[2.218,2.224]$ for $L=1024$, suggesting a rather rapid
convergence for these larger systems and a scaling dimension $\Delta_\phi \approx 0.61$. However, it should also be noted that the window of apparent
power law behavior shifts to larger distances with increasing $L$, which shows that the corrections to the leading power law continue to play some role
and may still affect the effective exponent even for $L=1024$.

\subsubsection{Scaling corrections}
\label{sub:delphifits}

To further investigate the stability of the value of $\Delta_\phi$, we also perform an analysis including scaling corrections, using fully converged
$C(r)$ results for $L=256$, $384$, and $512$. In order to properly account for the enhancement of the slowly decaying correlations close to (and even quite
far away from) $r=L/2$, because of the periodic boundary conditions, we should not just fit to a sum of power laws but use a form that respects the necessary
condition $C(r)=C(L-r)$. For a quantum system in one spatial dimension, the proper CFT geometry is a periodic chain (a ring), where the conformal distance
for a system of length $L$ is given by the chord length
\begin{equation}
x = (L/2)\sin{(r\pi/ L)},
\label{confdist}
\end{equation}
for lattice distances $r=0,1,\ldots,L-1$. In two spatial dimensions, the proper conformal geometry is a sphere. On the torus, we still find it useful to define
the distance $x$ as above with $r$ on the line $(r,0)$, which guarantees reflection symmetry about $r=L/2$ of whatever function is fitted for $x \in [0,L/2]$.
In the context of our 2D system, we will refer to $x$ as the proxy conformal distance.

The dominant corrections to the asymptotic critical form $C(x) \sim x^{-2\Delta_\phi}$ would normally arise from conventional corrections, with the leading one
often written as a factor $x^{-\omega}$ with $\omega \equiv |y_i|$ the smallest of all $y_i=3-\Delta_i$ values with $\Delta_i>3$ for irrelevant perturbations. There will
also be contributions from descendants of the $\phi$ operator. Because of lattice symmetry, only those with Lorentz spin $\ell=0$ mod $4$ (with scaling dimensions
$\Delta_\phi +2, \Delta_\phi +4,\ldots$) are present and they produce power-law contributions by themselves and from cross-terms with the primary. Thus, contributions
$x^{-2\Delta_\phi}$, $x^{-(2\Delta_\phi+2)}$, $x^{-(2\Delta_\phi+4)}$ etc.~are expected and the correction $x^{-\omega}$ (and higher-order ones with larger $\omega$ values)
multiply all these collectively. Moreover, we have to account for the fact that there is also weak long-range order, which we do by adding a constant $m^2$.

Fitting the spin and dimer correlations, we find that the form $m^2+ax^{-2\Delta_\phi}+bx^{-(2\Delta_\phi+2)}$ with $\Delta_\phi \approx 0.61$ produces reasonable fits
to the data except for the shortest distances, and by including also $cx^{-(2\Delta_\phi+4)}$ several shorter distances can also be included. The leading irrelevant
correction (i.e., arising from an intrinsic perturbation in the lattice model) typically has $\omega \approx 1$, e.g., in classical O($N$) models
\cite{hasenbusch10,poland19} but such corrections appear to be missing here. It is possible that $\omega \approx 2$ and therefore $x^{-(2\Delta_\phi+\omega)}$
cannot be distinguished in fits from the expected term $x^{-(2\Delta_\phi+2)}$. In the fuzzy sphere calculation \cite{zhou23} the only identified symmetry-allowed
(in the square-lattice $J$-$Q$ models) perturbations with scaling dimension less than $\approx 5$ (i.e., corresponding to $\omega \alt 2$) are those also studied
here in Sec.~\ref{sec:symm}; the SO($5$) violating operators with scaling dimensions $\Delta_4 \approx 3.7$ and $\Delta_{t'} \approx 4.7$. The former
corresponds to $\omega \approx 0.7$, but the fact that we do not observe any such correction implies that this perturbation must have a very weak effect on the
correlation functions. The larger dimension $\Delta_{t'}$ corresponds to $\omega \approx 1.7$, which is likely close enough to $2$ to not be detectable in the presence of
other similar contributions. Overall, the fact that $\Delta_\phi$ extracted from both spin and dimer correlations are the same within small error bars (quantified
below) suggests that intrinsic SO($5$) violations are not important here (though clearly detectable in finite-size scaling of specific quantities in Sec.~\ref{sec:symm}).
The leading correction should then indeed arise from the $x^{-(2\Delta_\phi+2)}$ mixed primary and descendant contribution. Fitting to
$m^2+ax^{-2\Delta_\phi}+bx^{-(2\Delta_\phi+\omega)}$, the correction exponent $\omega$ indeed comes out close to $2$ and when fixing $\omega$ to different
values close to $2$ there is no significant change in the resulting $\Delta_\phi$.

\begin{figure*}[t]
\includegraphics[width=160mm]{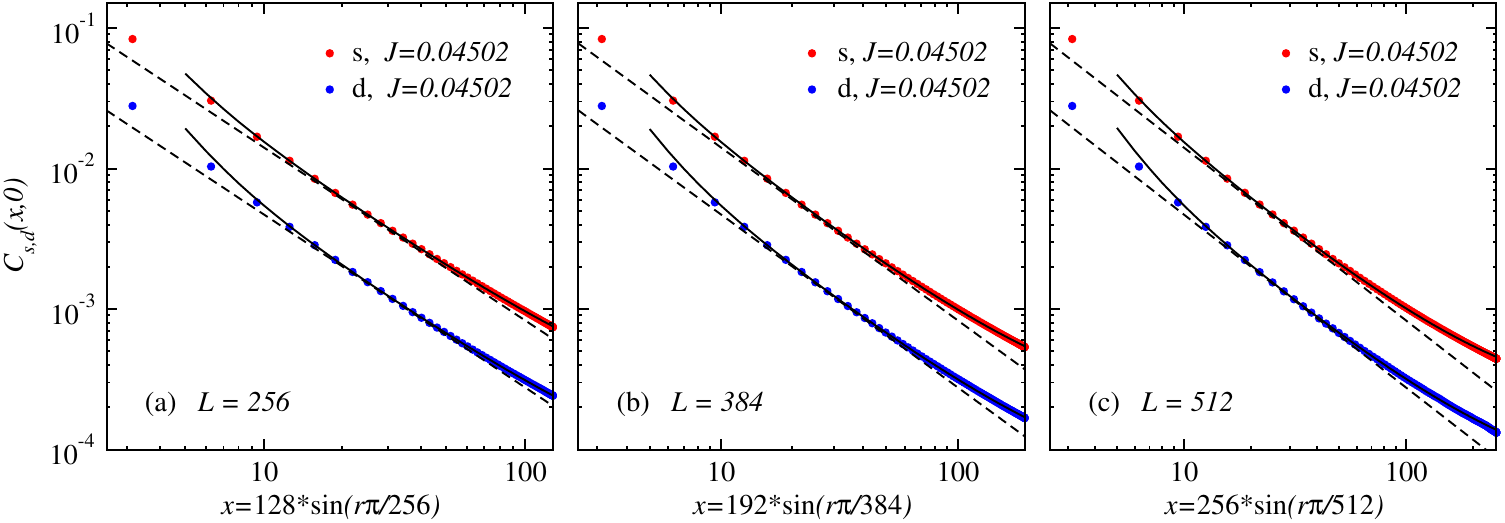}
\caption{Spin and dimer correlation functions graphed versus the proxy conformal distance, Eq.~(\ref{confdist}), for three different system sizes;
(a) $L=256$, (b) $L=384$, and (c) $L=512$ at the estimated transition point. The solid curves (which for large $x$ are hard to discern because of their close
match with the data points) are fits to Eq.~(\ref{cxfitform}) with $\Delta_\phi=0.61$ and $c=0$ (i.e., excluding the second correction term)
in all cases. The dashed lines indicate the slope corresponding to the leading term $\propto x^{-2\Delta_\phi}$.}
\label{corfit}
\end{figure*}

While remaining agnostic as to the actual source of the corrections to the leading power law, given the above we here discuss results obtained with the
fitting form
\begin{equation}
C(x)=m^2 + \frac{a}{x^{2\Delta_\phi}} + \frac{b}{x^{2\Delta_\phi+2}} + \frac{c}{x^{2\Delta_\phi+4}},
\label{cxfitform}
\end{equation}
without any explicit further correction in the form of a factor $(1+dx^{-\omega})$. This form still cannot be completely correct, not just
because of neglected higher-order terms but because additional terms beyond $m^2$ should be expected in the ordered coexistence state. Given that the system
at the transition point has emergent SO($5$) symmetry, we should expect contributions to the correlation functions from the four associated Goldstone modes of
the weakly ordered coexistence state (which is not symmetry-broken on the finite lattices we study), up to some large system size beyond which the SO($5$) symmetry
is violated due to perturbations that are irrelevant at the critical point but relevant in the ordered coexistence state (often referred to as dangerously irrelevant
operators). Even after this crossover, the AFM component of the coexistence state has exact SO($3$) symmetry and two associated Goldstone modes. The leading Goldstone
contributions decay as $r^{-1}$, i.e., {\it slower} than the critical correlations that we have assumed to be dominant above. The Goldstone contributions arise from
the transverse fluctuations of the order parameter and its amplitude should be of the order of the stiffness constant (helicity modulus) of the SO($5$) or SO($3$)
order parameter.

We analyze the spin stiffness $\rho_s$ at the transition point in Sec.~\ref{sec:fsexample} and find a very small value in the thermodynamic limit; as expected
given the small ordered moment of the coexistence state. We therefore assume that the Goldstone contribution will not significantly impact the fitted value of
$\Delta_\phi$ as long as we restrict the fit to relatively short distances, thus largely avoiding the complications arising from the crossover into the expected
complicated large-$r$ behavior. The constant $m^2$ in Eq.~(\ref{cxfitform}) is necessary in order to obtain good fits even with only short-distance data included;
it should be interpreted as a way to approximately include the effects of the long-range order at short to intermediate distances, as was also argued in
Ref.~\cite{wang21}. We have nevertheless also explored fits with an additional $x^{-1}$ Goldstone term, which we will further comment on below.

We here use interpolated data at our best estimate of the transition point, $(J/Q_2)_c=0.04502(1)$, which again falls between the ratios $0.04500$ and
$0.04505$ that bracket the transition based on the correlation functions above in Sec.~\ref{sec:corandderiv} but is obtained from extrapolating cumulant
crossings in Sec.~\ref{jqcscaling}. We only consider even distances $r$ to avoid small even-odd effects seen in Fig.~\ref{sdcor}.

We first discuss fits including also the longest distances. With only the leading correction and constant included, i.e., $c=0$ in Eq.~(\ref{cxfitform}), we can
obtain visually good fits to Eq.~(\ref{cxfitform}) for all $r\ge 10$ ($x\agt 15$), as shown in Fig.~\ref{corfit} for both spin and dimer correlations and system
sizes $L=256$, $384$, and $512$. The proxy conformal distance indeed works very well, suggesting that Eq.~(\ref{confdist}) should be close to an ideal CFT distance
definition also in two spatial dimensions. All of these fits deliver values of the scaling dimension $\Delta_\phi$ close to $0.61$. Deviations from
Eq.~(\ref{cxfitform}) when this function is extended below the shortest distance included in the fitting procedure are obvious, but all the data points can be
fitted well once the second correction, with exponent $2\Delta_\phi+4$ in Eq.~(\ref{cxfitform}), is included (fits not shown here). The value of $\Delta_\phi$
changes only marginally when the second correction is included.

\begin{figure}[t]
\includegraphics[width=80mm]{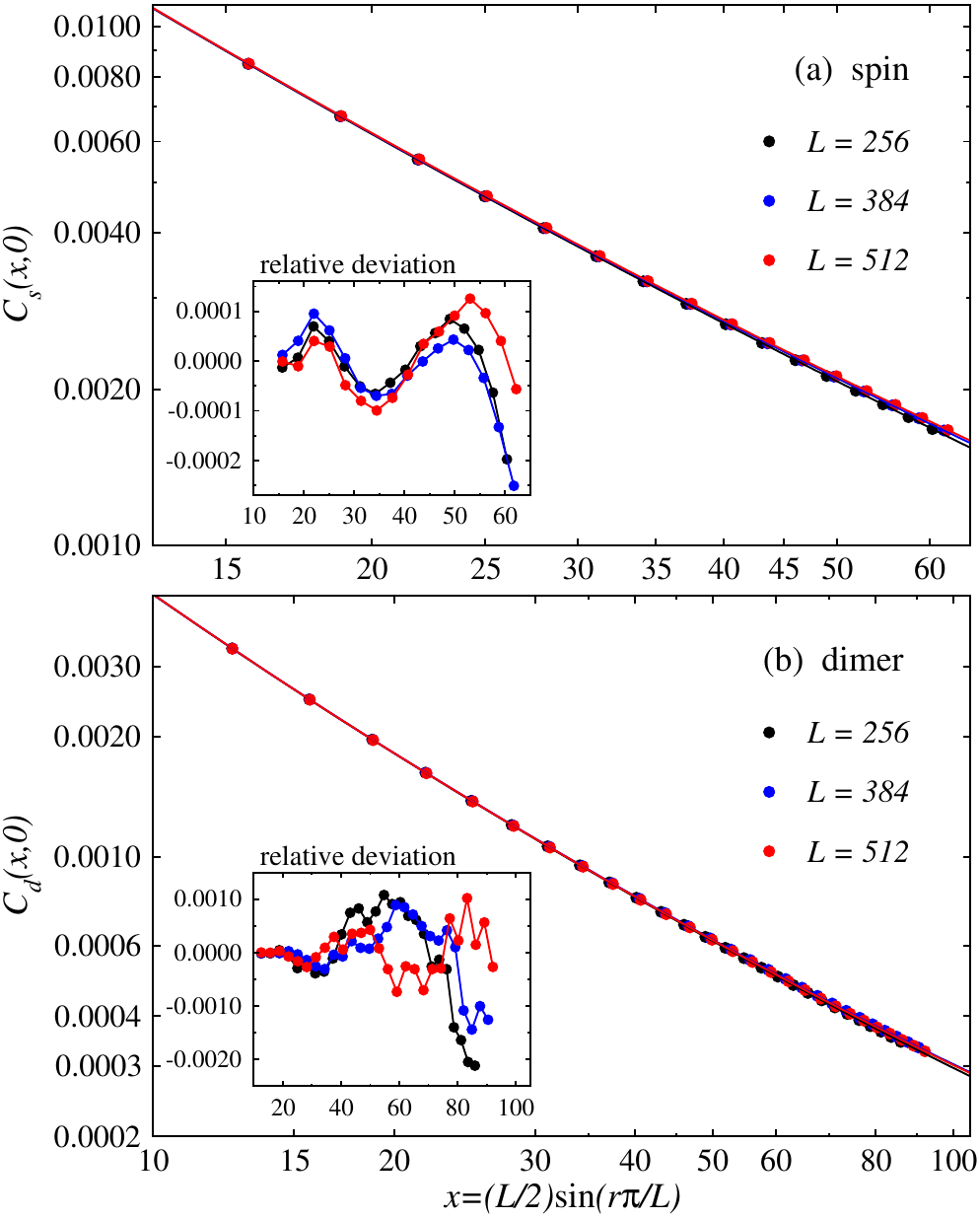}
\caption{Spin (a) and dimer (b) correlation function vs the proxy conformal distance $x$ for the same system sizes as in Fig.~\ref{corfit} but for
  limited ranges of distances where the full fitting function Eq.~(\ref{cxfitform}) can reproduce the QMC data very closely. In (a) the spin correlations are
  shown for $r \in [10,40]$, where fits to the Eq.~(\ref{cxfitform}) have relative deviations $(C_{\rm QMC}-C_{\rm fit})/C_{\rm fit}$ from the data
  points of order $\approx 10^{-4}$ or less, as shown in the inset. The best-fit scaling dimensions are $\Delta_\phi = 0.601$ for $L=256$, $0.605$ for
  $L=384$, and $0.606$ for $L=512$. In (b), the dimer correlations are shown for $r \in [10,60]$, where the relative deviations from the fitted
  functions (inset) are typically $10^{-3}$ or less. The scaling dimensions here are $\Delta_\phi = 0.592$ for $L=256$, $0.606$ for $L=384$, and $0.607$
  for $L=512$.}
\label{corfit2}
\end{figure}

While the fits in Fig.~\ref{corfit} look good for $x$ in the fitting range,
they are actually of very poor statistical quality. Because data at different distances are highly
correlated, it is necessary to compute the full covariance matrix and use it in the definition of the goodness of the fit $\chi^2$. The dominant source
of covariance---the eigenvector corresponding to the largest eigenvalue---is a common, almost uniform fluctuation. The eigenvalue of this mode is
typically (for large systems) more than an order of magnitude larger than the second largest eigenvalue. Therefore, even a statistically substandard fit can
look visually good, with the fitted function falling very close to the center of the data points, as in Fig.~\ref{corfit}, even though there are
statistically significant deviations when covariance is properly taken into account.

Henceforth we include data covariance and assure statistical soundness by monitoring the reduced-$\chi^2$ value defined with the eigenvalues of the
covariance matrix (in many cases using only the largest few eigenvalues, corresponding to clearly non-singular directions).
In Fig.~\ref{corfit2} we use smaller ranges of distances in order to eliminate distortions by potential shortcomings of the functional form Eq.~(\ref{cxfitform})
at both short and long distances. Here we also include the second correction, and in all cases we still obtain $\Delta_\phi \approx 0.61$. However, in the
case of the spin correlations, where we use the range of lattice distances $r \in [10,40]$ in Fig.~\ref{corfit2}(a), the statistical quality of the fits is
still not good, having reduced-$\chi^2$ values close to $50$ for $L=256$ and $384$ and about $8$ for $L=512$, despite the fact that the relative deviation
of the fitted function from the data points is typically less than $10^{-4}$, as shown in the inset of Fig.~\ref{corfit2}(a). These remarkably small deviations
are an order of magnitude below the naive statistical errors of the data points, defined using only the diagonal elements of the covariance matrix. The
deviations versus $x$ show very similar patterns for all three system sizes, demonstrating that they are not purely statistical in nature but arise from
additional corrections not taken into account by Eq.~(\ref{cxfitform}). Slightly changing the second correction from the form $x^{-(2\Delta_\phi+2)}$ to
$x^{-(2\Delta_\phi+\omega)}$ with $\omega \approx 2$ does not alter the overall shape of the relative deviations versus $r$. Including an overall multiplicative
correction $(1+dx^{-\omega})$ improves the quality of the fit somewhat, but at the same time makes the target $\Delta_\phi$ value much more uncertain.
In an absolute sense, the deviations from the fit without the further correction are very small and cannot appreciably distort the extracted value of $\Delta_\phi$,
but they still cause formally poor $\chi^2$ values because of the very small eigenvalues of the covariance matrix (beyond the largest eigenvalue corresponding
to the unison fluctuation). Below, we will still achieve full statistical soundness by further restricting the range of $r$ in the fit.

The dimer correlations, shown in Fig.~\ref{corfit2}(b) for distances $r \in [10,60]$, have larger overall statistical errors and show less effects of
neglected higher-order corrections. The $\chi^2$ values are good for all system sizes. While the deviations of the data from the fit have some similarities
for small $x$, and also for larger $x$ in the case of $L=256$ and $L=384$, they look quite different for $L=512$ at larger $x$, where the statistical errors
are larger and the systematic deviations are therefore likely masked by the noise. For a given $L$, the deviations are still not independent for different $x$
values, because of the covariance effects. Repeating the fitting procedure many times with Gaussian noise added to the data transformed to the eigenbasis of the
covariance matrix (with the magnitude of the noise set by each eigenvalue), we can also obtain statistical errors (one standard deviation) on the scaling
dimension. The final results from the dimer correlations with $r \in [10,60]$ are $\Delta_\phi=0.592(2)$ for $L=256$, $0.606(3)$ for $L=384$, and $0.606(5)$
for $L=512$.

In the case of the spin correlations, to achieve statistically good fits and compute meaningful error bars, we further reduce the range of distances to
$r \in [14,28]$, i.e., using only eight data points in the fit to the five-parameter form Eq.~(\ref{cxfitform}). While the number of degrees of freedom is
now very small (3), the fits are statistically sound and deliver the following values of the scaling dimension: $\Delta_\phi = 0.601(2)$ for $L=256$,
$0.605(2)$ for $L=384$, and $0.606(5)$ for $L=512$, all of which are consistent with the results obtained for the larger range $r\in [10,40]$ in
Fig.~\ref{corfit}. Overall, the statistical errors of $\Delta_\phi$ are very similar to those above from the dimer correlations. Though the raw dimer
correlations have significantly larger statistical errors, these errors are compensated by the fact that a wider range of points can be included in
fits with acceptable $\chi^2$ values. In fact, the range of the dimer correlations can be further extended and still allow good fits with only marginal
changes in $\Delta_\phi$, but to stay on the safe side in light of potential Goldstone effects we restrict the analysis to $r \in [12,60]$
in Fig.~\ref{corfit2}(b).

The four scaling dimensions extracted from the spin and dimer correlations for $L=384$ and $512$ are mutually consistent, while they are slightly smaller
for $L=256$. The size dependence is much weaker than what is observed in the derivative of the spin correlation function versus the original lattice
distance $r$ in Fig.~\ref{d1024}. The $\Delta_\phi$ value estimated from the $L=1024$ derivative in Fig.~\ref{d1024}(c) is fully consistent with the
results from the correlation functions at shorter distances for $L=384$ and $512$.

If a Goldstone contribution $ex^{-1}$ is added to the fitted function, the amplitude $e$ becomes unrealistically large (given the small ordered moment and associated
$\rho_s$ value) and the scaling dimension $\Delta_\phi$ becomes much larger, above $2$, to compensate
for the large $x^{-1}$ contribution. With our extracted value $\Delta_\phi \approx 0.61$, the forms $x^{-1}$ and $x^{-2\Delta_\phi}$ are very close to each other,
and it is not surprising that they cannot be separated in an unconstrained fit. We can also add the $e/x$ term while keeping $\Delta_\phi\approx 0.61$ and
fitting only at long distances. Then the amplitude $a$ of $x^{-2\Delta_\phi}$ becomes smaller to compensate for a rather large $e$ (with $e/a \approx 0.5$ in
the case of the spin correlations). However, this kind of fit may not be meaningful, because the $1/x$ term is only one aspect of the Goldstone effects.
Higher powers of $1/x$ are also expected \cite{reger88,sandvik10c} and the crossover between the critical form and the asymptotic form in the presence
of long-range order is likely complicated and beyond quantitative analysis with the present data and understanding. Fitting to a single power law plus a
constant, $C(x) = ax^{-1-\eta} + m^2$, for large systems we find that $\eta$ approaches $0$ from above when the fitting window is moved to larger $x$ values,
as also found previously for the loop model \cite{nahum15a}. We find similar behavior with $x=L/2$, and for large $L$ such data can also be fitted to a
polynomial with leading $1/L$ behavior, as in Fig.~\ref{corL_jq2}.

\begin{figure}[t]
\includegraphics[width=80mm]{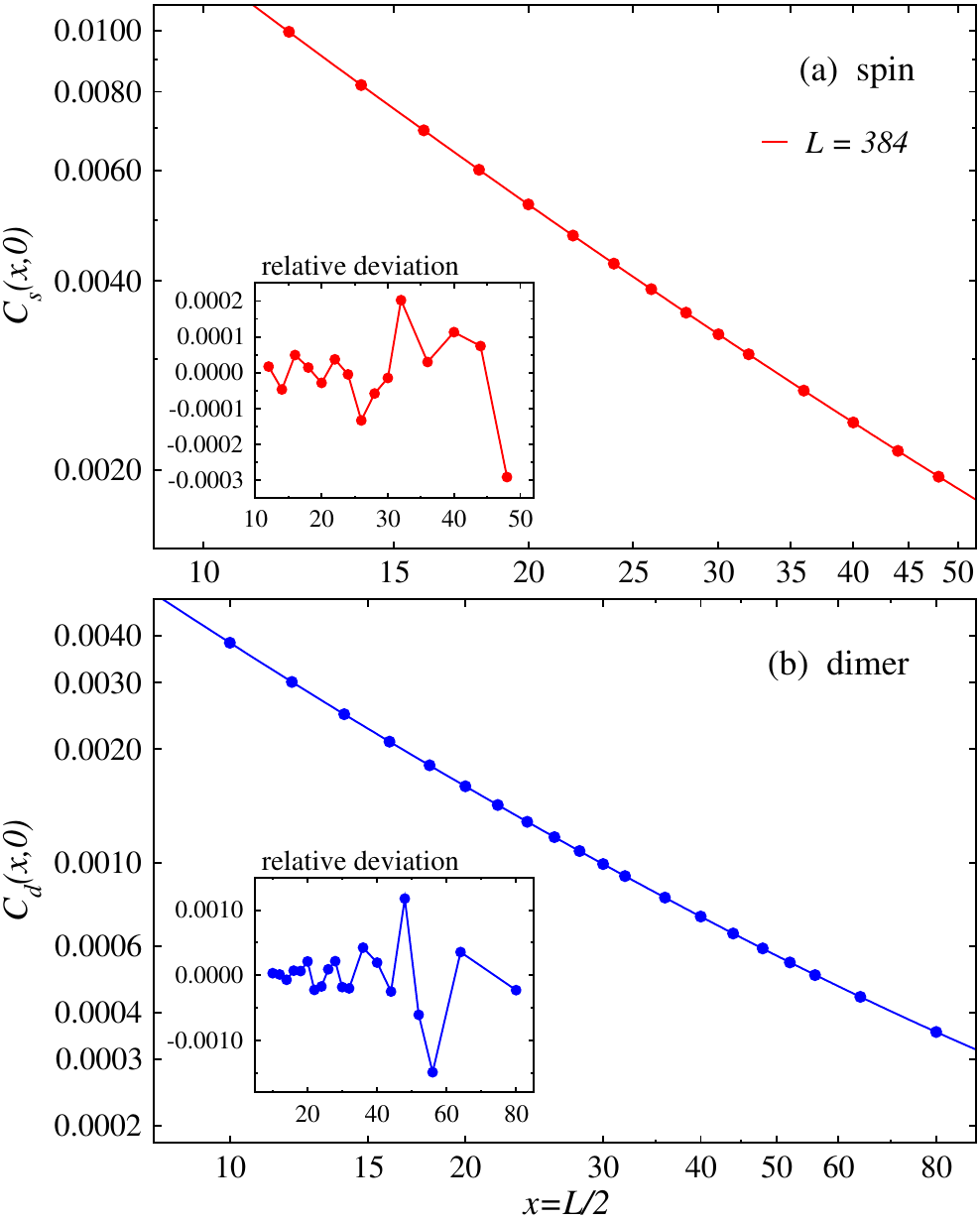}
\caption{Spin (a) and dimer (b) correlation functions vs distance $r=L/2$ along with fits to Eq.~(\ref{cxfitform}) with $x=L/2$. The ranges of data correspond
to windows of $L$ in which statistically good fits are obtained. The relative deviations from the fitted form, shown in the insets, are of the order of the
statistical errors. The scaling dimensions extracted from these fits are $\Delta_\phi = 0.606(3)$ in (a) and $0.613(3)$ in (b).}
\label{corfitL}
\end{figure}

The restriction of $x$ to rather short distances in our final analysis should ensure negligible impact of the Goldstone modes and crossover effects, and
this conclusion is also supported by the derivatives in Figs.~\ref{d1024}(a) and \ref{d1024}(b) for larger distances. In further support of not including
the $1/x$ term, in Fig.~\ref{corL_jq4} there are no significant changes in the approximate power-law behaviors for the three different models up to
$r \approx 100$, even though the $J$-$Q_4$ model has much larger ordered moments. Thus, we conclude that the expected Goldstone contributions must only play a
significant role in the correlation functions of the $J$-$Q_2$ model for distances beyond those used in Figs.~\ref{corfit} and \ref{corfit2}.

To complete the analysis, we next study the correlation functions at distance $r=L/2$ for a range of moderate system sizes $L$, where the critical correlations
dominate. In this case, we may also expect multiplicative corrections of the form $L^{-|y|}$ from the allowed irrelevant operators with scaling dimension $\Delta>3$
and $y=3-\Delta$. Again, we do not find any significant effects of leaving these out, obtaining good fits to Eq.~(\ref{cxfitform}) with $x=L/2$. The value of the
scaling dimension $\Delta_\phi$ is also close to the results of
the previous fits for $r \ll L$, where the irrelevant scaling corrections should be absent for large $L$ (which is supported by the very small differences
between results for $L=256$, $384$, and $512$). In this case the data for different $L$ are of course statistically independent and $\chi^2$ is defined with only
the individual standard deviations. In Fig.~\ref{corfitL} we show results for spin and dimer correlations obtained with windows of $r=L/2$ where good
$\chi^2$ values can be obtained. The deviations from the fitted function in this case are naturally of the order of the error bars of the data points
(on average about one standard deviation, corresponding to a statistically good $\chi^2$ value). The resulting scaling dimensions are $\Delta_\phi = 0.606(3)$
and $0.613(3)$ from spin and dimer correlations, respectively.

\begin{figure*}[t]
\includegraphics[width=150mm]{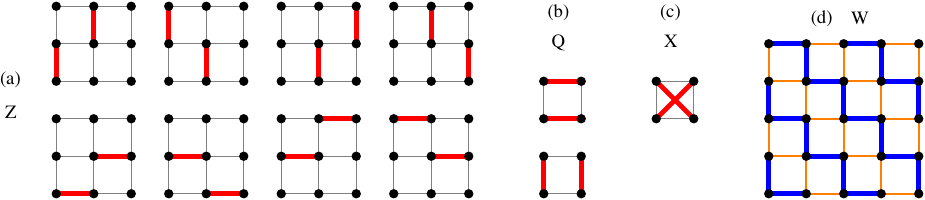}
\caption{Operators used in correlation functions for extracting relevant scaling dimensions \cite{zhao20}. In (a), (b), and (c), red
thick bars represent bond operators ${\bf S}_i\cdot {\bf S}_j$ on nearest-neighbor sites $i$ and $j$. The $Z$ and $Q$ operators are defined as sums of
products of two of these bond operators within $3\times 3$ (a) and $2\times 2$ (b) spin cells. The $X$ operator in (c) is a single product of the bond
operators on the diagonals of a plaquette. In (d), the W operator is defined in a cell of $5\times 5$ spins, with thick blue and thin orange lines
representing individual nearest-neighbor operators $\pm {\bf S}_i\cdot {\bf S}_j$, which are summed over the $5\times 5$ cell.}
\label{zqwops}
\end{figure*}

The suitability of the fitting form Eq.~(\ref{cxfitform}) is clearly supported by the stability of our results for $\Delta_\phi$ with respect to different system
sizes and fitting windows, and also by the good agreement between results obtained with $r=L/2$ and $r \ll L$. The fact that the results obtained from
spin and dimer correlations are statistically indistinguishable at the level below $1\%$ also lends further credibility to the SO($5$) scenario, according
to which $\Delta_\phi$ must be the common scaling dimension for the AFM and VBS order parameters. For the final common value reported in Table \ref{dtable},
we take the average of the spin and dimer results (for which the mutual correlation is rather weak; in fact the amplitudes of the raw correlation functions
are somewhat anti-correlated) for $L=512$; $\Delta_\phi = 0.607(4)$. This value is also statistically consistent with the two values from the $r=L/2$ fits and
the derivatives for $L=512$ and $L=1024$ in Fig.~\ref{d1024}. The difference between our final value and the input value $\Delta_\phi=0.63$ used in the CFT
bootstrap calculation \cite{chester24} could very plausibly account for the minor discrepancies between the other exponents seen in Table \ref{dtable}
\cite{chester24b}. It is also interesting to note that our $\Delta_\phi$ value is reasonably close to that obtained in the fuzzy sphere calculation,
where this scaling dimension is rather stable as a function of the number of orbitals \cite{zhou23}.

\section{Relevant perturbations}
\label{sec:scaledims}

Here we present results for correlation functions of operators that deliver the scaling dimensions $\Delta_t$ and $\Delta_j$ in Table \ref{dtable}.
In addition, we will present results for the ``pseudo dimension'' $\Delta_*$, which we believe is not a true level in the CFT spectrum but presumably
is a manifestation of the perturbed CFT slightly inside the first-order line. In principle, we should also be able to extract the singlet dimension
$\Delta_s$ from the same correlation function that has the leading decay exponent $2\Delta_t$ (i.e., assuming $\Delta_s > \Delta_t$). However, it is
difficult to separate its contribution from other potential corrections with similar exponents. The value of $\Delta_s$ in Table \ref{dtable} will instead
be extracted from the scaling of an SO($5$) indicator giving the exponent $\nu_*$ in Sec.~\ref{sec:symm} together with its predicted pseudocritical relationship
to $\Delta_s$ in Eq.~(\ref{nustarform}) derived in Sec.~\ref{sub:derivedeltastar}. The scaling behavior of the coexisting order parameters on the first-order
line gives a consistent but less precise value (standard deviation of a few percent) in Sec.~\ref{sec:jq2q6}.

\subsection{Operators and correlation functions}

Given a critical Hamiltonian $H_c$, we are interested in the stability of the fixed point to perturbations.
We here denote a generic extensive perturbing operator by
\begin{equation}
V = \sum_r \epsilon_rV_r,
\label{vopdef}
\end{equation}
where $V_r$ are local operators on individual spins or groups of spins centered at lattice position $r$,
and $\epsilon_r$ are phases that typically would be all unity or correspond to some regular
pattern or ``wave''. Upon adding $V$ to $H_c$ at small strength $\lambda$, $H \to H_c + \lambda V$, the system can either remain critical
or flow (here with increasing system size) to some other fixed point. Whether or not the perturbation is irrelevant or relevant in this sense
depends on its scaling dimension $\Delta_V$, or, equivalently, the corresponding exponent (eigenvalue in an RG transformation)
\begin{equation}
y_V=d+z-\Delta_V = 3 - \Delta_V
\label{yexpdef}
\end{equation}
that controls the flow of the strength of the perturbation with the length scale.

In finite-size scaling, the length scale is the system size $L$ and the scaling function (of some quantity exhibiting critical scaling when
$\lambda=0$) contains an argument $\lambda L^{y_V}$, which can be interpreted as the renormalized (coarse grained) strength of the perturbation
at length scale $L$ \cite{fisher72,binder81a,barber83}. Thus, the perturbation is relevant (the argument grows with $L$) if $y_V > 0$ and irrelevant
(the argument approaches zero in the thermodynamic limit) if $y_V < 0$. The length scale corresponding to a relevant perturbation diverges upon
approaching the critical point as $\xi_V \sim \lambda^{-\nu_V}$, with $\nu_V=y_V^{-1}$.

The most direct way to numerically extract a scaling dimension $\Delta_V$ in the absence of the perturbation ($\lambda=0$) is to compute the two-point connected
correlation function
\begin{equation}\label{cvconnected}
C_V(r)=\langle V_0V_r\rangle - \langle V_0\rangle^2,
\end{equation}
of the local operator $V_r$. The asymptotic power-law decay $C(r) \sim r^{-2\Delta_V}$ is governed by the (leading)
scaling dimension, as previously stated for the order parameters in Eq.~(\ref{cretaform}). Here the correlation function should be considered in
combination with the phase factors $\epsilon_r$ in Eq.~(\ref{vopdef}), and what matters in the relationship to $y_V$ in
Eq.~(\ref{yexpdef}) is $C(r)\epsilon_r$ integrated over space (and imaginary time, which when $z=1$ just gives one more factor of $L$). When considering
spatially uniform perturbations, i.e., $\epsilon_r =1 ~\forall ~r$, we are targeting the uniform component of the corresponding correlation function.
The previously studied critical order parameters are associated with phases taken into account in Eq.~(\ref{cafmdef}).

In this section we will study correlation functions of the operators illustrated in Fig.~\ref{zqwops}. The $Z$, $X$, and $W$ operators were previously
studied in Ref.~\cite{zhao20} and the $Q$ operator was studied in Ref.~\cite{sandvik20} (with a slightly different definition in terms of
singlet projectors instead of just ${\bf S}_i \cdot {\bf S}_j$). Here we provide more extensive analysis with improved PQMC data and discuss the results
in light of the scaling dimensions recently obtained within the CFT bootstrap calculation \cite{chester24} and the fuzzy sphere \cite{zhou23}.

A given lattice operator will in general contain all the CFT operators consistent with its symmetries. For a system with a CFT description, the correlation
function should therefore be an infinite sum of algebraic decays $a_{ij}r^{-(\Delta_i+\Delta_j)}$ (and possibly phases that we leave out here), with the
amplitudes $a_{ij}$ corresponding to the particular lattice operators chosen. The primary CFT operators are orthogonal, and two primaries $a$ and $b$ will
therefore only produce contributions $r^{-2\Delta_a}$ and $r^{-2\Delta_b}$. There are also no cross terms from operators in different conformal towers.
In general one would expect the decay corresponding to the smallest CFT scaling dimension to be clearly observable and in some cases it may be possible to
also extract additional larger scaling dimensions from the corrections to the leading form. In this section we will focus on relevant operators, leaving the
case of irrelevant perturbations of the SO($5$) symmetry to Sec.~\ref{sec:symm}, where we will use linear response techniques not relying on correlation
functions.

The models we study cannot be brought exactly to the critical point, and we observed large effects of the long-range ordered coexistence state in
Sec.~\ref{sub:corr}. We were able to reduce these effects by taking derivatives. However, we will find here that connected correlation functions of
operators not directly related to the order parameter are much less impacted by the ultimately first-order transition, which is natural since they do not
approach nonzero constants at large distances. Because of the overall faster decays, there are also less distorting effects of periodic boundary
conditions.

\subsection{Fully symmetric operator}
\label{sub:symop}

In the multicritical point scenario, the two tuning parameters required to reach the critical point of a lattice model must both control operators with
the same microscopic symmetries as the Hamiltonian; no explicitly symmetry-breaking operators should be introduced. The previously studied correlation
function of the $Z$ cell operator in Fig.~\ref{zqwops}(a) exhibited a leading decay $C_Z \sim r^{-2\Delta_Z}$ with $\Delta_Z \approx 1.40$ \cite{zhao20}
that we can now identify as the scaling dimension of the CFT $t$ operator; $\Delta_Z\equiv\Delta_t$.

All previous works had assumed that standard analysis of Binder cumulants or data collapse of the order parameters would deliver the exponent $\nu=\nu_t$
corresponding to the single correlation length expected within the original DQCP scenario. As already summarized in Sec.~\ref{sec:lattmodels},
large finite-size drifts had been observed in this exponent in many different works, but the two most extensive studies, of the $J$-$Q$ model \cite{sandvik20}
and the 3D classical loop model \cite{nahum15b}, both achieved apparent convergence toward $\nu \approx 0.45$, or $\Delta =3-1/\nu \approx 0.80$. A power
law decay consistent with this value of $\Delta$ was further obtained \cite{sandvik20} from the correlation function of an interaction energy (similar to
that of the $Q$ operator studied below in Sec.~\ref{sub:darkop}). It was therefore natural to assume that $\Delta_Q \approx 0.80$ and $\Delta_Z \approx 1.40$
reflected a CFT with two relevant symmetric operators, and, under the assumption of emergent SO($5$) symmetry, that $\Delta_Q$ and $\Delta_Z$ should be
identified as $\Delta_t$ and $\Delta_s$, respectively. However, in light of the CFT bootstrap calculation \cite{chester24} (and in hindsight also
Ref.~\cite{sreejith19}), it now seems more plausible that $\Delta_Z \approx 1.40$ is actually $\Delta_t$, while the anomalous scaling dimension extracted
from $C_Q$ (which also contains the VBS order parameter) is the rather mysterious pseudoscaling dimension (a manifestation of the weak first-order
transition) that we call $\Delta_*$ and for which we obtain an improved value in Sec.~\ref{sub:darkop}.

Given the very large value $\Delta_s \approx 2.36$ of the relevant SO($5$) singlet in the CFT \cite{chester24} and our slightly smaller value in
Table \ref{dtable}, it is not surprising that it could have been missed in previous lattice calculations. If indeed $Z$ also contains this CFT
operator, it should produce a $r^{-2\Delta_s}$ correction to the slower $r^{-2\Delta_t}$ decay of the Z-Z correlation function. However, we also expect
a conventional correction from irrelevant fields present in the model, of the form $r^{-(2\Delta_t+\omega)}$, where $\omega \approx 2$ was found in
Sec.~\ref{sub:delphifits}. If this correction is present also in the Z-Z correlations, the decay exponent $2\Delta_t+\omega \approx 4.8$ is very close
to the independent estimate of the exponent $2\Delta_s \approx 4.6$ from the relevant singlet obtained in Sec.~\ref{sub:so5}. It will unfortunately
then not be possible to separate these two corrections by fitting $C_Z$.

The value of the exponent that we call $2\Delta_t+\omega$ with $\omega \approx 2$ could also be $\Delta_t+\Delta'_t=2\Delta_t+2$, with $\Delta'_t$ being the
lowest $\ell=0$ descendant of $\Delta_t$ (i.e., its second descendant),
thus identical to the exponent with correction $\omega=2$. The second decay exponent that we extract here could
in actuality be an effective exponent approximating contributions from three different sources (from the descendant $\Delta'_t$, $\Delta_s$, or
$\omega \approx 2$). The fact that this effective exponent is much larger than $2\Delta_t$ nevertheless allows us to obtain a reliable value of $\Delta_t$.
Analyzing both $J$-$Q_2$ and $J$-$Q_3$ data, we reinforce the universal value $\Delta_t \approx 1.42$. Since the boundary enhancements of the $Z$ correlation
function are not significant, and also because we focus our quantitative analysis on $r=L/2$, there is no need to use the proxy conformal distance
Eq.~(\ref{confdist}) here.

\begin{figure}[t]
\includegraphics[width=80mm]{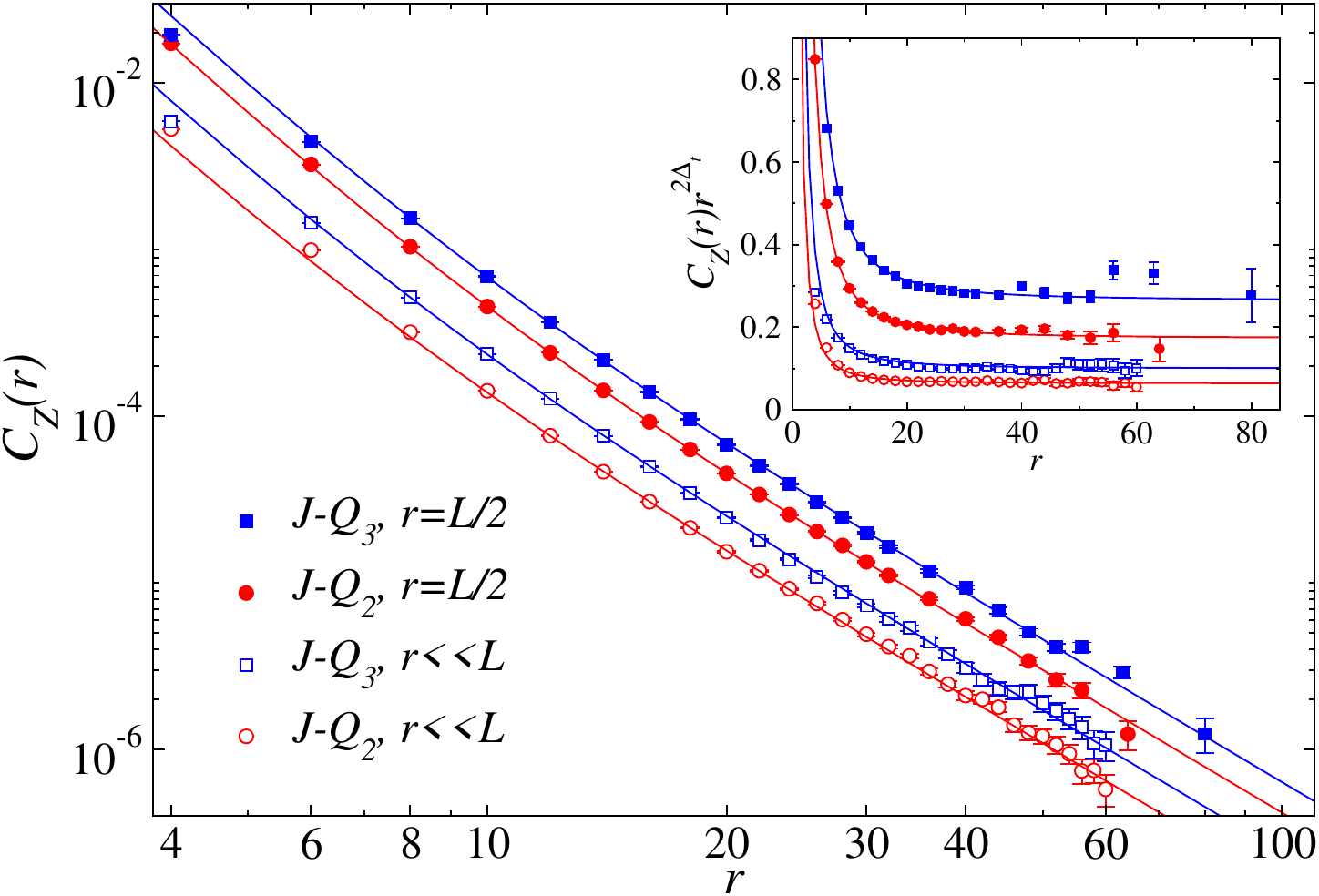}
\caption{Correlation function of the $Z$ operator illustrated in Fig.~\ref{zqwops}. Red circles and blue squares show results for the near-critical $J$-$Q_2$
and $J$-$Q_3$ models, respectively, and in both cases results are shown for distances $r$ much shorter than the system size (results averaged over
lattices with $L \ge 256$) and for $r=L/2$ in a range of system sizes $L$. The curves show fits to Eq.~(\ref{czfitform}) with fixed $\Delta_t=1.417$ and
$\omega = 2$ in all cases, with only the coefficients $a$ and $b$ optimized. The inset shows the results multiplied by $r^{-2\Delta_t}$.}
\label{zcor}
\end{figure}

The definition of $Z$ on a $3\times 3$ lattice cell with the sum of eight four-spin operators (two singlet projectors) in a staircase arrangement in
Fig.~\ref{zqwops}(a) guarantees that the correlation function does not contain any potential contamination from scaling dimensions corresponding to symmetry-breaking
singlet patterns \cite{sandvik20}. The 90$^\circ$ rotation and reflection symmetries of the cell also imply that only CFT operators with Lorentz spin $\ell=0$ will
contribute to their correlation functions (strictly $\ell =0$ mod $4$, but the higher ones will be invisible because of their very large scaling dimensions).
Fig.~\ref{zcor} shows results for both the $J$-$Q_2$ and $J$-$Q_3$ models versus the separation $r$ between the cells,
in either case for $r \ll L$ (lattices much larger than the maximum separation beyond which the results become too noisy) and also at $r=L/2$ up to $L$
exceeding $100$. In the case of $r \ll L$, to improve the statistics we have averaged results for different lattice sizes $L \ge 256$ that do not show
any differences beyond statistical errors. All these data sets indeed exhibit asymptotic decay with an exponent close to $2.8$ and a secondary power law with
exponent close to $4.8$.

The data points for $r=L/2$ in Fig.~\ref{zcor} represent independent stochastic outcomes, unlike the correlated $r$ points for the $r \ll L$ sets,
and we therefore use the former to determine the best-fit exponents and statistical errors. For the $J$-$Q_2$ model, for which we have the best data,
we fit to the function
\begin{equation}
C_Z(r)=\frac{a}{r^{2\Delta_t}}\left (1+\frac{b}{r^{\omega}}\right ),
\label{czfitform}
\end{equation}
and obtain a good $\chi^2$ value already starting from $r=L/2=6$, with $\Delta_t=1.417(7)$ and $\omega = 2.0(2)$. If we interpret the term with exponent
combination $(2\Delta_t+\omega)/2$ in Eq.~(\ref{czfitform}) as the singlet scaling dimension instead, we would have the reasonable value $\Delta_s = 2.40(2)$,
though in reality the second term should also contain a very similar contribution from $2\Delta_t+2$, as discussed previously. The leading contribution
governed by $\Delta_t$ is very stable with respect to reducing the number of system sizes included in the fit (with the statistical error of course
increasing) and we present the value given above as our final result for this important relevant scaling dimension in Table \ref{dtable}. This fitting
result for the $J$-$Q_2$ model at $r=L/2$ is also fully consistent with the $r$ dependence for $r \ll L$ as well as with analogous results for the $J$-$Q_3$ model,
all shown in Fig.~\ref{zcor}, where we fixed the decay exponents to the same values as above to explicitly show the good consistency between all the data sets.
Given the uncertainty in the interpretation of the subleading contributions to $C_Z(r)$, we do not report any value of $\Delta_s$ from this analysis.

Since the transitions in these models are eventually first-order, the algebraic decays seen in Fig.~\ref{zcor} cannot continue for ever but must eventually
reflect phase coexistence. Given that the two phases should eventually, when the SO($5$) symmetry has broken down, undergo separation in Hilbert space, the
correlation function should asymptotically approach a constant, reflecting large fluctuations from the finite system tunneling between the two phases.
In such a state, the generic connected energy correlation function $C_V$ should approach a constant, due to the expectation value $\langle V_0\rangle $ of the
local operator being different in the two phases (between which the system fluctuates) and, therefore, its square will not cancel the $r\to \infty$
value of $\langle V_0V_r\rangle$ in the definition Eq.~(\ref{cvconnected}). There are no signs of deviations from the power-law behavior up to the largest
distances studied here---not even for much larger distances than in Fig.~\ref{zcor}, where the results are too noisy to be useful for extracting the
exponents but do not indicate any obvious deviations from the fitted form.

\subsection{Conserved current operator}
\label{sub:current}

The $W$ operator illustrated in Fig.~\ref{zqwops}(d) was studied in Ref.~\cite{zhao20} in the context of a helically deformed VBS state. When the entire
lattice is modulated in the staircase fashion as in the $5\times 5$ cell depicted, the columnar VBS state acquires a twist that corresponds to helical
order, most likely with arbitrary (incommensurate) periodicity in the thermodynamic limit. The staircase modulation of the Heisenberg couplings
breaks the square-lattice symmetry, and it was shown that it induces nonzero winding number of the ground state wave function of the $J$-$Q_3$ model
expressed in the valence-bonds basis \cite{zhao20}, with the winding number directly related to the wavelength of the helical VBS order.

Many aspects of VBS states in $S=1/2$ systems can be understood based on their close relationship to classical dimer models; in some cases they can be
described by the same kind of height model, with a simple renormalized coupling constant \cite{tang11c,patil14}. Winding numbers generally correspond
to different topological sectors, which correspond to flux sectors when mapping to an effective electromagnetic theory \cite{chalker17}. The flux sectors
are associated with a conserved Noether current, which in the present case should not only be associated with the emergent U($1$) symmetry of the VBS
but also with the larger emergent SO($5$) symmetry. When applied as a macroscopic perturbation to a critical $J$-$Q$ model, the $W$ term induces the
conserved current, as observed in the form of the helical VBS in Ref.~\cite{zhao20}.

We note here that the conserved winding number, and associated topological sectors (hence the conserved current), has been explicitly demonstrated in $J$-$Q$
models also at the AFM--VBS transition, not only in the ordered VBS phase \cite{shao15}---where the conservation is essentially trivial, given that the valence
bonds are short with zero winding and the system is gapped. Its conservation at criticality, where the gap closes and bond-length distribution decays as a power
law of the length, is directly related to the emergent U($1$) symmetry of the VBS.

The CFT current operator that is contained in the $W$ operator must have scaling dimension $\Delta_j=2$, and a value close to $2$ was indeed found in
Ref.~\cite{zhao20}, $\Delta_W = 1.90(2)$, though the identification of $W$ as a current operator was not made. The conserved current has also been
studied using a correlation function in imaginary time \cite{ma19b}, but using real-space correlations of the $W$ operator seems an easier route. Though the
operator may appear complicated, it is not difficult to study its correlation function in QMC simulations. It is also not necessary to use a cell as
large as $5\times 5$ spins, but any odd length can be used and results for $3\times 3$ were also presented in Ref.~\cite{zhao20}. The larger cell has the
advantage of producing a correlation function with larger overall amplitude, making it easier to extract the power law. The correlations show an
interesting structure in the full 2D space $(r_x,r_y)$, as discussed in the Supplementary Material of Ref.~\cite{zhao20}. The strongest signal for
extracting the exponent is along the line $(x,-x)$, which we focus on here.

\begin{figure}[t]
\includegraphics[width=84mm]{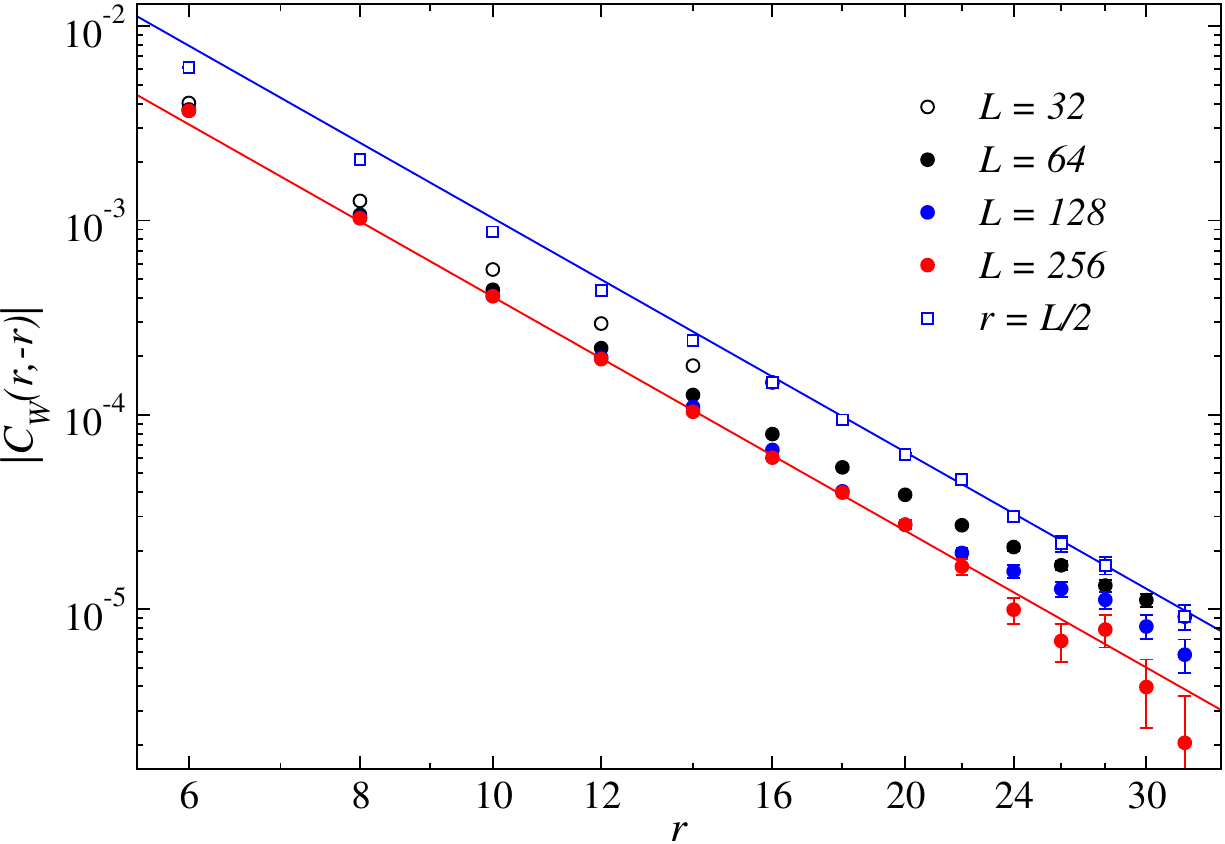}
\caption{Correlation function of the $W$ operator defined in Fig.~\ref{zqwops}(d) along the lattice line $(x,-x)$ on near-critical $J$-$Q_2$
  lattices of size $L=32$, $64$, $128$, and $256$, and also vs $r=L/2$ for other system sizes up to $L=56$. The solid lines show the $r^{-4}$ asymptotic
  decay expected for the conserved current operator ($\Delta_j=2$). A power-law fit $C_W \propto x^{-2\Delta_W}$ to the $L=256$ data for $r \ge 12$ gives
$\Delta_W = 2.01(3)$.}
\label{wcor}
\end{figure}

The previous result \cite{zhao20} for $\Delta_W$ was five error bars away from the expected result $\Delta_W=\Delta_j=2$, likely as a result of some remaining
corrections in short-distance data. We now have improved PQMC data and show the results in Fig.~\ref{wcor}. Looking at the $x$ dependence on
lattices of size $L=32$, $64$, $128$, and $256$, there is a systematic shift upward for $r$ approaching $L/2$, reflecting boundary
enhancement of these correlations. The trend for the distances where the finite-size effect becomes significant indicates that $L=256$ is sufficiently large
up to the largest $r$ for which we have results with reasonably small error bars. Fitting to the $x \ge 12$ data, we obtain $\Delta_W=2.01(3)$, and this
value is stable within error bars also when including $L=10$, which gives $\Delta_W=2.03(2)$. While the latter fit is also statistically sound, it may be
marginally affected by remaining short-distance corrections, noting that $\Delta_W$ further increases when additional small system sizes are included
(which leads to fits of poor quality).

The data at $x=L/2$ in Fig.~\ref{wcor} still delivers a scaling dimension close to $1.90$ \cite{zhao20} when small system sizes are included, even
when a scaling correction is included, while the data for larger distances are consistent with $\Delta_W=\Delta_j=2$. The data versus $r$ for $L=256$
exhibit less scaling corrections and we have entered our result from fitting these data for $r \in [12,32]$ as $\Delta_j$ in Table \ref{dtable}.

\subsection{Pseudo-scaling correlator}
\label{sub:darkop}

The $Q_n$ and $J$ operators of the Hamiltonian must also contain both the $t$ and $s$ field operators, but the $J$-$J$ and $Q$-$Q$ correlation functions
are complicated by the fact that they also contain the VBS order parameter. The dimer correlation function studied in Sec.~\ref{sec:order} is exactly a
correlator of two $J$ terms, though with phases corresponding to the modulation of singlet densities in a columnar state. The symmetric operators should be
detectable in the uniform component of the same correlation functions but their contributions are harder to extract because of the faster power law decays
(larger scaling dimensions)---we note that the value of $2\Delta_t \approx 2.8$ is not that far from $2\Delta_\phi+2 \approx 3.2$, the leading correction
exponent used in the fits in Fig.~\ref{corfit}(b). The lattice arrangement of the spins of the $Z$ operator in the previous section avoids this
complication because it does not contain the VBS order parameter.

As first noted in Ref.~\cite{sandvik20}, when the $Q$-$Q$ plaquette correlations are averaged over the four columnar states in the ordered
VBS phase the result vanishes on the line $(r,0)$ with $r$ odd. The same would be true also for a plaquette-ordered state, and then it will also hold
for all fluctuations between the columnar and plaquette patterns that are realized when the critical VBS develops emergent U($1$) symmetry. Thus,
on the $(r,0)$ line it would appear that the uniform component of the $Q$-$Q$ correlation function can be studied without trying to eliminate the
dominant VBS contributions. Indeed, the power-law decay corresponding to scaling dimension $\Delta_* \approx 0.80$ found in
Ref.~\cite{sandvik20} matched the expected value for a symmetric operator driving a phase transition with correlation-length exponent
$\nu_* = (3-\Delta_*)^{-1} \approx 0.455$. However, we now have a more reliable estimate of $\nu_*$ (Sec.~\ref{sub:so5}), $\nu_* \approx 0.416$,
which ruins the above relationship to $\Delta_*$, which still has a value close to that determined previously.

\begin{figure}[t]
\includegraphics[width=80mm]{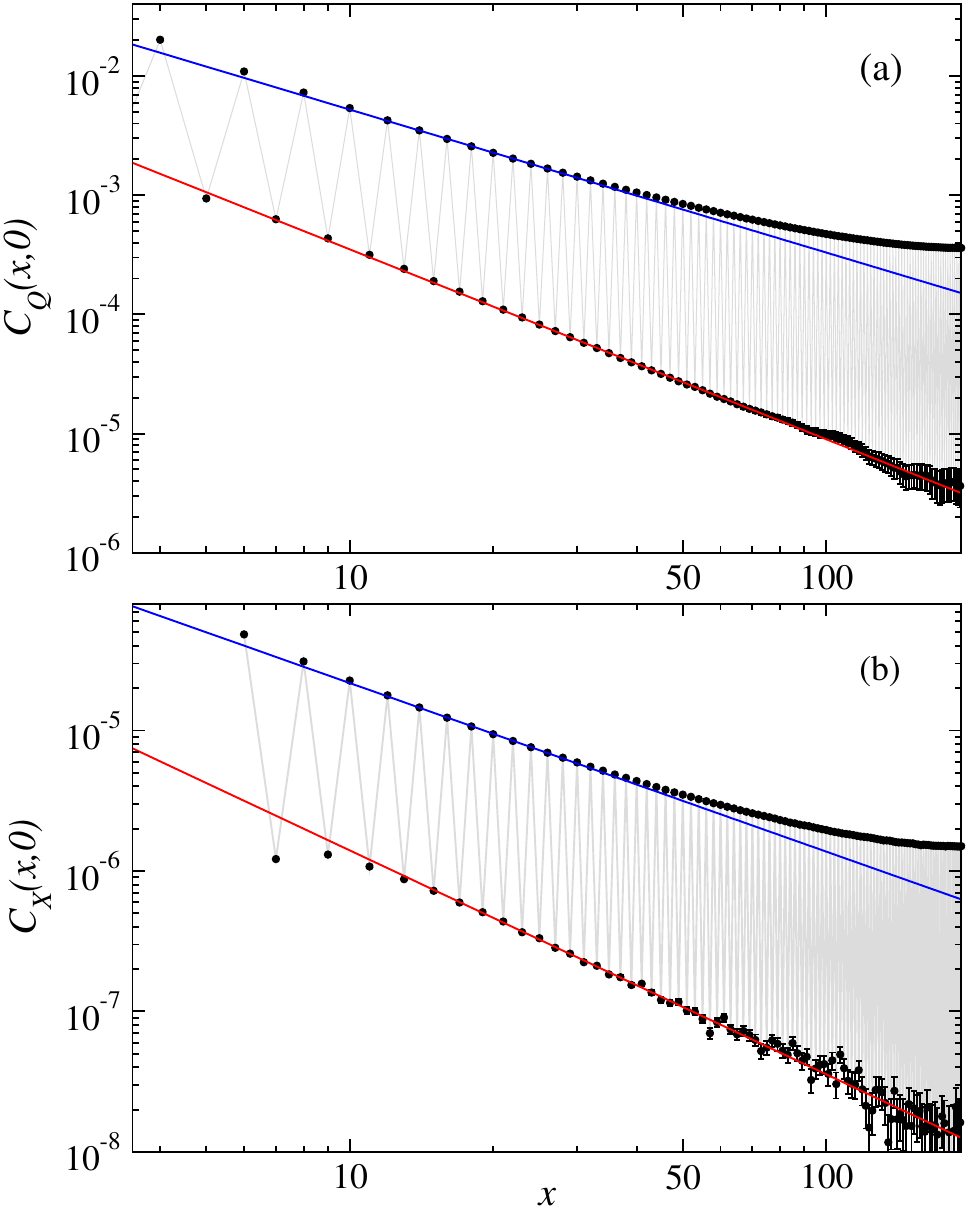}
\caption{Correlation function along a lattice axis of the four-spin $Q$ (a) and $X$ (b) operators in the near-critical $J$-$Q_2$ model
on a lattice of size $L=384$. The blue lines close to the even-$x$ (upper) branch show the power-law decay $x^{-2\Delta_\phi}$ of the
order parameter, with $\Delta_\phi=0.607$, while the red curves at the odd (lower) branch have the form $x^{-2\Delta_*}$ with $\Delta_*=0.80$.}
\label{qx384}
\end{figure}

In Fig.~\ref{qx384}(a) we show the $Q$-$Q$ correlation function $C_Q(r)$ computed using PQMC simulations of the $J$-$Q_2$ model on an $L=384$ lattice with a slightly
different definition of the $Q$ operator, with just ${\bf S}_i\cdot {\bf S}_j$ as the individual bond operators instead of the singlet projector
${\bf S}_i\cdot {\bf S}_j-1/4$
used in Ref.~\cite{sandvik20}, the latter being more practical in SSE simulations. The even and odd branches clearly show different power-law decays. On the
even branch, as shown with the blue line, the decay is consistent with the scaling dimension $\Delta_\phi \approx 0.61$ before the effects of the nonzero ordered
moment become significant, similar to Fig.~\ref{sdcor}. For the odd branch, the red line shows the decay with $\Delta_* = 0.80$, which is the value we
determine below by analyzing the decay at $r=L/2-1$ for a range of system sizes. This value obtained in the ground state with PQMC simulations is very close
to the result of Ref.~\cite{zhao20}, which was obtained with the SSE method with the inverse temperature scaled in proportion to $L$.

The $X$ operator illustrated in Fig.~\ref{zqwops}(c) consists of a single product of two-spin operators situated on the
diagonals of a plaquette. Its correlation function $C_X(r)$ shows a behavior similar to $C_Q$, with the same scaling dimensions used to draw the lines
in Figs.~\ref{qx384}(a) and \ref{qx384}(b), but the overall amplitude of $C_X$ is more than two orders of magnitude smaller. It is interesting to note that the
odd-$r$ branch does not show any significant enhancement of the correlations close to $r=L/2$, neither in $C_Q$ nor in $C_X$. Even though this
branch, unlike even $r$, does not sense the order parameter, one might still expect some boundary effect. The deviations from the power law are
likely just small in this case.

\begin{figure}[t]
\includegraphics[width=75mm]{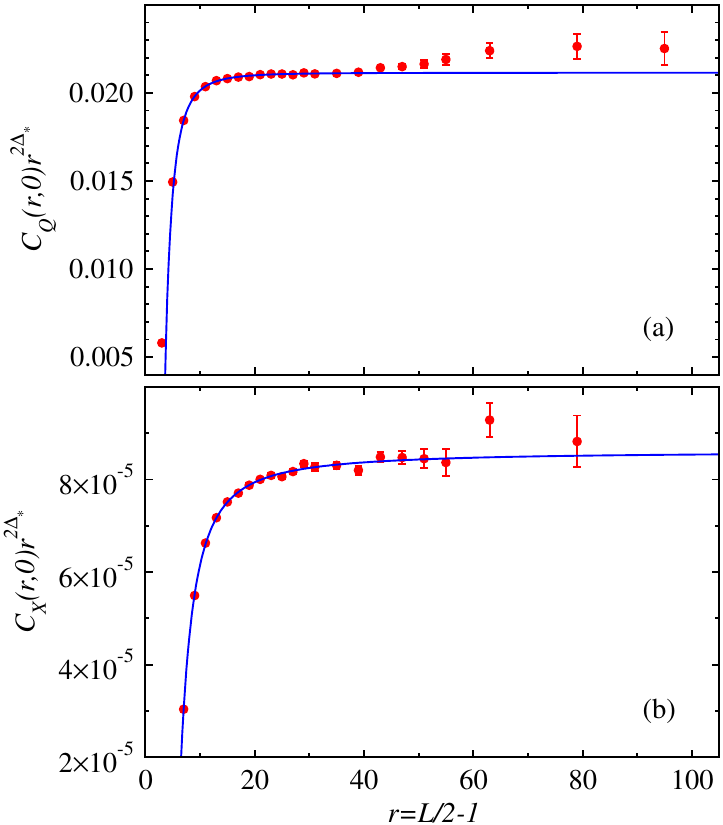}
\caption{Correlation function at distance $r=L/2-1$ of the $Q$ (a) and $X$ (b) operator computed with the near-critical $J$-$Q_2$ model. The values have been
multiplied by $r^{2\Delta_*}$ with the best-fit value $\Delta_*=0.80$. The blue curves show fits with one scaling correction in (a) and two corrections in (b),
as discussed in the text.}
\label{qxljq2}
\end{figure}

Our finite-size analysis of both the $Q$-$Q$ and $X$-$X$ correlations is illustrated in Fig.~\ref{qxljq2}, where we have multiplied the correlation functions
at $r=L/2-1$ by $r^{2\Delta_*}$ in order to make the corrections better visible. In the case of $C_Q$, we see violations of the scaling behavior for $r \agt 40$,
which we interpret as a crossover to the constant value expected (beyond some distance) in a first-order coexistence state. The multiplier $r^{2\Delta_*}$ then
causes the upward trend that is small but statistically significant in Fig.~\ref{qxljq2}(a). Such violations are not as apparent in $C_X$, though they may also be
present for $r\agt 50$, where the statistical errors are large. We fit the data only up to $r$ where the scaling violations do not seem significant. In light of
these results, it is interesting, and somewhat puzzling, to again note that the large-$r$ deviations from the leading power law are not apparent in Fig.~\ref{qx384}
for fixed $L=384$, with the decay exponent fully consistent with those extracted from $r=L/2-1$.

From $C_Q(L/2-1)$ we obtain $\Delta_* = 0.800(3)$ by fitting data for $r \in [9,39]$ to two power laws. The exponent of the correction is $4.7(2)$,
which matches $2\Delta_s$ in Table \ref{dtable} within the rather large error bar, but this exponent may again also be interpreted as $2\Delta_t+\omega$
with $\omega \approx 2$. When a third power law is added with fixed exponent equal to $2\Delta_t$, the amplitude of this term becomes very small. Since
$Q$ (summed without phases over the entire lattice) is the operator driving the AFM--VBS transition, the $t$ field must of course be present to some extent
in its correlation function, but apparently very weakly so---on the line considered here and along other lines. It is possible that the contributions from
the order parameter and $\Delta_*$ are overwhelmingly large, unlike in the $Z$ operator where these contributions are excluded.

In the case of the $X$-$X$ correlations, three power laws are required to describe the data. Unconstrained fits deliver exponents with large error bars and
we instead fix the corrections to plausible values; the first one at $2\Delta_t$ and the second one in the range $[4.5,4.8]$, the latter again corresponding to
either $2\Delta_t+\omega$, $2\Delta_s$, or some average of these. The fits have good $\chi^2$ values for $r \in [9,51]$ and produce
$\Delta_* \in [0.799,0.800]$, fully consistent with the value from $C_Q$. We therefore have a high confidence in the value of $\Delta_*$.

It is noteworthy that the contribution from the imposed $2\Delta_t$ decay appears to be a significant correction to the $X$-$X$ correlations, unlike the
case in the $Q$-$Q$ correlations. This can be explained by the much smaller overall amplitude of $C_X$, by about two orders of magnitude in both
Fig.~\ref{qx384} and Fig.~\ref{qxljq2}, which could mean that, relatively speaking, the $\Delta_t$ contribution is now much larger.

\begin{figure}[t]
\includegraphics[width=75mm]{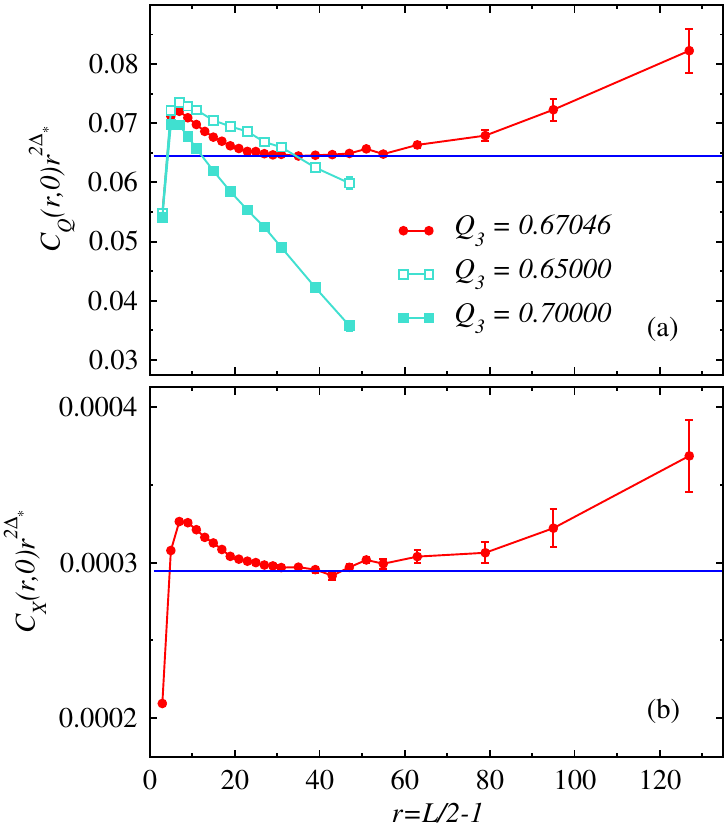}
\caption{Correlation function at distance $r=L/2-1$ of the $Q$ (a) and $X$ (b) operators in the near-critical $J$-$Q_3$ model ($Q_3=0.67046$, $J=1-Q$)
as well as inside the AFM ($Q_3=0.65$) and VBS ($Q_3=0.67$) phases. The results have been multiplied by $r^{2\Delta_*}$, with $\Delta_*=0.80$ having
the same value as in the $J$-$Q_2$ analysis in Fig.~\ref{qxljq2}.}
\label{qxljq3}
\end{figure}

The well defined common power law behavior controlled by $\Delta_*$ in both Fig.~\ref{qx384} and Fig.~\ref{qxljq2} indicates that we are not here
dealing merely with some slow gradual crossover to the first-order behavior, but truly a specific pseudocritical behavior governed by an exponent that does
not appear directly in the CFT spectrum. We believe that $\Delta_*$ and $\nu_*$ must be related in some way, as two manifestations of the weakly perturbed
CFT on the first-order line. The putative relationship $1/\nu_*=3-\Delta_*$ is clearly violated, with $1/\nu_* \approx 2.40$ extracted in Sec.~\ref{sub:so5w},
and it is not clear what the exact relationship is. A perturbative study of the SO($5$) CFT could hopefully give the answer.

In Fig.~\ref{qxljq3} we show results for the near-critical $J$-$Q_3$ model. The same critical decay as in the $J$-$Q_2$ model is manifested
up to distances $r\approx 50$, but with different sign of the leading correction. In this case it is even more difficult to fit the corrections,
and we only show a reference line to highlight a range of distances over which the decay controlled by $\Delta_*$ extracted from the $J$-$Q_2$ model
holds rather well. At large distances there is an even clearer upturn of the correlations from the critical form than in the $J$-$Q_2$
model. This more prominent precursor to the first-order behavior in the $J$-$Q_3$ model should again be a consequence of the slightly stronger
first-order nature of its transition. Note that the behavior is very different away from the transition point, as illustrated in Fig.~\ref{qxljq3}(a)
with data slightly inside the AFM and VBS phases. In general, energy-type correlators should be expected to decay rapidly inside ordered phases,
as we find, though we do not know the exact expected form.

\begin{figure*}[t]
\includegraphics[width=160mm]{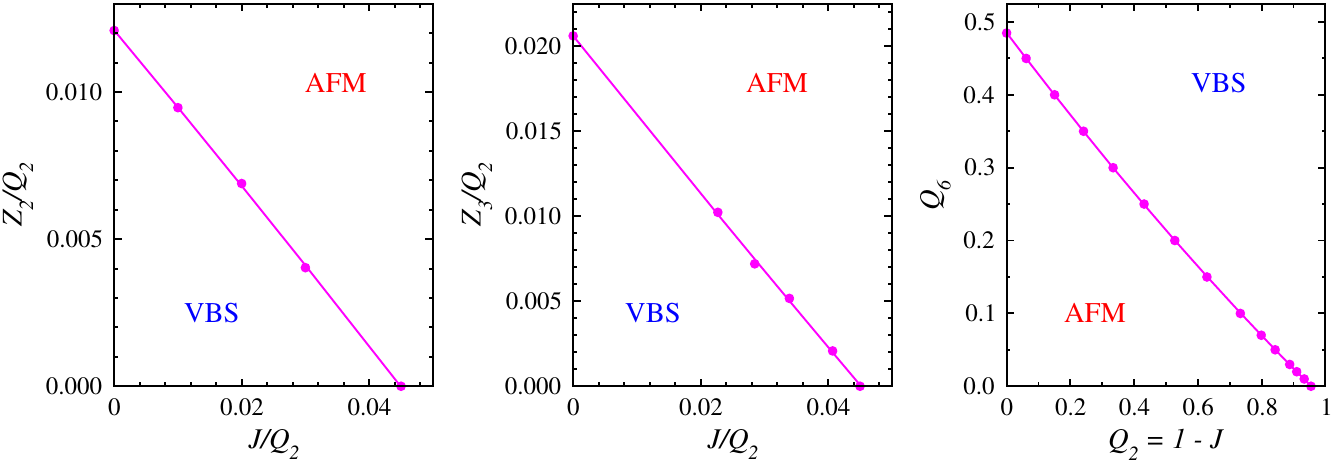}
\caption{Phase diagrams of models with two tuning parameters; (a) $J$-$Q_2$-$Z_2$, (b) $J$-$Q_2$-$Z_3$, and (c) $J$-$Q_2$-$Q_6$.
Because the $Z_n$ terms enhance the size of the AFM phase, $J/Q_2$ is used as the horizontal axis in (a) and (b). In (c), $Q_2=1-J$ is used
since the $Q_6$ term enhances the VBS phase. The curves are quadratic fits to the data points.}
\label{jq2q6phases}
\end{figure*}

An important aspect of our interpretation of the pseudo dimension $\Delta_*$ is that it is not a member of the CFT spectrum and therefore cannot be present
exactly at the critical point. The natural way in which ``fading'' of the effect characterized by $\Delta_*$ can take place upon approach to the critical point is that
the overall amplitude of contributions to correlation functions from $\Delta_*$ diminish gradually when the transition takes place closer to the critical point.
Comparing Figs.~\ref{qxljq2} and \ref{qxljq3}, it is apparent that the amplitude of both odd-branch correlations is more than three times larger in the $J$-$Q_3$
model, despite the fact that the values on the even-$x$ branch (in Fig.~\ref{qx384} for the $J$-$Q_2$ model) differ by only about 10\%.
Thus, the amplitude of the $\Delta_*$ correlator of the more critical $J$-$Q_2$ model has faded off to about $0.3$ of its value in the $J$-$Q_3$ model.
Fading is also manifested in the correlation length crossover schematically illustrated in Fig.~\ref{xi}, which will be demonstrated with data in
Sec.~\ref{sec:dark}.

\section{Scaling on the AFM-VBS coexistence line}
\label{sec:jq2q6}

As we demonstrated in Sec.~\ref{sub:corrL}, the $J$-$Q_n$ models exhibit increasingly strong first-order discontinuities with $n$ at their AFM--VBS
transitions. We here add a third interaction to the $J$-$Q_2$ model in order to have two continuous tuning parameters and realize the line of first-order
transitions shown schematically in Fig.~\ref{phases}. We study the evolution of the phase transition extending from that of the pure $J$-$Q_2$ model with
increasing coupling $Z_2$, $Z_3$, or $Q_6$. The $Z_2$ and $Q_6$ interactions are illustrated in Fig.~\ref{jqterms}(c) and Fig.~\ref{jqterms}(e), respectively,
and $Z_3$ is a direct generalization of $Z_2$ with one more singlet projector in the product, forming a staircase with three treads.

All results in this section were obtained using the SSE method, with the inverse temperature taken equal to the system size $L$. We define
both order parameters, $M_z$ (AFM) and $D=(D_x,D_y)$ (VBS), using the diagonal $S^z_i$ operators,
\begin{subequations}
\begin{eqnarray}
  M_z&& = \frac{2}{N}\sum_{i=1}^N S^z_i (-1)^{x_i+y_i},\label{mzdef} \\
  D_e&& = \frac{1}{N}\sum_{i=1}^N S^z_iS^z_{e(i)}(-1)^{e_i},~~~e \in \{x,y\},~~ \label{dxydef}
\end{eqnarray}
\label{m1defs}
\end{subequations}
where $e(i)$ in Eq.~(\ref{dxydef}) denotes the neighbor of site $i$ in the positive $e\in \{x,y\}$ direction and $e_i \in \{x_i,y_i\}$ are the corresponding
lattice coordinates. A factor $2$ is included in the definition of $M_z$ in Eq.~(\ref{mzdef}) because the two order parameters are then of similar magnitude
in their respective ordered phases. While this normalization is not crucial, when analyzing finite-size crossing points of the squared order parameters
versus a tuning parameter, the $L$ dependence is expected to benefit from a roughly equal scale of the two order parameters. We here denote the expectation
values of the squared order parameters by $m^2_z = \langle M^2_z\rangle$ and $m^2_d = \langle D^2\rangle$.

\subsection{Phase diagram of the $J$-$Q_2$-$Q_6$ model}

We have used both the criterion $m^2_z=m^2_d$ and analogous Binder cumulant (defined below) crossing points to extract the phase diagrams shown in
Fig.~\ref{jq2q6phases}, with the two criteria producing compatible results. We use $J/Q_2$ as the control parameter on the horizontal axis for the
$J$-$Q_2$-$Z_n$ models, switching to $Q_2=1-J$ for the $J$-$Q_2$-$Q_6$ model. In all three cases, we then have a phase boundary starting on the vertical
axis and dropping down to zero versus the horizontal tuning parameter.

Like any generic lattice operator obeying all the symmetries of the Hamiltonian (when summed over all lattice translations), the added interactions
can be expected to contain both the $s$ and $t$ fields of the CFT. Thus, we expect the introduction of $Z_2$, $Z_3$, or $Q_6$ into the $J$-$Q_2$
model to affect the way in which the phase transition is crossed, which corresponds to moving the dashed line in the schematic phase diagrams in
Fig.~\ref{phases}. With the two generic tuning parameters $g$ and $k$ introduced in Fig.~\ref{xi}, as opposed to the orthogonal CFT scaling fields
$s$ and $t$, we can take $g \in \{J,Q_2\}$ and $k \in \{Z_2,Z_3,Q_6\}$ (shifted so that $g=k=0$ is the critical point) in the models studied here.
We will also use the notation $\delta$ for a generic distance to the critical point in Hamiltonian parameter space, along with $s$ and $t$ used in
expressions where both specific distances are required.

In the case of $Q_6$, we know that the transition is first-order for $Q_2=0$, continuing the trend of increasingly strong discontinuities found for $J$-$Q_n$
models with $n$ up to $4$ in Sec.~\ref{sub:corrL}. We therefore expect an increasing $Q_6$ value in the $J$-$Q_2$-$Q_6$ model to move the system toward
larger $s$ values at the AFM--VBS transition, i.e., shifting the dashed line upward in Fig.~\ref{phases}. In the case of the $Z_2$ and $Z_3$ interactions, when
$Q_2=0$ there is no phase transition versus $Z_2/J$ (for the positive values of $J$ and $Z_2$ that can be accessed with QMC simulations), while increasing
$Z_3/J$ eventually leads to a strongly first-order transition into a staggered VBS ground state \cite{sen10}. When starting from the VBS phase of the
$J$-$Q_2$ model and turning on one of the $Z_n$ interactions, as on the left side of Figs.~\ref{jq2q6phases}(a) and \ref{jq2q6phases}(b), both $Z_3$ and
$Z_2$ rapidly destroy the VBS order, bringing the system into the AFM phase.

In the $Q_2$-$Z_3$ model, further increasing the $Z_3$ coupling beyond the values shown in the phase diagram in Fig.~\ref{jq2q6phases}(b) eventually leads
to a subsequent transition from the AFM phase into the staggered VBS phase \cite{sen10}. An extended phase diagram including
both columnar and staggered VBS phases, and either a direct transition between the two or an intervening AFM phase, was studied in Ref.~\cite{zhao20b}.
Here we will not consider the staggered VBS phase but observe that both $Z_2$ and $Z_3$ suppress the columnar VBS phase, in contrast to the clearly VBS
enhancing $Q_6$ interaction.

To investigate how the AFM--VBS transition evolves on the first-order line, we first consider the $J$-$Q_2$-$Q_6$ model. Here the changes with
increasing $Q_6$ are very clearly observable because the coexisting order parameters in the $J$-$Q_6$ model are robust when $Q_6 \agt 0.1$, much larger
than those in the plain $J$-$Q_2$ model. However, as we will see, $m_z^2$ and $m_d^2$ on the coexistence line still only reach a small fraction of their
maximum values deep inside the respective ordered phases, even up to the maximum value $Q_6 \approx 0.5$ that we can reach.

\subsection{Scaling of the Binder cumulant}

For testing scaling with an exponent $\nu_s=(3-\Delta_s)^{-1}$, we first study the dimensionless Binder cumulants on the coexistence line.
For a system of size $L$ at distance $\delta$ from the critical point, a generic cumulant $U$ should asymptotically be governed by a scaling
function $u(x)$, with the argument $x=\delta L^{1/\nu}$,
\begin{equation}
U(\delta,L)=u(\delta L^{1/\nu}).
\label{ufssform}
\end{equation}
For a system on the AFM--VBS coexistence line, i.e., $(t=0,\delta=s)$, $\nu=\nu_s$ should be expected. Otherwise, on approach to the critical point along
some other line where both $s$ and $t$ are nonzero and $\delta=\delta(s,t)$, there should be a second scaling argument in Eq.~(\ref{ufssform});
\begin{equation}
U=u(sL^{1/\nu_s},tL^{1/\nu_t}).
\label{ufssformw}
\end{equation}
Since $\nu_t < \nu_s$, the behavior for large systems will be governed by the second argument. Scaling of $U$ with
$\nu=\nu_s$ can asymptotically be observed only on the coexistence line, which is what we will pursue here.

For scaling purposes, the distance of a system from the critical point on the first-order line can be taken as $\delta = Q_6 - Q_{6c} \propto s$ (close
enough to the critical point so that any nonlinear correction in the field $s$ can be neglected), with some yet to be determined critical value $Q_{6c}<0$.
A negative $Q_{6c}$ follows from our expectation that increasing $Q_6$ above $0$ will bring the system further away from the critical point than
the very nearly critical $J$-$Q_2$ model. We also implicitly assume that this multicritical point exists in the real model space (i.e., it corresponds to a
unitary CFT), though it is located outside the region $Q_6\ge 0$ accessible with sign-free QMC simulations.

The most practical way to define a size dependent Binder cumulant at system size $L$ is to use the common value of the AFM and VBS cumulants when they cross
each other versus $J$ at fixed $Q_6$. To define the crossing point in an optimal way, the symmetry of the order parameters should be properly taken into account.
When defining the AFM and VBS cumulants, in the past the symmetries of the individual order parameters have been employed in the standard way in
the definitions so that $U_{\rm AFM} \to 1$ and $U_{\rm VBS}\to 0$ with increasing $L$ in the AFM phase and $U_{\rm AFM} \to 0$ and $U_{\rm VBS}\to 1$ in the
VBS phase. However, on the coexistence line close to the critical point (up to some large system size), we expect a combined order parameter with emergent
SO($5$) symmetry (which we will show further support for in Sec.~\ref{sub:so5}). When studying the transition, it may therefore be better to define the two
cumulants under the assumption that the AFM and VBS order parameters represent three and two components, respectively, of a five-component order parameter
with SO($5$) symmetry, which for our purposes again is no different from O($5$).

The Binder ratio is defined as
\begin{equation}
R = \frac{\langle X^4\rangle}{\langle X^2\rangle^2},
\label{bratio}
\end{equation}
where $X$ denotes an $N$-component vector $(x_1,\ldots,x_N)$. In the ordered phase $X \to 1$ trivially, while in the disordered phase the value
is easily obtained using Gaussian integrals. The cumulant with the desired values $0$ and $1$ in the two ordered phases is of the form $U = a(1-R/b)$
with $a$ and $b$ depending on $N$. When it is in practice difficult to compute the full ratio $R$ including all $N$ components of $X$ as in
Eq.~(\ref{bratio}), one can also use some subset of the components---typically those corresponding to diagonal observables in the basis used---which
is exactly what we do here. The coefficients $a$ and $b$ then have to be adjusted accordingly.

First consider the standard definitions. We here include only the $z$ component of the 3-component AFM order parameter $M$ and define the ratio
\begin{subequations}
\begin{equation}
R_{\rm AFMz} = \frac{\langle M_z^4\rangle}{\langle M_z^2\rangle^2},
\label{bratioafm}
\end{equation}
where $M_z$ is defined in Eq.~(\ref{mzdef}) and is evaluated at a given ``time slice'' of an SSE configuration. The expectation values can be averaged
in the usual way over many slices for better statistics. In the VBS case we use both components of the dimer order parameter defined in Eq.~(\ref{dxydef});
\begin{equation}
R_{\rm VBS} = \frac{\langle (D_x^2+D_y^2)^2\rangle}{\langle D_x^2+D^2_y\rangle^2}.
\label{bratiovbs}
\end{equation}
\end{subequations}
The properly normalized cumulants are given by
\begin{subequations}
\begin{eqnarray}
&& U_{\rm AFMz} = \frac{5}{2} \left (1-\frac{R_{\rm AFMz}}{3} \right ), \\
&& U_{\rm VBS} = 2 \left (1-\frac{R_{\rm VBS}}{2} \right ).
\end{eqnarray}
\label{udefs1}
\end{subequations}

The adjustments necessary under the assumption of SO($5$) symmetry again only require trivial Gaussian integrals.
In terms of the Binder ratios defined in Eqs.~(\ref{bratioafm}) and (\ref{bratiovbs}), the SO($5$) cumulants are
\begin{subequations}
\begin{eqnarray}
&& U_{\rm AFM,SO(5)} = \frac{35}{10} \left (1-\frac{R_{\rm AFMz}}{3} \right ), \\
&& U_{\rm VBS,SO(5)} = \frac{35}{10} \left (1-\frac{R_{\rm VBS}}{2} \right ).
\end{eqnarray}
\label{udefs2}
\end{subequations}
Asymptotically, for $L \to \infty$, the coexistence state does not host the higher symmetry, due to the dangerously irrelevant perturbations discussed in
Sec.~\ref{sub:so5}, which become relevant in the presence of long-range order. Nevertheless, the length scale above which the SO($5$) symmetry breaks down
will be large, at least for small values of $Q_6$. Even after the symmetry has broken down, crossing points of $U_{\rm AFM}$ and $U_{\rm VBS}$ will still
represent the phase boundary properly, though likely with some crossover affecting the form of the scaling function.

\begin{figure}[t]
\includegraphics[width=84mm]{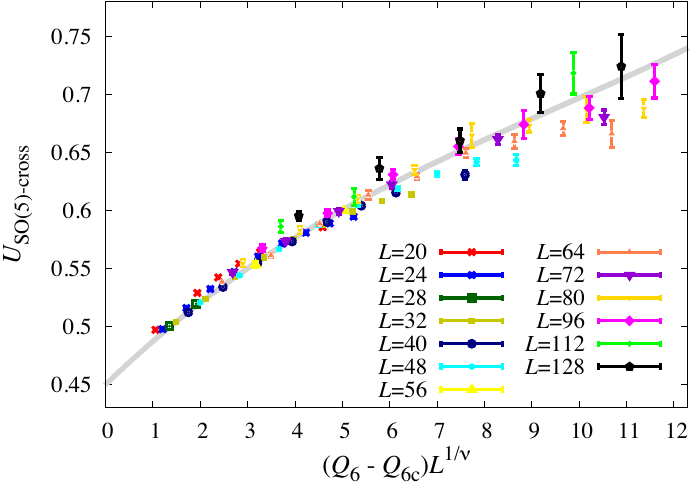}
\caption{Data collapse analysis of the common value of the SO($5$) variants of the AFM and VBS Binder cumulants, Eqs.~(\ref{udefs2}), at their
  finite-size crossing points. The SO($5$) singlet exponent was fixed at $1/\nu_s=3-\Delta_s=0.727$, corresponding to $\Delta_s$ in
  Table \ref{dtable}, and only the critical point was optimized to $Q_{6c}=-0.07$ for best data collapse. The grey curve shows a polynomial
  fitted to the data points $U(Q_6,L)$ for which the finite size deviations are judged to be mainly statistical (excluding all sizes $L \le 32$ as
  well as other points that fall significantly below the fitted function).}
\label{jq2q6binder}
\end{figure}

We here simply test data collapse of the crossing values of the two SO($5$) cumulants with our value of $\Delta_s$ in Table \ref{dtable}, which corresponds
to $1/\nu_s=0.727$ in Eq.~(\ref{ufssform}). For fixed $Q_6$, we extract crossing values $U(Q_6,L)$ with error bars from polynomial fits to several (of order 10) data points in the neighborhood of the transition versus $Q_2$, i.e., scanning horizontally in Fig.~\ref{jq2q6phases}. Results graphed versus the
scaled distance to the critical point are shown in Fig.~\ref{jq2q6binder}, where $Q_{6c}= -0.07(2)$ is the only adjustable parameter for optimizing the data
collapse. Though the collapse is not perfect, the deviations from an ideal common scaling function are largely what would be expected from scaling corrections,
e.g., the way the data for the smallest sizes shown here, $L=20$ and $L=24$, deviate somewhat from the other points for the scaling variable in the range
roughly from 1 to 4. There is also a systematic peeling-off of data from the emerging scaling function in the form of a slower growth with the scaling
argument as $Q_6$ increases for given $L$.

The scaling function to which the data should asymptotically collapse for sufficiently large $L$ is normally analytic in the scaling variable.
In Fig.~\ref{jq2q6binder} we have therefore fitted a cubic polynomial to the set of points that do not appear to be affected significantly by
finite-size corrections. The points for the three largest system sizes deviate somewhat from the fitted scaling function. This flaw is most likely
related to the eventual breakdown of the emergent SO($5$) symmetry, which will take place for any point on the first-order line above some system size
that depends on the distance to the critical point (as we will demonstrate explicitly in Sec.~\ref{sub:so5}). Since the violation of the SO($5$) symmetry
is shifted to larger system sizes upon moving closer to the critical point, a valid scaling function (one that has the correct asymptotic form on
approach to the critical point) can be defined by using only system sizes that are below the scale of SO($5$) breakdown, as we have done.

Considering the single adjustable parameter and no obvious flaws beyond likely scaling for the smallest and very largest ($L>100$) systems, these
results support the scenario of a critical end point of a first-order transition. It should also be noted again that the exponent $\nu_s$ is unusually
large, roughly twice the value at 3D O($N$) transitions with small $N$ \cite{poland19}. The good data collapse without adjusting any exponent therefore
confirms a very different and unconventional universality class.

\subsection{Scaling of coexisting order parameters}

As a further test, we extrapolate the order parameters to infinite size, first defining a common value $m^2$ as the crossing value $m^2_z(Q_2,L)=m^2_d(Q_2,L)$
at fixed $Q_6$. We have also tried to extrapolate the individual AFM and VBS order parameters to infinite size separately, but such a procedure is very
challenging because of the uncertainties in the location of the coexistence line. We can again in principle use the cumulant crossing points, but doing
so we find that the VBS order parameter sometimes is nonmonotonic in $1/L$, likely for similar reasons as discussed in Sec.~\ref{sub:corrL}
(Fig.~\ref{corL_jq4}). The crossing values $m^2(L)=m^2_z(L)=m^2_d(L)$ are ideal in this sense, guaranteeing that the coexistence line is followed in
a completely systematic way, with the order parameters balanced at a consistent mix as a function of the system size.

It should be noted that the coexistence line is fundamentally broadened in a finite system, in the present case likely representing a window of
asymptotic size $\propto L^{-(d+z)}$ with $d=2$ and $z=2$, the latter expected when the transition involves continuous symmetry breaking \cite{zhao19}.
Moving within this window will change the relative fraction of AFM and VBS order
from values close to zero to order one and vice versa. For system sizes where the SO($5$) symmetry remains at the transition, this change in
order parameters corresponds directly to the deformation of the five-sphere when the field $t$ crosses $0$. For larger systems, where the SO($5$)
symmetry should be violated, the coexistence state realized in QMC simulations will eventually (for very large system sizes that are not reached here)
correspond to the fraction of configurations (strictly speaking the fractions of equal-time slices of the space-time configurations) with either
type of order. In both cases, the change in relative fraction of the order parameters with the tuning parameter $Q_2$ will then be roughly linear
around the value where $m^2_z(L)=m^2_d(L)$.

\begin{figure}[t]
\includegraphics[width=84mm,clip]{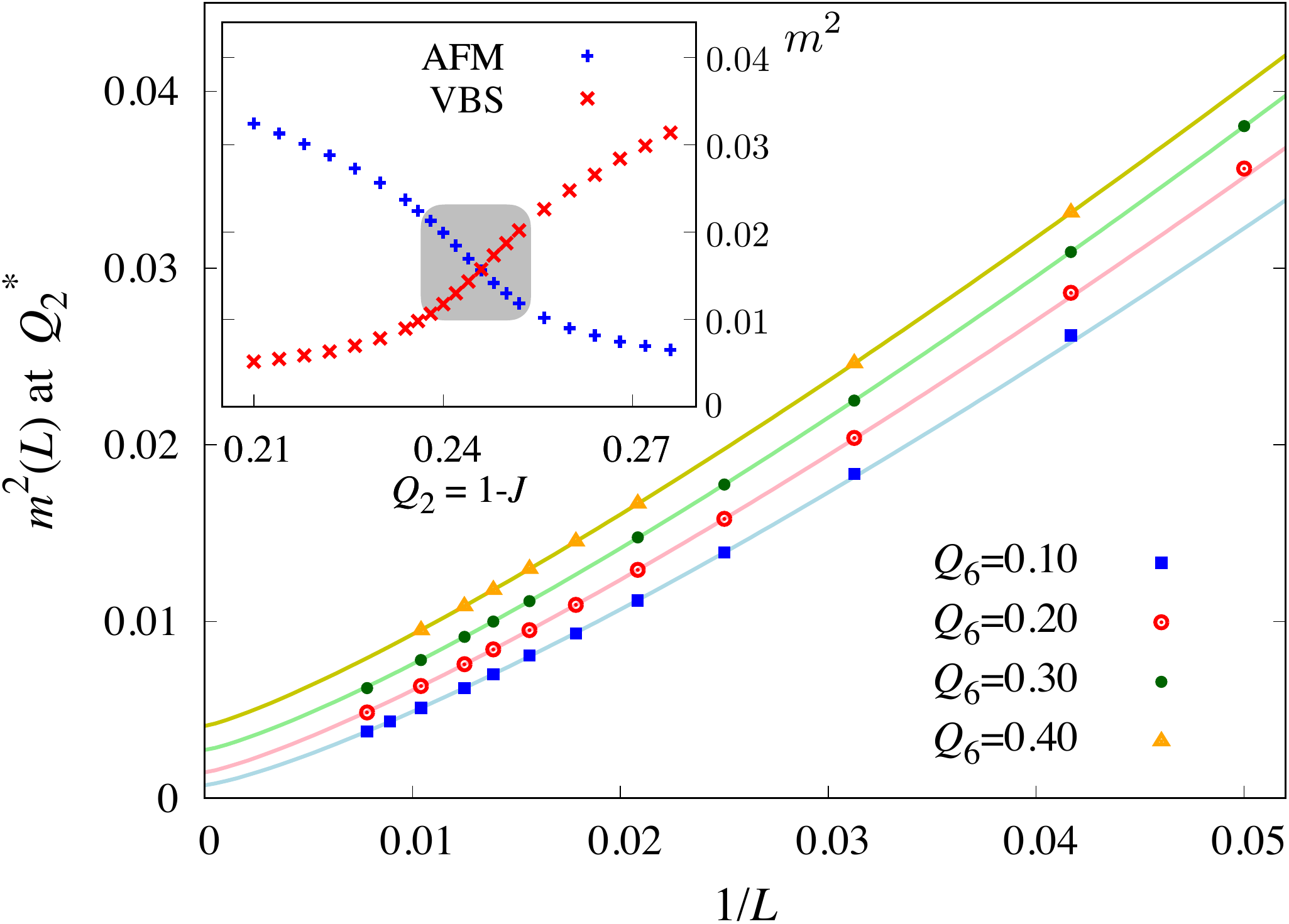}
\caption{The common value of the squared AFM and VBS order parameters on the coexistence line for several values of $Q_6$. Error bars are much
smaller than the symbol sizes in all cases. The definition of the crossing value $m^2(L)$ is illustrated in the inset for $Q_6=0.35$, $L=48$.
The crossing point $Q_2^*(L)$ is extracted from polynomials fitted to the data in the crossing region (roughly within the shaded square),
where the order parameters are nearly linear. The crossing values are extrapolated in the main
graph, using a constant plus a power law, shown with the curves.}
\label{m2method}
\end{figure}

Thus, the crossing method should provide a good definition of the center of the coexistence line, provided that the two order
parameters have roughly the same scale. In practice, it is not possible to perfectly normalize the order parameters for an unambiguous center of the
coexistence window, but, as shown in the inset of Fig.~\ref{m2method}, our normalizations in Eq.~(\ref{m1defs}) accomplish this balancing well. A good
test for balance of the order parameters close to an idealized mid-point of the transition region is to confirm that the crossing point is located
in the regime of the nearly linear dependence on the tuning parameter, which is the case in the inset of Fig.~\ref{m2method}.

As illustrated in the main part of Fig.~\ref{m2method}, we find that the common value $m^2(L)$ exhibits monotonic smooth behavior for all $Q_6$ values
studied. All these systems are still nearly critical, i.e., for large system sizes the order parameters tend to very small values relative to their
maximum attainable values. To fit the data, we therefore use a constant (representing the infinite-size squared order parameter) plus a power
law, representing the near-critical decay of correlations.

As already discussed in the case of $Q_6=0$ in Sec.~\ref{sec:order} (in that case for the correlation functions at $r=L/2$, which also extrapolate to the
squared order parameter), this fitting form cannot be asymptotically completely correct. We there focused on distances where the long-range order can be
taken into account by only adding a constant to the critical decay form of the correlation functions. Here we instead focus on the way the long range order
approaches its asymptotic $L \to \infty$ value, and the power-law form with optimized exponent (that is smaller than the critical exponent $1+\eta$ but
larger than $1$) should be regarded as a way to approximately account for the crossover from the critical form to the asymptotic $1/L$ form \cite{reger88}
for a system with Goldstone modes.

While polynomials including $1/L^2$ and possibly higher-order terms may also seem appropriate, such fitting produces a high level
of scattering (beyond the statistical errors) in the results versus $Q_6$, which is not the case with the constant plus power law. The polynomial form
should work well only when the systems are completely dominated by the Goldstone modes \cite{reger88,sandvik10c}, which is not yet true for the accessible
system sizes here. Though the eventual crossover to polynomial form implies some errors in our extrapolated $m^2$ values, if this crossover takes place
only for much larger system sizes (which is the case here, judging from the results in Sec.~\ref{sub:corrL}), there is not much room remaining in
$1/L$ for any drastic changes in the intercepts, as long as the true $m^2$ value is also not too small. To be clear, the balancing act here is rather
delicate, and it is not a priori guaranteed that the procedure can deliver meaningful results.

\begin{figure}[t]
\includegraphics[width=85mm]{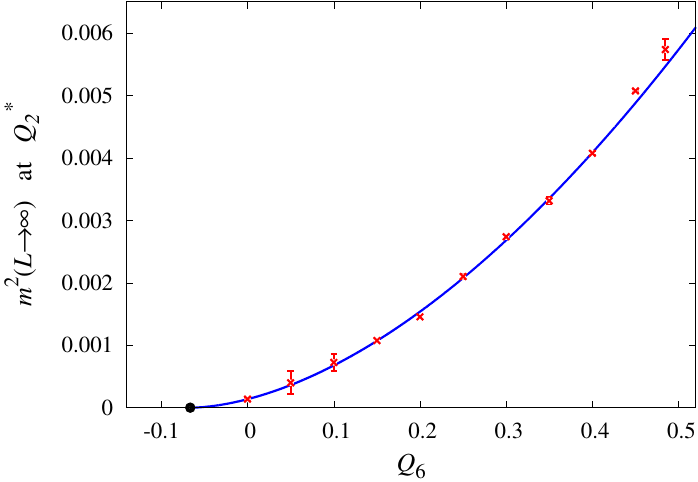}
\caption{Infinite-size extrapolated (as illustrated in Fig.~\ref{m2method}) common value of the squared AFM and VBS order parameters at their finite-size
crossing points, corresponding to the coexistence line in Fig.~\ref{jq2q6phases} for several values of $Q_6$. The dependence on $Q_6$ (excluding the two
points for largest $Q_6$) is fitted to the form $m^2 \sim (Q_6-Q_{6c})^{2\beta}$, resulting in the critical point $Q_{6c} = -0.07(2)$ and exponent
$\beta=0.87(5)$.}
\label{jq2q6_m2}
\end{figure}

The extrapolated order parameter
in the entire range of QMC accessible $Q_6$ values is shown in Fig.~\ref{jq2q6_m2}. In the case of $Q_6=0$ we have data for much larger system
sizes than in the other cases, and the extrapolated value therefore has smaller error bars than the nearby $Q_6>0$ points. For larger $Q_6$
the extrapolations are more stable, as the corrections to the $L \to \infty$ constant (Fig.~\ref{m2method}) are smaller. However, the simulations also
become more demanding for larger $Q_6$, and we are then limited to smaller systems and larger statistical errors on the raw data.

The squared order parameter in the thermodynamic limit should grow as $\delta^{2\beta}$, here with $\delta=Q_6-Q_{6c}$. In Fig.~\ref{jq2q6_m2} we show a fit
with $\beta$ and $Q_{6c}$ as free parameters, leading to $\beta=0.87(5)$ and $Q_{6c}=-0.07(2)$, the latter value (and error) coinciding with the critical point
extracted from the optimized cumulant collapse in Fig.~\ref{jq2q6binder}. Here we have excluded the two points with largest $Q_6$ values in Fig.~\ref{jq2q6_m2}.
Including one or both of these points, the $\beta$ value increases somewhat (not beyond the error bars) but the goodness of the fit $\chi^2$ becomes
larger than statistically acceptable. In general, the critical form governed by $\beta$ should only apply in the close neighborhood of the transition.
In the case at hand, $m^2$ is very small throughout the range of QMC accessible $Q_6$ values, and one may therefore expect the critical form to still
apply at the largest $Q_6$ values (which it almost does). However, it is possible that the curve $m^2(Q_6)$ for the largest $Q_6$ values begins to signal
the gradual breakdown of the SO($5$) symmetry with increasing $Q_6$.

The exponent $\beta$ is related to scaling dimensions of the critical point according to
\begin{equation}
\beta = \frac{1}{2}(1+\eta)\nu_s = \frac{\Delta_\phi}{3-\Delta_s} = 0.836(8),
\end{equation}
where the numerical value is obtained with our values of $\Delta_\phi$ and $\Delta_s$ in Table \ref{dtable}. Thus, the fitted form in Fig.~\ref{jq2q6_m2}
agrees within one error bar with what is expected under the multicritical scenario. Using the values of the scaling dimensions from the CFT bootstrap calculation
gives $\beta \approx 0.95$. If we instead use our value $\beta=0.87(5)$ and solve for the singlet scaling dimension, we obtain $\Delta_s=2.30(4)$,
which is consistent with our value in Table \ref{dtable} that we will obtain in an alternative way later.

\begin{figure}[t]
\includegraphics[width=80mm]{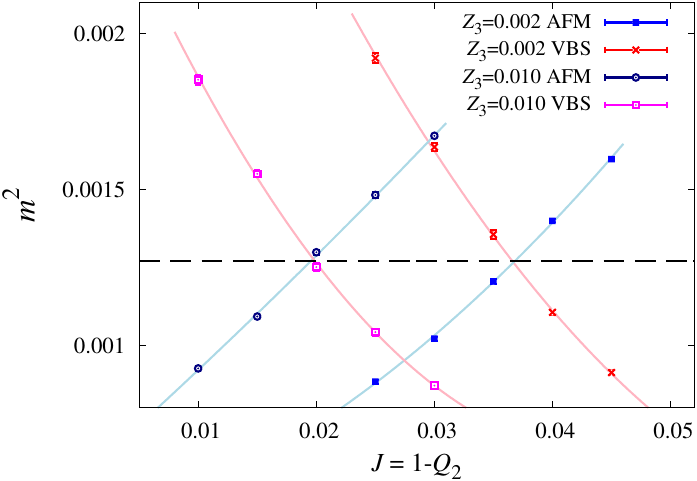}
\caption{The two squared order parameters defined in Eqs.~(\ref{m1defs}) graphed vs the coupling $J$ in the $L=80$ $J$-$Q_2$-$Z_3$ model at $Z_3=0.002$
  and $0.01$, the latter value being situated about halfway up to the maximum value on the phase boundary in Fig.~\ref{jq2q6phases}(b). As shown with the
  dashed horizontal line, the crossing value of the order parameters does not change appreciably with $Z_3$.}
\label{z3cross}
\end{figure}

With $Q_6$ replaced by one of the $Z_n$ interactions, we also expect to see the same kind of evolution of $m^2$ and the Binder cumulant on the first-order line.
However, as shown in Fig.~\ref{z3cross} for the $Z_3$ case, we observe almost no change in $m^2$ when $Z_3$ increases, and this applies also with the $Z_2$
interaction. This insensitivity of the order parameters, even though the phase boundary shifts significantly, may suggest that there is very little overlap of
the $s$ field of the CFT with the $Z_n$ operators (at least for $n=2,3$). It should be noted, however, that the maximum transition values of $Z_2$ and $Z_3$
in Figs.~\ref{jq2q6phases}(a) and \ref{jq2q6phases}(b), i.e., at $J=0$, are small. Therefore, the impact of the $s$ field is also limited trivially by these
small values of $Z_n$ on the first-order lines. In the case of $Q_6$, its maximum transition value is above $0.5$ in Fig.~\ref{jq2q6phases}(c), more than an
order of magnitude larger than the maximum $Z_2$ and $Z_3$ values. The singlet content of the $Q_6$ operator is presumably much larger as well, which we
conjecture to be a consequence of the large spatial extent of this operator.

As a way to compare with the relatively large dependence on $Q_6$ in the $J$-$Q_2$-$Q_6$ model, in Fig.~\ref{jq2q6_m2} the value of $m^2$ changes only
marginally, in absolute terms, from $Q_6=0$ to $0.02$ (the maximum value of $Z_3$), and the changes are most notable for large system sizes (larger
than those we have used for the $J$-$Q_2$-$Q_6$ model with small $Q_6$). This comparison suggests that the combination of less singlet (likely much
less) content in the $Z_n$ operators and the smaller range of the control parameter should be sufficient to explain the differences between the models.
It would still be useful to study $J$-$Q_2$-$Z_n$ models more extensively.

\section{Emergent symmetries}
\label{sec:symm}

The emergence of a symmetry higher than that of the Hamiltonian at a critical point requires that all perturbations $V$ of said symmetry are irrelevant, i.e.,
in the case at hand, have scaling dimensions $\Delta_V$ larger than $d+z=3$. Scaling dimensions of various perturbations can be computed using correlation functions
of corresponding local operators, as in Sec.~\ref{sec:scaledims}. However, if a system already contains an irrelevant perturbation of a given symmetry,
as is essentially guaranteed in the case of emergent symmetries (except at some accidentally fine-tuned point), there is a more direct and efficient
method: investigating the size dependence of a global (system integrated) quantity $V$ that is zero if the symmetry is exact and nonzero in a
finite system because of inherent perturbations. The coarse grained perturbation should decay with increasing size at the rate $L^{-|y_V|}$, where
$y_V=3-\Delta_V$, which is much more favorable for numerical purposes than analyzing a very fast decaying correlation function $C(r) \propto r^{-2\Delta_V}$
with $2\Delta_V > 6$.

This method of exploiting intrinsic perturbations to extract irrelevant scaling dimensions was applied in Ref.~\cite{shao20} to 3D classical clock
models i.e., the XY model with a nonzero $\mathbb{Z}_q$ symmetric field added (for $q \ge 4$), or by restricting the spins to $q$ ``clock'' angles
(for $q\ge 5$). In Sec.~\ref{sub:u1} we use the same method to study the emergent U($1$) symmetry of the VBS order parameter, determining the scaling dimension
$\Delta_4 \approx 3.72$ of the leading lattice-induced $\mathbb{Z}_4$ perturbation. In this case the perturbation is unavoidable, being truly intrinsic
to the model system, in contrast to being added at arbitrary strength to the Hamiltonian as in the clock models.

We further examine the expected relationship between $y_4=3-\Delta_4$ of the deformation of the U($1$) symmetry and the exponent $\mu_4$ governing its
length scale, $\xi'_t \sim t^{-\mu_4}$, in the near-critical VBS phase. We write this relationship for general number of angles $q$, with $q=4$ for
the square-lattice VBS:
\begin{equation}
\mu_q = \nu \left (1+ \frac{|y_q|}{p} \right ).
\label{nuqprime}
\end{equation}
Here $p=2$ in the classical $q$-state clock models \cite{chubukov94,okubo15,shao20} while a crossover from $p=2$ to
$p=3$ was found in a quantum clock model in Ref.~\cite{patil21}. Our results for the $J$-$Q_3$ model (which we use in order to reach sufficiently
deep into the VBS phase) are also consistent with $p=3$, though we cannot exclude $p=2$.

In Sec.~\ref{sub:so5} we study the larger SO($5$) symmetry of the combined five-component AFM and VBS order parameters. Here a key result is that the most
obvious SO($5$) perturbation inherent to the model, i.e., the $\mathbb{Z}_4$ lattice-induced perturbation with scaling dimension $\Delta_4$, is not causing the
same degree of deformation of the mixed distribution $P({\bf D},{\bf M})$ as in $P(D_x,D_y)$. While the perturbation of the U($1$) symmetry of the VBS
is of the form $D_x^2D_y^2$, which contains the rank-$4$ CFT operator with scaling dimension $\Delta_4 \approx 3.72$, the deformation involving both the AFM and
VBS sectors is richer because of the lack of exact symmetry between them. The AFM components of course always retain their exact SO($3$) symmetry and the symmetry
between $D_x$ and $D_y$ is also exact. Therefore the dominant deformation involving both sectors is of the leading form $D^2M^2$, which can still be probed in the
probability distribution of individual components, specifically $P(D_x,M_z)$. We find that the dominant source of deformation here has scaling dimension
$\Delta_{2,3} \approx 4.7$, which we then argue is actually $\Delta_{t'}$, i.e., the main source of the deformation is the next primary in the same symmetry
sector ($\ell=0$, rank-$2$, even parity) as the relevant $t$ operator driving the AFM--VBS transition. The expected contributions from the rank-$4$ operator
are very small, possibly completely absent. In contrast, the rank-$4$ lattice-induced perturbation of the form $D_x^2D_y^2$ causes the leading deformation
of the emergent VBS U($1$) and has scaling dimension $\Delta_{4} \approx 3.7$.

The finite-size scaling of the distribution $P(D_a,M_b)$ also offers a way to extract the exponent $\nu_*$ in Eq.~(\ref{nustarform}) to high precision, which is what
allows us to obtain the best estimate of $\Delta_s$, entered in Table \ref{dtable}. Further, in Sec.~\ref{sub:o5def} we construct an effective O($5$) model and
compare its analogous $D^2M^2$-deformed distribution with $P(D_x,M_z)$, thus showing that the inherent effective interaction in the $J$-$Q$ models must be repulsive.
This fact will be crucial when deriving the pseudocritical exponent $\nu_*$ in Sec.~\ref{sec:dark}.

\subsection{${\mathbb Z}_4$ perturbation and U($1$) crossover}
\label{sub:u1}

Following Ref.~\cite{shao20}, we treat the components $(D_x,D_y)$ of the VBS order parameter in analogy with the magnetization $(m_x,m_y)$ of
the 3D $q$-state clock models. For details of the method and tests of consistency with the conventional approach based on correlation functions, we refer
to Ref.~\cite{shao20}. We use the $J$-$Q_3$ model in the analysis here, because it has a stronger $\mathbb{Z}_4$ deformation at criticality and
can be pushed deeper into the VBS phase. The $J$-$Q_2$ model exhibits very similar behaviors but the statistical errors are larger.

Based on instances $(D_x,D_y)$ of the VBS order parameter generated either in the converged ground state in PQMC simulations or in SSE
simulations at $T \propto L^{-1}$, we define an angular order parameter as
\begin{equation}
  \phi_q = \langle \cos(q\theta)\rangle,
  \label{phiqdef}
\end{equation}
where the angle (which we can always shift to the $[0,\pi/2]$ quadrant because of lattice symmetries) is
\begin{equation}
\theta = \cos^{-1}(|D_x|/r),~~~r=(D^2_x+D^2_y)^{1/2}.
\label{thetadef}
\end{equation}
This angle is not defined if $D_x=D_y=0$, but such QMC configurations become extremely rare with increasing system size, and when
they do occur we simply do not include them in the averaging. In principle we could avoid this issue altogether by instead of Eq.~(\ref{phiqdef}) using,
e.g., $\langle r\cos(q\theta)\rangle/\langle r\rangle$, or with some other power of $r$. Here we follow Ref.~\cite{shao20} and obtain good results
with Eq.~(\ref{phiqdef}).

In a system with perfectly U($1$) symmetric order parameter, clearly $\phi_q=0$ for all $q$. In the 3D classical $q$-state clock models
the clock perturbation is irrelevant for $q\ge 4$, and the scaling dimension $\Delta_q$ increases
with increasing $q$. In a quantum magnet on the square lattice, the lattice itself induces a perturbation of a presumed U($1$) symmetric coarse-grained VBS
order parameter corresponding to $q=4$ by the fact that the local singlets form on the lattice links with coordination number $4$. When the angle $\theta$
is defined for the VBS order parameter on the entire lattice, the coarse-graining scale is the lattice length $L$, and we therefore expect scaling of the
$q=4$ angular order parameter of the form $\phi_4 \sim L^{-|y_4|}$, with $y_4=3-\Delta_4 < 0$. This is the scaling dimension that we target here.

In our PQMC simulations, we use the definition
\begin{equation}
D_e=\sum_{i=1}^N (-1)^{e_i} {\bf S}_i \cdot {\bf S}_{e(i)},~~e\in \{x,y\}
\label{dxdydef1}
\end{equation}
of the components of the VBS order parameter,
where $e(i)$ denotes the nearest neighbor of site $i$ in the $x$ or $y$ lattice direction, and $e_i$ is either the $x$ or $y$ coordinate of $i$.
The off-diagonal quantity ${\bf S}_i \cdot {\bf S}_{j}$ for any $i,j$ has a simple expression in the valence-bond basis used, being simply $-3/4$ or $0$,
for sites $i$ and $j$ being, respectively, in the same loop or different loops in the transition graph representing overlap of the bra and ket states
propagated by a sampled string of operators drawn from the Hamiltonian (Sec.~\ref{sub:qmc}) \cite{sandvik10c}.
In SSE simulations, we instead use the diagonal definition, Eq.~(\ref{dxydef}), in the basis of $S^z$ spins, taken at any ``slice'' in the periodic propagation
dimension (which is directly related to imaginary time \cite{sandvik92}). The corresponding angle $\theta$ is then extracted in each case and the
average $\phi_4$ is taken over slices.

We can also average all $S^z_i$ over imaginary time first and define $(D_x,D_y)$ using such averages $\bar S^z_i$, which in SSE correspond exactly to
imaginary time integrals;
\begin{equation}
\bar S^z_i = \frac{1}{\beta}\int_0^\beta d\tau S^z_i(\tau).
\label{tauint1}
\end{equation}
We then use $\bar S^z_i$ in Eq.~(\ref{dxydef}) to define the instance of the order parameter $(D_x,D_y)$ from which $\theta$ is extracted according
to Eq.~(\ref{thetadef}), which corresponds to a full space-time definition of the global angle. The scaling dimension extracted from Eq.~(\ref{phiqdef})
should be the same as with the space-only definition.

Yet another way to define $\theta$ in the angular order parameter is to use $Q_n$ operators to define $(D_x,D_y)$. The $Q_n$ operator summed over space
with the appropriate $(\pi,0)$ or $(0,\pi)$ phase detects the VBS order parameter as discussed in the context of the $Q$-$Q$ correlation function
in Sec.~\ref{sec:dark}. For the present purpose we use a space-time definition, because of the simplicity within the SSE approach of time integrals
of operators in the Hamiltonian---they are obtained by counting operators in the sampled operator strings \cite{sandvik92,sandvik96}.
Thus, for a given SSE configuration, we can compute
\begin{equation}
\bar O_i = \frac{1}{\beta}\int_0^\beta d\tau O_i(\tau),
\label{tauint2}
\end{equation}
where the operators $O_i$ of interest here are the local $Q$ interactions ($Q_2$, $Q_3$, etc., depending on the model considered). Since we here use
the $J$-$Q_3$ model, we should also take into account the lattice orientation of a given operator. Therefore, to evaluate this definition of
$D_e$, $e \in \{x,y\}$, we count only the set $q_e$ of $Q_3$ operators with $e$-oriented singlet projectors;
\begin{equation}
D_e=\sum_{i \in q_e} (-1)^{e_i} \bar Q_{3,i} ,~~e\in \{x,y\}.
\label{dxdydef3}
\end{equation}
Then again we extract the angle $\theta$ corresponding to $(D_x,D_y)$ and use it in Eq.~(\ref{phiqdef}).

\begin{figure}[t]
\includegraphics[width=80mm]{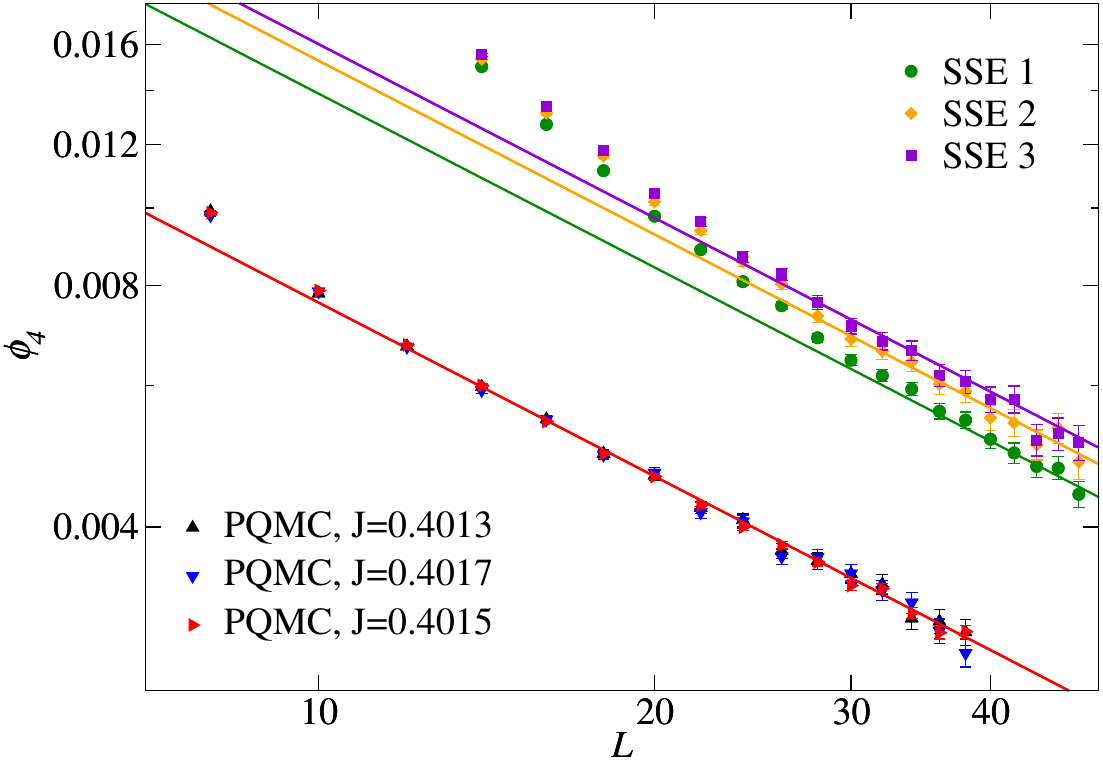}
\caption{Size dependence of the critical angular order parameter $\phi_4$ of the $J$-$Q_3$ model (with $J+Q_3=1$). PQMC results based on the off-diagonal
definition Eq.~(\ref{dxdydef1}) of the VBS order parameter are shown for three different couplings close to the AFM--VBS transition. The results are
so similar that the $J=0.4015$ points cover the other results for many of the system sizes, and where the other points are seen they are still statistically
indistinguishable. SSE results based on the three different definitions discussed in the text are shown for $J=0.4015$, with the legends corresponding to:
1) the diagonal definition of the VBS order parameter Eq.~(\ref{dxydef}), 2) the same with $S^z_i$ replaced by the
time average $\bar S^z_i$ in Eq.~(\ref{tauint1}), and 3) with the $Q_3$ operator time average, Eq.~(\ref{dxdydef3}). The red line is a fit to a power law
of the $J=0.4015$ PQMC data, delivering the exponent $y_4=-0.723(11)$. The line drawn through the SSE data points has the same slope.}
\label{phi4}
\end{figure}

\subsubsection{Scaling dimension $\Delta_4$}

Results for the size dependence of $\phi_4$ with all four definitions of $\theta$ are shown in Fig.~\ref{phi4}. We include PQMC data for three different
couplings close to the transition point of the $J$-$Q_3$ model, in order to demonstrate the insensitivity of the results to small variations of $J$ close
to $J_c$; among these, $J_c$ is closest to $J=0.4015$, which is also the coupling used in the SSE simulations. The PQMC results show very good power law scaling
already from $L=14$, and taking $L=16$ as the smallest size in a fit we obtain $y_4=-0.723(11)$, corresponding to the value $\Delta_4=3.723(11)$ entered in
Table \ref{dtable}. The SSE results with the three different $\theta$ definitions are fully consistent with the above value of $y_4$, but the finite-size
corrections are much larger for the smaller systems. In this case we therefore do not carry out independent fitting but just show consistency with the same
asymptotic power-law form for the larger system sizes. The presumed leading VBS deforming lattice operator $D_x^2D_y^2$ also contains rank-$2$ contributions
in addition to the leading rank-$4$ operator; its U($1$) charge-$2$ nematic component $D_x^2-D_y^2$ is forbidden because of the exact $D_x \leftrightarrow D_y$
(i.e., lattice $C_4$) symmetry. Moreover, the charge-$0$ contributions are not detected by $\phi_4$. The visible corrections in Fig.~\ref{phi4}, which we do not
analyze further here, must therefore have some other origin.

Our value of $\Delta_4$ is somewhat smaller than the CFT and fuzzy sphere results in Table \ref{dtable}, though here again we point out that the
bootstrap results depend to some extent on the exact value of the input parameter $\Delta_\phi$. In the case of the fuzzy sphere, the results are likely
not yet fully converged in the number of orbitals included (or possibly the coupling used may not be located exactly at its critical value).

\subsubsection{Emergent U($1$) scale in the VBS phase}
\label{sec:emergentu1}

While the ${\mathbb Z}_4$ perturbation is irrelevant at the critical point, it becomes relevant inside the VBS phase; the ground state is a four-fold degenerate
columnar pattern with tunneling paths corresponding to ${\mathbb Z}_4$ symmetry for finite systems (see Ref.~\cite{takahashi20} for a discussion of the
role of tunneling paths between the degenerate states) \cite{levin04}. Such a perturbation is often referred to as dangerously irrelevant \cite{amit82}.

The length scale on which the emergent U($1$) symmetry is manifested inside the VBS phase should diverge faster than the conventional correlation length
on approach to the critical point \cite{senthil04a,senthil04b,levin04}. The exponent $\mu_q$ governing this U($1$) scale has been the subject of much work,
and some controversy, in the context of the 3D clock models \cite{oshikawa00,lou07,okubo15,leonard15}. There is now overwhelming numerical evidence
\cite{okubo15,shao20} for Eq.~(\ref{nuqprime}) with $p=2$, a relationship first (to our knowledge) derived in Appendix B of Ref.~\cite{chubukov94}.
The derivation of this relationship involves the linearly dispersing gapless Goldstone mode of the U($1$) phase that the system flows toward
(in a RG sense) before the clock perturbation becomes effectively large and brings the system to the ultimately $\mathbb{Z}_q$ breaking fixed point. While
$p=2$ works well for the classical 3D clock models, results for a 2D quantum clock model suggested a crossover from $p=2$ to $p=3$ with increasing systems
size \cite{patil21}. The value $p=3$ had also previously been argued for in the case of the clock models \cite{lou07}. In
Ref.~\cite{patil21} the crossover to $p=3$ was also demonstrated in 3D classical clock models on lattices with $L\times L\times L_z$
spins when $L_z \gg L$, which in the quantum to classical mapping corresponds to the ground state of the 2D quantum clock model.

Like in the clock models, we expect Eq.~(\ref{nuqprime}) with $\nu \to \nu_t$ to hold also in the VBS phase of the $J$-$Q_n$ models close to the
critical point (i.e., for values of $n$ for which the VBS--AFM transition is near critical). Thus, there should be an exponent $\mu_4$ related to $\nu_t$
(the relevant correlation length exponent when tuning the system into the VBS phase) and $y_4=3-\Delta_4$ according to Eq.~(\ref{nuqprime}). We next test
this relationship with both $p=2$ and $p=3$.

\begin{figure}[t]
\includegraphics[width=70mm]{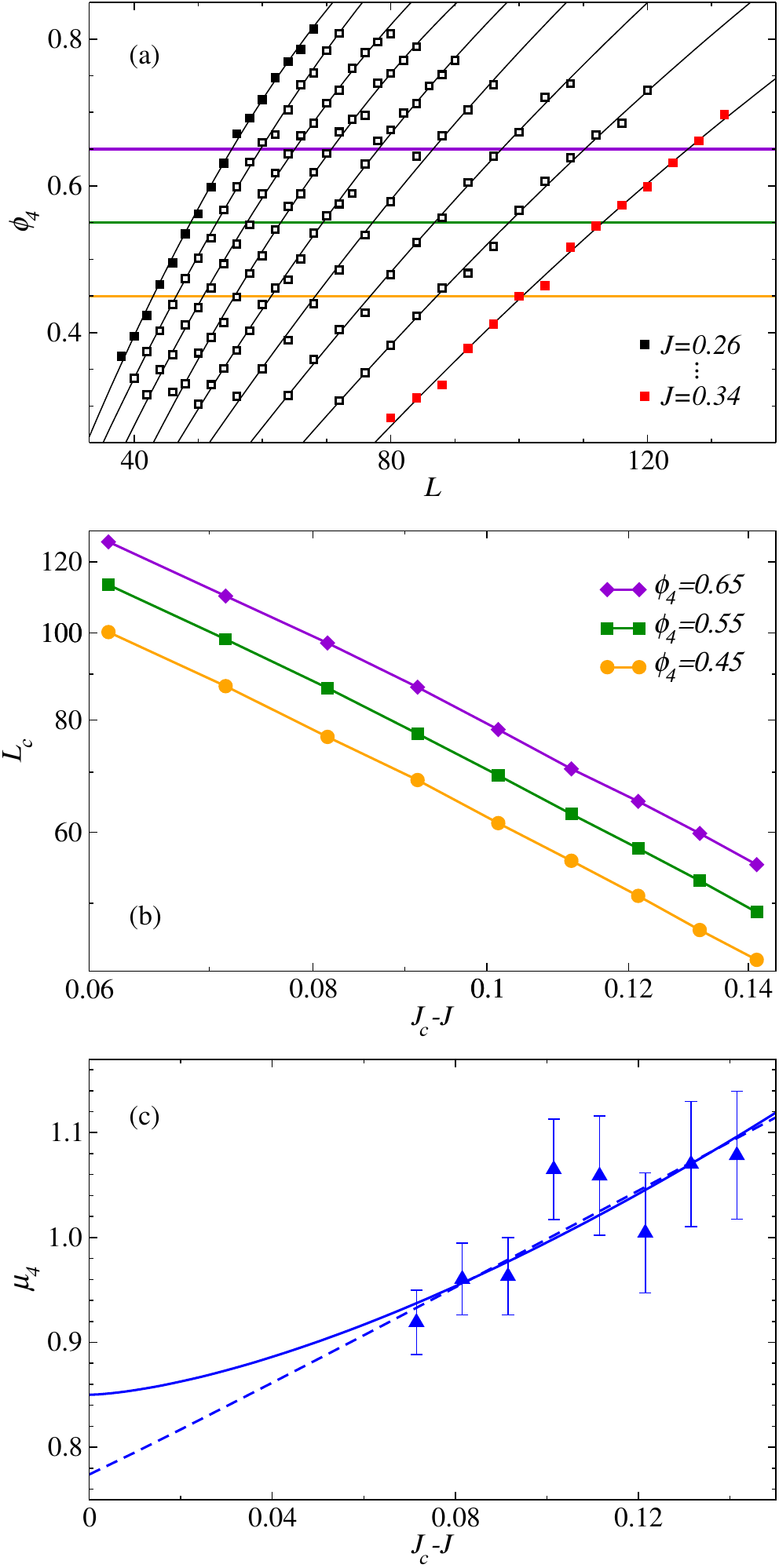}
\caption{(a) SSE results ($\beta=L$) for the size dependence of $\phi_4$ (using $\tau$-integrated $Q$ operator definition) for several values of $J$ inside the VBS
phase of the $J$-$Q_3$ model, with $J+Q_3=1$. The horizontal lines are used to define crossing sizes $L_c(J)$ for three different values of $\phi_4$. (b) The
crossing sizes vs the distance to the critical value $J_c \approx 0.4015$ extracted from the three lines in (a). (c) Local exponent extracted
from pairs of points at adjacent $J$ values in (b), using the scaling form $L_c \propto (J-J_c)^{-\mu_4}$. The asymptotic value of $\mu_4$ should be given by
Eq.~(\ref{nuqprime}) with $\nu_t=(3-\Delta_t)^{-1}$ from Table \ref{dtable}, $|y_4| \approx 0.72$ from the analysis in Fig.~\ref{phi4}, and with either
$p=2$ or $p=3$. The data points have therefore been fitted to $\mu_4(L) = \mu_4(\infty) + (J-J_c)^{\omega}$ with two values of the asymptotic exponent;
$\mu_4(\infty)=0.85$ and $\mu_4(\infty)=0.77$, corresponding, respectively, to $p=2$ and $p=3$ in Eq.~(\ref{nuqprime}).
The exponent $\omega$ reflects a scaling correction.}
\label{nu4}
\end{figure}

SSE and PQMC calculations again deliver consistent results; here we show SSE results for the definition of the VBS order parameter used to extract $\theta$
based on the $\tau$-integrated $Q$ operators, Eq.~(\ref{tauint2}). Fig.~\ref{nu4}(a) shows $\phi_4$ versus the system size for several values of the coupling $J$
of the $J$-$Q_3$ model in the regime where $\phi_4$ is already approaching $1$. Its scaling dimension is therefore $0$, which is where Eq.~(\ref{nuqprime})
should apply \cite{okubo15,shao20}. From these data, we can extract a length scale versus $J$ by finding the intersection point with a line at fixed
$\phi_4$, with three such lines drawn in Fig.~\ref{nu4}(a). The resulting length scales $L_c$ corresponding to the three lines are graphed versus $J-J_c$
in Fig.~\ref{nu4}(b). Here the behavior is not yet the ultimately expected pure power law, $L_c \sim L^{-\mu_4}$. To study the evolution of $\mu_4$ systematically,
we therefore define a size-dependent exponent by assuming the power-law form to apply between adjacent-$J$ points (which are equally spaced in $J$) in
Fig.~\ref{nu4}(b). The result for this floating exponent, where we have averaged results corresponding to all three fixed-$\phi_4$ lines, is graphed
versus $J-J_c$ in Fig.~\ref{nu4}(c).

According to Eq.~(\ref{nuqprime}) with $\nu \to \nu_t \approx 0.63$ from Table \ref{dtable} and $|y_4| \approx 0.72$ extracted above, we should have
$\mu_4 \approx 0.85$ if $p=2$ and $\mu_4 \approx 0.77$ if $p=3$. To check the consistency with these values we simply assume it to be asymptotically correct
and fit the data with a single adjustable power law correction in Fig.~\ref{nu4}(c). Even without such fits, the results are clearly statistically consistent
with Eq.~(\ref{nuqprime}) with either $p=2$ or $p=3$, given the large error bars. An independent reliable fit to the data with a constant plus power-law
correction is not meaningful. The fit is marginally better with $p=3$ but not enough so to rule out $p=2$.

The way the data approach $\mu_4$ in the expected range is reassuring, and there is no reason to doubt that emergent U($1$) symmetry crossover in the
near-critical VBS phase is governed by the same mechanisms as the emergent symmetry in the classical 3D clock models, only with different values of the exponents
involved. It would be interesting to improve the results further by going to larger system sizes and reducing the error bars. These calculations are very
time consuming, however.

Note that $y_4 \approx -0.11$ in the $q=4$ clock model \cite{shao20}, i.e., this clock perturbation is much closer to being relevant than the lattice deformation
of the U($1$) symmetry at the AFM--VBS transition. Most of the works on the clock models were done for $q=6$ \cite{shao20,patil21}, where $|y_4|$ is larger and all
the crossovers can be studied in greater detail.

Here we also point out that studies of $J$-$Q$ models on the honeycomb lattice, with coordination number and VBS degeneracy $q=3$, have indicated that there
are stronger violations of the U($1$) symmetry on this lattice \cite{pujari13,pujari15}. The quantity $\phi_3$ does not decay to zero but appears to approach
a constant, which would imply that the perturbation is marginal. It is also possible that the $\mathbb{Z}_3$ perturbation is weakly relevant, so that larger
lattices would be required to observe a clearly growing $\phi_3$. In the presumed SO($5$) CFT, the $q=3$ perturbation is indeed clearly relevant
\cite{nakayama16,chester24,zhou23}, but recent QMC calculations of the entanglement entropy have nevertheless demonstrated consistency with the SO($5$)
CFT expectations \cite{kumar26}.

\subsection{SO$(5)$ $\to$ SO$(3)$ $\times$ U$(1)$ perturbation}
\label{sub:so5}

The U($1$) symmetry of the VBS order parameter is just one aspect of the emergence of the higher SO($5$) symmetry. We next study another inherent deformation
of the SO($5$) sphere, using quantities involving one component each of the AFM and VBS order parameters in order to characterize the asymmetry between them.
We choose the $z$ spin component $M_z$ of the staggered magnetization and one of the components, $D_x$ or $D_y$, of the VBS order parameter and use these to test
for SO($5$) symmetry in the full five-component vector $(M_x,M_y,M_z,D_x,D_y)$.
The diagonal $M_z$ AFM order parameter was defined in Eq.~(\ref{mzdef}). For the VBS order parameter, we again use the definition in Eq.~(\ref{dxydef})
and Eq.~(\ref{dxdydef1}) with the PQMC and SSE methods, respectively. We can then extract an angle $\theta$ as in Eq.~(\ref{thetadef}) based on the components
$(M_z,D_x)$ or $(M_z,D_y)$ and study the angular order parameters $\phi_q$ in Eq.~(\ref{phiqdef}) for different values of $q$. However, to stay closer to previous
works studying emergent SO($5$) symmetry, we will here proceed with other related quantities.

Nahum et al.~\cite{nahum15b} considered generalizations of $\phi_q$:
\begin{equation}
\phi_{q,a}=\langle r^a\cos(q\theta)\rangle,
\label{phila}
\end{equation}
where $\theta$ is the angle between an AFM and a VBS component and $r$ is a normalized length of the vector that in our notation has the components
$M_z/\langle M_z^2\rangle^{1/2}$ and $D_x/\langle D_x^2\rangle^{1/2}$. They noted that, for $a=4$, these expressions can be converted into simple forms
only involving expectation values of powers of the order parameters; specifically, for $q=2$ and $q=4$ they used
\begin{subequations}\label{f2f4def}
\begin{eqnarray}
  F_2 & = & \frac{\langle M_z^4\rangle}{\langle M_z^2\rangle^2} - \frac{\langle D_x^4\rangle}{\langle D_x^2\rangle^2}, \label{f2def} \\
  F_4 & = & \frac{\langle M_z^4\rangle}{\langle M_z^2\rangle^2} + \frac{\langle D_x^4\rangle}{\langle D_x^2\rangle^2}
   - 6\frac{\langle M_z^2D_x^2\rangle}{\langle M_z^2\rangle\langle D_x^2\rangle}, \label{f4def}
\end{eqnarray}
\end{subequations}
which are equivalent to $\phi_{4,2}$ and $\phi_{4,4}$, respectively, in Eq.~(\ref{phila}). Both these quantities must vanish if the system has exact
SO($5$) [or O($5$)] symmetry, and the nonzero values inside the SO($3$) symmetric AFM phase and the ${\mathbb Z}_4$ or U($1$) symmetric (depending
on the length scale) VBS phase can also be easily computed.

We here consider the $J$-$Q_2$, $J$-$Q_3$, and $J$-$Q_4$ models only at and very close to their transition points but have also confirmed the correct
asymptotic values of $F_2$ and $F_4$ inside the phases. In the expressions above we can substitute $D_x$ for $D_y$, but when considering a single component
of the staggered magnetization we only have diagonal access to $M_z$ in both SSE and PQMC simulations (though we also have access to $M^2$ and $M^4$
through loop estimators, which we will not make use of here).

Nahum et al.~showed that the 3D loop model has very small $F_2$ and $F_4$ values at the transition point, vanishing within statistical errors that were
relatively large for large systems \cite{nahum15a}. However, size dependences were not studied in detail. Here we use the size and coupling
dependence to gain further insights into the emergence and eventual violation of SO($5$) symmetry.

\begin{figure}[t]
\includegraphics[width=80mm]{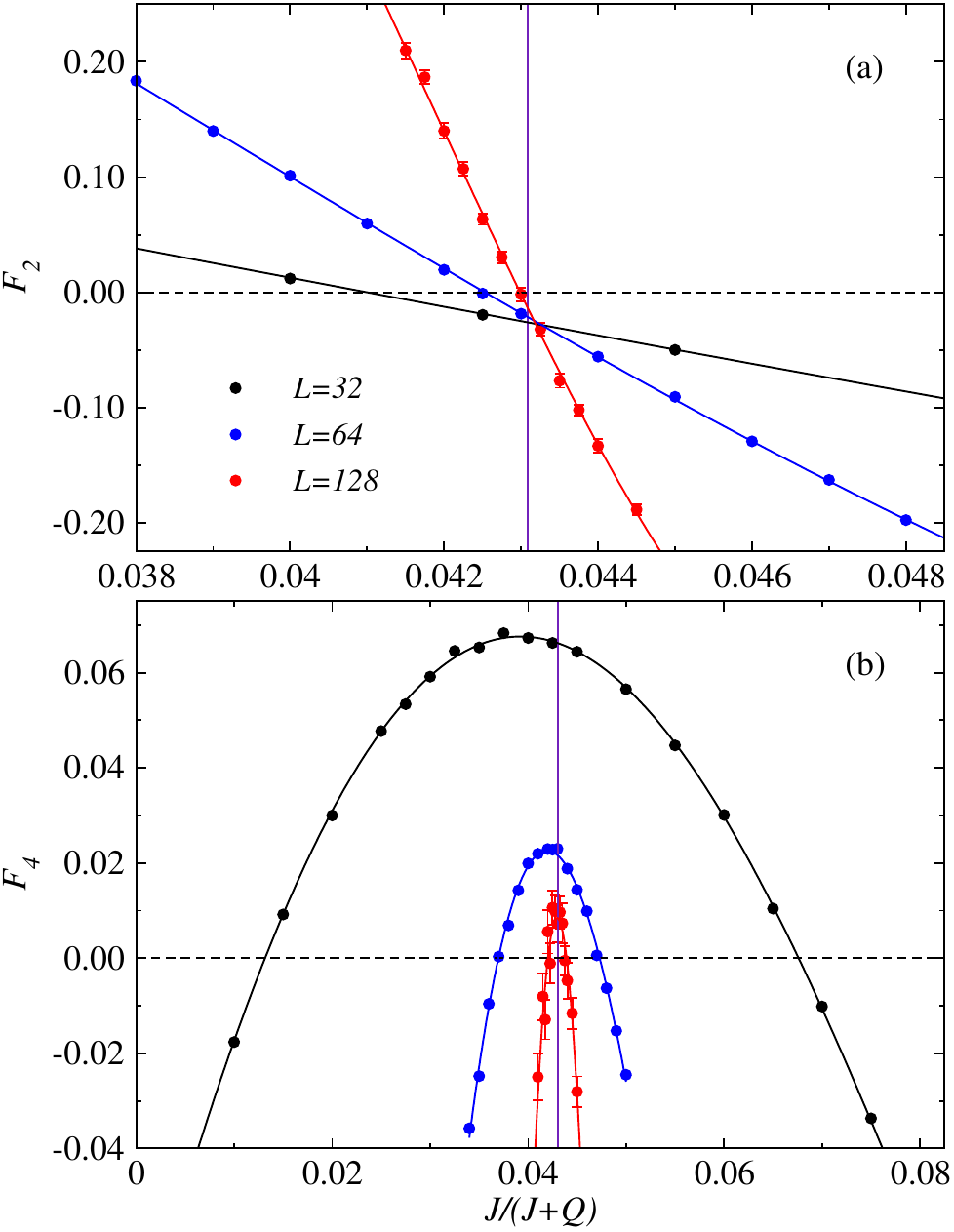}
\caption{The SO($5$) diagnostic functions $F_2$ and $F_4$, Eqs.~(\ref{f2f4def}), computed with SSE simulations ($\beta=L$, $J+Q_2=1$) vs the coupling ratio $J/(J+Q_2)$.
Results are shown for three different system sizes to illustrate how $F_2$ crosses $0$ close to the transition while $F_4$ exhibits a nonmonotonic behavior,
crossing $0$ twice. The curves show polynomial fits. The interpolated $J$ values at which $F_2=0$ and $F_4=0$ are graphed vs the inverse
system size in Fig.~\ref{zeros}.}
\label{f2f4}
\end{figure}

\begin{figure}[t]
\includegraphics[width=80mm]{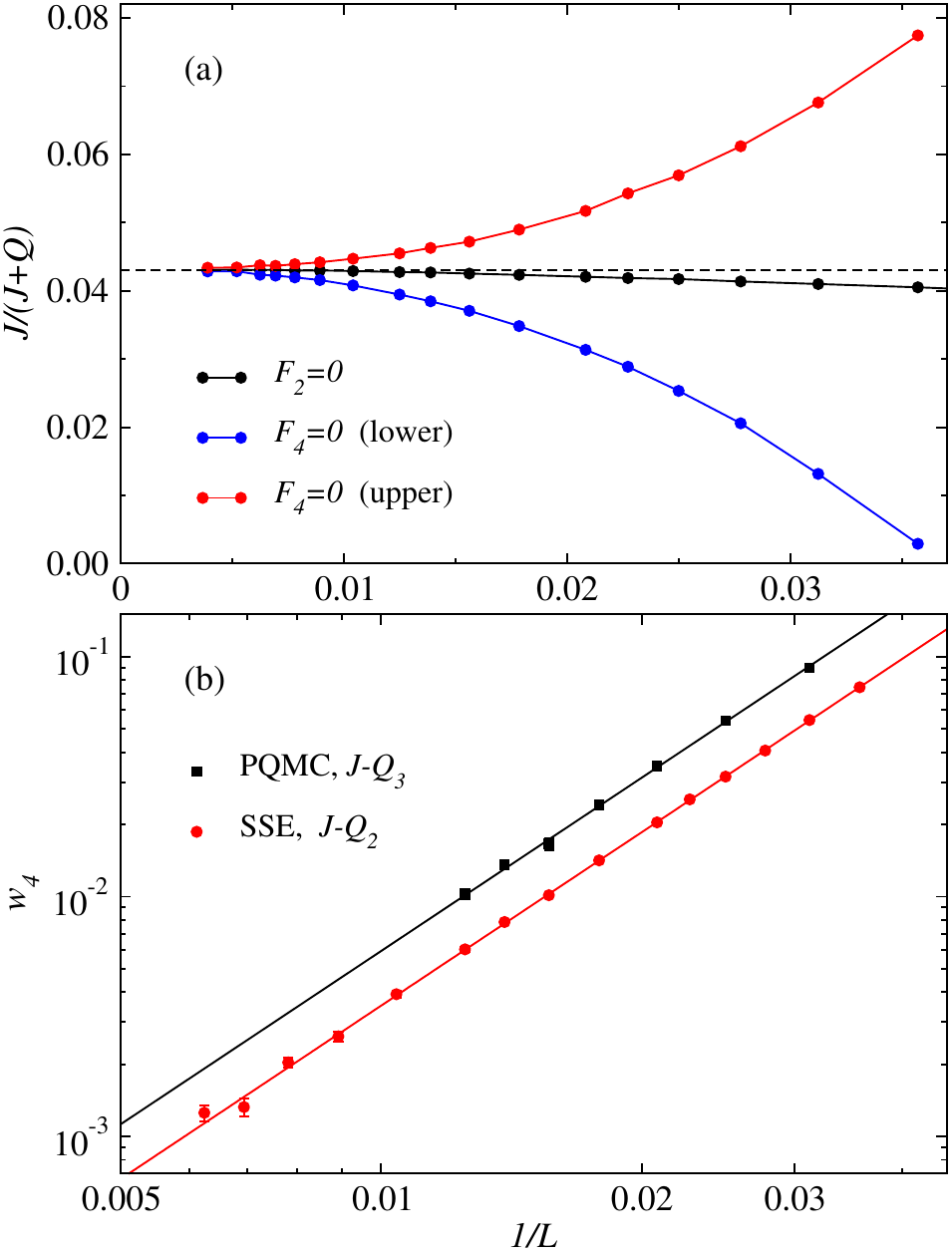}
\caption{(a) Location of the zeros of the functions $F_2$ and $F_4$, extracted by fitting polynomials to SSE data such as those in Fig.~\ref{f2f4} of
the $J$-$Q_2$ model. The horizontal dashed line shows the location of the extrapolated AFM--VBS transition---at $J_c/Q_2=0.04502$, here
rescaled to $J_c/(J_c+Q_2)$ pertaining to these simulations carried out with $J+Q_2=1$. (b) Log-log plot of the size of the window between
the upper and lower $F_4=0$ points are shown with red circles for the $J$-$Q_2$ model in (a). The black squares show analogous results for
the $J$-$Q_3$ model obtained with PQMC simulations. The straight lines show the scaling form $w_4 \sim L^{-1/\nu_*}$ with $1/\nu_* = 2.402$.}
\label{zeros}
\end{figure}

Figure \ref{f2f4} shows results for both $F_2$ and $F_4$ in the $J$-$Q_2$ model for three different lattice sizes. In the AFM phase,
for $L \to \infty$ we must have $F_2 \to -6/5$, while in the VBS phase $F_2 \to 2$ because of the ${\mathbb Z}_4$ symmetry of $(D_x,D_y)$ and there is an
intermediate scale at which $F_2 \to 3/2$ because of the emergent U($1$) symmetry \cite{nahum15b}. Thus, we expect $F_2$ to change signs close to
the transition point, eventually approaching $0$ at the transition if there is emergent SO($5$) symmetry. The point at which $0$ is crossed
indeed moves toward the transition point in Fig.~\ref{f2f4}(a), which is shown in detail in Fig.~\ref{zeros}(a) with results for $L$ up to $256$. The quantity
$F_4$ is negative inside both phases and, as seen in Fig.~\ref{f2f4}(b), is positive within a window of couplings that shrinks as $L$ increases,
as also demonstrated explicitly in Fig.~\ref{zeros}(a). Note that the $F_4=0$ points always bracket the $F_2=0$ point. Fig.~\ref{zeros}(b)
shows how the bracket shrinks as a power law with increasing system size; we will discuss the meaning of the corresponding exponent
further below.

\begin{figure}[t]
\includegraphics[width=80mm]{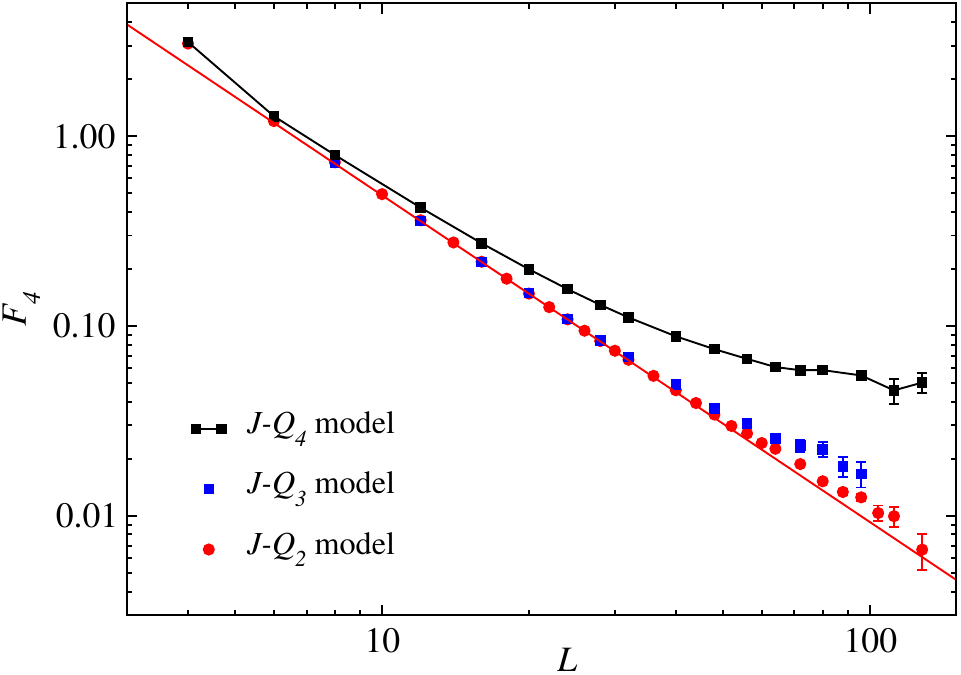}
\caption{Size dependence of $F_4$ computed with PQMC simulations in the near-critical $J$-$Q_2$, $J$-$Q_3$, and $J$-$Q_4$ models. The
line drawn through the $J$-$Q_2$ data for $L \alt 50$ has slope $-1.71$, corresponding to a scaling dimension $\Delta_{2,3} \equiv \Delta_{t'} \approx 4.71$. Deviations
from the critical form for the largest systems (essentially all sizes in the case of the $J$-$Q_4$ model) indicate the length scale at which the
emergence of SO($5$) symmetry is affected by the first-order transition, or, possibly, initial crossover to a decay governed by $y_4 \approx -0.72$. Note
that the $J$-$Q_3$ points completely cover many of the $J$-$Q_2$ points because of the close coincidence of the values.}
\label{jq234_f4}
\end{figure}

In analogy with the quantity $\phi_4$ studied in Sec.~\ref{sub:u1}, $F_4$ at the transition point is a quantitative measure of the violation of the
SO($5$) symmetry (provided that this symmetry really is emergent) at a scale given by the system size $L$. Thus, $F_4$ should scale as $L^{-|y|}$,
where $y=3-\Delta$, with $\Delta$ the smallest of scaling dimensions of the inherent irrelevant SO($5$) perturbations visible to the $F_4$ definition.
At first sight, one might expect $\Delta = \Delta_4$, but, as shown in Fig.~\ref{jq234_f4}, the exponent governing $F_4$ of the $J$-$Q_2$ model is much larger
in magnitude; $y_{2,3} \approx -1.71 \approx y_4-1$ (where the subscripts $2,3$ refer to the number of components of the two order parameters) for system sizes
from $L=6$ to $L \approx 60$. The rather small deviations from the fitted line occur for slightly smaller system sizes for the $J$-$Q_3$ model, and for the
$J$-$Q_4$ model there is no range of $L$ with power-law decay at all, only a gradual flattening out. Knowing that the first-order discontinuities of the
AFM--VBS transition become stronger for $J$-$Q_n$ models as $n$ increases, the large-$L$ behaviors of the $J$-$Q_2$ and $J$-$Q_3$ data in Fig.~\ref{jq234_f4}
are plausibly explained as the initial, barely detectable stage of the breakdown of the SO($5$) symmetry to ${\rm SO}(3)\times{\rm U}(1)$ and finally to
${\rm SO}(3)\times \mathbb{Z}_4$, as expected in the coexistence state. There could also in principle be some effects of crossover from behavior $y_{2,3}$
to $y_4$ dominance, but it is clearly not possible to determine the exact source of the small (especially for the $J$-$Q_2$ model) deviations from
$y_{2,3}$ scaling in Fig.~\ref{jq234_f4} just based on the data.

To understand the at first sight unexpected behavior of $F_4$, note that the SO($3$) symmetry of the AFM order parameter is exact and likewise the symmetry between
$D_x$ and $D_y$ is exact, whether in the regime of emergent U($1$) symmetry or just $\mathbb{Z}_4$. Therefore all perturbations of the form $\lambda_{ab}M^2_aD^2_b$
with $a \in \{x,y,z\}$ and $b \in \{x,y\}$ must share the same value of the coupling $\lambda_{ab}$. Thus, the dominant intrinsic SO($5$) perturbing interaction involving
both order parameters is really of the form $M^2D^2$, which is different from the $D^2_xD^2_y$ perturbation that is detected by the VBS distortion measure $\phi_4$.
We therefore have to consider the possible CFT operator content of $M^2D^2$ more carefully and compare with that of $D^2_xD^2_y$.

Both interactions are quartic in the number of order parameters and, therefore, contain $\ell=0$, parity-even CFT operators of rank-$0$ (singlet), rank-$2$, and
rank-$4$. The singlet corresponds to the $s$ field with scaling dimension $\Delta_s$ and, by definition, does not deform the SO($5$) sphere. The VBS symmetry indicator
$\phi_4$ explicitly detects the effects of a U($1$) charge-$4$, rank-$4$ operator, which should also in principle be present in $M^2D^2$. However, its overlap with
the intrinsic deformations in the model must be very small, to explain the much faster decay with exponent $y=y_{2,3}$. This dominant perturbation must then originate
from a rank-$2$ contribution, at least for system sizes up to $L \approx 60$ (and likely beyond, as we will argue below). One operator in the symmetry sector in
question is the $t$ field tuned by $J/Q_n$ (with scaling dimension $\Delta_t \approx 1.42$), which vanishes at the critical point. We then conclude that the $F_4$ decay
is dominated by the second primary operator $t'$ in the same rank-$2$ sector as $t$, with the observed scaling dimension $\Delta_{t'} \equiv \Delta_{2,3} \approx 4.71$.
In the fuzzy-sphere approach, Zhou et al.~indeed identified a primary CFT operator with matching quantum numbers and scaling dimension $\Delta_{t'} \approx 4.35$
\cite{zhou23}. Although the values do not agree perfectly, considering the significant drifts of some of the scaling dimensions with the number $n_f$ of orbitals in
the fuzzy sphere, our symmetry arguments strongly support this matching of the two calculations and our improved value of the scaling dimension $\Delta_{t'}$.

The results in Fig.~\ref{jq234_f4} were all obtained with PQMC simulations very close to the transition point, but SSE results for the $J$-$Q_2$ model, like those
in Fig.~\ref{f2f4} interpolated to the transition point, fall almost exactly on top of the PQMC results (with larger error bars). Because of the finite-size drifts,
one might question the reliability of the value of $y_{t'}$. The statistical errors are, however, very small for system sizes up to $L\approx 50$, and the extracted
exponent from statistically good fits in the range from $L=16$ to $32$ (to minimize corrections both for small and large $L$) is stable with respect to exactly
which system sizes are included. Our final estimate from this analysis is $y_{t'} = -1.711(4)$.

The corresponding dangerously irrelevant length-scale of SO($5$) symmetry in the coexistence state should be given by Eq.~(\ref{nuqprime}), now with
$\nu \to \nu_s$ and $|y_q| \to |y_{t'}|$. Using the scaling dimensions in Table \ref{dtable}, we obtain a very large exponent, $\mu_s \approx 2.8$,
for the exponent governing the divergence of the SO($5$) length scale upon approaching the critical point. In principle we could test this value in
the same way as we tested the expected value of $\mu_4$ in Sec.~\ref{sub:u1}, by monitoring $F_4$ versus the system size for points on the coexistence
line that we determined in the phase diagram of the $J$-$Q_2$-$Q_6$ model in Sec.~\ref{sec:jq2q6}. However, such a program is currently computationally
too expensive and we have not pursued it.

Though eventually $F_4$ may be governed by $y_4$, we assume that $y_{t'}$ is to be used in Eq.~(\ref{nuqprime}), and also in the $\nu_*$ relationship
Eq.~(\ref{nustarform}), either because the corresponding intrinsic perturbation is very small and does not come into play before the ultimate onset of the
first-order behavior with convergent length scales, or possibly its amplitude completely vanishes. One would indeed expect operators of higher rank to have
smaller overlap with ``smooth'' intrinsic perturbations, which would be a generic explanation for $F_4$ to be dominated by the rank-$2$ operator up to some
length scale. We will see in the next section that the length scale on which the rank-$4$ operator possibly takes over must be much larger than the size
$L \approx 60$, where small deviations from the fit to the $J$-$Q_2$ data first become apparent in Fig.~\ref{jq234_f4}.

\subsection{Length scale of SO($5$) perturbation}
\label{sub:so5w}

Returning to the $J$-$Q_2$ data in Fig.~\ref{f2f4}, we next investigate the zeros of the $F_2$ and $F_4$ functions in Fig.~\ref{zeros}(a) more carefully.
Here we observe that the $J$ point where $F_2(J,L)=0$ moves toward the previously extracted transition point with increasing $L$, and this point is
always (for any $L$ studied) between the two points where $F_4(J,L)=0$. We have already seen above that $F_4 \to 0$ at the infinite-size critical
point (though in the models, which are not exactly at the critical point, $F_4$ will eventually flow away from $0$ for large system sizes), and we have
also confirmed that $F_4$ at its $L$ dependent peak exhibits the same scaling governed by the exponent $y_{t'} \approx -1.71$. Thus $F_4 \to 0$
as $L$ grows in the entire region between the couplings where the function vanishes. This behavior is of course qualitatively necessary for
there to be emergent SO($5$) symmetry at the critical point, and the above results reinforce the notion that such a critical point is almost realized
in the $J$-$Q$ models in the QMC accessible part of their phase diagrams.

The region between the zeros of $F_4(J,L)$ can be taken as a practical definition of the finite-size region where the system hosts an emergent SO($5$)
symmetry that is gradually violated when moving out from the transition point, which can be taken as the point where $F_2=0$ or the peak location of $F_4$.
An important quantitative question is then how this region shrinks with increasing $L$. The width $w_4$ of the SO($5$) region defined by the zeros of $F_4$
is graphed versus the inverse system size in a log-log plot in Fig.~\ref{zeros}(b) for both the $J$-$Q_2$ and $J$-$Q_3$ models. In both models we observe the
same kind of decay by a single power law, $w_4 \sim L^{-a}$, with apparently very small corrections. From the $J$-$Q_2$ results we can extract the exponent
$a=2.402(6)$, which is a stable value with respect to the number of points included in the fit. Here we point out that the smallest system size included
in Fig.~\ref{zeros}(b) is the first size for which the two zeros of $F_4$ are both within the QMC accessible region $J \ge 0$ (as is in Fig.~\ref{f2f4}).
The fit to the single power law is of good statistical quality already starting from this $L$.

\begin{figure}[t]
\includegraphics[width=75mm]{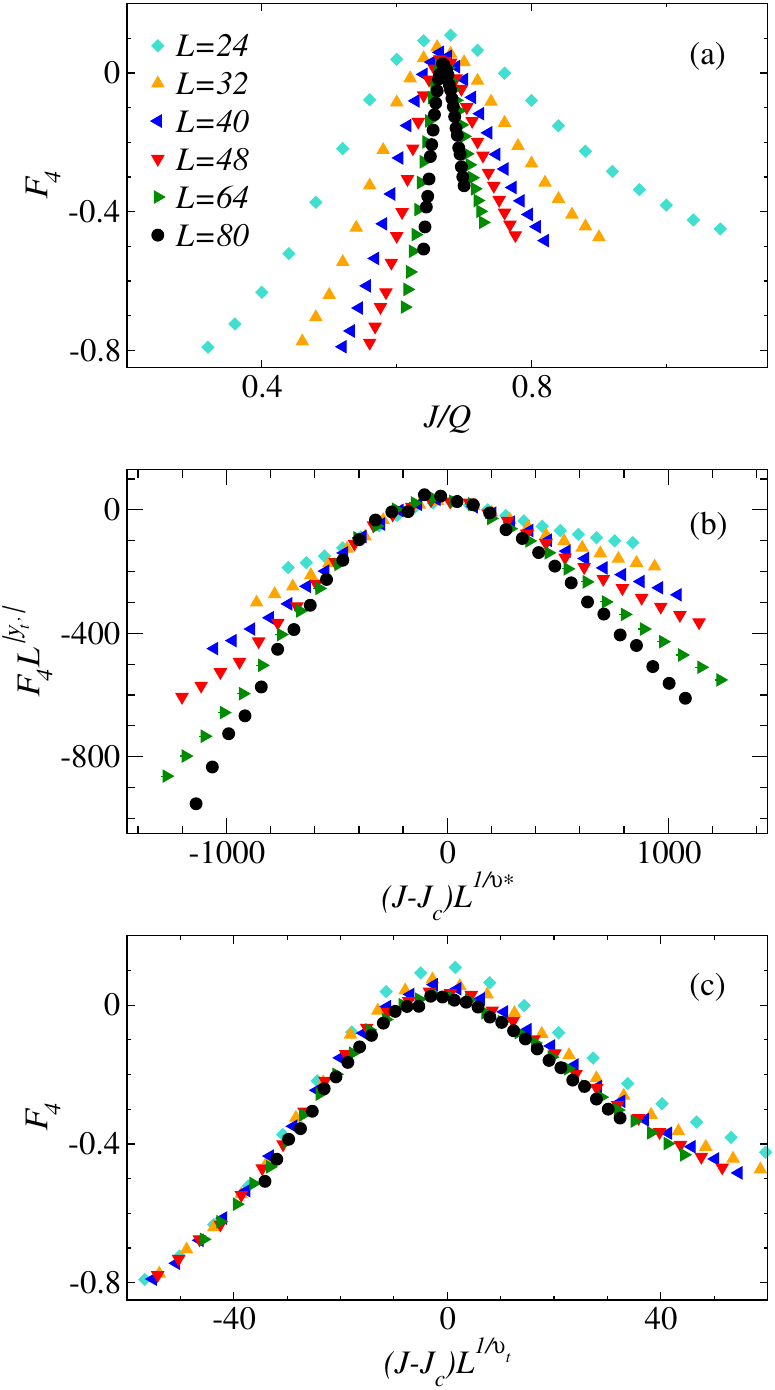}
\caption{Scaling analysis of the function $F_4$ from PQMC simulations of the $J$-$Q_3$ model. (a) Raw data in the neighborhood of the VBS--AFM
transition vs $J$ ($Q=1$) for different system sizes. (b) Scaling collapse in the neighborhood of the maximum, where the magnitude of $F_4$ decays as
$L^{-|y_{t'}|}$ with $y_{t'} \approx -1.71$, according to Fig.~\ref{jq234_f4}, and its dependence on $J$ is controlled by $1/\nu_*=2.40$ according
to the scaling observed in Fig.~\ref{zeros}(b). The data collapse extends to larger values of the scaled coupling $(J-J_c)L^{1/\nu_*}$ with increasing
system size. (c) Data sufficiently far away from the maximum---the points peeling off from the common scaling function in (b)---are collapsed by
scaling $(J-J_c)L^{y_t}$ with $y_t \approx 1.58$. Here $F_4$ is dimensionless, i.e., there is no rescaling of the value of $F_4$.}
\label{f4collapse}
\end{figure}

The power-law forms of $F_4(J_c,L)$ and $w_4$ suggest that $F_4(J,L)$ obeys the scaling form
\begin{equation}
F_4 = L^{-|y_{t'}|}f(\delta L^{a}),
\label{f4scaleform}
\end{equation}
in the close neighborhood $\delta=J-J_c$ of the transition at $\delta=0$. In Fig.~\ref{f4collapse}(a) we show PQMC data for the $J$-$Q_3$ model over a
wide range of the tuning parameter and system sizes, which we replot in Fig.~\ref{f4collapse}(b) in the manner that should lead to data collapse
if Eq.~(\ref{f4scaleform}) holds. Systematic collapse indeed builds up as the system size increases, with $f(x)$, $x=\delta L^{a}$ forming a close to
parabolic shape and with data peeling off from this form at values of $|x|$ that increase with $L$. As shown in Fig.~\ref{f4collapse}(c), the peeled off
data instead exhibit data collapse when the rescaled variable is $x=\delta L^{1/{\nu_t}}$, then with no rescaling of the value of $F_4$.

The asymptotic value of $F_4$ inside both ordered phases is negative of order one. Thus, its scaling dimension in the ordered phases
should be taken as zero, as we do in Fig.~\ref{f4collapse}(c). It is also natural that the scaling here is controlled by the exponent $\nu_t$ corresponding
to the tuning parameter, with $J$ largely driving the $t$ field. The different scaling seen very close to the transition in Fig.~\ref{f4collapse}(b)
reflects the emergence of the SO($5$) symmetry in the overall decay of $F_4$ to zero, $F_4 \propto L^{-|y_{t'}|}$, but the scaling variable $x=\delta L^{a}$
with $a=2.40$ does not have any apparent explanation from the CFT, matching neither $1/\nu_t$ nor $1/\nu_s$. The value is very close to $1/\nu_\phi=3-\Delta_\phi$,
with the order parameter scaling dimension $\Delta_\phi$ from Table \ref{dtable}. However, $1/\nu_\phi$ controls the behavior when an order parameter
field is added to the Hamiltonian, i.e., breaking its symmetries explicitly. There are no such symmetry breaking terms in the Hamiltonian, and therefore
the exponent $a$ cannot be $1/\nu_\phi$.

The only plausible explanation we can find for the value of $a$ is to identify it with the pseudocritical length scale illustrated in Fig.~\ref{xi}. It was
previously believed that $1/\nu \approx 2.20$ \cite{shao16,sandvik20} with $\nu$ the misidentified conventional correlation length exponent $\nu \approx 0.45$ in
both the $J$-$Q_2$ \cite{shao16,sandvik20} and the classical loop model \cite{nahum15a}. However, the extrapolations of the quantities used to obtain this
value of $\nu$ are challenging, as large system sizes are required for convergence of, e.g., Binder cumulant slopes, and the statistical errors grow
with $L$. As we will see in Sec.~\ref{sec:dark}, $1/\nu = 2.40$ also fits high quality data very well but this exponent should actually be identified as
$1/\nu_*$. Thus, we identify the exponent $a$ in Eq.~(\ref{f4scaleform}) as $1/\nu_*$. It is then very interesting to note that this exponent describes the $w_4$
window [and the scaling function $f(\delta L^{1/\nu_*})$ in Eq.~(\ref{f4scaleform})] up to the largest system sizes studied, $L=256$, even though some signs
(though rather weak) of an impending breakdown of the overall decay governed by $y_{t'}$ begin to appear already at $L \approx 60$ for the $J$-$Q_2$ model in
Fig.~\ref{jq234_f4}. This seems to imply that the latter crossover to some slower decay is indeed the result of the coexistence state, that the $w_4$ window
is much less sensitive to the first-order transition, and that the rank-$2$ operator is governing $1/\nu_*$ in Eq.~(\ref{nustarform}) up to the largest length
scale reached here without contamination by a possible slower $y_4$ decay originating from the rank-$4$ operator.

In Sec.~\ref{sub:derivedeltastar} we will derive an expression for $1/\nu_*$, already given in two different forms in Eqs.~(\ref{nustardeltas}) and (\ref{nustarform}),
by explicitly evoking the dangerously irrelevant interaction that is primarily responsible for violations of the SO($5$) relation between the AFM and VBS amplitude on
the first-order line, which we take as the rank-$2$ operator with $\Delta_{t'}\approx 4.71$. The relationship is then fully consistent with the value of $\Delta_s$ that
we found in Sec.~\ref{sec:jq2q6} using scaling on the first-order line and also with the bootstrap value of the exponent in Table \ref{dtable}. However, if we use $y_4$
in place of $y_{t'}$ in Eq.~(\ref{nustardeltas}), the predicted value of $\nu_*$ comes out much smaller, far from consistent with the behavior of $F_4$. This fact as well
supports the identification of the rank-$2$ part of the mixed AFM and VBS SO($5$) deformation as being solely responsible for the pseudocritical length
scale $\xi_*$.

Beyond Eq.~(\ref{f4scaleform}), a more complete scaling form of $F_4$ involves also $\nu_t$ and $\nu_s$
\begin{equation}
F_4 = L^{-|y_{t'}|}f(\delta L^{1/\nu_*},\delta L^{1/\nu_t},\delta L^{1/\nu_s}),
\label{f4scaleform2}
\end{equation}
which we will discuss and motivate in a more general context of finite-size scaling in Sec.~\ref{sub:darkscale}.
Since $1/\nu_* > 1/\nu_t > 1/\nu_s$, there is a regime where $\delta L^{1/\nu_*}$ is much larger than the other two arguments but still relatively small
so that the function $f$ can be Taylor expanded in this argument---this regime corresponds to the near parabolic shape formed by the collapsed data
in Fig.~\ref{f4collapse}(b). When also the argument $\delta L^{1/\nu_t}$ becomes large $\delta L^{1/\nu_*}$ can be considered as approaching infinity
and the function therefore crosses over to a form $f(\delta L^{1/\nu_t})$, as seen in the data collapse plot Fig.~\ref{f4collapse}(c). The third argument
$\delta L^{1/\nu_s}$ will also to some extent influence the behavior further away from $\delta=0$. In Sec.~\ref{sec:dark} we will present several examples of
physical observables that can be described by similar functions of three arguments, and where all of them are needed to fully describe numerical data.

\begin{figure}[t]
\includegraphics[width=75mm]{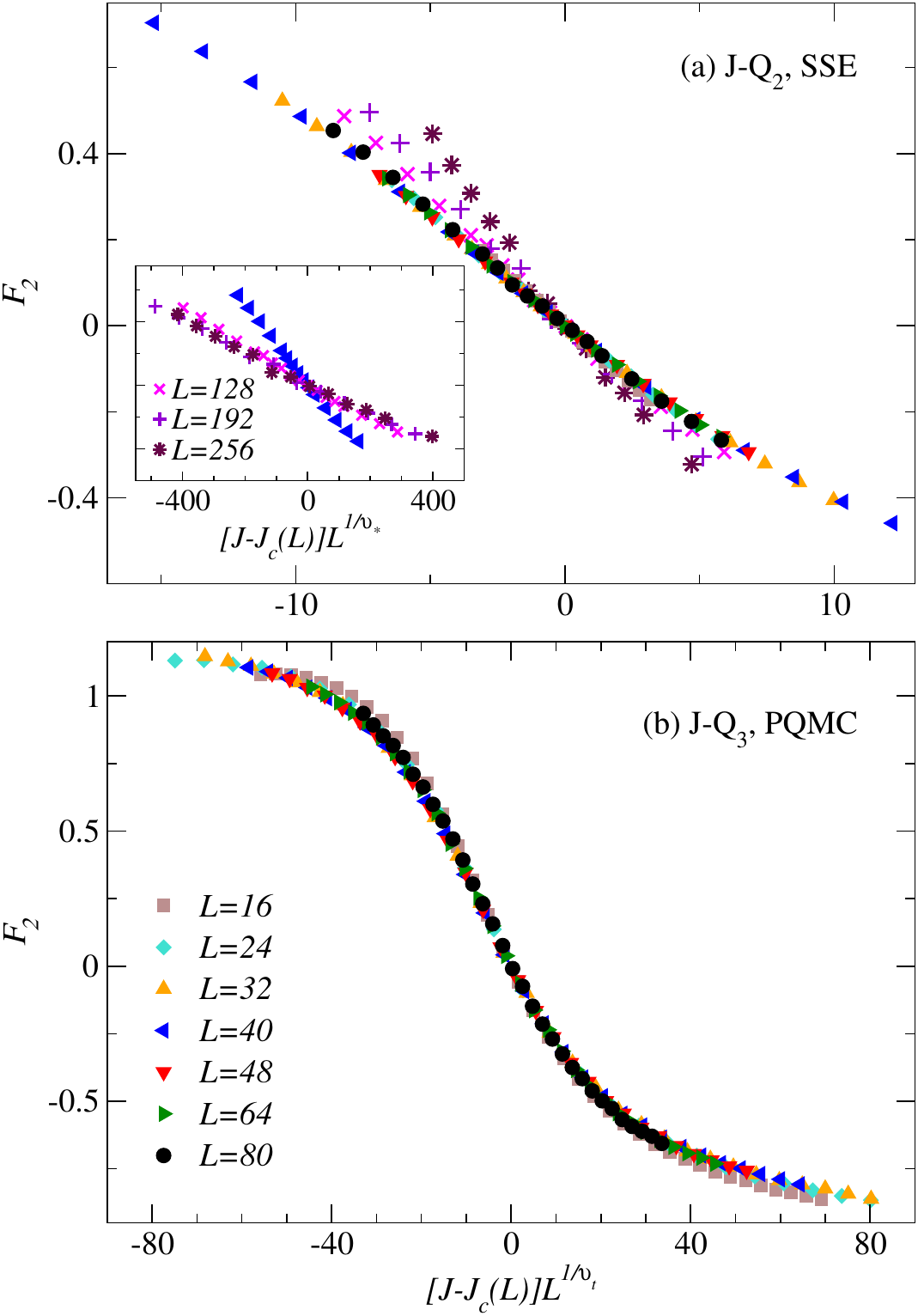}
\caption{The function $F_2$ of the $J$-$Q_2$ (a) and $J$-$Q_3$ (b) models, obtained with SSE ($Q_2=1$, $\beta=L$) and PQMC simulations,
respectively. The distance $J-J_c(L)$ to the size-dependent critical point is defined here by $F_2=0$ from Fig.~\ref{zeros}(a) (and similar PQMC results)
and is rescaled by $L^{1/\nu_t}$, with $1/\nu_t=3-\Delta_t \approx 1.58$. For common system sizes the plot symbols in (a) and (b) match, while the symbols
for the three largest sizes in (a) are defined in the inset.}
\label{f2collapse}
\end{figure}

To conclude the discussion of the emergent SO($5$) symmetry, in Fig.~\ref{f2collapse} we show a data collapse analysis for the function $F_2$ obtained
with both PQMC and SSE calculations (using different windows on the $x$ axis). This analysis confirms that $F_2$ should be regarded as dimensionless, i.e.,
no scaling of the function values are required to achieve data collapse. This may not be surprising, considering that $F_2$ in Eq.~(\ref{f2def}) is the
difference between two Binder ratios, both of which are dimensionless.

In Fig.~\ref{f2collapse}(b) and the main part of Fig.~\ref{f2collapse}(a) the distance from the transition point is scaled with the
conventional correlation length exponent $\nu_t$, which follows from the fact that $F_2$ measures the violations of the SO($5$) symmetry that are
``manually'' tuned by the parameter $J$, taking the system between the AFM phase with SO($3$) order parameter and the VBS phase with emergent U($1$)$=$O($2$)
symmetry by effectively deforming the five-sphere defining the order parameter. To effectively cancel a possible scaling correction, we have here used
the size dependent distance $J-J_c(L)$ from the critical point, with $J_c(L)$ defined as the point where $F_2(J,L)=0$, graphed for the $J$-$Q_2$ model
in Fig.~\ref{zeros}(a). The essentially perfect data collapse for system sizes up to $L\approx 100$ of a function manifestly probing the imposed
deformation of the SO($5$) order away from the transition lends further support to the interpretation of the exponent $\nu_t\approx 0.63$ as that of the
traceless symmetric tensor.

In Fig.~\ref{f2collapse}(a) we also observe clearly that the data for system sizes $L=128$ and above do not collapse to the common function. As shown in the
inset plot, for the large system sizes we instead see good data collapse when $J-J_c$ is scaled with $L^{1/\nu_*}$. This crossover between
behaviors controlled by $\nu_t$ and $\nu_*$ is a manifestation of the pseudocriticality illustrated in Fig.~\ref{xi}. In the case of Fig.~\ref{f2collapse}(a),
the larger sizes in finite-size scaling translate to systems closer to the transition, i.e., moving to the right on the logarithmic scale in Fig.~\ref{xi}.
It is important to note that $F_4$, in contrast to $F_2$, exhibits scaling with $\nu_*$ already from much smaller system sizes in Fig.~\ref{f4collapse},
which shows that the length scale $\xi_*$ exists already further away from the transition but is manifested in many physical observables only closer to
the transition, at larger scales.

\subsection{O(5) deformation model}
\label{sub:o5def}

Bruce and Aharony considered an O($N$), $N=N_A+N_B$, classical multicritical point separating O($N_A$) and O($N_B$) critical lines, where an attractive
interaction between the O($N_A$) and O($N_B$) order parameters $A$ and $B$ is presumed irrelevant \cite{bruce75}. While it is now known that such a multicritical
point is stable only for $N_A=N_B=1$ (i.e., the interaction is actually relevant for $N_A+N_B>2$) \cite{pelissetto02,eichhorn13}, the arguments of a coexistence phase
versus a first-order line and, in the former case, the shape of the boundaries of the coexistence phase between the pure A and B phases, are still formally correct
and useful. Moreover, while the scenario in practice is invalid for classical O($N$) models, a central aspect of our findings is that the multicritical DQCP
exists and corresponds to phase A being VBS ordered with $N_A=2$ and B is AFM ordered with $N_B=3$. However, unlike the case of attractive interactions that
cause the coexistence phase studied by Bruce and Aharony, we will show here that the interactions are repulsive, thus leading instead to the first-order line
extending from the multicritical point, as illustrated in Figs.~\ref{phases}(a), \ref{phases}(b), and \ref{defects}.

It is normally very difficult, even impossible, to distinguish a first-order line from the extremely narrow coexistence phase, as qualitatively drawn
in Fig.~\ref{phases}(c), in the neighborhood of the critical point. The shape of the coexistence boundaries were derived by Bruce and Aharony, and in
Sec.~\ref{sec:dark} we will generalize their arguments to define a crossover region in which the repulsive order-parameter interactions are important
and give rise to the emergent bubble length scale $\xi_*$. Here we will first present our arguments for detecting repulsive interactions through the form of
the SO($5$) deformation.

Inspecting the definition of $F_4$ in Eq.~(\ref{f4def}) and considering the fact that the denominators do not decay to zero with increasing system size in
a coexistence state (since both order parameters are nonzero by definition), it seems plausible that $F_4>0$ in the neighborhood of the transition, as seen
in Figs.~\ref{f2f4}(b) and \ref{jq234_f4}, corresponds to anti-correlation between AFM and VBS order parameters, relative to how they must relate to each other
for perfect SO($5$) symmetry with $F_4=0$. Anti-correlation clearly would correspond to effectively repulsive interactions that eventually must vanish in the
RG sense (as the interaction is irrelevant) at the critical point on large length scales [and note that $F_4$ stays positive also when the SO($5$) symmetry
breaks down for all the models in Fig.~\ref{jq234_f4}]. For attractive SO($5$) breaking interactions, we should instead have positively correlated order
parameters, thus $F_4 \to 0$ from below with increasing system size in the critical regime.

In order to confirm the ways attractive and repulsive interactions manifest themselves differently in $F_4$, and for additional insights into possible causes
of the symmetry violations, we here study a classical O($5$) deformation model of a five-component order parameter ${\bf v}=(v_1,\ldots,v_5)$,
where ${\bf D}=(v_1,v_2)$ and ${\bf M}=(v_3,v_4,v_5)$ correspond, respectively, to the VBS and AFM order parameters. Our unperturbed model is simply a 5-sphere
of radius $1$ with perfect O($5$) symmetry of ${\bf v}$. We then introduce perturbations by considering the following ``action'':
\begin{equation} \label{evdef}
S({\bf v}) = \lambda (D^2-M^2) + \kappa M^2D^2 + \mu D_1^2D_2^2.
\end{equation}
Here $\lambda$ tunes the system from VBS order for $\lambda < \lambda_c$ to $\lambda > \lambda_c$, with $\lambda_c=0$ if $\kappa=\mu=0$. This tuning parameter
plays the role of the $t$ field (the horizontal axis in Fig.~\ref{phases}) in the CFT or $J/Q$ in a $J$-$Q$ model. In general, we can define $\lambda_c$ as the
point where $F_2$ in Eq.~(\ref{f2def}) changes sign. The second term in Eq.~(\ref{evdef}) is the simplest form of
an interaction between the order parameters and we already analyzed it in the context of the SO($5$) CFT in Sec.~\ref{sub:so5}. The last term breaks the symmetry
between the two components of the VBS order parameter, similar to the lattice in the $J$-$Q$ model.

\begin{figure}[t]
\includegraphics[width=80mm]{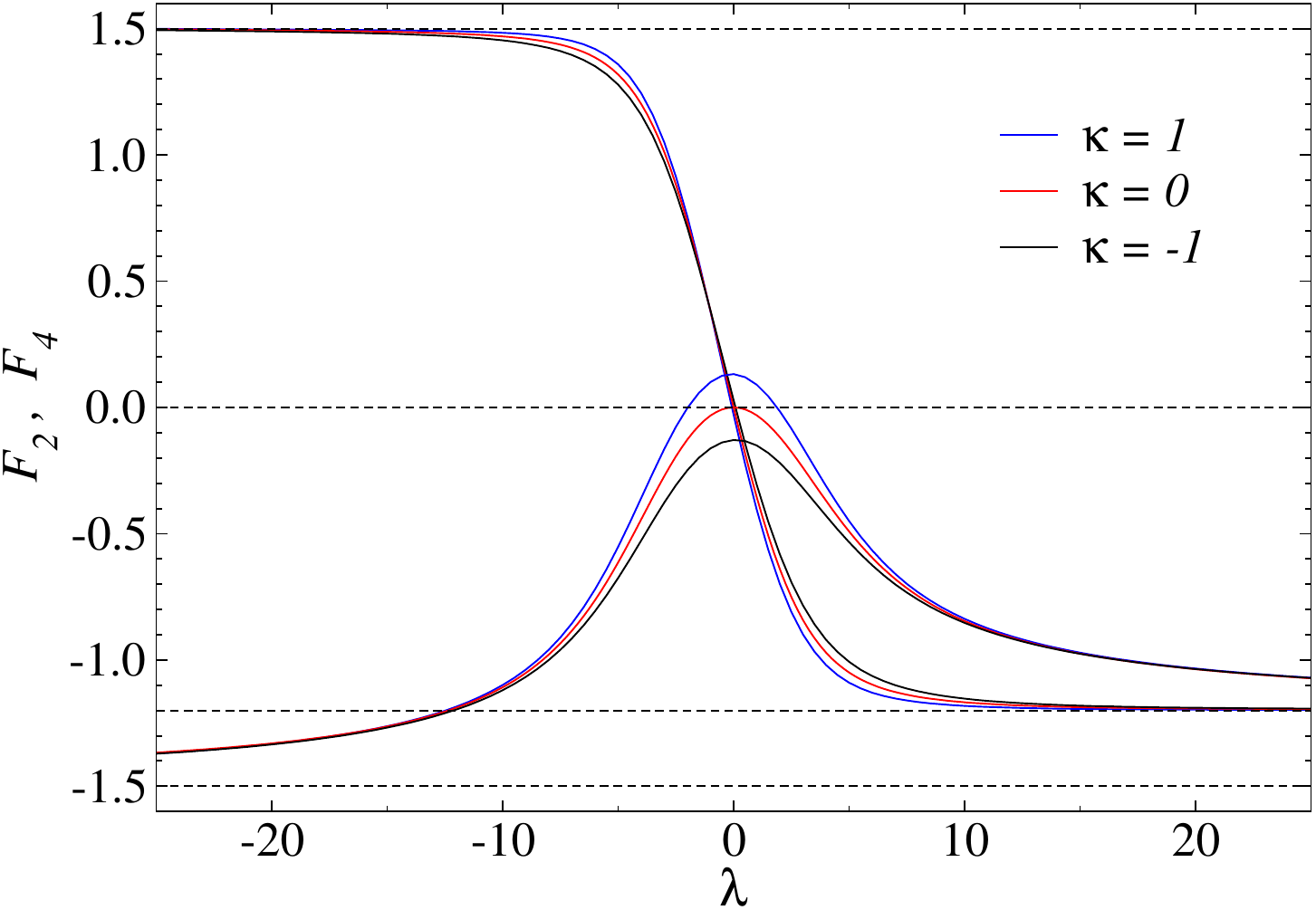}
\caption{Results for the O($5$) deformation model with $\mathbb{Z}_4$ coupling $\mu=0$. The O($5$) diagnostic quantities $F_2$ and $F_4$ defined
in Eqs.~(\ref{f2f4def}) are shown vs the parameter $\lambda$ that tunes the transition between the O($2$) and O($3$) ``phases''. Three different
values of the interaction parameter $\kappa$ are used; $\kappa \in \{0,\pm 1\}$. The dashed lines indicate $0$ and expected asymptotic values of
the quantities.}
\label{f2f4_5.eps}
\end{figure}

In order to mimic a non-uniform critical radial distribution we define the probability of ${\bf v}$ with a factor $r^2=\sum_i v_i^2$ in front of the
exponential;
\begin{equation} \label{pvdef}
P({\bf v}) = \frac{1}{Z}r^2{\rm e}^{-S({\bf v})},
\end{equation}
though the results presented below are qualitatively insensitive to the presence of this factor or the exponent $2$ used. We can now define physical
observables $G$ in the model by computing
\begin{equation} \label{gvpdef}
\langle G\rangle = \int dv^5 G({\bf v})P({\bf v}),~~~~ Z = \int dv^5 P({\bf v}).
\end{equation}
It should be possible to solve this model analytically, at least perturbatively for small values of the interaction parameters, but here we simply
sample the vector.

We first keep $\mu=0$ for the purpose of testing the effects on the O($5$) symmetry on the interaction strength $\kappa$. Results for $\kappa=0$
and $\kappa=\pm 1$ are shown in Fig.~\ref{f2f4_5.eps} and can be compared with those for the $J$-$Q_2$ model in Fig.~\ref{f2f4}, which includes results only very
close to the transition point. In Fig.~\ref{f2f4_5.eps} we show results for a larger range of the tuning parameter $\lambda$ in the O($5$) model in order to
confirm the known asymptotics. As expected, a repulsive interaction, $\kappa > 0$ is required for the $F_4$ peak to move above $0$ in the transition region.
Since here it is definitely true that it is the repulsion that causes $F_4>0$, we can also define the distance between the points at which $F_4=0$ as the
window in which the repulsion has a significant effect. In this simple model we of course only get trivial values of the exponents that we investigated in
the $J$-$Q$ models in Figs.~\ref{zeros} and \ref{jq234_f4}. The $F_4$ peak value (or the value at $\lambda=0$) decays linearly with $\kappa$ and the window
of $F_4>0$ is of size $\propto \kappa^{1/2}$. The decay versus system size in the spin models of course reflects a non-trivial renormalized
(coarse grained on scale $L$) effective interaction, and the scaling of the order parameters themselves, versus the length scale $L$.

\begin{figure}[t]
\includegraphics[width=80mm]{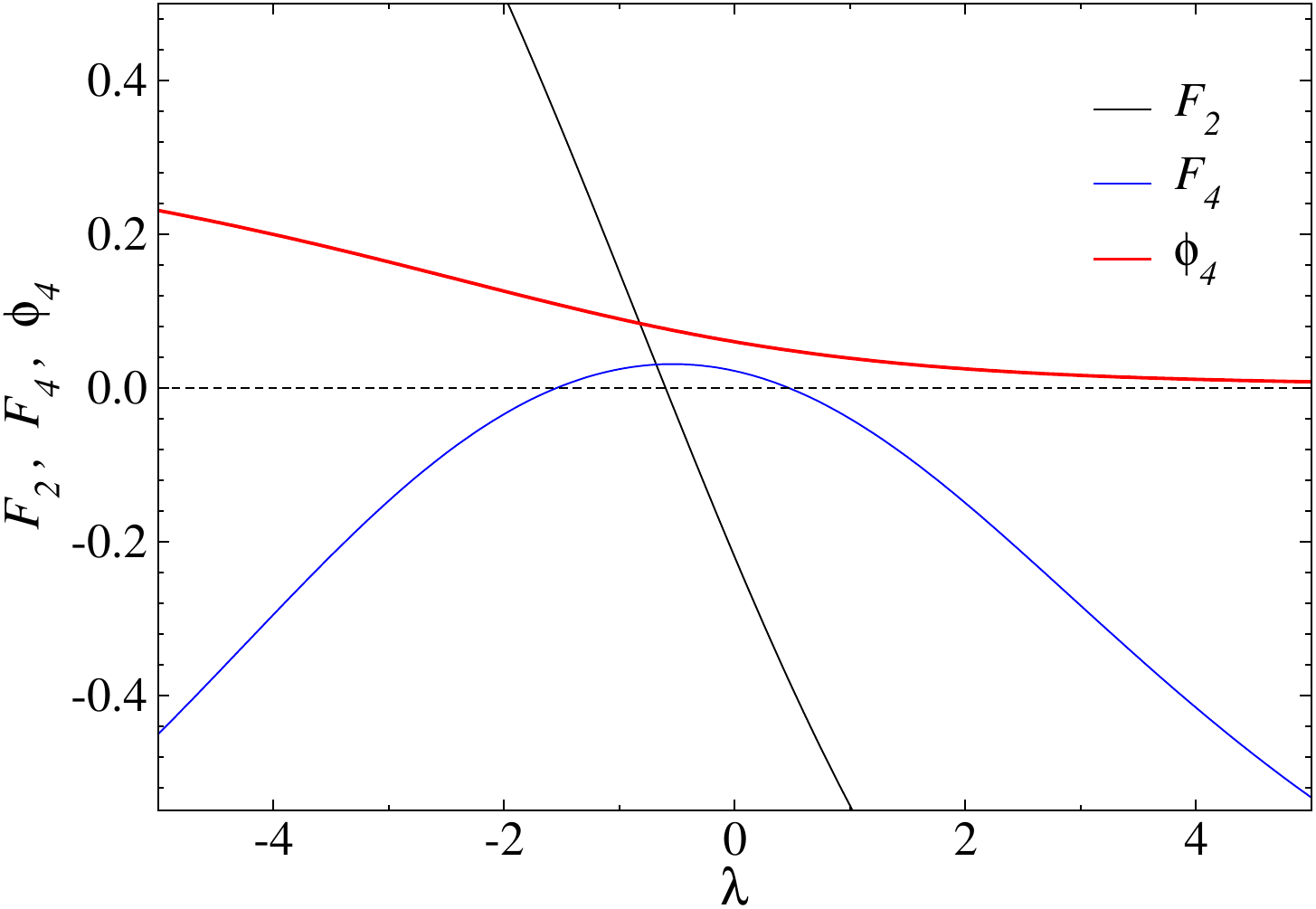}
\caption{Results for the O($5$) deformation model with $\mathbb{Z}_4$ coupling $\mu=8$ in the neighborhood of the phase transition.
Here the $\mathbb{Z}_4$ diagnostic function $\phi_4$ defined in Eq.~(\ref{phiqdef}) is shown in addition to the $F_2$ and $F_4$ functions.}
\label{f2f4p4_5}
\end{figure}

We next turn on the $\mu D_1^2D_2^2$ interaction in Eq.~(\ref{evdef}) and investigate the $\phi_4$ order parameter defined in Eq.~(\ref{phiqdef}). Note
that $D_x^2D_y^2 = [1-\cos(4\theta)]/8$, i.e., the quantity $\phi_4$ measures the expectation value of this same interaction operator. Though of course this
interaction leads to a nonzero $\phi_4$ (and $\phi_4 \propto \mu$), as shown in Fig.~\ref{f2f4p4_5} for a rather large value of $\mu$ to compensate
for the $1/8$ factor above, another interesting point is that a positive $F_4$ is also generated. This behavior should not necessarily be strictly
interpreted as $D_1^2D_2^2$ corresponding to a repulsive interaction between the VBS and AFM components, though $F_4>0$ does signal a similar effect of
anti-correlated order parameters.

In the many-body system as well we would expect the lattice by itself to not just cause the eventual crossover of the VBS order parameter symmetry
from O($2$) to $Z_4$, but in the process also induce other violations of the SO($5$) symmetry. However, and importantly, in the $J$-$Q$ models the length
scale of the O($2$) symmetry is very large, as we quantitatively demonstrated in Sec.~\ref{sub:u1}, with the exponent $\mu_4 \approx 0.80$ governing the
U($1$) length scale $\xi_{{\rm U}(1)} \sim t^{-\nu}$ in the VBS phase being larger than the conventional correlation-length exponent $\nu_t \approx 0.63$
and the bubble-size exponent $\nu_* \approx 0.42$. Moreover, the amplitude of the size dependent $\phi_4$ in Fig.~\ref{phi4} is also very small in the
PQMC case (where $\phi_4$ is on equal footing with the PQMC results for $F_2$ and $F_4$ in Figs.~\ref{zeros} and \ref{jq234_f4}). Since the deviations
from the U($1$) VBS symmetry are required before this distortion can have a secondary effect on the SO($5$) symmetry (e.g., as manifested in $F_4$),
the SO($5$) distortion with its faster decay (but rather large amplitude) in Fig.~\ref{jq234_f4} cannot be a direct byproduct of the
rank-$4$ lattice-induced perturbation, or any other member of the multiplet with scaling dimension $\Delta_4$, but of a different perturbation that we
have argued above is the second rank-$2$ primary operator.

\section{Pseudocritical scaling}
\label{sec:dark}

We have argued that the exponent $\nu_*$ describes a novel phenomenology of the weakly first-order AFM--VBS transition, in the form of the crossover of the
correlation length illustrated in Figs.~\ref{defects} and \ref{xi}, and also in finite-size scaling of an important indicator of emergent SO($5$) symmetry above
in Sec.~\ref{sub:so5}. No scaling dimension matching $\nu_*$ appears in the (approximate) CFT spectrum now available from the CFT bootstrap calculations
\cite{li18,chester24} and the fuzzy sphere model \cite{zhou23}. In light of the CFT results, the initial observation of $\xi_*$ in the form of a correlation
length with anomalously small growth appears puzzling---reminding us of Rabi's famous question ``Who ordered that?'' when pondering Anderson's 1936 discovery
of (as it turned out to be) the muon \cite{nyt}.

The emergent pseudocriticality actually does fit perfectly into the SO($5$) multicritical scenario, though it is not directly observable exactly at the
critical point, only in its neighborhood of small $s\not=0$. The main manifestations of the pseudocriticality are in the evolution of observables on approach
to the critical point, as illustrated with the emergence of the bubble scale $\xi_*$ in Fig.~\ref{defects}. This scale corresponds directly to the exponential
decay rate of VBS correlations in the AFM phase and vice versa, and, therefore, is easy to extract. Here in Sec.~\ref{sub:derivedeltastar} we will derive the
expression Eq.~(\ref{nustardeltas}) relating $\nu_*$ to conventional scaling dimensions of the underlying CFT and then in Sec.~\ref{sub:darkscale} we deduce
the crossover points $g_1^*(k)$ and $g_2^*(k)$ (with $g_2^* < g_1^*$) between which bubble size grows as $g^{-\nu_*}$. In Sec.~\ref{sec:cumslopes} we demonstrate the
finite-size crossover from $\xi_t$ to $\xi_*$ in the $J$-$Q_2$ model and also analyze the correlation length computed on large lattices with the 3D classical loop
model in Ref.~\cite{nahum15b}. The crossover there had been noted by Nahum et al., but not explained, and here we go further and show that the largest values of
the correlation length versus the distance $g$ to the critical point agree with our $\xi_* \sim g^{-\nu_*}$ with $\nu_*$ obtained from the $F_4$ analysis
in Fig.~\ref{f2f4}(b).

\subsection{Emergent length scale}
\label{sub:derivedeltastar}

Bruce and Aharony \cite{bruce75} considered a system with two order parameters, ${\bf \phi}_A$ and ${\bf \phi}_B$, in Landau theory including
an interaction term $v\phi_A^2\phi_B^2$. They further assumed that the joint order parameter with a total of $N$ components
develops an emergent O($N$) symmetry at an isolated critical point in the space of a parameter preserving this symmetry and one that violates it; like
the $s$ and $t$ field strengths in our phase diagrams in Fig.~\ref{phases}. Within this scenario, a first-order line forms, as in Fig.~\ref{phases}(b),
if the two order parameters are mutually repulsive, i.e., $v>0$. For attractive order parameters, $v<0$, the coexistence phase forms instead, as
in Fig.~\ref{phases}(c).

Though we do not have a coexistence phase in our case, as demonstrated in Sec.~\ref{sub:o5def}, for the moment our discussion will be framed around
such an imagined phase, in order to derive useful results that also will apply once we have abandoned the extended coexistence phase in favor of a coexistence
line. The key result of Ref.~\cite{bruce75} for our purpose here is that the shape of the phase boundaries of either side ($t<0$ and $t>0$) of the
coexistence phase (in our case an imagined coexistence phase) is given by
\begin{equation}
s_c(t) \sim |t|^{1/\psi},
\label{sct}
\end{equation}
where the exponent $\psi$ depends on the crossover exponent as well as the scaling dimension of the leading irrelevant perturbation of the O($N$) symmetry.
According to Eq.~(5.11) of Ref.~\cite{bruce75}, $\psi=\phi_g-\phi_v$, which is translated to the situation at hand here as $\phi_g \to y_t=1/\nu_t=3-\Delta_t$
and $\phi_v \to y_{t'} = 3-\Delta_{t'}$. Thus, the exponent controlling the observed behavior of the phase boundaries (if they were actually to exist) would be
\begin{equation}
\psi = y_t - y_{t'} = \Delta_{t'} - \Delta_t.
\label{psiyty4}
\end{equation}
As discussed in Secs.~\ref{sub:so5w} and \ref{sub:o5def}, we cannot prove that $y_{t'}$ should be used here (after we generalize to repulsive interactions)
instead of $\Delta_4$, and in principle $\Delta_4$
might control the behavior on even larger length scales, i.e., there would be yet another crossover before the growth of the bubble scale halts when the
$\xi_t$ converges to the finite $\xi_s$ as illustrated in Fig.~\ref{xi}. However, the crossover may also formally take place after the final convergence of all the
length scales, in which case it would play no role at all. We see no signs of such crossover, and, given that we obtain remarkable agreement with all numerical
results when using $\Delta_{t'}$, we conjecture that the operator responsible for the irrelevant violation of SO($5$) symmetry here is the first irrelevant
rank-$2$ CFT primary operator discussed in Sec.~\ref{sub:so5}.

Using Eq.~(\ref{sct}), we can also obtain the width $w_{\rm co}$ of the coexistence phase in the $t$ field direction as a function of the $s$ field:
\begin{equation}
w_{\rm co}(s) \sim s^\psi.
\label{deltats}
\end{equation}
From this expression we can obtain an effective finite-size width of the coexistence phase at the critical point, $s\to 0$, in the
standard way by expressing $s$ using the correlation length $\xi_s \sim s^{-\nu_s}$ and replacing $\xi_s$ by $L$:
\begin{equation}
w_{\rm co}(L) = L^{-\psi/\nu_s},
\label{deltatl}
\end{equation}
which of course vanishes for $L \to \infty$. If a length scale diverges as $t^{-\nu}$ with some $\nu>0$, then the value of $t$ for which this length
reaches the system length $L$ scales as $L^{-1/{\nu}}$. Applying these arguments in the reverse, from Eq.~(\ref{deltatl}) we can identify an exponent
\begin{equation}
\nu_*=\frac{\nu_s}{\psi}
\label{nustarnusnut}
\end{equation}
governing a divergent length scale $\xi_* \sim t^{-\nu_*}$, which, when using the form of $\psi$ in Eq.~(\ref{psiyty4}), leads to Eq.~(\ref{nustardeltas})
or, equivalently, Eq.~(\ref{nustarform}).

Another way to arrive at the same result is to consider the $s$ correlation length $\xi_s \sim s^{-\nu_s}$ in combination with a proposed
new length scale $\xi_* \sim |t|^{-\nu_*}$. If there is a singular boundary in the $(s,t)$ plane given by Eq.~(\ref{sct}), then its functional form should
correspond to $\xi_* \sim \xi_s$, i.e., $s_c(t)=|t|^{\nu_*/\nu_s}=|t|^{1/\psi}$, so that $\psi=\nu_s/\nu_*$. Such an argument applies generally to the shape of a
transition line emerging from a critical point when a relevant perturbation is turned on, which would correspond to $\nu_s$ and $\nu_t$ being known and
the exponent $\psi$ is the one sought. Here instead the general form of $\psi$ in Eq.~(\ref{psiyty4}) is known, and we propose a related length scale
controlled by $\nu_*$, which according to the above steps is given by Eq.~(\ref{nustarnusnut}). The unusual aspect here is that $\psi$ in Eq.~(\ref{psiyty4})
involves not only $\Delta_t$ but also $\Delta_{t'}$, because of the emergent symmetry with its dangerously irrelevant perturbation.

Note that, with the form of $\psi$ in Eq.~(\ref{psiyty4}), the coexistence phase shrinks when $\Delta_{t'}$ increases, i.e., when the SO($5$) perturbation becomes more
irrelevant. Thus, if the symmetry is exact there will be no coexistence phase, only a line with SO($5$) symmetry (and spontaneous symmetry breaking).
For reasonable values of the scaling dimensions, the coexistence phase is extremely narrow in the $t$ direction for small $s>0$, and, as pointed out already in
Ref.~\cite{bruce75}, it can easily be mistaken for a direct first-order transition.

In our present case we do not have a coexistence phase. Why then are the above results relevant to our case? The reason is that the scaling forms do not rely on the
sign of the interactions between the order parameters; only the physical interpretation does. For repulsive interactions, the form of $s_c$ versus $t$ in Eq.~(\ref{sct})
should not be interpreted as a phase boundary, but as a crossover value of $s$ for given $t$ above which the repulsive interaction becomes strong enough (relative to the
$t$ and $s$ scales) to be the main source of gapping the pseudo-Goldstone modes corresponding to the components of the AFM order parameter inside the VBS phase or
vice versa. Similarly, the key result in Eq.~(\ref{deltatl}) should be interpreted as the window of $t$ values around $t=0$ in which the order parameter repulsion has
detectable consequences in a finite system. Thus, all the above arguments that lead to Eq.~(\ref{deltatl}), and then to Eq.~(\ref{nustardeltas}), still formally apply
though the physical situation is quite different. In the repulsive case, the pseudo-Goldstone modes develop anomalously large gaps corresponding to smaller bubble
size, as discussed qualitatively in Sec.~\ref{sub:crossover} and illustrated in Fig.~\ref{defects}. The exponent $\nu_*$ describes the impeded growth of the bubble
size $\xi_*$.

A practically useful aspect of the exponent $\nu_*$ is that its numerical estimate from the finite-size scaling of the SO($5$) window extracted
from $F_4$ in Fig.~\ref{zeros} has rather small statistical errors. If correct, the form Eq.~(\ref{nustarform}) can therefore be used to obtain a
good estimate of $\Delta_s=3-1/\nu_s$, which we were not able to extract from the correlation function that gave us $\Delta_t$ in Sec.~\ref{sub:symop}.
The resulting value $\Delta_s=2.271(3)$ has been entered in Table \ref{dtable} and it also is consistent with the value of the critical exponent $\beta$
governing the growth of the coexisting order parameters on the first-order line, Fig.~\ref{jq2q6_m2}, which corresponds to $\Delta_s=2.30(4)$.

\subsection{Scaling crossover}
\label{sub:darkscale}

We now consider tuning a system on a line where the $s$ and $t$ fields are parametrized by $g$ as in Fig.~\ref{xi}; $s=s(g)$, $t=t(g)$. A scaling function
governing a dimensionless observable when the line crosses the multicritical point should be of the form $f(g L^{y_s},g L^{y_t})$, while for a dimensionful
quantity a power of $L$ also multiplies $f$. Eventually for $g \to 0$, the larger of the two arguments will dominate the scaling behavior, which in our
case means $f(g L^{y_s},gL^{y_t}) \to f(gL^{y_t})$, except on the coexistence line ($t=0$) where $f(sL^{y_s})$ applies. If instead the first-order line is
crossed at a small value of $s=s_0$ ($k>0$ in Fig.~\ref{xi}), the scaling function also will depend on $y_*$;
\begin{equation}
f=f(sL^{y_s},tL^{y_t},ts_0^bL^{y_*}),
\label{fform3}
\end{equation}
where we could replace $s_0$ by $s$ but we here keep $s_0$ as an indication that we have in mind traversing the transition versus $t$ with $s$ almost
constant and crossing $t=0$ at $s=s_0$. The factor $s_0^b$ with $b>0$ has been introduced to account for the diminished effects of the pseudocritical scaling
when $s_0\to 0$. While we do not know the value of the exponent $b$, we can obtain a bound from the fact that pseudocriticality has been observed on rather
large length scales \cite{nahum15a,shao16,sandvik20}, as we will also show with improved numerical results below in Sec.~\ref{sec:cumslopes}.

We would like to know the range of length scales over which the bubble size dominates, i.e., in the thermodynamic limit the range $[\xi_*(g_1^*),\xi_*(g_2^*)]$
in Fig.~\ref{xi}. We consider a corresponding finite-size version first and cross the transition versus $t$ at some small $s>0$. For the three inverse
correlation length exponents we have $y_s < y_t < y_*$, and there will therefore be an $(s,t)$ region where the correlation length is controlled by $t$
and the first argument of the scaling function in Eq.~(\ref{fform3}) is unimportant (effectively $s$ can be taken as $0$ but we have to keep $s_0>0$). Further,
since $y_* \approx 2.4$ is significantly larger than $y_t \approx 1.6$, above some large $L$ the function crosses over to effectively one of a single argument,
$f \to f(ts_0^bL^{y_*})$, disregarding for now the fact that this scaling form must break down when $L$ becomes sufficiently large for the final crossover
to finite length scales at the first-order transition (inside the dashed fan in Fig.~\ref{defects}). However, when $s_0^b$ is small, there will also be some range
of smaller system sizes for which $tL^{y_t}$ dominates in Eq.~(\ref{fform3}). Balancing the two arguments, in a scaling sense, produces the crossover length
scale. From $L^{y_t} \sim s_0^bL^{y_*}$ we obtain the finite-size crossover length, and by replacing $L$ by the generic correlation length $\xi$ in the thermodynamic
limit we can write the first crossover length in Fig.~\ref{xi} as
\begin{equation}
\xi(g_1^*) \sim {s_0}^{-b/(y_* - y_t)}.
\label{xicross1}
\end{equation}

The upper crossover length $\xi(g_2^*)$ in Fig.~\ref{xi} will roughly correspond to the saturation length---the final converged bubble size at the first-order transition.
With the bubble growth ceasing when $\xi_t \to \xi_s \sim s^{-{\nu_s}}$ and $\xi_*=\xi_t^{\nu_*/\nu_t}$, the second crossover length is given by
\begin{equation}
\xi(g_2^*) \sim {s_0}^{-{\nu_s\nu_*/\nu_t}},
\label{xicross2}
\end{equation}
and the entire window in which the pseudoscaling holds
\begin{equation}
\xi_* \in [{s_0}^{-b/(y_* - y_t)},{s_0}^{-{\nu_s\nu_*/\nu_t}}].
\label{darkrange}
\end{equation}
Provided that $\nu_s\nu_*/\nu_t > b/(y_* - y_t)$, the range diverges when $s \to 0$, and it is in this sense that $\xi_*$ can be a truly divergent length scale
even if it is completely invisible exactly at the multicritical point, where the first crossover never takes place. The fact that we can observe the crossover
suggests that the above condition holds, i.e., that the unknown ``fading exponent'' $b$ in Eq.~(\ref{fform3}) satisfies $0 < b < (y_* - y_t)\nu_s\nu_*/\nu_t$, or,
using the numerical values of the scaling dimensions, $0 < b \alt 1.70$, which is quite a substantial range of values.

\subsection{Tests of the correlation length}
\label{sec:cumslopes}

A common way to extract a generic correlation length exponent $\nu$ is to consider the finite-size scaling form of a dimensionless singular
quantity. The dependence of such a quantity on the distance $\delta$ to the critical point is exemplified in the case of the Binder cumulant in
Eq.~(\ref{ufssform}). This form is often used to collapse data for a range of $\delta$ values and system sizes, as we did with the Binder cumulant in the
coexistence state with $\nu=\nu_s$ in Fig.~\ref{jq2q6binder}. If the exponent is not known, it can be determined by adjusting the
value of $\nu$ for optimal collapse onto a scaling function when $L$ is sufficiently large for scaling corrections to be negligible. We have already seen above that a
scaling function with three arguments is in principle needed in the present case when crossing the first-order transition, Eq.~(\ref{fform3}). This
function crosses over into a function $f_*(tL^{y_*})$ with a single argument $y_* \approx 2.4$ for large $L$, while for smaller $L$ the argument
$ts_0^b$ can be small relative to just $t$ and then $f \approx f_t(tL^{y_t})$ with $y_t \approx 1.58$.

Instead of performing a full data collapse analysis, the presence of two scaling regimes, separated by a smooth crossover, can be tested more
systematically (without having to make choices on what data points should be included) by analyzing the derivative of the cumulant with respect to
the tuning parameter. Using $J$ as the tuning parameter in the $J$-$Q$ models, $\delta = J-J_c$, the scaling form Eq.~(\ref{ufssform}) of the
cumulant implies that its derivative with respect to $J$ scales as
\begin{equation}
\frac{dU(J)}{dJ} \vert_{\delta=0}~ \propto L^{1/\nu},
\label{uderiv}
\end{equation}
at the critical point. This way of extracting $\nu$ from the derivative was used in Refs.~\cite{nahum15a}, \cite{shao16}, and \cite{sandvik20}.

We now generalize the above conventional form of the cumulant slope $U'$ to multiple length scales. Using the scaling form of a generic dimensionless
quantity in Eq.~(\ref{fform3}), where the fields $s$ and $t$ depend linearly on the tuning parameter $J$, we expect
\begin{equation}
U'(J_c) = aL^{1/{\nu_*}} + bL^{1/{\nu_t}} + cL^{1/{\nu_s}}.
\label{uderiv_abc}
\end{equation}
We will test this three-length scaling form with data for the $J$-$Q_2$ and $J$-$Q_2$-$Q_6$ models.

The derivatives can either be obtained from a set of closely spaced
numerical values of the cumulants fitted to polynomials, by taking the derivative of the polynomials, or directly from an estimator in the SSE simulations.
Typically, we find the latter to produce results with smaller statistical errors. In either case, results are interpolated at the crossing point of the AFM
and VBS cumulants, or at the infinite-size critical point if it is known to sufficient precision. Alternatively, the maximum slopes (which occur at slightly
different $J$ values for the two cumulants) can be analyzed and should scale in the same way. Here we use the derivatives at the cumulant crossing point.

\begin{figure}[t]
\includegraphics[width=80mm]{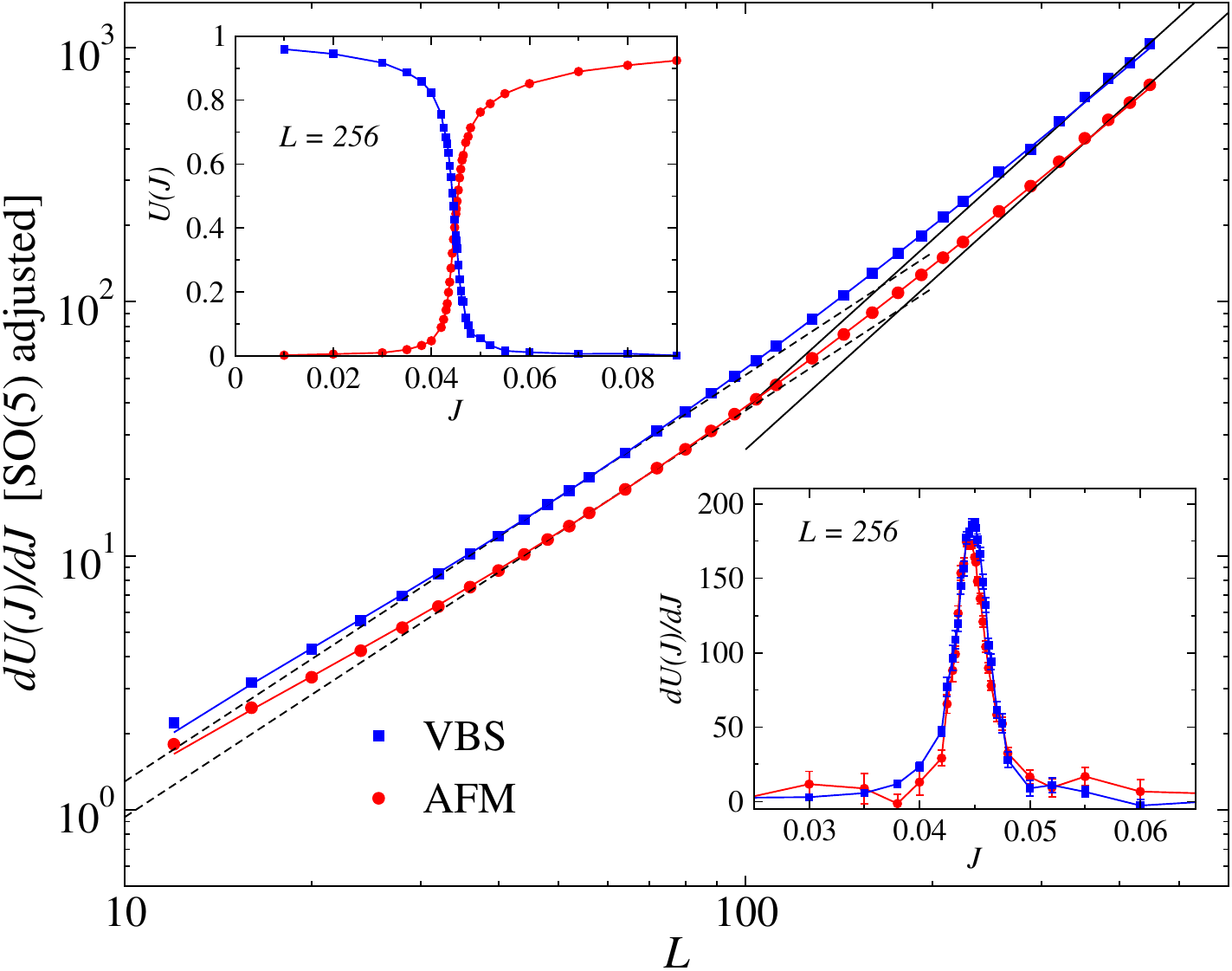}
\caption{System size dependence of the SO($5$) adjusted derivatives of the AFM and VBS Binder cumulants of the near-critical $J$-$Q_2$ model
($Q_2=1$) obtained with SSE simulations at inverse temperature $\beta=0.42L$.
The solid red and blue curves closely matching the data points are fits to Eq.~(\ref{uderiv_abc})
including $L\ge 16$ data (excluding only the $L=12$ points). The dashed black lines show the predicted form when the growth of the correlation length is
governed purely by the exponent $\nu_t$ (slope $y_t=1.58$) and the solid black lines where $\nu_*$ fully dominates (slope $y_*=2.40$). The top-left
inset shows raw data for the SO($3$) (AFM) and U($1$) (VBS) definitions of the cumulants, Eqs.~(\ref{udefs1}). The derivatives in the main
figures have been interpolated to the size-dependent point where the two SO($5$) adjusted cumulants cross each other. The bottom-right
inset shows the derivatives obtained directly from an estimator in the simulations.}
\label{slopes_jq2}
\end{figure}

Fig.~\ref{slopes_jq2} shows results for the $J$-$Q_2$ model, with insets showing raw data for the cumulants and the derivatives directly produced by
the SSE simulations for each value of $J$. In practice, we find no significant differences between results obtained with the conventional cumulant
definitions in Eqs.~(\ref{udefs1}) and the SO($5$) versions in Eq.~(\ref{udefs2}), beyond some small shifts of the size dependent transition points
and the values of the cumulants and their derivatives. The data presented below are for the SO($5$) definitions, Eqs.~(\ref{udefs2}), with the
derivatives evaluated at the $L$ dependent point at which the two cumulants cross each other. We have also analyzed data at our most precisely determined
infinite-size transition point $J/Q_2=0.04502(1)$ (obtained in Sec.~\ref{jqcscaling}) and in addition studied the size behavior of the maximum slope (peak values
in the right-bottom inset of Fig.~\ref{slopes_jq2}). Both these definitions produce only small overall shifts of the data points relative to those in
Fig.~\ref{slopes_jq2}.

For system sizes roughly in the range $L=30$ to $100$, the results in Fig.~\ref{slopes_jq2} are well described by the exponent $\nu_t$, while for larger
sizes the behavior crosses over to a significantly faster growth with $L$. The expected form with $1/\nu_* \approx 2.40$ is fully consistent with the data
only for the largest few system sizes (up to $L=448$) but fits to the predicted form Eq.~(\ref{uderiv_abc}), shown as the continuous curves in Fig.~\ref{slopes_jq2},
describe the data well for all sizes $L \in [16,448]$. This good fit is not trivial, since only the three factors $a,b,c$ in Eq.~(\ref{uderiv_abc})
are adjustable.

\begin{figure}[t]
\includegraphics[width=85mm]{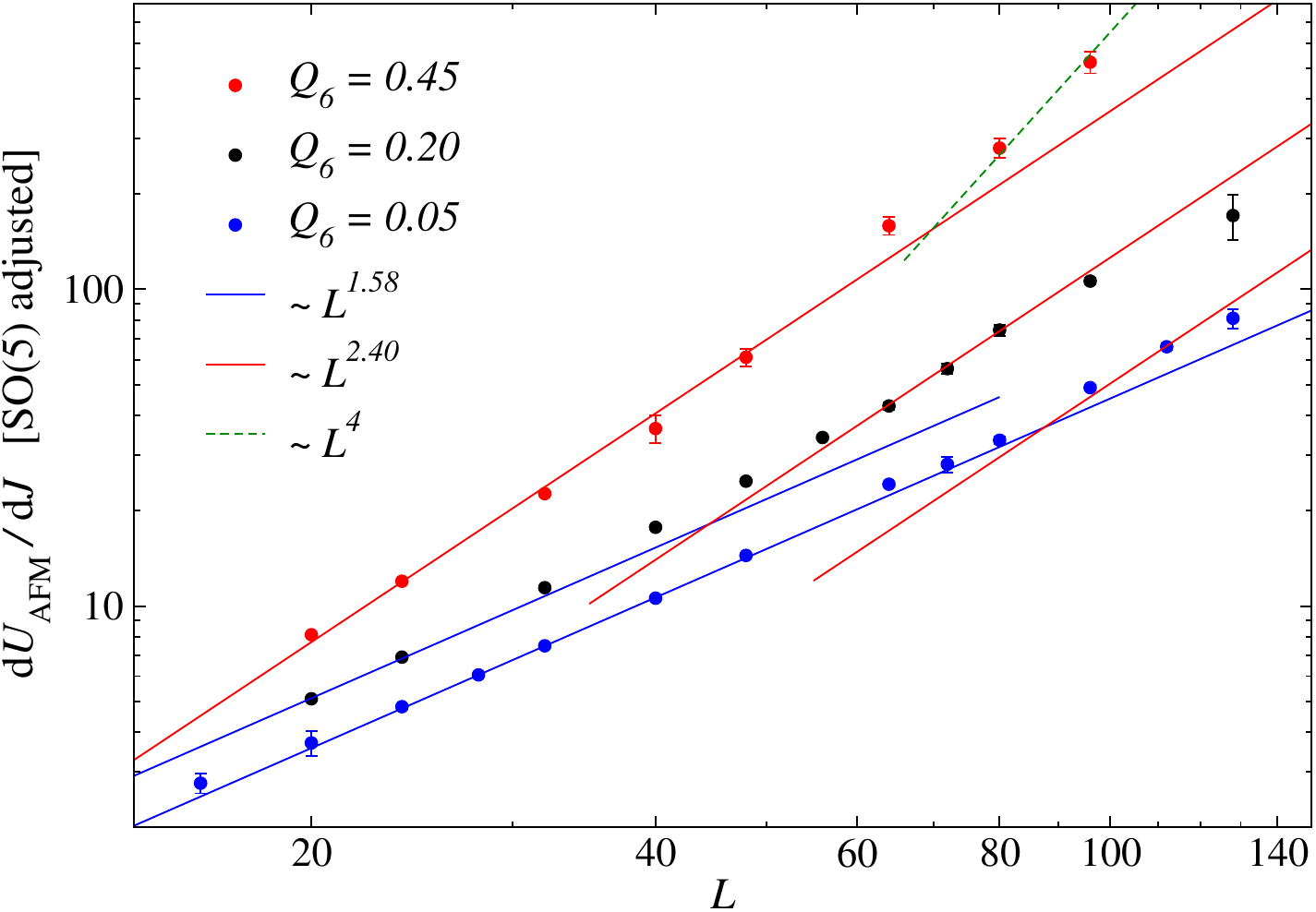}
\caption{Size dependence of the derivatives of the SO($5$) AFM and VBS cumulants of the $J$-$Q_2$-$Q_6$ model at the points where the two cumulants cross each
  other vs $Q_2$. Results are shown close to the critical point ($Q_6=0.05$), near the maximal $Q_6$ value accessible without QMC sign problems ($Q_6=0.45$), and
  one point in between ($Q_6=0.20$). The lines have slopes $1/\nu_t = 1.58$ (blue lines) and $1/\nu_* = 2.40$ (red lines), where, in support of our expectations,
  the former has a larger range of applicability for small $Q_6$ and the latter should describe the slopes in the pseudocritical scaling regime before the crossover
  to asymptotic first-order behavior with much larger slope $d+z'=4$ (green dashed line) \cite{zhao19} that the largest-$L$ points for $Q_6=0.45$ may already
  approach.}
\label{jq2q6uslope}
\end{figure}

Next we study the cumulant derivatives in the $J$-$Q_2$-$Q_6$ model, in order to examine the expected shift of the crossover from $\nu_t$ to
$\nu_*$ scaling as we move deeper into the first-order line, as schematically illustrated in Fig.~\ref{xi}, where the generic parameter $k$
now corresponds to $Q_6-Q_{6c}$. Fig.~\ref{jq2q6uslope} shows results for three values of $Q_6$, with the slopes extracted at the cumulant crossing
points as before in Fig.~\ref{slopes_jq2} at $Q_6=0$. In this case we only show results obtained from the AFM cumulants, but the VBS results look
very similar. The expected shift of the crossover is seen clearly. At $Q_6=0.05$, the behavior is still similar to the case $Q_6=0$ in
Fig.~\ref{slopes_jq2}, with the data points up to $L \approx 80$ falling close to the line corresponding to $\nu_t$ scaling. We do not have data
for very large $L$ here, but the early stage of crossover to $\nu_*$ scaling is seen at $L\approx 80$, somewhat smaller
than at $Q_6=0$ in Fig.~\ref{slopes_jq2}. At the point furthest inside the first-order line, $Q_6=0.45$, there is no regime for smaller sizes where
the scaling is governed cleanly by $\nu_t$; instead, for $L\agt 30$ the behavior is already well described by $\nu_*$. Thus, when moving further
into the first-order line, the manifestations of the bubble scale are significant already on smaller length scales, as in the schematic illustration
Fig.~\ref{xi} for increasing $k$. Finally, at a point between these extremes, $Q_6=0.2$, we explicitly see the crossover taking place
at $L \approx 40$.

A final crossover not seen clearly in Fig.~\ref{jq2q6uslope} is expected that reflects the conventional aspects of the first-order transition. For a
finite system, the scaled distance $\delta L^{a}$ to the first-order point has a large exponent; $a=d$ classically and likely $a=4$ in the case
at hand \cite{zhao19}, which implies an ultimately faster large-$L$ growth of the cumulant derivatives with $L$ than $L^{1/\nu_*}$. We expect this
first-order finite-size scaling to hold also in the presence of two physical length scales, $\xi_s$ and $\xi_*$, at the transition, and there are
indeed signs of an onset of this faster growth for the largest systems at $Q_6=0.45$ in Fig.~\ref{jq2q6uslope}.

To further elaborate on the ultimate first-order finite-size scaling, the effective correlation-length exponent reflects two different phases occupying different
parts of the Hilbert space. Since a finite system can still tunnel between these subspaces, through the formation of system spanning domain walls, there
are fluctuations on the length scale $L$ that are not present in the thermodynamic limit. It is well known from the classical literature that these
finite-size fluctuations result in a scaling argument $\delta L^{1/d}$, i.e., $\nu=1/d$ in analogy with the correlation-length exponent at a critical
point \cite{binder81b,vollmayr93,iino19}. In the quantum case we should have $1/\nu=d+z'$, with $z'=2$ when at least one of the coexisting phases is
long-range ordered and associated with a tower of Anderson rotor states (with states at energies $1/L^d$) in a finite-size system in which the spin-rotational
symmetry is not broken \cite{zhao19}. For the $J$-$Q_2$ model this final stage of the transition cannot be observed with systems of practical size (Fig.~\ref{slopes_jq2}),
and even for the $J$-$Q_2$-$Q_6$ model at $Q_6=0.45$ (Fig.~\ref{jq2q6uslope}) the slope taken with the largest 2-3 points corresponds to $1/\nu \approx 3$. Values as large
as $1/\nu \approx 3.5$ were observed in a different model with emergent O($4$) symmetry \cite{zhao19}, where the first-order tendency is larger.

\begin{figure}[t]
\includegraphics[width=80mm]{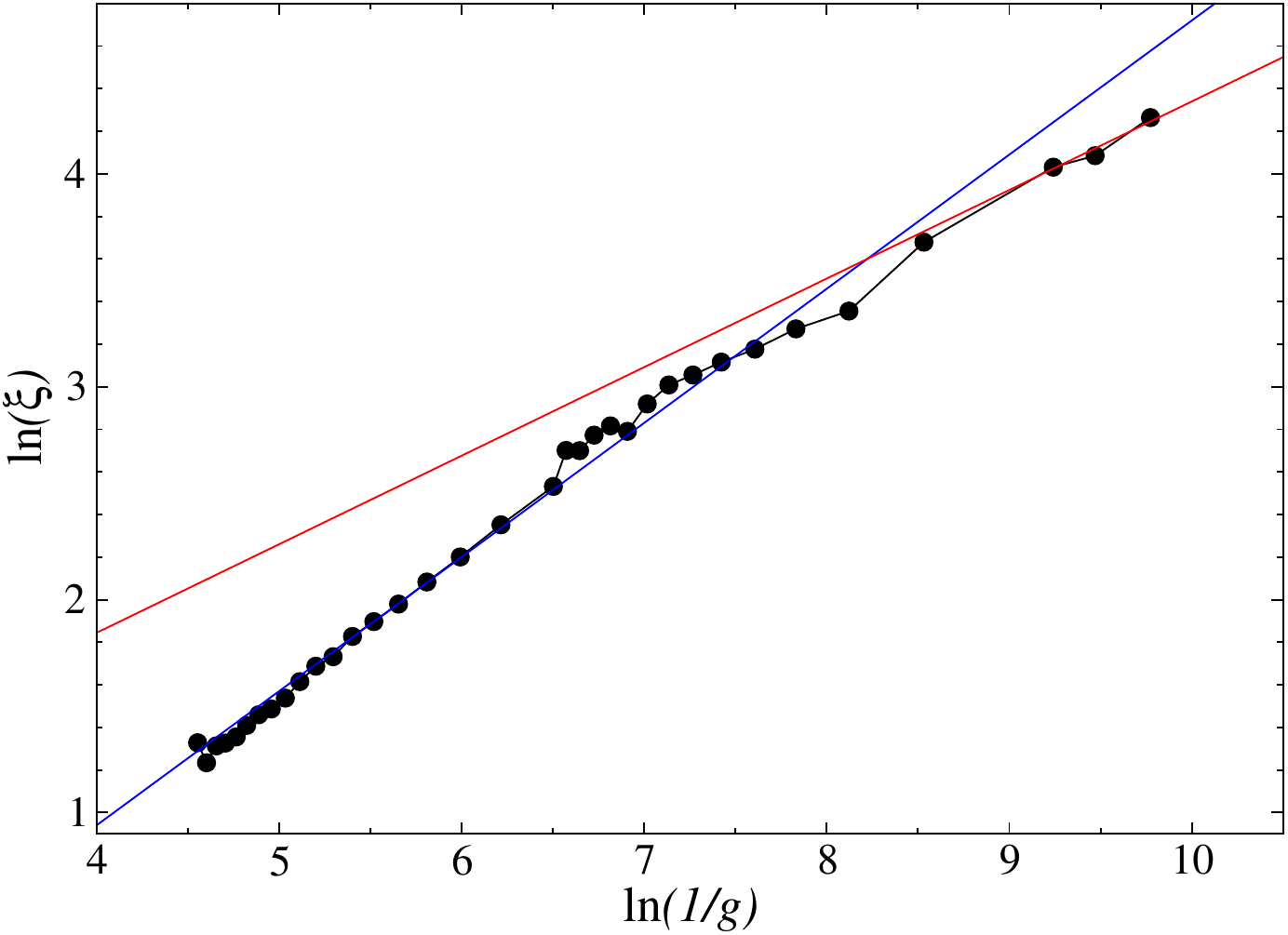}
\caption{Dependence of the logarithm of the correlation length on the logarithmic distance to the phase transition in the 3D loop model \cite{nahum15a},
with $g=|J-J_c|$. The circles are data reproduced by digitizing Fig.~18 of Ref.~\cite{nahum15a} (for error bars we refer to the original article)
and the blue and red lines show, respectively, the predicted scaling behavior with our exponent values $\nu_t = 0.632$ and $\nu_*=0.416$. The crossing
point of the red and blue lines corresponds to the crossover coupling $g_1^*$ in the schematic Fig.~\ref{xi}.}
\label{nahumxi}
\end{figure}

Here we have focused on the manifestation of the crossover in finite-size scaling of the exponent governing the correlation length. The physical bubble
size at the first-order transition in the thermodynamic limit will eventually saturate at a value set by the field $s$, as schematically illustrated in
Fig.~\ref{xi}. Though it is the conventional correlation length $\xi_t$ that converges to the finite $\xi_s \sim s^{-\nu_s}$, the growth of $\xi_*$ must
also halt at the same point because the ground state does not evolve further until the first-order transition is crossed. The bubble size thus converges
to a scale of order $\xi_s^{\nu_*/\nu_t}$ or $s^{-\nu_s\nu_*/\nu_t}$, which was already used in Sec.~\ref{sub:darkscale} and which is approximately $s^{-0.90}$
with our exponent values.

We have not computed the correlation length itself under the condition $\xi \ll L$, but exactly such a calculation for the 3D loop model was presented by
Nahum et al.~in Fig.~18 of Ref.~\cite{nahum15a} versus a distance $g$ to the transition. We expect the length $\xi$ studied there to be governed by the
conventional exponent $\nu_t$ for some $g>g_1^*$ but then cross over to the slower growing bubble scale with exponent $\nu_*$ for $g < g_1^*$. For convenience,
we have re-graphed their data in Fig.~\ref{nahumxi} along with lines showing the predicted scaling with the inverse of the distance $g=|J-J_c|$ to the transition
point. The behaviors $\xi \sim g^{-{\nu_*}}$ and $\xi \sim g^{-{\nu_t}}$ with our values of $\nu_*$ and $\nu_t$ are matched remarkably well for systems close to and far away,
respectively, from the transition. Like in Fig.~\ref{slopes_jq2}, the largest length scale here is barely enough to clearly follow the behavior $\xi \sim g^{-\nu_*}$
without any influence from $\nu_t$, but the crossover to the slower growth is still compelling, considering that only the amplitudes of the fitted lines are
adjustable in Fig.~\ref{nahumxi}. The behavior conforms with our assertion in Fig.~\ref{xi}, apart from the missing crossover to the eventual first-order behavior,
which just means that the saturation value has not yet been reached at the minimum value of $g$ (and going to still smaller $g$ values would necessitate very large
system sizes in order to satisfy $L \gg \xi_*$). This previously unexplained behavior serves as further confirmation of the pseudocritical crossover, as well as
the close similarity of the $J$-$Q$ and loop models.

\section{Anomalous finite-size scaling}
\label{sec:anomalous}

The crossover behavior of the correlation length will impact also other physical observables---those that are directly related to the correlation
length, in a way that will be made more precise in this section. In Ref.~\cite{shao16} it was already realized that a scaling form involving two
length scales could explain some of the anomalous behaviors that had been observed in the lattice models. While those insights remain technically valid,
the physical interpretation of the roles of the different length scales has to be revised in light of our current understanding of the exponent $\nu_*$.
Previously this exponent was just called $\nu$, as it was assumed to be associated with the single relevant symmetric
operator within the original DQCP scenario. Another exponent $\nu'>\nu$ was extracted that appeared to fit within the conventional DQCP scenario and
its larger U($1$) scale in the near-critical VBS phase, which was predicted \cite{senthil04a,senthil04b,levin04} to also set the scale of spinon
deconfinement. The numerical values of the exponents determined in Ref.~\cite{shao16} were $\nu = 0.446(8)$, extracted from cumulant slopes, and
$\nu'= 0.585(18)$, from the scaling of the critical VBS domain wall energy density as well as an observable explicitly probing the size of a bound
state of two spinons. These values agree reasonably well with our improved results (which benefit from larger systems and in many cases also smaller
statistical errors) when we replace $\nu \to \nu_*$ and $\nu' \to \nu_t$.

The quantitatively most important result of Ref.~\cite{shao16} was that the finite-size scaling forms for many physical observables depend on the ratio
$\nu/\nu'$. As we will see below, this ratio, now identified as $\nu_*/\nu_t$, is indeed of crucial importance in finite-size scaling of a broad class of
observables, but with $\nu_*$ governing the bubble scale and $\nu_t$ the conventional correlation length when approaching the SO($5$) critical point.
If indeed $\nu' \equiv \nu_t$, as it appears,
the scale of spinon deconfinement is actually the correlation length and should then, presumably, exhibit crossover between $\nu_t$ and $\nu_*$ scaling
when approaching the weak first-order line. Since it is hard to extract the size of a bound state for large systems---only lattices of size up to $L=64$
were used in Ref.~\cite{shao16}---the crossover scale was likely not reached (though some weak tendencies are seen in Fig.~1 of
Ref.~\cite{shao16}). We will not revisit the spinon scale directly here, but will discuss our updated view of the extended scaling hypothesis
where $\nu_*/\nu_t$ appears. We first demonstrate the impact of more than one length scale in the drift of the AFM--VBS transition point in
Sec.~\ref{jqcscaling}, then formulate an extended finite-size scaling ansatz for other observables in Sec.~\ref{sec:fsform}, and present
an example of this ansatz applied to the spin stiffness in Sec.~\ref{sec:fsexample}.

\subsection{Crossover scaling of the transition point}
\label{jqcscaling}

\begin{figure}[t]
\includegraphics[width=80mm]{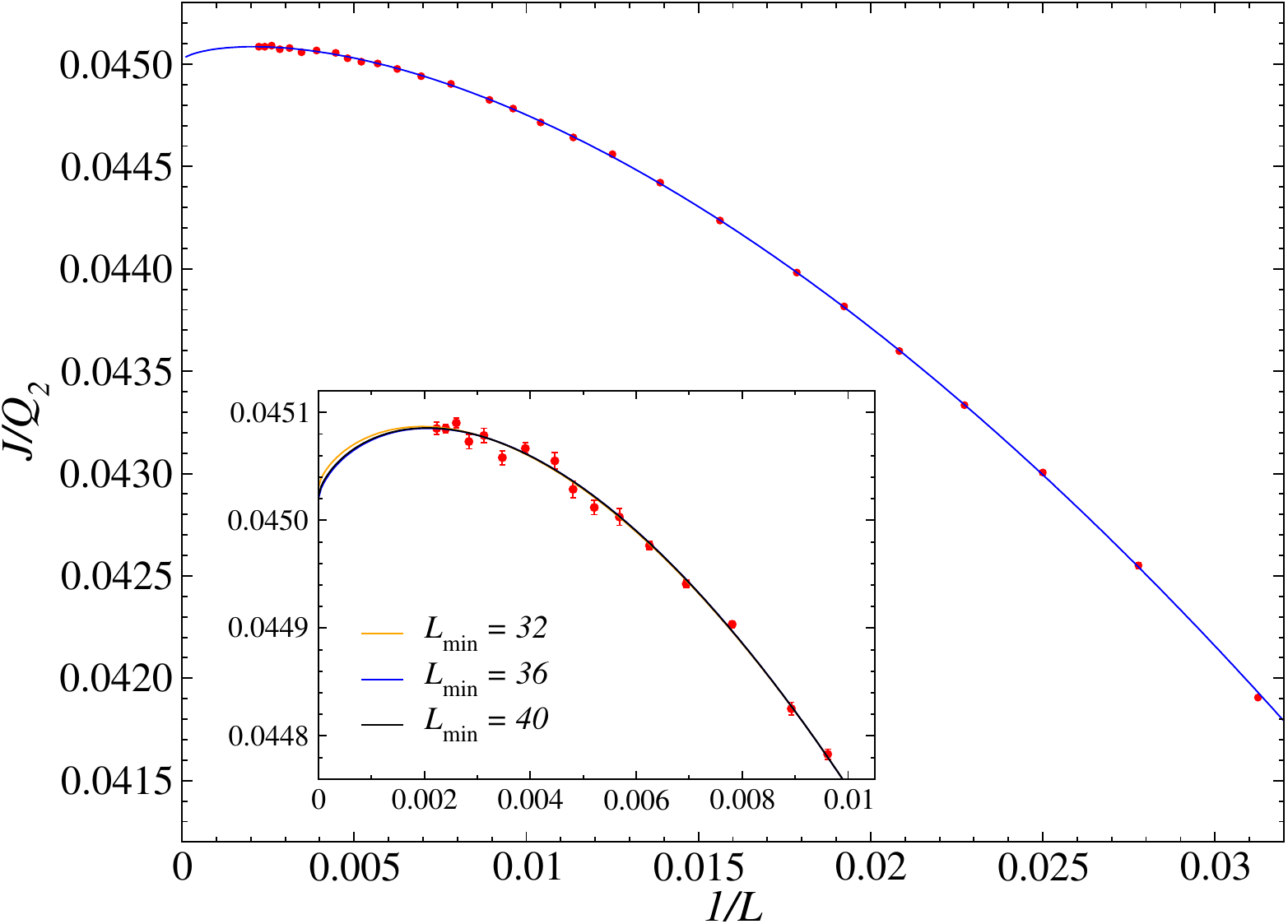}
\caption{Crossing values $J/Q_2$ of the SO($5$) versions of the AFM and VBS Binder cumulants defined in Eqs.~(\ref{udefs2}) for the $J$-$Q_2$ model,
obtained from the same data sets as in Fig.~\ref{slopes_jq2}. The curve is a fit to the form Eq.~(\ref{jcfitform}). The inset zooms in on the data
for $L > 100$, with the three almost indistinguishable curves corresponding to different minimum system sizes included in the fit.}
\label{cpoint}
\end{figure}

Based on the long-distance correlation functions of the $J$-$Q_2$ model in Fig.~\ref{corL_jq2}, the AFM--VBS transition should be located somewhere
between $J/Q_2=0.04500$ and $J/Q_2=0.04505$, which is slightly smaller than the value $J_c/Q_2=0.04510(2)$ obtained by extrapolating the conventional cumulant
crossings in Ref.~\cite{sandvik20}. Results based on the SO($5$) crossing points and improved SSE data are shown in Fig.~\ref{cpoint}. At first sight,
it again looks like the $L \to \infty$ crossing point tends to about $J_c/Q_2=0.04510$, at odds with the results in Fig.~\ref{corL_jq2}, which indicates
that the system at that coupling ratio is marginally inside the AFM phase. We therefore analyze the cumulant crossing points in more detail next.

Instead of fitting the data versus $x=1/L$ to the form $f(x)=f_0 + ax^b+cx^d$ with five adjustable parameters (as was done in Ref.~\cite{sandvik20}),
we here use a constrained form that should be expected in the presence of the three length scales governed by the already determined exponents
$\nu_s$, $\nu_t$, and $\nu_*$:
\begin{equation}
J_c(L) = J_c(\infty) + a L^{-1/\nu_s} + bL^{-1/\nu_t} + cL^{-1/\nu_*},
\label{jcfitform}
\end{equation}
with only $J_c(\infty)$ and the coefficients $a,b,c$ taken as fitting parameters. This form can be motivated by first considering a system approaching
the critical point exactly, where $\nu_*$ does not come into play. On the line $t=0$, the condition of the two cumulants being equal to each other
corresponds to some function $h(x)=h(sL^{1/\nu_s})=0$, which has some solution $x=x_0$. Thus, the size dependent critical point scales as
$s_c \sim L^{-1/\nu_s}$. Similarly, on the line $s=0$ we have $t_c \sim L^{-1/\nu_t}$. For a system tuned to the critical point on some line $\delta$
such that $s=d_s\delta$ and $t=d_t\delta$, with $d_s,d_t$ arbitrary nonzero coefficients, both powers come into play for a finite-size
shift of the form $\delta_c \sim aL^{-1/\nu_s} + bL^{-1/\nu_t}$. In Eq.~(\ref{jcfitform}) we have posited that a similar term originating from the
bubble scale appears when approaching the weak first-order line with small $s \not= 0$.

Since $1/\nu_s \approx 0.73$ is much smaller than the other exponents, it will eventually control the finite-size corrections for large systems. In fact, we find
that the coefficients $a$ and $b$ have different signs, leading to nonmonotonic behavior that is not yet obvious with the SSE $L \le 448$ data at hand but
is a very stable feature of the fit. As shown in the inset of Fig.~\ref{cpoint}, for three data sets including different minimum system sizes, the fitted
function consistently has a maximum at $L \approx 500$. The extrapolated value is stable (also when excluding additional small systems) and now agrees
very well with the previous conclusion in Sec.~\ref{sec:order} (from the long-distance spin and dimer correlation functions) that the critical coupling ratio
should be between $J/Q_2=0.04500$ and $J/Q_2=0.04505$. Based on the statistical quality of the fits and the variations among different $L_{\min}$, we conclude that
the critical point is at $(J/Q_2)_c = 0.04502(1)$. The third correction in Eq.~(\ref{jcfitform}) is actually not very important for system size $L \agt 50$, since
the exponent $1/\nu_*$ is much larger than the other two exponents. Leaving out this term still produces a statistically good nonmonotonic fit and critical point
consistent with the value stated above.

It should be noted that fitting two independent power laws does not produce the same exponents as the known ones imposed above. In fact, the best fit has
two exponents very close to each other, both around $1.9$, and with opposite signs of their coefficients. This behavior in itself indicates that the best
fit mimics a different form. While the fit with fixed exponents $1/\nu_s$ and $1/\nu_t$ is not completely optimal (since other exponents give marginally lower
$\chi^2$ values), it is still statistically good, as can even be judged visually in the inset of Fig.~\ref{cpoint} (and $\chi^2/{\rm dof} \approx 1$). The
known exponents thus provide a good explanation for the nonmonotonic crossover behavior, the presence of which can also be inferred from our other indications
(Sec.~\ref{sec:order}) of a transition point falling between $J/Q_2=0.04500$ and $J/Q_2=0.04505$. The extrapolated transition point $(J/Q_2)_c \approx 0.04502(1)$
also agrees very well with an estimate based on the AFM and VBS correlation lengths (in a combination that apparently cancels a correction) \cite{sandvik10b}.

A similar nonmonotonic behavior of crossing points was observed in a dimerized 2D Heisenberg model \cite{ma18b}, where the behavior is caused
by a conventional scaling correction. An even more intricate (and harder to extrapolate) nonmonotonic behavior was observed in crossing points of the
conventional [i.e., not SO($5$)] version of the Binder ratio Eq.~(\ref{bratioafm}) at a different ratio $\beta/L$ \cite{shao16}. In principle there could
be some further change in behavior for even larger $L$ than in Fig.~\ref{cpoint}. However, the very good agreement between the transition point extracted
here and the bracketing values obtained from long-distance spin correlation in Sec.~\ref{sub:corrL} suggests that any further change in $(J/Q_2)_c$ must
be very small. Note that not all quantities are affected equally by the presence of more than one length scale, since the coefficients of the corresponding
terms in the scaling function are specific to a given quantity \cite{ma18b}.

\subsection{Extended finite-size scaling hypothesis}
\label{sec:fsform}

In standard finite-size scaling \cite{fisher72,binder81a,barber83}, a physical observable $A$ with scaling behavior $A \sim \delta^\kappa$ (with $\kappa$
denoting a generic critical exponent) in the thermodynamic limit is first rewritten using the correlation length $\xi\sim \delta^{-\nu}$, i.e.,
$A \sim \xi^{\kappa/\nu}$. Then at criticality $\xi$ is simply replaced by the system size $L$. Thus, $A(L) \sim L^{-\kappa/\nu}$, which is valid in a window
of size $L^{-1/\nu}$, where the correlation length exceeds $L$. Finite-size scaling effectively corresponds to the flow of observables with the length
scale on which they are probed, motivating the sometimes used term ``phenomenological renormalization'' \cite{fisher72}.

The easiest way \cite{shao16} to construct a modified finite-size behavior in the presence of the bubble scale $\xi_*$ is to posit that it should
be used in place of $\xi$ above, i.e., $A \sim \xi_*^{\kappa/\nu}$. Here we have only changed the relevant length scale that appears, without changing
$\kappa/\nu$ (which is related to the scaling dimension of $A$ exactly at the critical point). Since $\xi_* \sim \xi^{\nu_*/\nu}$, we can also write
$A \sim \xi^{(\kappa/\nu)(\nu_*/\nu)}$. If we now replace $\xi$ by $L$ we obtain $A(L) \sim L^{-(\kappa/\nu)(\nu_*/\nu)}$, or, more specifically when the AFM--VBS
transition is crossed by tuning the $t$ field and $\nu$ corresponds to $\nu_t$,
\begin{equation}
A(L) \sim L^{-(\kappa/\nu)(\nu_*/\nu_t)},
\label{alnustarform}
\end{equation}
Here the exponent indeed differs by a factor $\nu_*/\nu_t$ from the conventional finite-size exponent $\kappa/\nu_t$. The form should be valid at the
transition point roughly in the range of system sizes corresponding to the window $[g_2^*(k),g_1^*(k)]$ of a tuning parameter $g$ in the schematic drawing
of the crossover behavior in Fig.~\ref{xi}.

The physical interpretation of the steps leading to Eq.~(\ref{alnustarform}) was \cite{shao16} that the shorter length scale $\xi_*$ ceases to grow when
the larger scale $\xi$ reaches $L$. If that is so, then it would appear that the two length scales are present at the same time (with the longer scale $\xi_t$
having been replaced by $L$ in finite-size scaling), which is not clear from the different finite-size analyses we have carried out and from the results
for $\xi$ versus the distance to the transition by Nahum et al.~\cite{nahum15a} (data reproduced here in Fig.~\ref{nahumxi}). Here we have explained this
simultaneous presence of the length scales as originating from irrelevant repulsive interactions between the two order parameters, as illustrated in
Fig.~\ref{defects}. The length scales physically correspond to the typical scale of amplitude fluctuations ($\xi_t$) and the bubbles ($\xi_*$). The shorter
bubble scale should in practice be more prominently present in most observables, since a bubble in one phase represents an order parameter of the other phase;
thus directly impacting the AFM and VBS correlation lengths.

A key question now is to what observables anomalous scaling involving $\nu_*$ applies. Clearly, a form like Eq.~(\ref{alnustarform}) cannot apply to all
operators, because the exponent $\kappa/\nu_t$ is the scaling dimension of the operator $A$ and we have presented numerous examples in the previous sections
of scaling dimensions extracted from system size dependence. For instance, in Sec.~\ref{sub:symop} we extracted the relevant scaling dimensions
$\Delta_t$ from the correlation function of the $Z$ operator at distance $r=L/2$. If this correlation function $C_Z(L)$ takes the form where
the exponent $2\Delta_t$ is modified by the factor $\nu_*/\nu_t$, then we would have to compensate by multiplying the exponent from the fit by the
inverse ratio to obtain the true values of the two scaling dimensions. This would completely ruin the good agreement in Table \ref{dtable}
between the AFM--VBS transition and the SO($5$) CFT and the fuzzy sphere model. Moreover, we considered both $r=L/2$ and $r\ll L$ in Fig.~\ref{qxljq2},
with full agreement between the decay exponents. A very strong confirmation of the correctness of the procedure is that the scaling dimension of
the conserved current comes out very close to $\Delta_j=2$ in Fig.~\ref{wcor}.

The bubble size $\xi_*$ should be at the heart of the matter, and we propose that only those quantities are affected that relate explicitly to the order
parameters through the conventional correlation lengths $\xi_t$ and $\xi_s$. Such quantities would typically involve integration over the lattice and there
should then also be similar anomalous scaling behaviors away from the transition point in the thermodynamic limit, related to the crossover of the correlation
length illustrated in Fig.~\ref{xi}. For example, a squared order parameter (here VBS or AFM) scales with the distance to the critical point as $m^2=\delta^{2\beta}$,
where $2\beta = 2\Delta_\phi\nu$, with $\nu=\nu_t$ if only the field $t$ is tuned at $s=0$ and $\nu=\nu_s$ if $s$ is tuned with $t=0$. Thus, $m^2$ depends explicitly
on an inverse correlation length, $\xi_s$ or $\xi_t$. Another example is the spin stiffness $\rho_s$, which scales as $\rho_s \sim \delta^{z\nu}$. This quantity as
well is directly related to the (AFM) long-range order (as a response to a twist of the order parameter distributed over the entire system); since $z=1$ here, the
scaling should be identical (in terms of exponents, not the shape of the full scaling function) to that of the inverse correlation length. Then it is also plausibly
affected by the crossover behavior. After presenting the general formalism here, we will study the important spin stiffness $\rho_s$ as an example below
in Sec.~\ref{sec:fsexample}.

The inverse correlation length of a system with a single tuned relevant field can be written as
\begin{equation}
\xi^{-1} = \delta^\nu f(\delta L^{1/\nu}),
\label{invxi1}
\end{equation}
where $\delta^\nu$ is the form for $L \to \infty$ and the function $f(x)$, $x=\delta L^{1/\nu}$, accounts for the finite-size cut-off. The finite-size
form $\xi^{-1} \sim L^{-1}$ is obtained from the necessary condition (to eliminate $\delta$ at criticality) that $f(x) \to x^{-\nu}$ when $x \to \infty$,
i.e., for $L \to \infty$ at $x \not=0$. An alternative, more common form of Eq.~(\ref{invxi1}) for finite-size scaling including $\delta \not= 0$ is
\begin{equation}
\xi^{-1} = L^{-1} f(\delta L^{1/\nu}),
\label{invxiL}
\end{equation}
with a different form of the scaling function $f(x)$. In practice, this form holds up to some value of $x$, beyond which there is a finite-size
peel-off from the scaling function, with that peel-off value increasing with $L$. Here we will start from a generalized form of Eq.~(\ref{invxi1})
with more than one scaling argument and derive a corresponding critical form with more than one power of $L$.

Generalizing Eq.~(\ref{invxi1}) to the case at hand, we have already argued that a scaling function depending on all the exponents $\nu_s$, $\nu_t$,
and $\nu_*$ is needed; Eq.~(\ref{fform3}). Further, we have in mind a situation where the bubble scale dominates the behavior for small values of
$s$ in the thermodynamic limit and therefore start from the form
\begin{equation}
\xi^{-1} = \delta^{\nu_*} f(\delta L^{1/{\nu_*}},\delta L^{1/{\nu_t}},\delta L^{1/{\nu_s}}).
\label{invxi2}
\end{equation}
Here we have replaced all the factors in front of $L$ by just $\delta$, but it should be kept in mind that different arguments will dominate
the behavior depending on how $s$ and $t$ are tuned by $\delta$. The last scaling argument should really be $(\delta+s_0) L^{1/{\nu_s}}$, with $s_0$ the
value of $s$ when we cross $t=0$ ($\delta=0$). Then the system will eventually cross over to a first-order form that we do not consider here. The question
is whether there is some range of $L$ where Eq.~(\ref{invxi2}) can describe the finite-size behavior when the transition is weakly first-order,
i.e., when $s_c$ is very small and there is a large range of length-scales governed by $\xi_*$ in the thermodynamic limit, as in Fig.~\ref{xi} for small $k$.

In the example of the cumulant derivative $U'$ in Sec.~\ref{sec:cumslopes} we already saw how the scaling function in Eq.~(\ref{fform3}) can, effectively,
produce three different power laws. Since $U$ is a dimensionless quantity, there is no power of $\delta$ in front of the scaling function and
Eq.~(\ref{uderiv_abc}) is produced by the derivative. Such a form was able to reproduce the data in Fig.~\ref{slopes_jq2} very well from small to
large systems. For the inverse correlation length of interest now, the asymptotic power of the arguments in Eq.~(\ref{invxi2}) must be
$-1/{\nu_*}$ in order to eliminate the overall $\delta$ dependence, thus producing the form
\begin{equation}
\xi^{-1}(L) = aL^{-1}+bL^{-\nu_*/{\nu_t}}+cL^{-\nu_*/{\nu_s}}.
\label{invxi3}
\end{equation}
This form is valid under the assumption that each argument, when large, produces its own power law to remove the $\delta$ dependence at criticality.
While we have not proved this property, Eqs.~(\ref{invxi2}) and (\ref{invxi3}) are certainly mutually consistent. We will also see below that
Eq.~(\ref{invxi3}) is sufficient to describe numerical data without any cross terms or even more complicated crossover behavior.

According to Eq.~(\ref{invxi3}), a conventional finite-size scaling term $L^{-1}$ will be produced in addition to the anomalous $L^{-\nu_*/{\nu_t}}$
term of the original two-length proposal \cite{shao16}. The anomalous term decays slower and may dominate the behavior for large system sizes. In the
last term, $L^{-\nu_*/{\nu_s}}$ decays even slower but here the coefficient $c$ should be small if the crossing value $s_0$ of the $s$ field is small.
Thus, it may only become important for very large system sizes, where the other terms have essentially decayed away. When the last term does become
important, we also expect the first-order behavior to take over. In practice, only the $L^{-1}$ and $L^{-\nu_*/{\nu_s}}$ terms will likely reflect pseudocritical scaling.

\subsection{Spin stiffness}
\label{sec:fsexample}

The spin stiffness $\rho_s$ provided the first hint of anomalous scaling in the $J$-$Q_2$ model \cite{jiang08} and has frequently been used to argue for a
first-order transition \cite{jiang08,chen13}. It is computed in SSE (and other QMC) simulations in terms of winding number fluctuations (of spin currents in
systems with periodic boundary conditions) \cite{pollock87,sandvik97}. The conventional critical form when $z=1$ is $\rho_s \sim L^{-1}$, arising from
$\rho_s \sim \delta^{\nu_t}$ when crossing the transition by tuning $t$ at $s=0$ in the thermodynamic limit. Thus, the modified $L$ dependent scaling form is
identical to that for $\xi^{-1}$ in Eq.~(\ref{invxi3}),
\begin{equation}
\rho_s(L) = \rho_s(\infty)+aL^{-1}+bL^{-\nu_*/{\nu_t}}+cL^{-\nu_*/{\nu_s}},
\label{rhoslform}
\end{equation}
except for the added constant $\rho_s(\infty)$, which reflects an expected nonzero value for $L \to \infty$ because of the weak AFM order in the
coexistence state, in analogy with how the non-vanishing order parameter itself was handled in correlation functions in Sec.~\ref{sec:order}.
While the contribution originating from the $s$ field close to criticality may play some role, the crossover to the ultimate constant
form for $L \to \infty$ at the transition (or to other forms away from the transition) is likely more intricate. Moreover, even with high precision data
it is not possible to reliably extract the relative contributions of the constant and the $L^{-\nu_*/{\nu_s}}$ term. We therefore leave out the $s$
contribution here.

\begin{figure}[t]
\includegraphics[width=80mm]{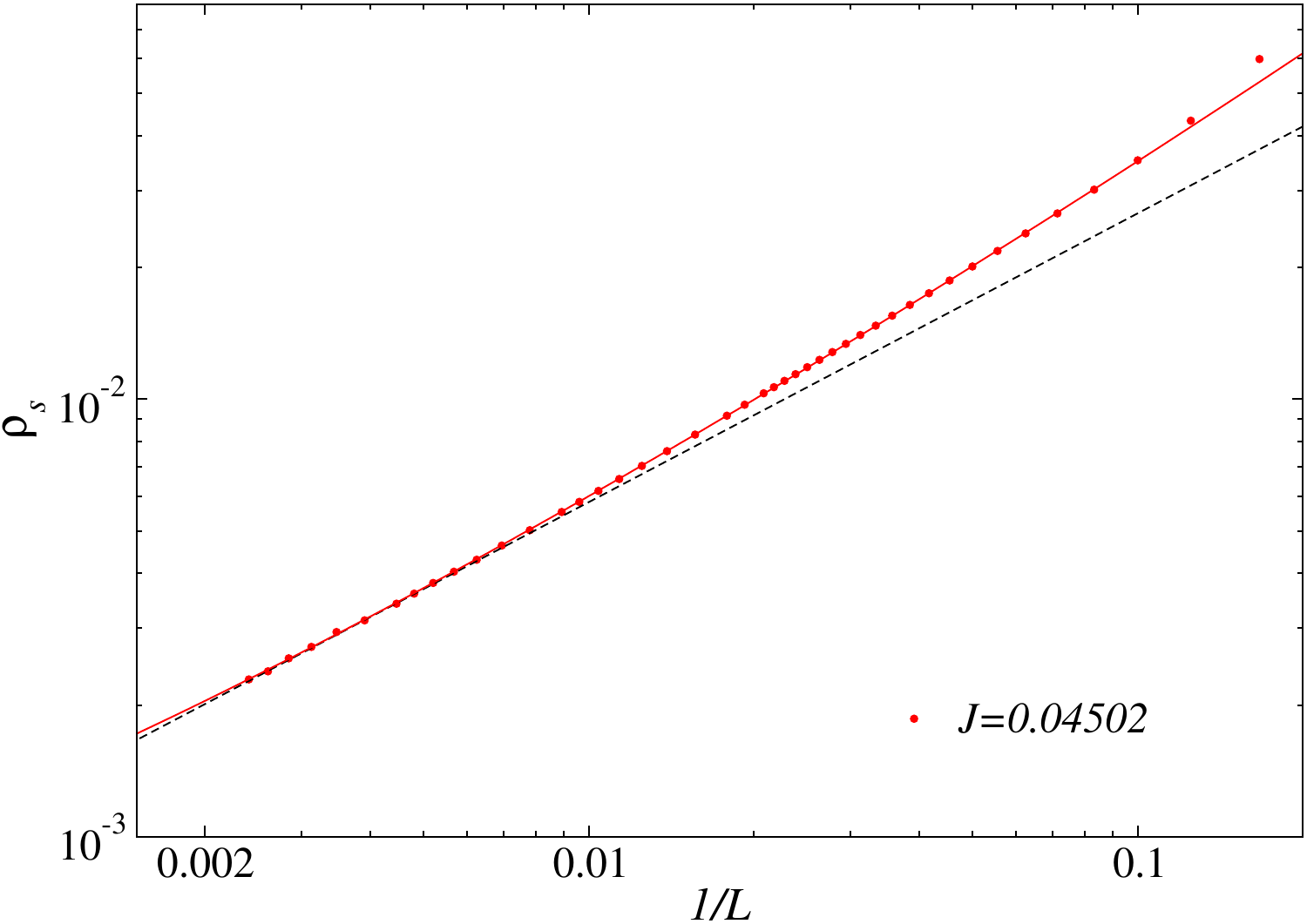}
\caption{Spin stiffness of the $J$-$Q_2$ model at the transition point, taken as $J=0.04502$ ($Q_2=1$). The red curve shows a fit to
Eq.~(\ref{rhoslform}) with $c=0$ and the exponent fixed at $\nu_*/{\nu_t}=0.66$. The dashed line shows the single
power law $\propto L^{-\nu_*/{\nu_t}}$, with the factor adjusted for best fit to the large-$L$ data.}
\label{rhos}
\end{figure}

We fit to SSE data for the $J$-$Q_2$ model generated at inverse temperature $\beta=0.42L$,
with the factor $0.42$ being approximately the inverse of the critical spin wave velocity \cite{suwa16}, thus representing cubic space-time geometry
(which is not a prerequisite for scaling but may reduce some corrections). We have data for large systems, up to $L=448$, at two coupling ratios, $J=0.0450$ and
$J=0.0451$ ($Q_2=1$), and interpolated in those data sets to the estimated transition point $J=0.04502$.

The data are graphed versus $1/L$ in Fig.~\ref{rhos} and have been fitted to the expected form for $L \ge 12$, shown with the red curve. The dashed
black line shows the form $L^{-\nu_*/{\nu_t}} = L^{-0.66}$, where the exponent is slightly smaller than in Ref.~\cite{shao16} ($\nu_*/{\nu_t} \approx 0.715$).
While this term indeed governs the large-$L$ data very well, the $L^{-1}$ term is important to match the results for the smaller systems and also to improve
the fit overall before the asymptotic behavior sets in. The $L^{-1}$ term corresponds almost exactly to a correction term with optimized exponent
used in Ref.~\cite{shao16}. Thus, the results of the two fits are almost the same (but here we also have data for larger systems), with the main
difference being the added constant $\rho_s(\infty)$ in Eq.~(\ref{rhoslform}), which has a value $\approx 0.0003$ in our fit and plays only a
minor role in the available range of system sizes. There are still, not unexpectedly, further corrections for the smaller system sizes
that are not included in Eq.~(\ref{rhoslform}). Overall, the extended scaling ansatz works well here.

\section{Conclusions and Discussion}
\label{sec:discuss}

The many past efforts to elucidate the nature of the DQCP and its adjacent phases can be likened to the ancient Indian parable of a group of blind men
examining an elephant, an entity that they have never before encountered. Feeling different parts of the animal without access to the whole, they
present their individual findings and theories but, given the conflicting observations, they cannot reach any consensus on what they are dealing
with. In the case of the putative DQCP, many models and methods have been applied to study the AFM--VBS transition in different ways. Individually, most of
these studies present plausible and coherent scenarios for exotic critical phenomena, but, taken together, no consistent unified picture has emerged.
Here we have reconstructed the entire elephant---a multicritical point with emergent SO($5$) symmetry at the end of a line of weak first-order
transitions---which explains previously irreconcilable observations and scenarios.

\subsection{Multicritical SO(5) point}
\label{sub:disc1}

We have analyzed our numerical results in a space of two scaling fields (contained in ratios of model parameters)
and extracted scaling dimensions compatible with results for an SO($5$) CFT. The CFT results of main interest here \cite{li18,chester24}
are strictly speaking only boundaries of allowed values in the space of scaling dimensions. However, it is believed that the true exponents should fall at or
very close to the boundaries, as is the case for O($N$) models \cite{poland19}. Our relevant scaling dimensions listed in Table \ref{dtable} fall within about
1-2\% from the boundaries, read off of graphs available for the combinations $(\Delta_\phi,\Delta_s)$ and $(\Delta_\phi,\Delta_t)$ \cite{li18}. While $\Delta_4$,
as well as the other dimensions, is in less good agreement with Ref.~\cite{chester24}, the discrepancies can likely be explained by the input value
$\Delta_\phi=0.63$ used in that calculation. Our value is lower, $\Delta_\phi= 0.607(4)$, and once this is taken into account the agreement is improved
for all scaling dimensions \cite{chester24}. Overall, the remarkably good agreement between two very different calculations represents strong support for the
multicritical SO($5$) CFT.

The multicritical point is inaccessible (with our current models) in QMC simulations, because of sign problems (e.g., a negative $Q_6$ interaction
would be needed, according to Fig.~\ref{jq2q6_m2}). However, in our view the point is not fundamentally inaccessible, in contrast to the complex CFT scenario
\cite{wang17,zhou23}. Methods such as DMRG and machine learning approaches can be used to study models like the $J_1$-$J_2$-$Q_2$
Hamiltonian, but on much smaller lattices than those accessed here or with infinite-size approaches that require other cut-offs. Useful results may
still be obtained, because the scaling behavior will not be contaminated by the effects from the weak first-order transition. Based on previous
results for frustrated spin models \cite{gong14,morita15,wang18,ferrari20,nomura21,yang22,wang22,liu22,liu23,viteritti23,corboz26}, which can be
continuously deformed into the $J$-$Q$ model studied here, there is also most likely a gapless QSL phase adjacent to the parameter regime accessible to
efficient (sign problem free) QMC simulations. The simplest scenario then is that the phase diagram is of the type in Fig.~\ref{phases}(b), where the
multicritical point (which then is bicritical) is also the tip of the QSL phase. A QSL connected to a direct AFM--VBS transition was actually observed
in a tensor-network study of the $J_1$-$J_2$-$J_3$ model \cite{liu23}, but it was not possible to determine the nature of the special point and
the direct transition.

In principle, there could also be a line of generic DQCPs, as in Fig.~\ref{phases}(a), with the spin liquid only opening for still smaller $s$ values.
However, $N=2$ is a special case \cite{senthil06} of the more general SU($N$) theory \cite{senthil04a}, being the only case for which the
emergent SO($5$) symmetry is expected---at least for large $N$, where the anomalous dimensions for the spin and dimer order parameters become different,
thus prohibiting a unified order parameter. The multicritical scenario without generic DQCP transition may therefore also be unique to this case.
The generic line of DQCPs between the first-order end point and the tip of the spin liquid would then have to emerge for $N>2$. It is already known that
some first-order signatures weaken with increasing $N$ \cite{song23b} and some of the critical exponents have been extracted for different values of $N$
\cite{beach09,kaul12,block13,harada13}. With our new insights on the nature of the transition for $N=2$ and the associated complications in extracting
scaling dimensions, further work on SU($N$) spin models with $N>2$ is also called for.

\subsection{Emergent pseudocriticality and excitations}
\label{sub:disc2}

Beyond the nature of the critical point that we have now resolved, we have also demonstrated that the emergent SO($5$) symmetry leads to previously not
understood pseudocritical scaling behaviors when the weak first-order transition is approached. While previous works found evidence of two length scales and
related anomalous scaling behaviors \cite{shao16}, here we have clarified the origin of the emergent length scale and its consequences. We have
presented concrete evidence for the presence of an effectively repulsive irrelevant interaction between the AFM and VBS order parameters, which
leads to a specific deformation of the SO($5$) symmetry that we have observed. We have argued that the CFT operator responsible for the observed
behavior is the lowest irrelevant rank-$2$ CFT operator $t'$, with all other quantum numbers the same as the traceless symmetric tensor $t$ driving
the AFM--VBS transition. Such an operator with reasonably similar scaling dimension was identified with the fuzzy sphere approach \cite{zhou23} but not ascribed
any particular significance.

The symmetry deformation leads to two different length scales of fluctuations of the symmetry-breaking phases in the neighborhood of the weak (nearly critical)
first-order transition. The size of defects (bubbles) formed by fluctuations of one of the ordered phases into the other phase is governed by the pseudocritical
exponent $\nu_*$, for which we have derived a relationship, Eq.~(\ref{nustardeltas}), to scaling dimensions of the multicritical point. This relationship is
obeyed by our numerical results for the $J$-$Q$ models. The conventional correlation length is associated with the amplitude fluctuations of the order parameter,
and there are also two (in the AFM phase) or one (in the VBS phase) gapless Goldstone modes left. The various crossovers and symmetry considerations are
summarized in Fig.~\ref{defects}. It should be noted that the length scale $\xi_*$ is different from the conventional length scale generated inside an ordered
phase (specifically here in the VBS phase) by a dangerously irrelevant perturbation, which we have also discussed and in some cases studied quantitatively.
Importantly, while the conventional emergent U($1$) and SO($5$) length scales are larger than the conventional correlation lengths governed by the exponents
$\nu_t$ (when crossing between the AFM and VBS phases) and $\nu_s$ (on the coexistence line), the pseudocritical length $\xi_*$ is actually smaller;
$\nu_* < \nu_t < \nu_s$. The pseudocritical crossover is also different from the conventional ``walking'' scenario of pseudocriticality
\cite{wang17,gorbenko18a,gorbenko18b,ma20,nahum20,he21}, where no emergent length scale $\xi_*$ has been discussed.

The pseudocriticality can be regarded as a ``beyond Landau'' phenomenon, since it relies on the emergent SO($5$) symmetry. Classical $N$-component spin models cannot
have emergent O($N>2$) (unless all relevant perturbations are tuned away) \cite{pelissetto02,eichhorn13}. Thus, the proposed topological $\theta$ term
\cite{senthil06,tanaka05} is likely needed in the field theory, which in the most well known incarnation would be the 2+1 dimensional WZW $k=1$ theory with
SO($5$) symmetry (which has to be properly regularized, e.g., within the fuzzy sphere approach \cite{zhou23} or with lattice simulations \cite{chen26}). It has
been believed that this theory has no relevant SO($5$) singlet, contrary to what we have found here at the AFM--VBS transition. Interestingly, the emergent
bubbles are somewhat akin to the topological defects in the original DQCP proposal, where a defect in one phase can be considered as having the order of the
other phase \cite{senthil04a}, though the origin of this phenomenon is different in our scenario and requires multicriticality with an irrelevant
repulsive interaction between the two order parameters.

We have not investigated excitations explicitly here, but the amplitude and bubble excitations of course correspond to vacuum fluctuations of excitations;
particles whose masses can in principle be extracted from correlation functions in imaginary time, from which spectral functions can also be extracted
\cite{shao23}. Such calculations have already been employed with $J$-$Q$ models to investigate the expected spectrum of (almost) deconfined spinons
\cite{ma18,yang25}, whose dispersion relation turned out to be in remarkable agreement with the fermionic $\pi$-flux state, one of the effective descriptions
of the AFM--VBS transition that is also related to the SO($5$) WZW theory \cite{tanaka05}. More recently, spin-charge separation was detected
for a hole injected into the host system \cite{yang25} and this type of fractionalization can also be described within the $\pi$-flux framework.

An important question now is how the bubble excitations discussed here fit into the picture of deconfined spinons. On the VBS side of the weak first-order
transition, the AFM bubbles can be understood as bound states of spinons \cite{shao23}, with $\xi_*$ being the size of such a state. In order to account
for the four states of two almost deconfined spinons, there must be a triplet and a singlet bubble, and these two types will only become completely degenerate
exactly at the multicritical point (where the concept of bubble excitation also breaks down). On the AFM side, the VBS bubbles carry U($1$) charge $q = \pm 1$
(vorticity, assuming only minor effects of the ultimate $\mathbb{Z}_4$ symmetry) and the total spin $S$ label does not apply because of the weak SO($3$)
symmetry breaking (though nevertheless the VBS bubbles can be regarded as $S=0$). It is not clear to what degree all these excitations (including the amplitude
modes and the remaining gapless Goldstone modes) can be separated in QMC calculations with analytic continuation, though steps in this direction have
already been taken in studies at the very nearly critical transition point \cite{ma18,yang25,liu26}. In order for the separation of the length scales (and thereby the
mass scales) to be sufficiently large, the best approach may be to stay very close to the first-order line, where $\xi_t$ and $\xi_*$ have converged, but ideally
not exactly at the transition point. Since the system size $L$ also has to be larger than both of the two length scales, the plain $J$-$Q_2$ or $J$-$Q_3$ models are
likely not ideal. The separation of the amplitude-mode and bubble scales should be clearer in $J$-$Q_n$ models with larger $n$ or extended models like $J$-$Q_2$-$Q_6$.
In the case of the latter, Fig.~\ref{jq2q6uslope} shows that $\xi_*$ can apply already from $20$ (or even less), but the larger $\xi_s$ ($\xi_t \to \xi_s$ when
converged) may still make it challenging to satisfy $L \gg \xi_s$ (to approach the thermodynamic limit). In any case, the scenario presented here should
motivate such calculations.

\subsection{Higgs-like mass generation}
\label{sub:disc3}

A very interesting aspect of the two different massive excitations is the functional resemblance to the Higgs mechanism in particle physics, in which
a SO($4$) Higgs field and a unified electroweak SU($2$)$\times$U($1$) gauge field constitute different gauge-coupled sectors. The Higgs potential drives the
spontaneous symmetry breaking of the SO($4$) field, producing the massive Higgs boson (amplitude mode), and the gauge coupling causes the three would-be
Goldstone modes to combine with the gauge fields, forming the massive $W^\pm$ and $Z$ bosons and leaving the gapless U($1$) photon. The analogy between the
near-multicritical SO($5$) quantum magnet with its amplitude mode (Higgs), bubble excitations (massive soliton-like configurations of the five-vector),
and remaining Goldstone (photon) mode(s) is actually closer to alternatives to the standard Higgs mechanism. Such theories often start from a unified field
with higher symmetry instead of the separate Higgs field \cite{bellazzini14,schmaltz05,hill03}, similar to the SO($5$) critical order parameter of the quantum
magnet, and the spontaneous symmetry breaking of this field leads to the same Higgs, $W^\pm$ and $Z$ bosons, and photon as in the standard theory, but with
possibly different mass scales.

In our scenario for the weak first-order AFM--VBS transition the irrelevant repulsive order-parameter interaction remaining when
the relevant $t$ field has been tuned very small is what causes the separation between the Higgs and bubble scales. In the absence of this interaction,
they are both controlled by the exponent $\nu_t$ of the tuning field $t$, but in the presence of the interaction the energetically costly domain
walls between the order parameter components and the bubble components cause a larger bubble mass controlled by $\nu_* < \nu_t$. The separation of the scales
and the eventual convergence of the lengths (and their inverses corresponding to masses) very close to the first-order line is illustrated in
Fig.~\ref{defects}, where the number of bubble excitations is two on the AFM side (VBS bubbles) and three on the VBS side (AFM bubbles). There is a single
gapless Goldstone mode in the VBS phase and two in the AFM phase, though the one in the VBS phase is also strictly gapped because of its eventual
$\mathbb{Z}_4$ crossover. That gap is, however, smaller than the bubble gap, with the emergent U($1$) length scale being larger than $\xi_*$.

Although the analogy to the hierarchy problem in particle physics is clearly only suggestive, especially since the number of space dimensions is two instead of three
and the massive bubble excitations are scalars (at least at first sight) instead of $\ell=1$ vector bosons, it would be interesting to explore further whether the
crossover mechanism with the emergent length scale $\xi_*$ (and related mass scale of order $\xi_*^{-1}$ that we have not explored directly here), driven by a dangerously
irrelevant interaction (in an unconventional manner), could have some counterpart in field theories of interest in the context of mass hierarchies in particle physics.
Our scenario essentially corresponds to the WZW $k=1$ SO($5$) theory to which the irrelevant repulsive order-parameter interaction has to be added in order to generate
the pseudocritical $\xi_*$ scale governed by an effective exponent $\nu_*$ in a regime close to the transition. This is reminiscent of the soft mass scales that arise
in near-conformal (walking) Higgs alternatives, e.g. the light dilaton, where proximity to a fixed point sets a mass parametrically below the natural scale
\cite{yamawaki86,goldb08,appel10,aoki14,golterman16}. Whether the specific pseudocritical mass separation mechanism and structure that we have found
have counterparts there is an intriguing question.

To make the analogy with particle physics formally clearer (i.e., beyond the functional analogy discussed above), the original CP$^{1}$ gauge field theory of deconfined
quantum criticality \cite{senthil04a} could be considered. Since it is believed to be related to the SO($5$) WZW theory through the web of dualities \cite{wang17} it should
be possible to analyze the effects of the irrelevant interaction also there and elucidate the possible relationship to the electroweak gauge-field theories and
proposed alternative Higgs mechanisms in particle physics. The CP$^1$ theory may also reveal whether the bubble excitations couple to the emergent gauge field and
thereby carry hidden vector content, which would imply a more literal electroweak correspondence.

\subsection{$J$-$Q$ model extensions}
\label{sub:disc4}

It is also possible to study $J$-$Q$ type models with emergent O($4$) symmetry, either by having a two-fold degenerate plaquette-singlet (PS) phase
transitioning into an SO($3$) breaking AFM state \cite{qin17b}, or by four-fold degenerate VBS transitioning into an O($2$) AFM state \cite{lee19}. Previous
work on a ``checker-board'' $J$-$Q$ model with AFM--PS transition showed emergent O($4$) symmetry up to the largest system sizes studied ($L=96$)
even though the transition was rather obviously first-order \cite{zhao19} (a related classical loop model found similar behaviors \cite{serna19}).
An XY anisotropic (planar) $J$-$Q$ model with O($2$) deformed AFM state has also been studied \cite{qin17b,ma18}. It also has a first-order transition but
signatures of deconfined spinons were seen in the dynamic spectral functions \cite{ma18}. These models should be further studied in light of our results
presented here.

Ising symmetric versions of the $J$-$Q$ models have not been considered so far. Such models would be interesting in light of recent work on 1D analogues of the
DQCP separating two Ising-type ordered phases \cite{jiang19,roberts19,mudry19,huang19}. While these order-order
transitions fall into the well known Luttinger liquid classification and have been studied in that context previously
\cite{nakamura00,sengupta02,sandvik04}, 2D Ising-like $J$-$Q$ models could shed further light on the higher-dimensional DQCP phenomenon. In the 1D models
considered, the two Ising-type order parameters vanish continuously and there is emergent O($4$) symmetry at the common critical point. Recent work
\cite{romen26} has introduced an additional tuning parameter---a spin-phonon coupling---that eventually causes the line of critical points to terminate
at a tricritical point in the 4-state Potts universality class, after which the transition is of first order. This scenario is also very similar to the
symmetrically equivalent stripe-AFM transition in classical $J_1$-$J_2$ Ising model \cite{kalz11,jin12}, through a simple mapping between the phase diagram
and transitions by adding a field favoring vertical or horizontal stripes in the classical Ising model.

The Ising anisotropy contrasts with isotropic quantum spin chains with long-range
interactions \cite{sandvik10d,yang20}, which retain the critical WZW level $k=1$ phase of the standard Heisenberg chain (a 1D gapless QSL, sometimes
referred to as a quasi long-range ordered state in the context of this model) but also have an adjacent true long-range ordered AFM phase (circumventing
the Mermin-Wagner theorem) and a two-fold degenerate VBS phase. There is a direct AFM--VBS transition ending at the tip of the critical phase.
In one variant of the model---the frustrated $J_1$-$J_2$ chain with long-range unfrustrated interactions $J_{r>2}$ \cite{sandvik10d}---the direct AFM--VBS
transition is strongly first-order, while in a $J$-$Q$ chain with likewise unfrustrated long-range $J_r$ couplings \cite{yang20} it was believed to be
continuous. However, in light of the present knowledge of the 2D case, the direct transition likely also is weakly first-order, with only a bicritical
point with emergent O($4$) symmetry at the tip of the critical phase.

The introduction of Ising anisotropy in the long-range interacting 1D models should destroy the critical phase and lead to a similar diagram with
tricritical point as in the spin-phonon model. The QSL phase should also be destroyed by Ising anisotropy in 2D systems, and one can reasonably speculate
that systems like the Ising-anisotropic $J$-$Q$ models should then also have a line of direct AFM--VBS transitions replacing the extended QSL phase.
However, this Ising-AFM to VBS transition will unlikely be continuous in 2D, and there is then the possibility of two first-order lines meeting at a
special multicritical point, likely associated with a duality. A line of continuous transitions can perhaps not be completely excluded, as a higher-dimensional
analogy of ``classical deconfined criticality'' at finite temperature in 2D \cite{mudry25}, or possibly a weak first-order transition in the complex CFT
scenario could apply here.

An important question is whether the generic DQCP could appear with electronic model Hamiltonians \cite{gazit18,christos23,bonetti26} [i.e., beyond the fuzzy
sphere approach of directly regularizing the SO($5$) field theory]. A candidate is the transition
between an AFM ground state and a $d$-wave superconductor, which was found to be continuous within the resolution of early QMC studies \cite{assaad97} and for
which further evidence of excitations expected at the DQCP was presented very recently \cite{chen26}. At finite doping, the AFM--VBS DQCP may evolve into an exotic
spin-charge separated ``holon metal'' \cite{kaul07}, and the spectral function of a single injected fermionic hole in a $J$-$Q$ model is indeed consistent
with spin-charge separation \cite{yang25}. More detailed comparisons between the AFM--$d$-wave and AFM--VBS transitions are called for, especially
as concerns the question of generic versus multicritical DQCP in the fermionic models, which would require high-precision values of the scaling dimensions
also in those models.

\subsection{Recent developments}
\label{sub:disc5}

After the first version of this paper appeared online (arXiv:2405.06607v1) two significant works addressing the complex CFT
\cite{zou25} and the multicriticality DQCP scenarios \cite{li26} for the DQCP have appeared. Ref.~\cite{zou25} constructed
a non-Hermitian version of the planar $J$-$Q$ model and demonstrated that the first-order discontinuity could be increased by turning
on the imaginary coupling. However, as also noted by the authors, the results cannot be taken as a demonstration of the conventional $J$-$Q$ model
realizing the complex CFT scenario. Ref.~\cite{zou25} made significant progress on the numerical conformal bootstrap approach, going beyond the previous
SO($5$) works \cite{li18,chester24} in how much of the space of scaling dimensions can be excluded by the symmetries of the CFT. In particular,
a multicritical cone in the space was found where the several scaling dimensions considered all agree very closely with the results presented
here in Table \ref{dtable}. These calculations are also claimed to be inconsistent with the complex CFT scenario, but still roughly consistent with
the numerical fuzzy sphere results \cite{zhou23}---not just the scaling dimensions (which are less precise than those presented here) but also other
CFT data.

Ref.~\cite{li26} together with the results presented here can be taken as a rather definite proof of the multicritical SO($5$) CFT scenario, which now is
being further explored in the context of models that contain gapless QSL phases with SO($5$) tips in their phase diagrams (beyond the previous
works \cite{yang22,liu23}) \cite{feuer26,maity26}. It would still be useful to further improve the QMC calculations, e.g., implementing the recently
proposed operator covariance method \cite{sandvik24,chen26b} and constructing models with interactions beyond the $J$ and $Q_n$ terms \cite{flynn26}, which
may potentially reach closer to the multicritical point without sign problems. Even in cases where the sign problem cannot be avoided, the fact that the
multicritical point in the $J$-$Q_2$ model is so close to the sign free region (and perhaps even closer in models yet to be constructed) offers hope
that some QMC calculations may still be possible there. One approach would be to add a second PQMC stage after projecting to the transition in the sign-free
regime; technically an imaginary-time quench from the transition point of the $J$-$Q_2$ model $H_0$ into the ground state of a $J$-$Q_2$-$K$ model $H_K$,
where $K$ is an appropriate small term (e.g., negative $Q_6$, negative $J$, positive $J_2$, etc.), so that the ground state is now at or closer to the
multicritical point.

\begin{acknowledgments}
We thank Shai Chester, Yin-Chen He, Ribhu Kaul, Zi Yang Meng, Masaki Oshikawa, Slava Rychkov, Subir Sachdev, Martin Schmaltz, T. Senthil,
Ning Su, and Cenke Xu for valuable discussions. The work of J.T. was supported by the National Science Foundation Grant No.~PHY-2116246, JSPS KAKENHI Nos. JP25K17310,
JP25H01391, JP25H01388, JP25K24845, and JP25K24848. H.S. and W.G. were supported by the National Natural Science Foundation of China under Grants No.~12574251 and
No.~12574252, by National Key Projects for Research and Development of China under Grant No.~2021YFA1400400. A.W.S. was supported by the Simons Foundation under Grant
No.~511064 and by a Fellowship from the John Simon Guggenheim Memorial Foundation. This research was also supported in part by grant NSF PHY-2309135 to the Kavli
Institute for Theoretical Physics (KITP). Many of the numerical calculations were carried out on the Shared Computing Cluster managed by Boston University's
Research Computing Services.
\vskip3mm

$^\dagger$ These two authors contributed equally to this work.
\end{acknowledgments}

\end{document}